%% file: main.tex
\documentclass[12pt]{article}
\usepackage{amsmath}
\usepackage{amsthm}
\allowdisplaybreaks
\usepackage{graphicx,psfrag,epsf}
\usepackage{enumerate}
\usepackage{hyperref}
\usepackage{natbib}
\usepackage[labelformat=simple]{subcaption}

\usepackage{amsfonts}
\usepackage{bm}
\usepackage{url} % not crucial - just used below for the URL
\setcitestyle{authoryear,open={(},close={)}} 
\newcommand{\citeay}[1]{\citeauthor{#1}, \citeyear{#1}}
\usepackage{mathtools}
\usepackage{xcolor}
\usepackage{multirow}
\usepackage{algorithm2e}
\RestyleAlgo{ruled}
\SetKwInput{KwInput}{Input}
\SetKwInput{KwReturn}{Return}
\usepackage[ddmmyyyy]{datetime}
\newdateformat{monthyeardate}{\THEDAY{ }\monthname[\THEMONTH]{ }\THEYEAR}
\usepackage{bbm}
\usepackage{setspace}

\newcommand{\rc}[1]{{\color{black}#1}}

\newtheorem{proposition}{Proposition}

\begin{document}

\def\spacingset#1{\renewcommand{\baselinestretch}%
{#1}\small\normalsize} \spacingset{1}

\title{\bf Piecewise-linear modeling of multivariate geometric extremes}
\author{Ryan Campbell\thanks{
Corresponding author, \href{mailto:ryan.campbell@univ-cotedazur.fr}{ryan.campbell@univ-cotedazur.fr}}\hspace{.2cm}\\
Laboratoire J.A. Dieudonn\'{e}, Universit\'{e} C\^{o}te d'Azur, France\\
and \\
Jennifer Wadsworth \\
School of Mathematical Sciences, Lancaster University, UK}
\date{\monthyeardate\today}
\maketitle
\bigskip
\begin{abstract}
    A recent development in extreme value modeling  uses the geometry of the dataset to perform inference on the multivariate tail. A key quantity in this inference is the gauge function, whose values define this geometry. Methodology proposed to date for capturing the gauge function either lacks flexibility due to parametric specifications, or relies on complex neural network specifications in dimensions greater than three. We propose a semiparametric gauge function that is \emph{piecewise-linear}, making it simple to interpret and providing a good approximation for the true underlying gauge function. This linearity also makes optimization tasks computationally inexpensive. The piecewise-linear gauge function can be used to define both a radial and an angular model; this allows for a flexible fit to all data, either separately or jointly. We further expand the toolkit for geometric extremal modeling  through the estimation of high radial quantiles at given angular values via kernel density estimation. We apply the new methodology to air pollution data, which exhibits a complex extremal dependence structure.
\end{abstract}

\noindent%
{\it Keywords:} extrapolation, extremal dependence, geometric extremes, limit sets, piecewise-linear modeling.
%\vfill

\newpage
\spacingset{1} % DON'T change the spacing!

\section{Introduction}
\label{sec:intro-PWL}

\subsection{Multivariate geometric extremes}
In multivariate extreme value analysis, interest lies in characterizing the extremal dependence structure of random vectors. Let $\bm{X}=\left(X_1,\dots,X_d\right)^\top$ be a $d$-dimensional random vector with components $X_j$ representing measurements of a simultaneous process. For example, $\bm{X}$ may comprise measurements of $d$ different air pollutants at a single site, contemporaneous river flows at $d$ locations, or values of $d$ different stock returns. 
Such multivariate vectors can exhibit complex dependence structures, with some variables experiencing simultaneous extremes while others are smaller. Recently, the framework of geometric extremes has emerged as a tool for modeling extremes of potentially complex dependence structures \citep{wadsworth2024statistical, papastathopoulos2025statistical}.
When $\bm{X}$ has common light-tailed margins (e.g., exponential, Weibull, Gaussian, Laplace),
often achieved via a transformation, a useful geometric interpretation of multivariate extremes arises. 

For a wide variety of distributions, the scaled sample cloud of independent copies of light-tailed random vectors, $\left\{{\bm{X}_1}/{r_n},\dots,{\bm{X}_n}/{r_n}\right\}$, converges onto a \emph{limit set} $G$ \citep{davis1988almostsure, kinoshita1991convergence,balkema-nolde-2010}, where the scaling sequence $r_n$ depends on the margins.
The limit set can be characterized by the gauge function, $g$,
where $g:\mathbb{R}^d\rightarrow\mathbb{R}$ is 1-homogeneous, through the relation $G = \left\{\bm{x}\in\mathbb{R}^d : g(\bm{x})\leq 1\right\}$. The shape of $G$ is affected by both the margins and dependence. In order to study dependence features, the margins are fixed to a convenient form, typically standard exponential or Laplace. For both these choices, a suitable scaling factor is  $r_n=\log n$, in which case the coordinatewise supremum of $G$ is given by $\left(1,\dots,1\right)^\top$.
The coordinatewise infimum is $\left(0,\dots,0\right)^\top$ for exponential margins or  $\left(-1,\dots,-1\right)^\top$ for Laplace margins.
This means that $g(\bm{x})\geq\left\|\bm{x}\right\|_{\infty}$, where $\left\|\cdot\right\|_{\infty}$ is the max-norm.
When $\bm{X}$ has exponential or Laplace margins, density $f_{\bm{X}}(\bm{x})$, and the limit
\begin{equation}\label{eq:gauge-limit}
	g(\bm{x}) = \lim_{t\rightarrow\infty} {-\log f_{\bm{X}}(t\bm{x})}/{t},
\end{equation}
is continuous, then the scaled sample cloud converges onto a limit set with gauge function $g$ \citep{balkema-nolde-2010,nolde2022linking}. The boundary $\partial G$ of the limit set is given by the unit level set of $g$. \cite{nolde2014geometric} and \cite{nolde2022linking} show how $g$ can be used to describe the extremal dependence structure of known distributions, while \cite{wadsworth2024statistical} introduced methodology to perform inference with $g$. In contrast to alternative statistical methods for multivariate extremes, inference
based on this new geometric framework can capture highly complex extremal dependence
structures. For example, most statistical methodology based on multivariate regular variation \citep[e.g.,][]{resnick2007htpbook} is suited only to the case of all variables attaining simultaneous extremes. Recent attempts to estimate and model so-called \emph{extreme directions} \citep{mourahib2025penalized} partially address this, but provide no way to model sub-asymptotic dependence. The conditional extreme value model of \citet{heffernan2004conditional} enjoys somewhat greater flexibility, but, as explored in \citet{wadsworth2024statistical}, cannot handle certain more complex structures.
Therefore, estimating $g$, or equivalently $G$, is crucial to flexible multivariate extremal inference.

\begin{figure*}[t!]
    \centering
    \includegraphics[width=0.23\textwidth]{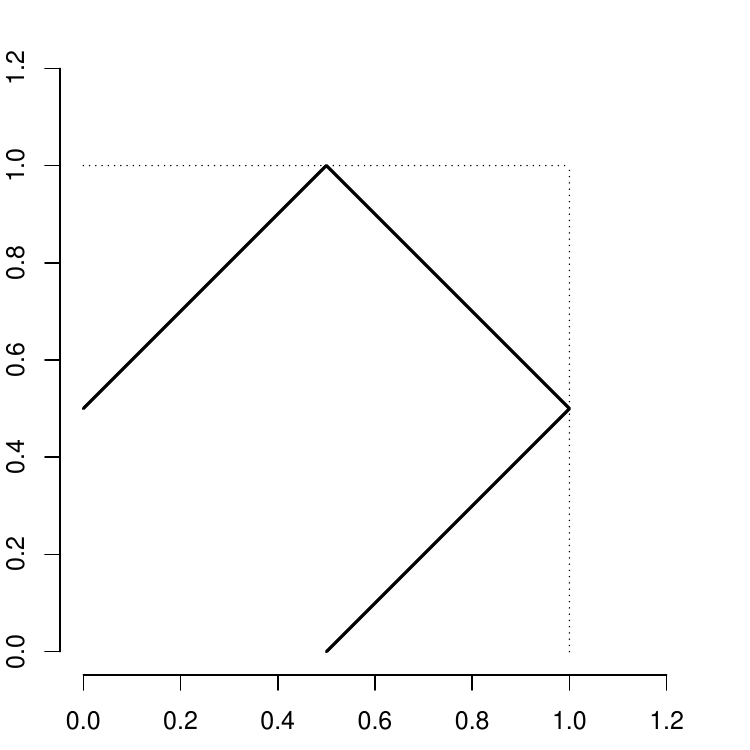}
    \includegraphics[width=0.23\textwidth]{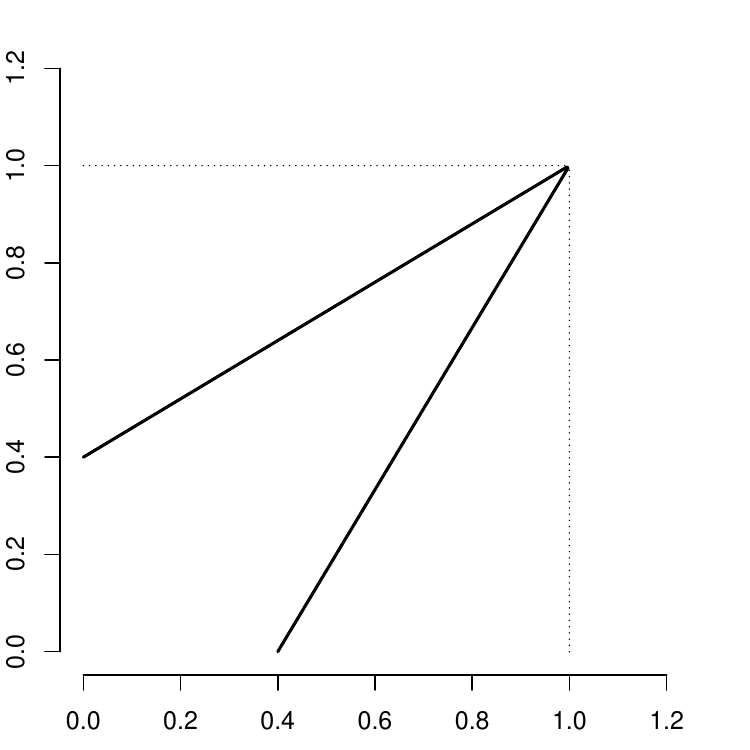}
    \includegraphics[width=0.23\textwidth]{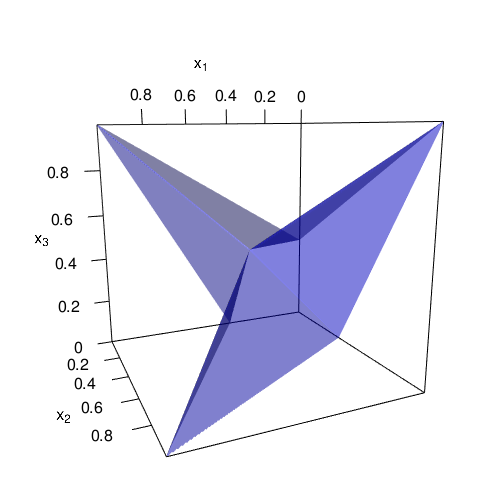}
    \includegraphics[width=0.23\textwidth]{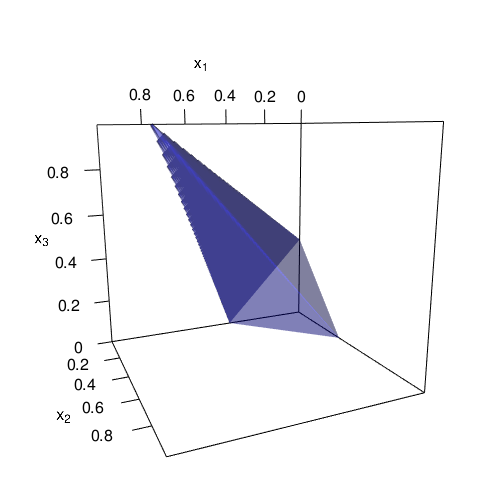}
	\caption[Illustration of limit set boundaries and their interpretation in terms of simultaneous extremes.]{Illustration of limit set boundaries and their interpretation in terms of simultaneous extremes. Left: Example where extreme events occur separately. Center-left: Example where extreme events occur together. Center-right: Example where only pairs of variables grow large together. Right: Example where all three variables grow large simultaneously.}
    \label{fig:d3-g-dep}
\end{figure*}

The limit set, and therefore the gauge function, provide us with a useful description of the extremal dependence structure of the random vector $\bm{X}$ by telling us which groups of variables exhibit simultaneous extremes while the remaining variables are of smaller order. Let $D=\left\{1,\dots,d\right\}$ and $C\subseteq D$.
We say that the variables in group $C$ can be simultaneously extreme while the others are smaller if there exists $\bm{z}^C$ such that $g(\bm{z}^C)=1$, where $z^C_{j}=1$ for all $j\in C$ and $z^C_j = \gamma_j$ for all $j\in D\setminus C$, for some $\gamma_j\in[0,1)$ in exponential margins or $\gamma_j\in[-1,1)$ in Laplace margins. Note that these are points of intersection of the limit set boundary with its bounding box $[0,1]^d$ or $[-1,1]^d$.
The collection of sets $C\subseteq D$ with $g(\bm{z}^C)=1$ is denoted by $\mathcal{C}$.
Each variable must be represented at least once in $\mathcal{C}$, since the coordinatewise supremum is $(1,\dots,1)^\top$.
Figure~\ref{fig:d3-g-dep} displays 
bivariate examples with $\mathcal{C}=\left\{\left\{1\right\},\left\{2\right\}\right\}$ and $\mathcal{C}=\left\{\left\{1,2\right\}\right\}$. Examples for $d=3$ are also displayed with $\mathcal{C}=\left\{\left\{1,2\right\},\left\{1,3\right\},\left\{2,3\right\}\right\}$ and $\mathcal{C}=\left\{\left\{1,2,3\right\}\right\}$.
Note that our definition of simultaneous extremes based on $g$ is slightly different to definitions that arise in the framework of multivariate regular variation (e.g., \cite{goix2017sparse}), although the two will often overlap theoretically and are essentially indistinguishable at a practical level. We therefore prefer this simple definition when working in the geometric framework. 

To perform statistical inference using the gauge function, it is useful to consider the radial-angular decomposition $\bm{X}\longmapsto\left(R,\bm{W}\right)=\left(\left\|\bm{X}\right\|,{\bm{X}}/{\left\|\bm{X}\right\|}\right)\in\mathbb{R}_+ \times \mathcal{S}_{d-1}$, where $\left\|\cdot\right\|$ is a norm and  $\mathcal{S}_{d-1}=\left\{\bm{x}\in\mathbb{R}^d : \left\|\bm{x}\right\|=1\right\}$. 
We emphasize here that this decomposition is made in light-tailed margins, and therefore is fundamentally different to radial-angular decompositions in heavy-tailed margins, which are central to ``classical'' multivariate extremes, in which multivariate regular variation plays a key role (see, for example, \citeay{beirlant2006statistics}, Chapter 8).
When appropriate, the $L_1$ norm is preferred for its simplicity, defined by $\left\|\bm{x}\right\|_1 = \sum_{j=1}^d |x_j|$.
In the radial-angular framework, we are interested in $\bm{X}=R\bm{W}$ when the conditional variable $R\mid\bm{W}$ achieves large values. 
\cite{wadsworth2024statistical} explain that, when working with exponential-tailed variables and the $L_1$ norm, the limiting behavior in equation \eqref{eq:gauge-limit} leads to the asymptotic distribution of $R\mid\left\{\bm{W}=\bm{w}\right\}$ with density
$
	f_{R\mid\bm{W}}(r\mid\bm{w}) \propto r^{d-1}\exp\left\{-rg(\bm{w})\left[1+o(1)\right]\right\}
$ as $r\rightarrow\infty$,
%\end{equation}
where $g(\bm{w})$ is the gauge function corresponding to the joint distribution of $(R,\bm{W})$ evaluated at $\bm{w}\in\mathcal{S}_{d-1}$. However, they showed that in a wide variety of examples, the same asymptotic form also holds with the $[1+o(1)]$ outside of the exponent, i.e., 
\begin{equation}
	\label{eq:trunc-gam-cond}
	f_{R\mid\bm{W}}(r\mid\bm{w}) \propto r^{d-1}\exp\left\{-rg(\bm{w})\right\}\left[1+o(1)\right] \,\,\,\,\text{as}\,\,r\rightarrow\infty.
\end{equation}
This suggests that a gamma model is asymptotically appropriate for large values of $R\mid\bm{W}$.
This limiting density is shown to hold very broadly, although for the multivariate Gaussian dependence structure, the shape parameter of the gamma distribution also depends on $\bm{w}$.
Nonetheless, \cite{wadsworth2024statistical} demonstrate that it is typically not problematic to assume that this parameter is constant.  
They model $R\mid\left\{\bm{W}=\bm{w},R>r^\star(\bm{w})\right\}$ using a left-truncated gamma distribution with rate parameter $g(\bm{w})$, where $r^\star(\bm{w})$ is a high threshold estimated to correspond to $r_\tau(\bm{w})$, the $\tau^\text{th}$ quantile of the distribution of $R\mid\left\{\bm{W}=\bm{w}\right\}$, for $\tau$ close to 1. We denote such an estimate by $\hat{r}_\tau(\bm{w})$ henceforth.
Using parametric forms of the gauge function $g$ derived from known copulas, they perform statistical inference with high accuracy, showing how to use their fitted model for inference on extremal probabilities through simulation of $\bm{X}\mid \left\{R>\hat{r}_\tau(\bm{W})\right\}$.
% Given $\tau\in(0,1)$ close to 1 and setting $r_\tau(\bm{w})$ as the $\tau^\text{th}$ quantile value of $R\mid\left\{\bm{W}=\bm{w}\right\}$, \cite{wadsworth2024statistical} model $R\mid\left\{\bm{W}=\bm{w},R>r_\tau(\bm{w})\right\}$ with a truncated gamma distribution with rate parameter $g(\bm{w})$, using parametric forms of the gauge function $g$ derived from known copulas to perform statistical inference with high accuracy, showing how to use this fitted model for inference on extremal probabilities through simulation of $\bm{X}\mid \left\{R>r_{\tau}(\bm{W})\right\}$.

\subsection{Semiparametric estimation of the gauge function}\label{intro:models}

For dimensions $d\geq 3$ in particular, the current suite of parametric models employed in \cite{wadsworth2024statistical} may not be sufficiently flexible for real datasets, where the dependence structure can be complicated. Therefore, the natural consideration is to develop semiparametric approaches for approximating $g$.
\cite{simpson2024estimating}, \cite{majumder2025semiparametric}, and \cite{papastathopoulos2025statistical} all aim to do this, the latter two in a Bayesian manner. The methods in \cite{simpson2024estimating} and \cite{majumder2025semiparametric} approximate $g$ in the bivariate case, and use these estimates to describe the underlying tail dependence structure. The framework of \cite{papastathopoulos2025statistical} is generalized for $d$-dimensions, but Bayesian inference using the integrated nested Laplace approximation (INLA) is suitable only for problems of dimension 2 or 3. 
\cite{simpson2024estimating} and \cite{papastathopoulos2025statistical} model exceedances $\left\{R-\hat{r}_\tau(\bm{w})\right\}\mid\left\{\bm{W}=\bm{w}, R>\hat{r}_\tau(\bm{w})\right\}$ using a generalized Pareto distribution in place of the truncated gamma distribution. When far enough into the tails, both choices should perform well, but the truncated gamma form may be more accurate at finite levels, especially for larger $d$.
Deep learning methods to estimate the shape of the limit set have been explored for statistical inference using feed-forward neural networks and generative modeling through normalizing flows in \cite{murphybarltrop2024MLgauges} and \cite{deMonte2025generative}, respectively, indicating potential for higher-dimensional inference.

\begin{figure}[t!]
    \centering
    \includegraphics[width=0.25\textwidth]{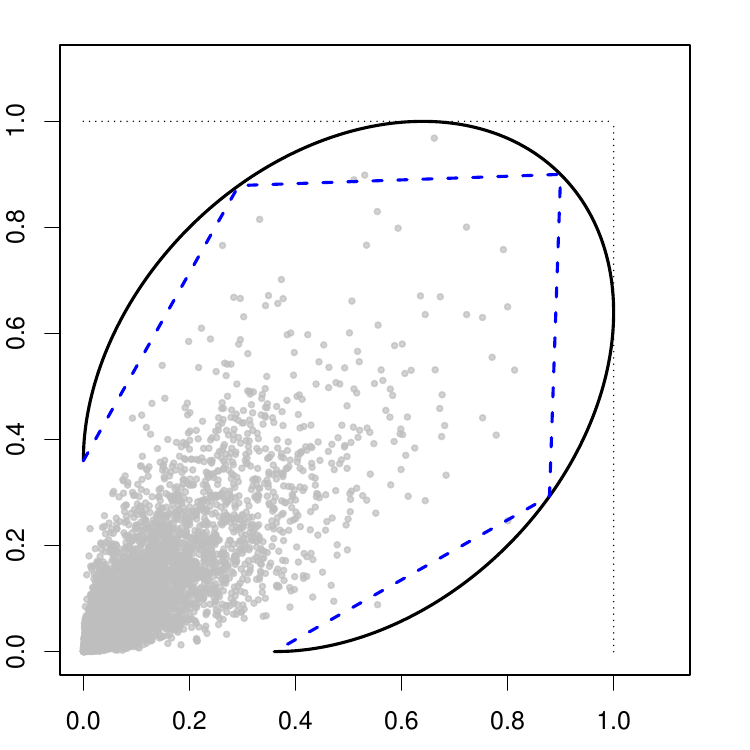}
    \includegraphics[width=0.25\textwidth]{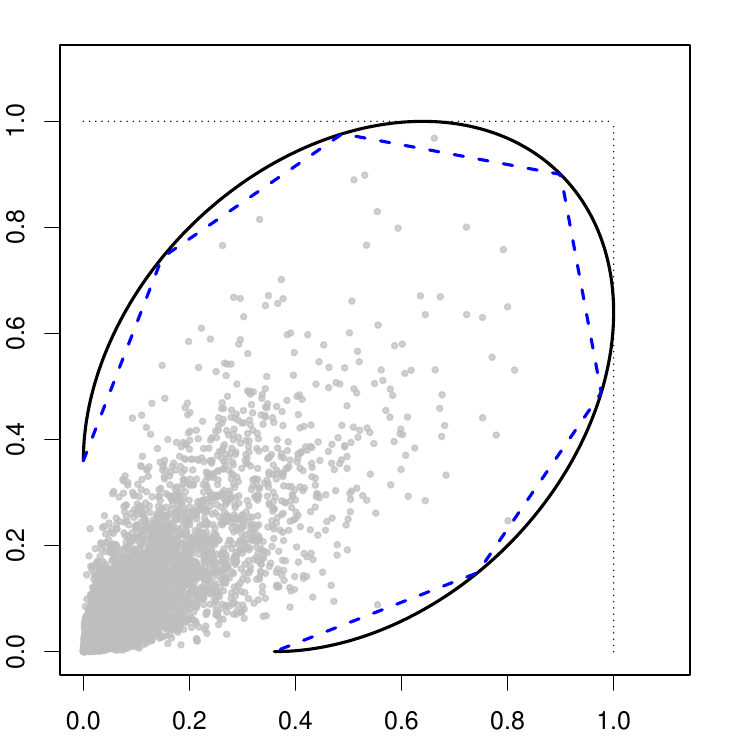}
    \includegraphics[width=0.25\textwidth]{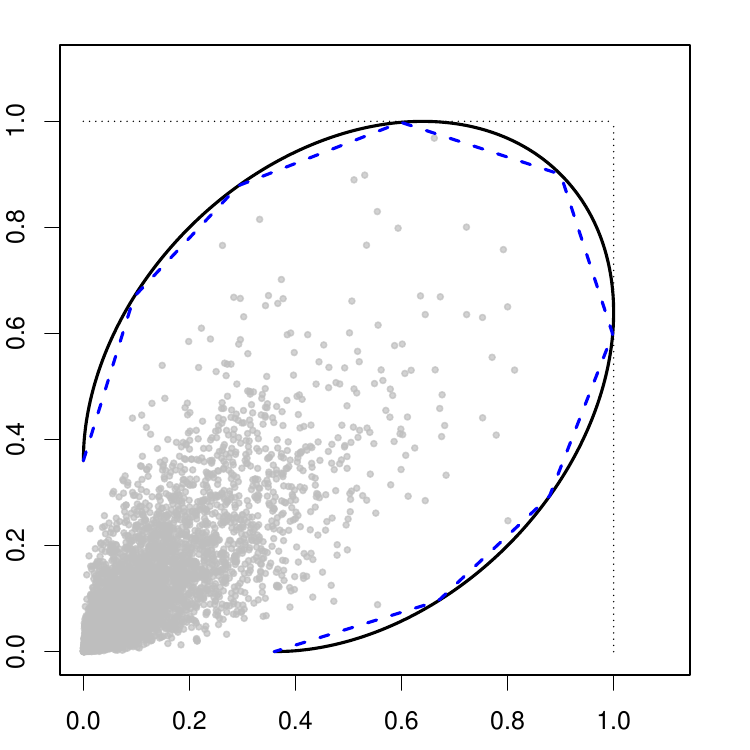}
	\caption{%Top row: 
	$\log n$-scaled bivariate Gaussian data in standard exponential margins, with true limit set boundary given by the solid line. The piecewise-linear limit set boundary is given by the dashed line using 5, 7, and 9 reference angles (left to right).
%	Bottom row: $\log ({n}/{2})$-scaled bivariate Gaussian data in standard Laplace margins. The piecewise-linear limit set boundary uses 8, 16, and 24 reference angles (left to right).
	}
    \label{fig:d2-gauss-pwl-ill}
\end{figure}

In this work, we present a simple and interpretable piecewise-linear representation of $g$. Given a choice of reference angles in $\mathcal{S}_{d-1}$, the parameters of $g$ define the distance from the origin to the boundary of $G$ at these reference angles.  The value of $g$ is given by linear interpolation between these points. When $d\geq3$, a triangulation is also required to define the linear interpolation; we use a Delaunay triangulation. The simple construction of our gauge function has numerous benefits. The main benefit is that a model for large radii and exceedance angles can be easily obtained with quick convergence to maximum likelihood estimates for parameters.
Figure~\ref{fig:d2-gauss-pwl-ill} demonstrates what our proposed limit set boundary would look like in the bivariate setting when working in exponential margins. The piecewise-linear limit set boundary represents a rough approximation of the truth, derived using the limit \eqref{eq:gauge-limit}, and becomes closer to the truth as the number of reference angles increases. 
The approximation error decreases with the distance between the reference angles. Under the mild assumption that $g$ is Lipschitz-continuous, we sketch a bound for this approximation error in Appendix~\ref{app:pwl-error-bounds}. However, increasing the number of reference angles increases the number of parameters to estimate leading to a typical bias-variance trade-off. This is addressed in our work using a regularization approach.

Piecewise-linear approaches have been used recently in different contexts for extreme value analysis. \cite{barlow2023penalisedpwl} define non-stationary shape and scale parameters of the generalized Pareto distributions in a piecewise-linear manner in the presence of covariates for univariate peaks-over-threshold modeling.
\cite{winterstein1993environmental}, \cite{huseby2013new}, and \cite{mackay2023pwlenvcont} use piecewise-linear interpolation of the $d$-dimensional domain to estimate and visualize environmental contours, which are multivariate sets used to approximate the occurrence of extreme events.

As in \cite{wadsworth2024statistical}, we model large radii using the truncated gamma distribution with rate parameter given by the gauge function value. 
This leads to a likelihood from which the parameters of the piecewise-linear gauge can be estimated.
When simulating from the distribution of $\bm{X}\mid \left\{R>\hat{r}_\tau(\bm{W})\right\}$, two things are required: the distribution of $\bm{W}\mid \left\{R>\hat{r}_\tau(\bm{W})\right\}$ and the distribution of $R\mid \left\{\bm{W}=\bm{w},R>\hat{r}_\tau(\bm{w})\right\}$. The second of these is given by the truncated gamma model, while for the first, \cite{wadsworth2024statistical} used the empirical distribution of $\bm{W}\mid \left\{R>\hat{r}_\tau(\bm{W})\right\}$. A natural progression, particularly for moderate dimensions, is also to estimate a semiparametric form for this.
We propose an angular density inspired by the homothetic case presented in \cite{balkema-nolde-2010}, with joint density equivalent to the one used in \cite{papastathopoulos2025statistical}. 
Specifically, a valid density over $\mathcal{S}_{d-1}$ is 
$
	f_{\tilde{\bm{W}}}(\bm{w}) = {\tilde{g}(\bm{w})^{-d}}/{\{d\text{vol}(\tilde{G})\}} \:,
$
where $\tilde{g}$ is a gauge function for the set $\tilde{G}$. We emphasize here that this density can be expressed and fitted independently of $R\mid\bm{W}$, so that the gauge function $\tilde{g}$ in the angular density need not correspond to the gauge function in equation \eqref{eq:trunc-gam-cond}. This is discussed in greater detail in Section~\ref{sec:lik}.
The major advantage to  our piecewise-linear setting is that the normalizing constant for this joint density has an explicit form, rendering its estimation simple.

When standardizing the margins of a given dataset, the choice of marginal distribution is a nuanced one. If negative dependence arises, then the limit definition of the gauge function in equation \eqref{eq:gauge-limit} may not hold on the axes in exponential margins, and Laplace margins are preferred for revealing greater structure. Furthermore, if the dataset has negative values in its original margins, then it may be more intuitive to use Laplace margins, as they preserve the domain to all orthants of $\mathbb{R}^d$. In many cases, real datasets are positive-valued and have no negative associations, and so exponential margins are a suitable choice.
While the methodology presented is suitable for data with a variety of light-tailed margins, we choose to model data in standard exponential margins for simplicity. A key advantage of this is that because the standardized data lies in the positive orthant, the $L_1$ norm simplifies to taking the sum of the components, so that $\mathcal{S}_{d-1}$ is the unit simplex $\{\bm{w}\in\mathbb{R}^d_+:w_j\geq 0, \sum_{j=1}^dw_j=1\}$, and working with the $(d-1)$-dimensional embedding is as simple as removing the $d^\text{th}$ element from the angular vector. In Supplement~\ref{supp:laplace}, we explain the differences required to work in standard Laplace margins.

The outline of the paper is as follows. In Section~\ref{sec:method}, we detail the piecewise-linear construction of a gauge function, and consider the calculation of the corresponding angular density normalizing constant.
As we only consider radial values above a threshold, a good estimate $\hat{r}_\tau(\bm{w})$ of the threshold $r_\tau(\bm{w})$ is needed. The empirical version presented in \cite{wadsworth2024statistical} and quantile regression techniques are not currently well-suited to higher dimensions. 
In Section~\ref{sec:quantiles}, we outline a method to estimate $r_{\tau}(\bm{w})$ via kernel density estimation of the joint distribution of $(R,\bm{W})$.
Section~\ref{sec:inference-PWL} covers how we fit our piecewise-linear models, including a regularization technique for when there are many parameters. We also consider diagnostics and probability estimation techniques using our models.
In Section~\ref{sec:sim}, simulation studies show that repeated fits of the piecewise-linear model are comparable to parametric models where knowledge of the true copula is exploited. 
Section~\ref{sec:pollution} details an application to extremes of four air pollutants, demonstrating that the piecewise-linear model can be used to perform statistical inference in dimensions where other semiparametric methods struggle.

In order to select the best hyperparameters for both threshold estimation and model penalization, as well as evaluating probability estimates, we assess our models on datasets of dimensions $d=2$ and $d=3$ generated from seven different multivariate distributions, all in standard exponential margins.
The distributions used to generate the data are as follows
\begin{enumerate}[(I)]
	\setlength\itemsep{1.9pt}
	\item \label{distn:log1} $d=2$ logistic with dependence parameter $\alpha=0.4$. $\mathcal{C}=\left\{\left\{1,2\right\}\right\}$.
	\item \label{distn:log2} $d=2$ logistic with dependence parameter $\alpha=0.8$. $\mathcal{C}=\left\{\left\{1,2\right\}\right\}$, but with weaker dependence than \eqref{distn:log1}.
	\item \label{distn:gauss} $d=2$ Gaussian distribution, correlation $\rho=0.8$. $\mathcal{C}=\left\{\left\{1\right\},\left\{2\right\}\right\}$.
	\item \label{distn:invlog} $d=2$ inverted logistic with dependence parameter $\alpha=0.7$. $\mathcal{C}=\left\{\left\{1\right\},\left\{2\right\}\right\}$.
	\item \label{distn:alog1} $d=3$ asymmetric logistic with dependence parameters $\alpha_{\left\{1,2\right\}}=\alpha_{\left\{1,3\right\}}=\alpha_{\left\{2,3\right\}}=0.4$. $\mathcal{C}=\left\{\left\{1,2\right\}, \left\{2,3\right\}, \left\{1,3\right\}\right\}$.
	\item \label{distn:alog2} $d=3$ asymmetric logistic, with dependence parameters $\alpha_{\left\{1\right\}}=\alpha_{\left\{1,2\right\}}=\alpha_{\left\{2,3\right\}}=0.4$. $\mathcal{C}=\left\{\left\{1\right\},\left\{1,2\right\}, \left\{2,3\right\}\right\}$.
	\item \label{distn:mix} $d=3$ equally-weighted mixture of asymmetric logistic and Gaussian. The Gaussian correlations are $\rho_{12}=\rho_{13}=\rho_{23}=0.6$ and asymmetric logistic dependence parameters $\alpha_{\left\{1,2\right\}}=\alpha_{\left\{1,2,3\right\}}=0.4$.  $\mathcal{C}=\left\{\left\{1\right\},\left\{2\right\},\left\{3\right\},\left\{1,2\right\}, \left\{1,2,3\right\}\right\}$.
\end{enumerate}
A catalog of these distributions is presented in more detail in Supplement~\ref{supp:copulas}.

\section{A piecewise-linear model}
\label{sec:method}

\subsection{The piecewise-linear gauge function}
\label{sec:radial-model}
\begin{figure}[t]
    \centering
    \includegraphics[width=0.3\textwidth]{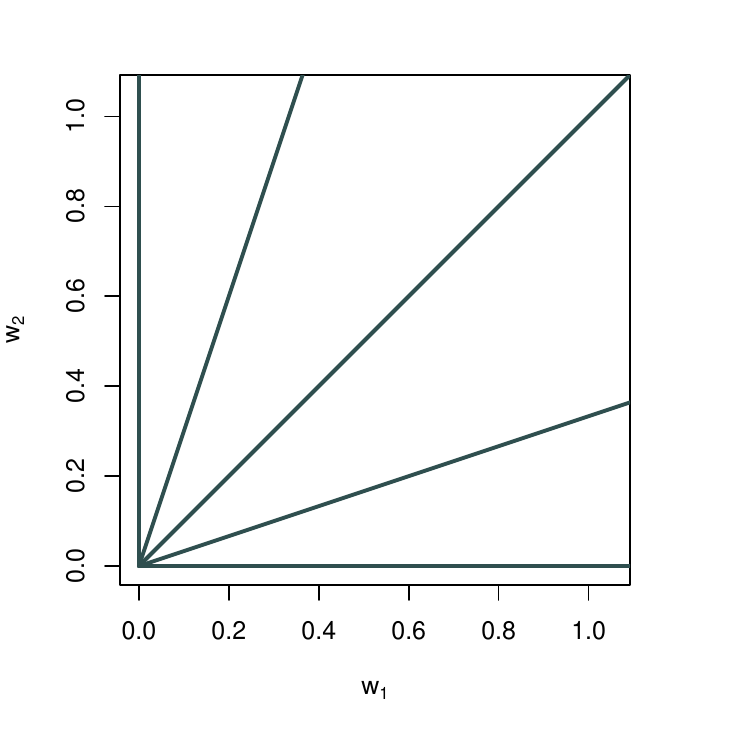}
    \includegraphics[width=0.3\textwidth]{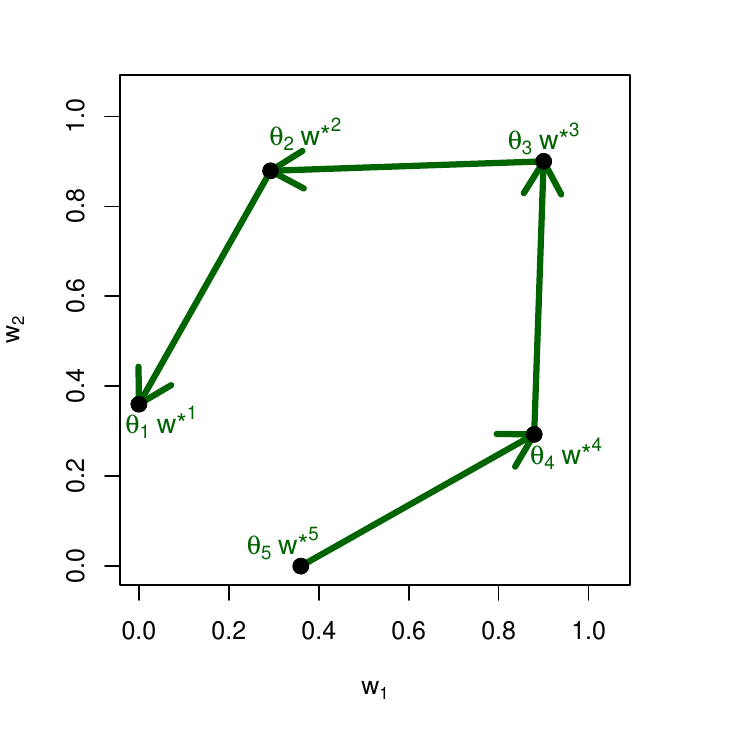}
%    \hspace{0.08\textwidth}
    \includegraphics[width=0.3\textwidth]{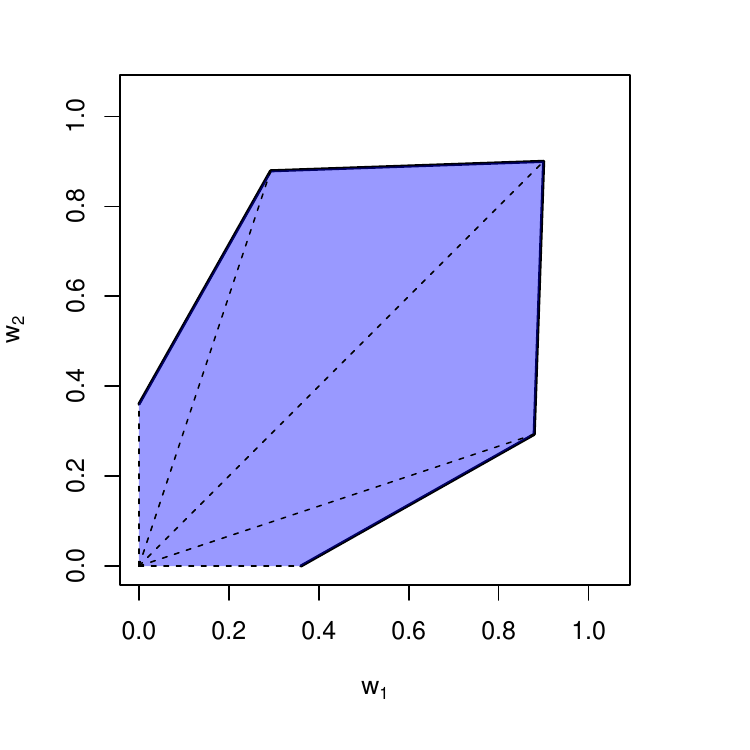}
	\caption[Piecewise-linear gauge function construction illustration for $d=2$.]{Piecewise-linear gauge function construction illustration for $d=2$. Left: Partition of $\mathcal{S}_1$ with reference angles $w^{\star 1}=0$, $w^{\star 2}=0.25$, $w^{\star 3}=0.5$, $w^{\star 4}=0.75$, and $w^{\star 5}=1$. Middle: Coplanar vectors $\bm{C}^{(1)}$, $\bm{C}^{(2)}$, $\bm{C}^{(3)}$, and $\bm{C}^{(4)}$ (green arrows) for pre-set parameter values $\theta_1=0.36$, $\theta_2=1.17$, $\theta_3=1.8$, $\theta_4=1.17$, and $\theta_5=0.36$. Right: Limit set at chosen parameter values. Solid line indicates the unit level set of the piecewise-linear gauge function at the chosen parameter and reference angle values. Dashed lines indicate the distances dictated by the parameter values.}
    \label{fig:d2-ill}
\end{figure}
Construction of the piecewise-linear (PWL) gauge function, denoted $g_{\small{\textsc{pwl}}}$, relies on segmenting the simplex $\mathcal{S}_{d-1}$ using $N$ nodes called \emph{reference angles}. A parameter value is assigned at each reference angle, defining the distance from the origin to the limit set boundary at that angle: for each reference angle $\bm{w}^{\star}\in\mathcal{S}_{d-1}$, the parameter corresponds to the value ${1}/{g_{\small{\textsc{pwl}}}(\bm{w}^{\star})}$. %, the distance from the origin to the unit level set boundary at angle $\bm{w}^{\star}$. 
Because the true gauge function $g$ satisfies $g(\bm{x})\geq \left\|\bm{x}\right\|_\infty$, a parameter at location $\bm{w}^{\star}$ has an upper bound of ${1}/{\left\|\bm{w}^{\star}\right\|_{\infty}}$.
Depending on the dimension, straight lines, planes, or hyperplanes are used to connect these limit set boundary values. The construction of $g_{\textsc{pwl}}$ will depend on coplanar vectors to obtain the necessary perpendicular vectors that define these geometric objects.

\sloppy To provide intuition into the general approach, we begin by illustrating for $d=2$. 
Taking the $L_1$ norm in our radial-angular decomposition, reference angles can be defined by scalar values in $[0,1]$.
Let $\left\{w^{\star1},\dots,w^{\star N}\right\}$ be an increasing sequence of reference angles such that $w^{\star1}=0$ and $w^{\star N}=1$.  
This partition of the interval $[0,1]$ has $N-1$ segments with vertices $\left\{w^{\star1},w^{\star2}\right\},\dots,\left\{w^{\star N-1},w^{\star N}\right\}$. 
Further let $\bm{\theta}=\left(\theta_1,\dots,\theta_N\right)^\top$ be $N$ positive parameter values with $\theta_k = {1}/{g_{\small{\textsc{pwl}}}(w^{\star k},1-w^{\star k})}$.
To define $g_{\small{\textsc{pwl}}}$, we consider coplanar vectors, $\bm{C}^{(k)} = \left(\theta_k w^{\star k}-\theta_{k+1} w^{\star k+1},\theta_k (1-w^{\star k})-\theta_{k+1} (1-w^{\star k+1})\right)^\top$, $k=1,\dots,N-1$, an example of which is displayed in Figure~\ref{fig:d2-ill}.
The line interpolating from $\theta_k\left(w^{\star k},1-w^{\star k}\right)^\top$ to $\theta_{k+1}\left(w^{\star k+1},1-w^{\star k+1}\right)^\top$ is 
${\left(C^{(k)}_2 x - C^{(k)}_1 y\right)}/{\left(C^{(k)}_2 \theta_k w^{\star k} - C^{(k)}_1 \theta_k (1-w^{\star k})\right)}$. Therefore the piecewise-linear gauge function is given by
\begin{equation}\label{eq:d2-pwl-gauge-N5}
	g_{\small{\textsc{pwl}}}(x,y;\bm{\theta})=\sum\limits_{k=1}^{N-1} \mathbbm{1}_{(w^{\star k},w^{\star k+1})}\left(\frac{x}{x+y}\right)\,\frac{A_k x - B_k y}{A_k \theta_k w^{\star k} - B_k \theta_k(1-w^{\star k})},
\end{equation}
for $(x,y)\in\mathbb{R}_+^2$, where $\mathbbm{1}_{A}(x)$ is an indicator function with value 1 if $x\in A$ and 0 otherwise, and $A_k =\theta_k (1-w^{\star k}) - \theta_{k+1}(1-w^{\star k+1})$, $B_k =\theta_k w^{\star k} - \theta_{k+1}w^{\star k+1}$ for $k\in\{1,\dots,N-1\}$.
Figure~\ref{fig:d2-ill} displays an example in which a piecewise-linear gauge function is used to approximate the gauge function corresponding to the bivariate Gaussian distribution, $g_{\small{\text{N}}}(x,y;\rho)=\left(1-\rho^2\right)^{-1}\left(x + y - 2\rho(xy)^{{1}/{2}}\right)$. The $N=5$ reference angles correspond to an equally-spaced mesh, and parameters are set to $\theta_k={1}/{g_{\small{\text{N}}}(w^{\star k},1-w^{\star k};\rho)}$.
The resulting limit set in Figure~\ref{fig:d2-ill} does not satisfy the coordinatewise supremum property because of the absence of a reference angle $w^\star$ such that $g_{\text{N}}(w^\star,1-w^\star;\rho)=\left\|(w^\star,1-w^\star)^\top\right\|_{\infty}$. When performing inference, we will develop an algorithm to ensure that limit set estimates can have the coordinatewise supremum $(1,\dots,1)^\top$.

In dimensions $d\geq 3$, denote the set of $N\geq d$ reference angles $\left\{\bm{w}^{\star 1},\dots,\bm{w}^{\star N}\right\}$, each lying in the simplex $\mathcal{S}_{d-1}$. We partition $\mathcal{S}_{d-1}$ using a Delaunay triangulation \citep{delaunay1934sphere} with a point set based on the $N$ reference angles. Given a set of $N$ $(d-1)$-dimensional reference angles, the Delaunay triangulation creates a partition of $\mathcal{S}_{d-1}$ comprised of $M$ regions with vertices given by these reference angles. These regions are constructed in such a way that their surface area (or volume) is maximized, leading to no insignificant segments.
While there is no direct one-to-one correspondence between the number of reference angles $N$ and the number of partitions $M$ in the resulting Delaunay triangulation, it is known that $M$ is between $O(N)$ and $O(N^{\left\lceil{(d-1)}/{2}\right\rceil})$ in general $d$-dimensions. When $d=3$, the stronger statement of $N-2\leq M \leq 2N-5$ holds \citep{deBerg2008delaunay}.
Each region of the Delaunay triangulation, $\triangle^{(k)}\subset\mathcal{S}_{d-1}$, $k\in\left\{1,\dots,M\right\}$, is defined by $d$ vertices, $\mbox{vert}(\triangle^{(k)}):=\{\bm{w}^{\star(k),1},\dots,\bm{w}^{\star(k),d}\}$.
%Further let $\textbf{e}=\left\{\bm{e}_1,\dots,\bm{e_d}\right\}$ be the elementary basis vectors of $\mathbb{R}^d$. For example, $\textbf{e}=\left(\bm{i},\bm{j},\bm{k}\right)^\top$ when $d=3$. 
Given the parameter values $\bm{\theta}=\left(\theta_1,\dots,\theta_N\right)^\top\in\mathbb{R}^N
_+$, we define $\bm{\theta}^{(k)}=(\theta^{(k)}_1,\dots,\theta^{(k)}_d)^\top$ as the parameters from $\bm{\theta}$ associated with the $d$ vertices of $\triangle^{(k)}$.
% and $\bm{\theta}^{(k)}\triangle^{(k)}$ as a bounded hyperplane with vertices $\theta^{(k)}_1\bm{w}^{\star (k),1},\dots,\theta^{(k)}_d \bm{w}^{\star (k),d}$. 
Define the $(d-1)\times d$ \emph{coplanar matrix} $\textbf{C}^{(k)}$ for triangulation segment $k\in\left\{1,\dots,M\right\}$ where the $i^\text{th}$ row is given by the vector
\begin{equation*}
	\textbf{C}_{i,\cdot}^{(k)} = 
		\theta^{(k)}_{1}\left(\bm{w}^{\star(k),1}\right)^\top - \theta^{(k)}_{i+1}\left(\bm{w}^{\star(k),i+1}\right)^\top\in\mathbb{R}^d\;;\;\; i=1,\dots,d-1.
\end{equation*}
For an arbitrary angle $\bm{w}\in\triangle^{(k)}$, $g_{\small{\textsc{pwl}}}(\bm{w};\bm{\theta}) = {\bm{n}^{(k)\top}\bm{w}}/{\bm{n}^{(k)\top}\theta^{(k)}_{1}\bm{w}^{\star{(k),1}}}$, where the normal vector $\bm{n}^{(k)}\in\mathbb{R}^d$ to the plane defined by vertices $\theta^{(k)}_1\bm{w}^{\star (k),1},\dots,\theta^{(k)}_d\bm{w}^{\star (k),d}$ is
\begin{equation}\label{eq:normalvec}
	\bm{n}^{(k)} = \sum\limits_{j=1}^d (-1)^{j+1} \det\left(\textbf{C}^{(k)}_{\cdot,-j}\right) \bm{e}_j\:.
\end{equation}
In equation \eqref{eq:normalvec}, $\textbf{C}^{(k)}_{\cdot,-j}$ is the matrix $\textbf{C}^{(k)}$ with the $j^\text{th}$ column removed, and $\bm{e}_j$ the $j^{\text{th}}$ standard unit vector.
%, a vector of length $d$ of zeros except for a 1 in the $j^{\text{th}}$ entry.
%Then add parameters.
%Then do it for all triangles.
Performing the summation over all regions in the Delaunay triangulation gives the proposed piecewise-linear gauge function 
\begin{equation}\label{eq:gauge-full-w}
	g_{\small{\textsc{pwl}}}(\bm{x};\bm{\theta})=\sum\limits_{k=1}^M \mathbbm{1}_{\triangle^{(k)}}\left({\bm{x}}/{\left\|\bm{x}\right\|}\right)\frac{\bm{n}^{(k)\top}\bm{x}}{\bm{n}^{(k)\top}\theta^{(k)}_{1}\bm{w}^{\star{(k),1}}}, \qquad \bm{x}\in\mathbb{R}_+^d,
\end{equation}
which is 1-homogeneous and continuous in $\bm{x}$.
Note that the formulation of the $d=2$ case in equation \eqref{eq:d2-pwl-gauge-N5} is covered by equation \eqref{eq:gauge-full-w}, but was described separately to give an intuition in the bivariate setting.

If the underlying gauge function $g$ is Lipschitz continuous, and the parameter values are chosen such that $\theta_k=1/g(\bm{w}^{\star k})$, then we have the approximation error bound
\begin{align*}
    |g(\bm{w})-g_{\small{\textsc{pwl}}}(\bm{w})| \leq (K_g+K_{\small{\textsc{pwl}}})\sup_{1\leq k\leq M}\sup_{(i,j) \in \mbox{vert}(\triangle^{(k)})}\|\bm{w}_i-\bm{w}_j\|,
\end{align*}
where $\mbox{vert}(\triangle^{(k)})$ is the set of vertices of $\triangle^{(k)}$, and $K_g, K_{\small{\textsc{pwl}}}$ are the Lipschitz constants of $g$ and $g_{\small{\textsc{pwl}}}$, respectively. See Appendix~\ref{app:pwl-error-bounds} for further details. This means that the approximation error generally decreases in line with the distance between vertices in the Delaunay triangulation, which in turn decreases as $N$ increases.

\begin{figure*}[t]
    \centering
    \includegraphics[width=0.3\textwidth]{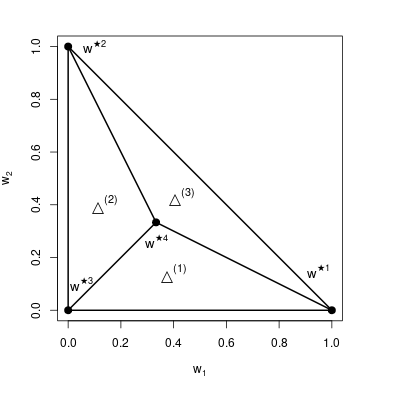}
    \includegraphics[width=0.3\textwidth]{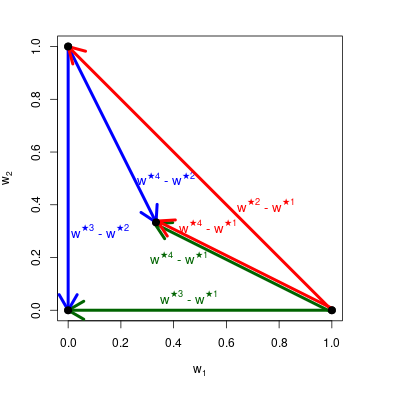}
    \includegraphics[width=0.3\textwidth]{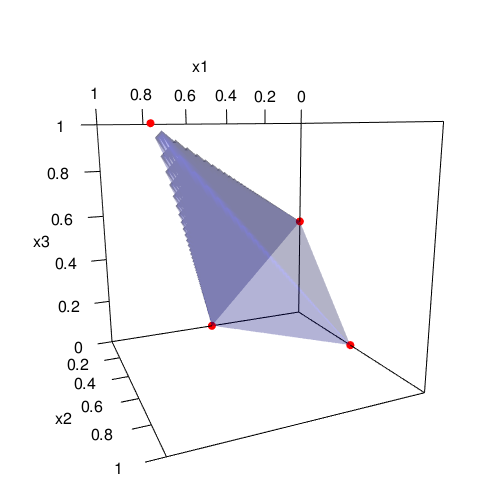}
	\caption[Piecewise-linear gauge function construction illustration for $d=3$.]{Piecewise-linear gauge function construction illustration for $d=3$. Left to right: Delaunay triangulation based on $N=4$ chosen reference angles; the resulting coplanar vectors; limit set boundary.}
    \label{fig:d3-ill}
\end{figure*}

\paragraph{Example:}
Suppose $d=3$ and $N=4$, with $\bm{w}^{\star 1},\bm{w}^{\star 2},\bm{w}^{\star 3}$ chosen to lie on each of the vertices, and $\bm{w}^{\star 4}$ in the center of the $\mathcal{S}_{2}$ simplex. As displayed in Figure~\ref{fig:d3-ill}, the resulting Delaunay triangulation gives $M=3$ regions: $\mbox{vert}(\triangle^{(1)})=\left\{\bm{w}^{\star 1},\bm{w}^{\star 3},\bm{w}^{\star 4}\right\}$, $\mbox{vert}(\triangle^{(2)})=\left\{\bm{w}^{\star 2},\bm{w}^{\star 3},\bm{w}^{\star 4}\right\}$, and $\mbox{vert}(\triangle^{(3)})=\left\{\bm{w}^{\star 1},\bm{w}^{\star 2},\bm{w}^{\star 4}\right\}$. Each region has two coplanar vectors that make up the following coplanar matrices of dimension $2\times 3$,
$$
	\textbf{C}^{(1)} = \begin{pmatrix}\theta_1\left(\bm{w}^{\star 1}\right)^\top - \theta_3\left(\bm{w}^{\star 3}\right)^\top\\
	\theta_1\left(\bm{w}^{\star 1}\right)^\top - \theta_4\left(\bm{w}^{\star 4}\right)^\top\end{pmatrix}, \,\,\,\,\textbf{C}^{(2)} = \begin{pmatrix}\theta_2\left(\bm{w}^{\star 2}\right)^\top - \theta_3\left(\bm{w}^{\star 3}\right)^\top\\
	\theta_2\left(\bm{w}^{\star 2}\right)^\top - \theta_4\left(\bm{w}^{\star 4}\right)^\top\end{pmatrix},$$
	$$
	\textbf{C}^{(3)} = \begin{pmatrix}\theta_1\left(\bm{w}^{\star 1}\right)^\top - \theta_2\left(\bm{w}^{\star 2}\right)^\top\\
	\theta_1\left(\bm{w}^{\star 1}\right)^\top - \theta_4\left(\bm{w}^{\star 4}\right)^\top\end{pmatrix},
$$
the rows of which are represented by the arrows in Figure~\ref{fig:d3-ill}. For $\bm{\theta}=\left(0.5,0.5,0.5,3\right)^\top$, the unit level set of $g_{\small{\textsc{pwl}}}$ is evaluated and is also displayed in Figure~\ref{fig:d3-ill}.

\subsection{Angular model}
\label{sec:angular-model}
Given that the geometric approach uses a pseudo-radial-angular decomposition, it is desirable to model the distribution of angles $\bm{W}\mid\left\{R>\hat{r}_\tau(\bm{W})\right\}$ with a flexible semiparametric model. This can reduce issues with the curse of dimensionality that can arise when using the empirical distribution of $\bm{W}\mid\left\{R>\hat{r}_\tau(\bm{W})\right\}$, and even in low dimensions may be helpful for ensuring the ability to estimate non-zero extremal probabilities. Note that if $r_\tau(\bm{w})$ corresponds to the exact $\tau$ quantile of $R|\{\bm{W}=\bm{w}\}$, then $\bm{W} \overset{d}{=}\bm{W}|\{R>r_\tau(\bm{w})\}$ \citep{papastathopoulos2025statistical}. However, since we use an estimate $\hat{r}_{\tau}(\bm{w})$ of $r_\tau(\bm{w})$, and it is unlikely to be perfect, we focus only on the distribution of exceedances. The form of $f_{\tilde{\bm{W}}}$ given in Section~\ref{sec:intro-PWL} arises as the exact angular density for a certain type of homothetic density \citep{balkema-nolde-2010}. Specifically, given a gauge function $\tilde{g}$, a valid joint density is $f_{\tilde{\bm{X}}}(\bm{x})={\exp\left\{-\tilde{g}(\bm{x})\right\}}/{\{ d!\text{vol}(\tilde{G})\}}$,
as considered in some examples of \cite{nolde2022linking}. The set $\tilde{G}$ is linked to $\tilde{g}$ through $\tilde{G}=\{\bm{x}\in\mathbb{R}^d: \tilde{g}(\bm{x} )\leq 1\}$. The margins of $\tilde{\bm{X}}$ are near-exponential sub-asymptotically, and are exactly exponential asymptotically. For $\tilde{R}=\|\tilde{\bm{X}}\|$ and $\tilde{\bm{W}}={\tilde{\bm{X}}}/{\|\tilde{\bm{X}}\|}$, we have $f_{\tilde{R}\mid\tilde{\bm{W}}}(r\mid\bm{w})={r^{d-1}\exp\left\{-r\tilde{g}(\bm{w})\right\}}/{\{ d!\text{vol}(\tilde{G})\}}$ and $f_{\tilde{\bm{W}}}(\bm{x})={\tilde{g}(\bm{x})^{-d}}/{\{d\text{vol}(\tilde{G})\}}$. This suggests that if the extremes of $\bm{X}$ are well-approximated by the density $f_{\tilde{\bm{X}}}$ then the angles $\bm{W}$ might be well-approximated by the density $f_{\tilde{\bm{W}}}$, where the gauge function $\tilde{g}$ is the same as the one in the gamma distribution of $R\mid\bm{W}$. However, when this approximation is poor, this still presents a way to construct a flexible model for $\bm{W}$ via a gauge function $\tilde{g}$ that can be parametrized independently of that used in the truncated gamma distribution.  \cite{papastathopoulos2025statistical} and \cite{deMonte2025generative} both use this form of model for $\bm{W}$ when performing probability estimation, the former in the Bayesian context where the gauge function is modeled by Mat\'{e}rn Gaussian random fields, and the latter in the normalizing flows framework.

\begin{figure*}[t!]
    \centering
    \includegraphics[width=0.3\textwidth]{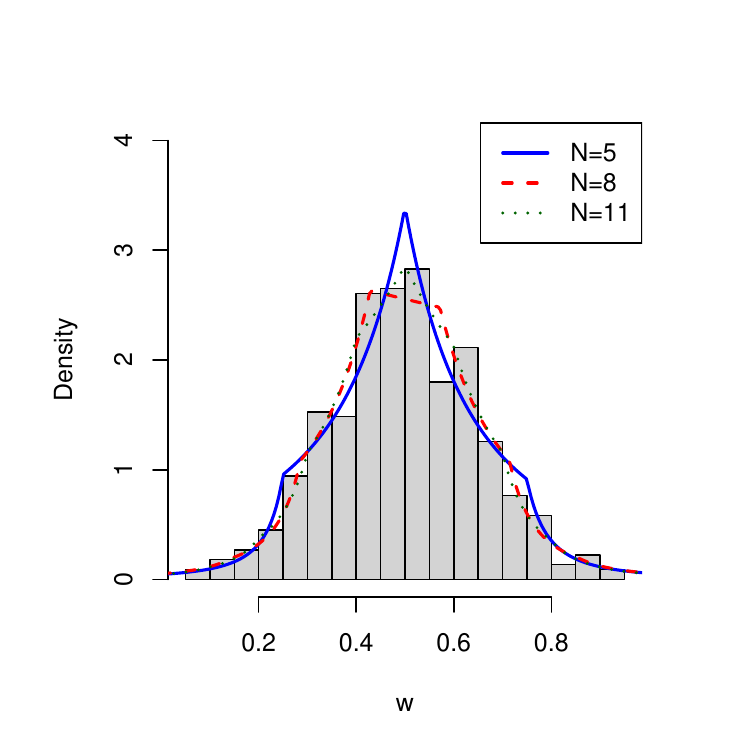}
    \includegraphics[width=0.3\textwidth]{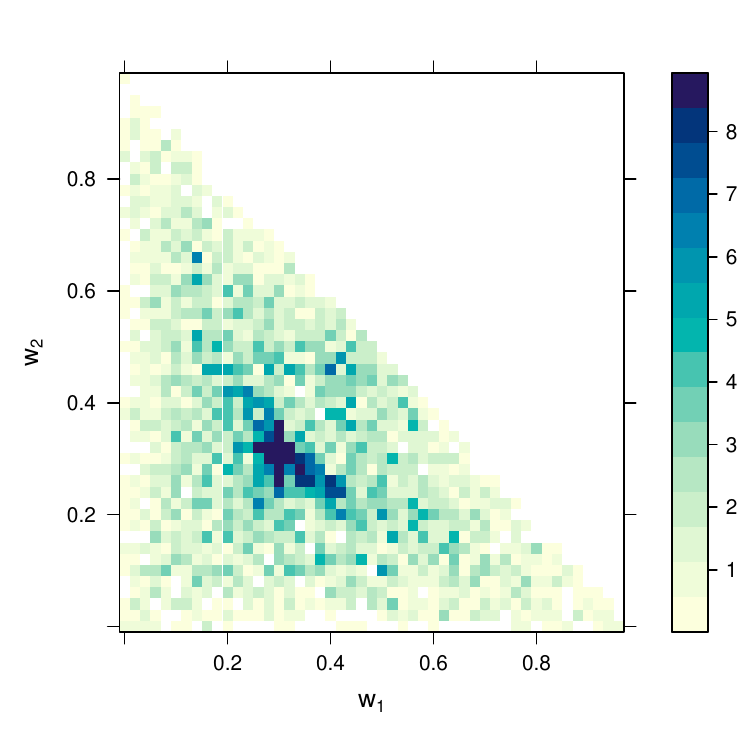}
    \includegraphics[width=0.3\textwidth]{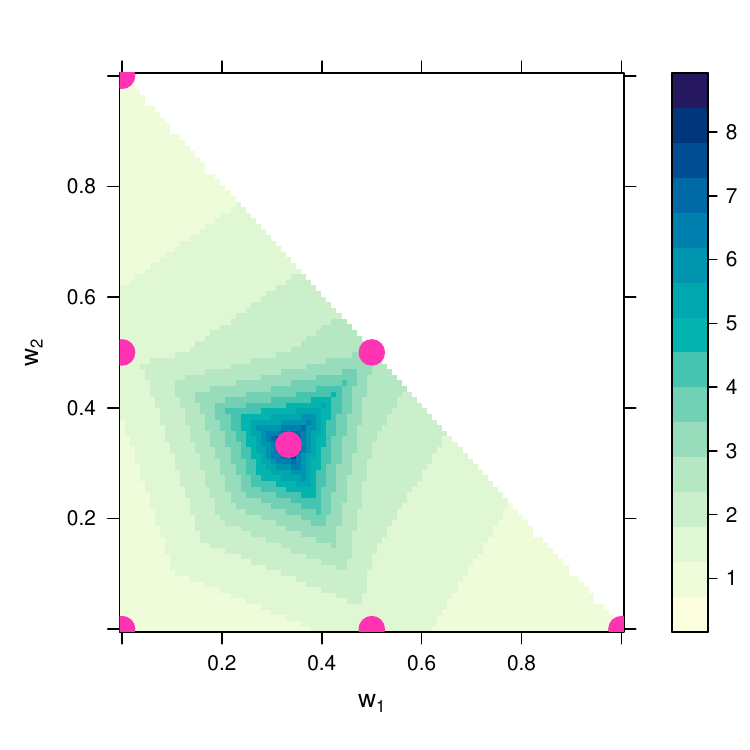}
    \caption{Left: Histogram of exceedance angles from bivariate Gaussian data, with fitted $f_{\tilde{\bm{W}}}(\bm{w})$ for $N=5,8,11$ reference angles. Center: Histogram of exceedance angles from a $d=3$ mixture model. Right: A fit of $f_{\tilde{\bm{W}}}(\bm{w})$ on this data with reference angles overlaid.}
    \label{fig:d2-d3-fW-ex-main}
\end{figure*}

When fitting the model given by $f_{\tilde{\bm{W}}}$ via maximum likelihood estimation, the computation of $\text{vol}(\tilde{G})$ needs to be done at every likelihood evaluation, which can be computationally expensive if the form of this volume is not explicit. Numerical integration methods may be possible, but can drastically slow down maximum likelihood estimation.
\cite{papastathopoulos2025statistical} estimate the volume during model fitting via a latent variable in the likelihood, and a numerical integration procedure is performed during posterior prediction.
Due to the piecewise-linear nature of $g_{\small{\textsc{pwl}}}$, the corresponding $\text{vol}(G_{\small{\textsc{pwl}}})$ is easily obtained, as the unit level set defines the union of $M$ $d$-dimensional faces. In particular, computing $\text{vol}(G_{\small{\textsc{pwl}}})$ is reduced to solving $M$ determinants of $d\times d$ matrices, whose columns correspond to the vertices of the $M$ piecewise-linear regions that make up the set $G_{\small{\textsc{pwl}}}$.
\begin{proposition}\label{prop:vol}
For the piecewise-linear gauge function $g_{\small{\textsc{pwl}}}$ in equation \eqref{eq:gauge-full-w}, the volume of the corresponding set $G_{\small{\textsc{pwl}}}$ is given by
\begin{equation*}
  \textnormal{vol}(G_{\small{\textsc{pwl}}}) =\frac{1}{d!} \sum\limits_{k=1}^{M} \left|\det\begin{pmatrix}
		\theta^{(k)}_1\bm{w}^{\star(k),1} & \theta^{(k)}_2\bm{w}^{\star(k),2} & \hdots & \theta^{(k)}_d\bm{w}^{\star(k),d}
	\end{pmatrix}\right|
\end{equation*}
\end{proposition} 
\noindent The proof of Proposition~\ref{prop:vol} is given in Appendix~\ref{app:vol-proof}.

We illustrate this construction with two examples. First, consider data from distribution \eqref{distn:gauss}, Gaussian dependence with exponential margins. 
Taking $\bm{W}={\bm{X}}/{\|\bm{X}\|}$, we model $\bm{W}$ with $f_{\tilde{\bm{W}}}(\bm{w})={g_{\small{\textsc{pwl}}}(\bm{w})^{-d}}/{\{ d\text{vol}(G_{\small{\textsc{pwl}}})\}}$ without any knowledge of the underlying joint distribution. After doing so, Figure~\ref{fig:d2-d3-fW-ex-main} shows a good fit of the estimated density using regularly-spaced reference angles over the empirical distribution of exceedance angles, with the fit improving as $N$ increases.
Secondly, consider distribution \eqref{distn:mix}, which has a difficult angular structure to capture. Nonetheless, using the piecewise-linear method with $N=7$ reference angles, the fitted model based on $f_{\tilde{\bm{W}}}$ aligns reasonably well with the empirical distribution, both shown in Figure~\ref{fig:d2-d3-fW-ex-main}.
Further diagnostics on samples obtained from the fitted density $f_{\tilde{\bm{W}}}$ using MCMC for these datasets are presented in Supplement~\ref{supp:angle-MCMC-performance}. We note that in all density fits we implement regularization techniques to ensure smoothness of parameter estimates across neighboring regions. This is covered in detail in Section~\ref{sec:ref-angle-choice}.

\section{High radial threshold estimation}
\label{sec:quantiles}
\subsection{Background}
\label{sec:thresholdbackground}
The truncated gamma distribution for $R\mid\bm{W}$ holds asymptotically, and therefore is fitted using datapoints which exceed a threshold.
To define this threshold, denote exceedance radii as observations from the distribution corresponding to $R\mid\left\{\bm{W}=\bm{w},R>r^\star(\bm{w})\right\}$. As mentioned above, we take this threshold to be an estimate of the quantile $r_{\tau}(\bm{w})$, corresponding to the solution to $F_{R\mid\bm{W}}(r_{\tau}(\bm{w})\mid\bm{w})=\tau$, for $\tau$ close to 1. This is a natural choice, and, as \cite{wadsworth2024statistical} point out, leads to an independent estimate of the gauge function through the relation $g(\bm{w})\approx {C_{\tau}}/{r_{\tau}(\bm{w})}$, thus can be helpful for model checking.
They consider two approaches to estimate $r_\tau(\bm{w})$. 
For $d=2$ they suggest additive quantile regression (AQR, \citeay{fasiolo2021fast}), but this could not be extended for $d>2$ using available methodology due to a lack of basis functions defined on the simplex. 
A second approach using an empirical binning method was instead implemented when $d\geq 2$. This requires splitting the simplex into overlapping bins and computing the empirical $\tau^{\text{th}}$ quantile of $R\mid\left\{\bm{W}=\bm{w}\right\}$ in each bin to which the angle $\bm{w}$ belongs.
The radial threshold value at new angles is computed using local means with threshold values already computed in the overlapping bins.
Such an approach is not ideal as $d$ increases, as very little data may be observed in certain bins. Furthermore, it provides a very rough approximation of $r_{\tau}(\bm{w})$.
\cite{papastathopoulos2025statistical} model $\log\left( r_{\tau}(\bm{w})\right)$ using a Mat\'{e}rn Gaussian random field, with implementation currently suitable for dimension $d\leq 3$.

Given the semiparametric nature of our proposed piecewise-linear approach to modeling  $R\mid\bm{W}$, a good estimate of the radial threshold across all angles in the simplex is needed to avoid regions of $\mathcal{S}_{d-1}$ about which little is known.
To overcome some of the issues that persist in current methods, we develop an approach based on kernel density estimation (KDE). The use of kernel-based methods for conditional quantile estimation is well-established, and relevant references include \citet{samanta1989non} and \citet{bhattacharya1990kernel}. A kernel-based approach gives smooth results, akin to the AQR approach. As with empirical estimation, the quality of kernel density estimation deteriorates in high dimensions. However, the empirical approach completely fails when there are too few angles in a bin, while the kernel approach does not, and we find that it at least produces reasonable estimates in moderate dimensions ($d\in\{4,5\}$).

\subsection{Threshold estimate}
\label{sec:threshestimate}
 Given a dataset comprised of $n$ $d$-dimensional observations, we first transform the margins to standard exponential. This is achieved via non or semiparametric estimation of the margins. The standardized datapoints $\bm{x}_1,\dots,\bm{x}_n$ are then transformed to radii $r_1,\dots,r_n$ and angles $\bm{w}_1,\dots,\bm{w}_n$.
We begin with the integral
\begin{equation}\label{eq:KDE-quants-1}
  F_{R\mid \bm{W}}(r\mid\bm{w})
  = \int\limits_0^r \frac{f_{R,\bm{W}}(\tilde{r},\bm{w})}{f_{\bm{W}}(\bm{w})}\,d\tilde{r}.
\end{equation}
Using the radial-angular observations $(r_i,\bm{w}_i)$, $i=1,\dots,n$, we then adopt kernel-based estimates for the densities $f_{R,\bm{W}}$ and $f_{\bm{W}}$:
\begin{align}
  &\widehat{f}_{R,\bm{W}}(r,\bm{w}) =\frac{1}{n} \sum\limits_{i=1}^n k_R\left(\frac{r-r_i}{h_R}\right)k_{\bm{W}}\left(\frac{\bm{w}-\bm{w}_i}{h_{\bm{W}}}\right)\frac{1}{h_R}\frac{1}{h_{\bm{W}}^{d-1}} \label{eq:KDE-quants-2}\\
  &\widehat{f}_{\bm{W}}(\bm{w}) =\frac{1}{n} \sum\limits_{i=1}^n k_{\bm{W}}\left(\frac{\bm{w}-\bm{w}_i}{h_{\bm{W}}}\right)\frac{1}{h_{\bm{W}}^{d-1}}\:, \label{eq:KDE-quants-3}
\end{align}
where $h_R,h_{\bm{W}}>0$, are bandwidths (or smoothing parameters), and $k_R,k_{\bm{W}}$ are kernels. 
Substituting \eqref{eq:KDE-quants-2} and \eqref{eq:KDE-quants-3} into \eqref{eq:KDE-quants-1}, 
\begin{align}
  &\widehat{F}_{R\mid \bm{W}}(r\mid\bm{w}) \notag \\
  &= \frac{\sum\limits_{i=1}^n k_{\bm{W}}\left(\frac{\bm{w}-\bm{w}_i}{h_{\bm{W}}}\right) \int\limits_{0}^r k_R\left(\frac{\tilde{r}-r_i}{h_R}\right) \frac{1}{h_R} d\tilde{r}}{\sum\limits_{i=1}^n k_{\bm{W}}\left(\frac{\bm{w}-\bm{w}_i}{h_{\bm{W}}}\right)}= \frac{\sum\limits_{i=1}^n k_{\bm{W}}\left(\frac{\bm{w}-\bm{w}_i}{h_{\bm{W}}}\right)\left[K_R\left(\frac{r-r_i}{h_R}\right)-K_R\left(-\frac{r_i}{h_R}\right)\right]}{\sum\limits_{i=1}^n k_{\bm{W}}\left(\frac{\bm{w}-\bm{w}_i}{h_{\bm{W}}}\right)}\:,\label{eq:fhatrgivenw}
\end{align}
where $K_R$ is the distribution function associated with the kernel density $k_R$. The right-hand side of~\eqref{eq:fhatrgivenw} is valid when $k_R$ is supported on $\mathbb{R}_+$. An estimate of $r_\tau(\bm{w})$ can be obtained by solving for $r$ in $\widehat{F}_{R\mid \bm{W}}(r\mid\bm{w}) = \tau$ using numerical inversion methods presented in \cite{brent2013algorithms}.
The choice of kernel and bandwidths will affect the smoothness of the estimate of $r_{\tau}(\bm{w})$. In practice, we employ the univariate Gaussian kernel for $k_R$ and its multivariate counterpart for $k_{\bm{W}}$, with identity correlation matrix. While the choice of Gaussian kernel for $R$ means that $k_R$ has negative support, in practice the right-hand side of~\eqref{eq:fhatrgivenw} still provides an accurate approximation for moderate to large $\tau$, since the relevant values of $r_i$ are typically $\gg h_R$. For smaller values of $\tau$, one can truncate the Gaussian at zero, use an alternative kernel, or replace $K_R$ with its empirical counterpart, which is common in the literature. The Gaussian kernel for $\bm{W}$ will also place mass outside of $\mathcal{S}_{d-1}$. One could again re-normalize to a different support, but this becomes more complicated for $d>2$. Nonetheless, we still find that it works adequately for this quantile regression task. In Supplement~\ref{supp:quants}, we compare the Gaussian kernel to the compact Epanechnikov kernel, finding that both can achieve similar quantile estimation accuracy.
Existing methods that do not place mass outside of the domain on interest can be found in, for example, \cite{de2013euclidean}, who use a beta kernel to estimate the so-called spectral measure on $\mathcal{S}_1$. However, for $d>2$ this would prove to be a computationally-expensive replacement for our approach, as numerical methods would be required for obtaining parameter estimates.

\subsection{Performance evaluation}
\label{sec:performanceevaluation}

As a performance diagnostic for threshold estimates, we propose a metric based on $K$-fold cross-validation of the check function, commonly used as the objective function in quantile regression \citep{koenker1978QRcheckfn}. For $K\in\mathbb{N}$, we split the data $\bm{x}_1,\dots,\bm{x}_n$ into a fitting set of length $n_{fit}=n-\lfloor{n}/{K}\rfloor$ and a evaluation set of length $n_{eval}=\lfloor{n}/{K}\rfloor$. On the $k^\text{th}$ set of fitting data, we obtain a radial threshold estimate $\hat{r}_{\tau}^{(k)}(\bm{w};\;\cdot\;)$, $k=1,\dots,K$. In KDE-based estimation, the obtained threshold values depend on the bandwidth $h_{\bm{W}}$. In the empirical procedure of \cite{wadsworth2024statistical}, it depends on the amount of bin overlap. Once obtained, the mean of the check function is evaluated on the evaluation data, resulting in the following score on the $K$ cross-validation partitions,
\begin{equation*}
	S(\cdot) = \frac{1}{K}\sum\limits_{k=1}^K \frac{1}{n_{eval}}\sum\limits_{i=1}^{n_{eval}}\left[r_i - \hat{r}_{\tau}^{(k)}(\bm{w}_i;\;\cdot\;)\right]\left[\tau - \mathbbm{1}_{(-\infty,0)}\left(r_i - \hat{r}_{\tau}^{(k)}(\bm{w}_i;\;\cdot\;)\right)\right].
\end{equation*}
The bandwidth or the amount of bin overlap are hyperparameters which control how smooth the radial threshold is. A smoothing parameter whose score $S(\cdot)$ is closest to zero is thought of as the optimal hyperparameter setting. 

In Supplement~\ref{supp:quants}, we compare KDE-based estimation to the empirical method described above. This shows little difference in the quality of threshold estimates between the methods for $d=2,3$, but we prefer the KDE approach for the ability to evaluate $\hat{r}_{\tau}(\bm{w})$ for any $\bm{w}\in\mathcal{S}_{d-1}$. 
From the same study on $d=2,3$ datasets, we found that optimal values of $h_{\bm{W}}$ when using the Gaussian kernel were often in the neighborhood of 0.05. 
We also show that varying the radial bandwidth made little difference in the performance of the threshold estimate, so we choose to fix $h_R=0.05$.

Adaptive bandwidths offer an alternative approach.
Rather than letting $h_{\bm{W}}$ be constant, we can allow it to vary with the angle $\bm{w}_i$, giving more bandwidth to regions of $\mathcal{S}_{d-1}$ with low density and less bandwidth to regions with less density. Given a set of angular observations $\bm{w}_1,\dots,\bm{w}_n$, there are numerous proposed methods for obtaining $h_{\bm{W}}(\bm{w}_i)$. In particular, the method outlined in \citet[Section~5.3.1]{silverman1986density} is conveniently available within the \texttt{R} package \texttt{pdfCluster} \citep{azzalini2014clustering}, and we can apply this to each component.
Our results indicate that the adaptive method performs comparably to manual selection of the constant $h_{\bm{W}}$ in dimensions $d=2,3$. However, its effectiveness diminishes in higher dimensions, likely due to being a univariate approach. Nevertheless, it remains a valuable tool for much of this work.

More direct methods for KDE-based threshold estimation exist. \cite{chaudhuri1991nonparametric} and \cite{takeuchi2006nonparametric} both outline methodology based on constructing $\hat{r}_{\tau}$ using a linear combination of $n$ kernel functions, then minimizing the check-function to estimate the $n-1$ weighting parameters. We considered this approach, but found it to be far more computationally intensive than our indirect inversion method.

\section{Inference}
\label{sec:inference-PWL}

\subsection{Model fitting}
\label{sec:lik}
We begin with the dataset $(r_i,\bm{w}_i), i=1,\ldots,n$ as outlined in Section~\ref{sec:quantiles}, and use these to obtain an estimate of $r_\tau(\bm{w})$ using the kernel density estimation approach from Section~\ref{sec:quantiles}.
Given that we have candidate densities for radial and angular components, we can choose to model $\bm{W}$ and $R\mid\left\{\bm{W}=\bm{w}\right\}$ separately or jointly using the following likelihood functions:
$$
\begin{aligned}
    & L_{\bm{W}}(\bm{\theta}_W;r_{1:n},\bm{w}_{1:n}) &&\propto \prod\limits_{i\,:\,r_i>\hat{r}_\tau(\bm{w}_i)}\text{vol}(G_{\small{\textsc{pwl}}}(\bm{\theta}_W))^{-1}g_{\small{\textsc{pwl}}}(\bm{w}_i;\bm{\theta}_W)^{-d}\notag\\
	& L_{R\mid\bm{W}}(\bm{\theta}_{R|W};r_{1:n},\bm{w}_{1:n}) &&\propto \prod\limits_{i\,:\,r_i>\hat{r}_\tau(\bm{w}_i)}\frac{f_{\small{\text{Ga}}}(r_i;d,g_{\small{\textsc{pwl}}}(\bm{w}_i;\bm{\theta}_{R|W}))}{1-F_{\small{\text{Ga}}}(\hat{r}_\tau(\bm{w}_i);d,g_{\small{\textsc{pwl}}}(\bm{w}_i;\bm{\theta}_{R|W}))}\notag\\
  	& L_{R,\bm{W}}(\bm{\theta}_{R,W};r_{1:n},\bm{w}_{1:n}) &&= L_{R\mid\bm{W}}(\bm{\theta}_{R,W})L_{\bm{W}}(\bm{\theta}_{R,W}),\notag
\end{aligned}
$$
where $f_{\small{\text{Ga}}}(\cdot;d,g(\bm{w}))$ and $F_{\small{\text{Ga}}}(\cdot;d,g(\bm{w}))$ are the density and distribution functions of the gamma distribution with shape parameter $d$ and rate parameter  $g(\bm{w})$. 
In this setting, there are two ways to fit a joint model for $(R,\bm{W})\mid\left\{R>\hat{r}_\tau(\bm{W})\right\}$. One can implement a joint approach by maximizing $L_{R,\bm{W}}$. This results in lower variability of parameter estimates of $g_{\small{\textsc{pwl}}}$ via the use of more data. However, potential bias can occur if $g_{\small{\textsc{pwl}}}$ is taken to be the same for the radial and the angular models when the joint tail density of the random vector $\bm{X}$ is not well represented by the homothetic form discussed in Section~\ref{sec:angular-model}. Instead a two step approach of maximizing $L_{\bm{W}}$ and $ L_{R\mid\bm{W}}$ separately, and having models for $\bm{W}\mid\left\{R>\hat{r}_\tau(\bm{W})\right\}$ and $R\mid\left\{\bm{W}=\bm{w},R>\hat{r}_\tau(\bm{w})\right\}$ can be implemented.
One may also wish to model $R\mid\left\{\bm{W}=\bm{w}\right\}$ alone if the empirical distribution of $\bm{W}$ is suitable for all estimation tasks.
Each of these settings is considered extensively in the simulation studies in Section~\ref{sec:sim}.

Maximizing these likelihoods leads to parameter values that do not guarantee the marginal condition on the limit set, that $\max (G_{\small{\textsc{pwl}}})=(1,\dots,1)^\top\in\mathbb{R}^d$. Algorithm~\ref{alg:bound-fit} in Supplement~\ref{supp:bounding-alg} provides an adjustment to the parameter estimation procedure to ensure this condition holds. 
In it, a piecewise-linear model is first fitted via maximum likelihood using the likelihood of interest, $L_{\bullet}$. The parameter(s) that correspond to locations where the limit set is at its largest value, but does not lie on the unit box boundary, is then divided by the fitted gauge function value at that location. These parameters are fixed and the likelihood of interest is then re-maximized with respect to the remaining parameters at the starting values given by the maximum likelihood estimates of the previous fit. This is repeated until $\max (G_{\small{\textsc{pwl}}})=(1,\dots,1)^\top\in\mathbb{R}^d$.
This bounding procedure is suitable when maximizing $L_{R\mid\bm{W}}$ and $L_{R,\bm{W}}$, as there is no such constraint in the angular model. Both unbounded and bounded gauges are fitted in simulation studies and compared for their bias in extremal probability estimates.
We additionally make note of a parameter redundancy in the angular model $f_{\tilde{\bm{W}}}(\bm{w})={g_{\small{\textsc{pwl}}}(\bm{w})^{-d}}/{\{ d\text{vol}(G_{\small{\textsc{pwl}}})\}}$, notably, $f_{\tilde{\bm{W}}}(\bm{w};c\bm{\theta})=f_{\tilde{\bm{W}}}(\bm{w};\bm{\theta})$ for any constant $c>0$. To remedy this when using $L_{\bm{W}}$, we fix $\theta_1=1$, and maximize over the remaining $N-1$ parameters. We further note our choice to fix the gamma shape parameter at the dimension $d$ to simplify model fitting, as the number of parameters is large in our piecewise-linear setting. In experiments, we found that estimating the shape parameter, as is done in \cite{wadsworth2024statistical}, led to little difference in terms of bias in extremal probability estimates.

\subsection{The reference angles and penalization}\label{sec:ref-angle-choice}

\subsubsection{Choosing the reference angles}

The choice of reference angles is important for the quality of the approximation to the underlying gauge function.
In essence, there are two ways of choosing the reference angles for $g_{\small{\textsc{pwl}}}$. The first is to strategically choose them where the underlying limit set boundary has a cusp or a change in direction.
This is hard to do in practice without using knowledge of the true  gauge function, but would lead to a model with fewer parameters. 
We opt for a second approach, setting a relatively fine mesh of reference angles. The result is more parameters than are perhaps needed, requiring some form of penalization during model fitting, to be discussed in Section~\ref{sec:gradpen}.
When $d=2$, we take an equally spaced mesh, ensuring a reference angle is placed at $w={1}/{2}$, which allows for capturing whether or not the variables exhibit simultaneous extremes. 
In practice, we found $N=11$ to be a good choice for approximating bivariate gauge functions. 
For $d=3$, we partition at the subset of nodes $\left\{0,{1}/{6},\dots,1\right\}^2$ which lie in $\mathcal{S}_{2}$, giving a triangulation of $\mathcal{S}_2$ with $N=28$ nodes, as displayed in Figure~\ref{fig:d3-DT} in Supplement~\ref{supp:sim}.
For $d\geq 4$, the grid-based approach leads to a very large number of angles. A sparser approach is to initially place reference angles at the edges $\bm{e}_j$, $j=1,\dots,d$ and the center $\left({1}/{d},\dots,{1}/{d}\right)^\top$ of $\mathcal{S}_{d-1}$, along with an angle at the center of all subfaces of $\mathcal{S}_{d-1}$, adding further angles if diagnostics indicate the need.

\subsubsection{Gradient-based penalization}\label{sec:gradpen}

\begin{figure}[t!]
    \centering
     	\includegraphics[width=0.24\textwidth]{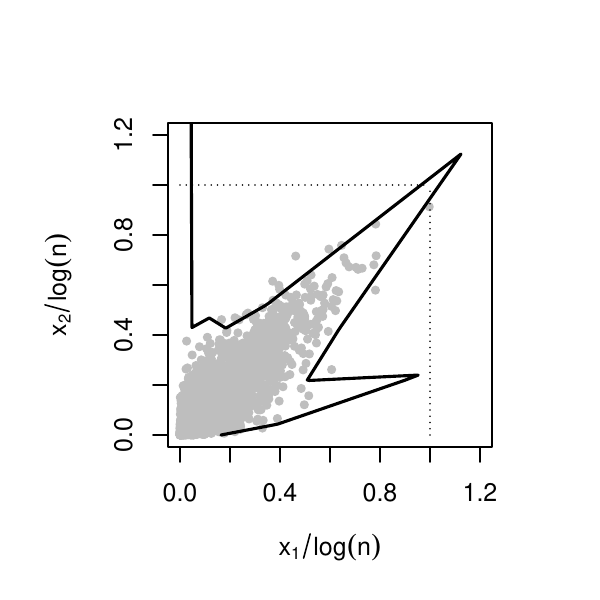}
     	\includegraphics[width=0.24\textwidth]{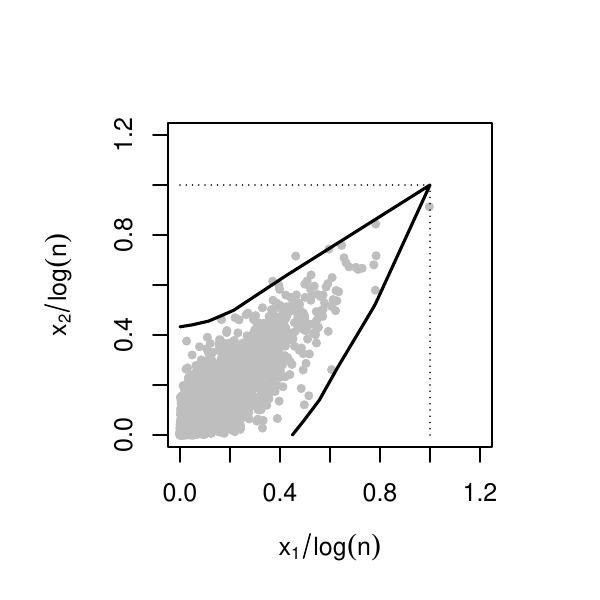}
     	\includegraphics[width=0.24\textwidth]{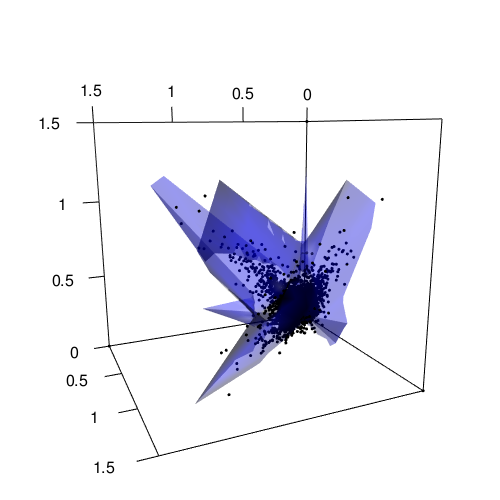}
     	\includegraphics[width=0.24\textwidth]{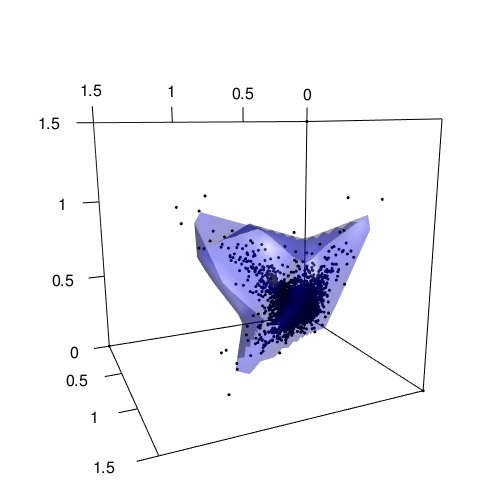}
    \caption[\protect{Left: $d=2$ fitted unit level set of $g_{\small{\textsc{pwl}}}$ using $N=11$ equally-spaced reference angles on a dataset generated from distribution \protect \eqref{distn:log1}. Right: $d=3$ fitted unit level set of $g_{\small{\textsc{pwl}}}$ using $N=28$ reference angles on a dataset generated from distribution \protect \eqref{distn:alog1}.}]{Left: $d=2$ fitted unit level set of $g_{\small{\textsc{pwl}}}$ using $N=11$ equally-spaced reference angles on a dataset generated from distribution \protect \eqref{distn:log1}. The first panel has no penalty; the second panel uses the gradient penalty with $\lambda=1$ and is bounded using Algorithm \protect \ref{alg:bound-fit} in Appendix \protect \ref{supp:bounding-alg}.
    Right: $d=3$ fitted unit level set of $g_{\small{\textsc{pwl}}}$ using $N=28$ reference angles on a dataset generated from distribution \protect \eqref{distn:alog1}. The first panel has no penalty; the second panel uses the gradient penalty with $\lambda=1$ and is bounded using Algorithm \protect \ref{alg:bound-fit} in Appendix \protect \ref{supp:bounding-alg}.}
    \label{fig:pen-ill}
\end{figure}

If we maximize the likelihoods from Section~\ref{sec:lik} with a large number of reference angles $N$, the amount of data contributing to each parameter may be small, leading to high variability in parameter estimates.
To remedy this, we propose penalizing the gradients of $g_{\small{\textsc{pwl}}}$ so that they do not vary too much on either side of the reference angle locations. 
Linearity makes the gradient of $g_{\small{\textsc{pwl}}}$ from equation \eqref{eq:gauge-full-w} simple to calculate:
\begin{equation*}
	\nabla g_{\small{\textsc{pwl}}}(\bm{x};\bm{\theta})=\sum\limits_{k=1}^M \mathbbm{1}_{\triangle^{(k)}}\left({\bm{x}}/{\left\|\bm{x}\right\|}\right)\frac{\bm{n}^{(k)}}{\bm{n}^{(k)\top}\theta^{(k)}_{1}\bm{w}^{\star{(k),1}}}.
\end{equation*}
As the gradients do not change within a segment of the triangulated simplex, we define $\nabla g^{(k)}_{\bm{\theta}}={\bm{n}^{(k)}}/{\bm{n}^{(k)\top}\theta^{(k)}_{1}\bm{w}^{\star{(k),1}}}$ to be the gradient of $g_{\small{\textsc{pwl}}}(\cdot;\bm{\theta})$ on $\triangle^{(k)}$. Further define  $\mathcal{I}_{\ell}$ as the collection of pairs of indices of neighboring segments in the Delaunay triangulation containing the vertex $\bm{w}^{\star \ell}$, where $\ell\in\{1,\dots,N\}$. Neighboring segments of the Delaunay triangulation are defined  as segments that have $d-1$ matching vertices.
For example, in the triangulation of $\mathcal{S}_2$ in Figure~\ref{fig:d3-ill}, we have $\mathcal{I}_1 = \left\{(1,3)\right\}$, $\mathcal{I}_2 = \left\{(2,3)\right\}$, $\mathcal{I}_3 = \left\{(1,2)\right\}$, and $\mathcal{I}_4 = \left\{(1,2),(2,3),(1,3)\right\}$ with $\left|\mathcal{I}_1\right|=\left|\mathcal{I}_2\right|=\left|\mathcal{I}_3\right|=1$ and $\left|\mathcal{I}_4\right|=3$.
Given a likelihood $L_{\bullet}$ from Section~\ref{sec:lik}, we add a penalty to give the objective function
\begin{equation}\label{eq:pen-fit}
	-\log L_{\bullet}(\bm{\theta};r_{1:n},\bm{w}_{1:n}) + \lambda \frac{1}{N}\sum\limits_{\ell =1}^N \frac{1}{|\mathcal{I}_{\ell}|}\sum\limits_{(i,j) \in \mathcal{I}_{\ell}} \|\nabla g^{(i)}_{\bm{\theta}}-\nabla g^{(j)}_{\bm{\theta}}\|_2^2\,.
\end{equation}
This penalty term can be interpreted as the average sum of squared differences between neighboring segment gradients at each node of the Delaunay triangulation. 
Figure~\ref{fig:pen-ill} illustrates the effect of this penalty on two and three-dimensional data.

We propose to select the penalty value $\lambda$ via $K$-fold cross-validation. As described in Section~\ref{sec:quantiles}, the dataset is split into a fitting and an evaluation set. For each value of $\lambda$ on a grid, we fit the model by minimizing \eqref{eq:pen-fit}, and evaluate the negative log-likelihood on the evaluation set,  repeating this $K$ times 
to yield $\text{CV}(\lambda) = K^{-1}\sum_{k=1}^{K} -\log L_{\bullet}(\widehat{\bm{\theta}}_{\lambda,k}; r^{(k)}_{1:n_{eval}}, \bm{w}^{(k)}_{1:n_{eval}})$. The parameter vector $\widehat{\bm{\theta}}_{\lambda,k}$ is the minimizer of \eqref{eq:pen-fit} evaluated at the $k^{\text{th}}$ fitting set, and the $\lambda$ value that minimizes $\text{CV}(\lambda)$ is said to be optimal. 
In Supplement~\ref{supp:pen-str}, we compute the median $\text{CV}(\lambda)$ value across repeatedly-generated datasets from distributions \eqref{distn:log1}--\eqref{distn:mix} with $K=4$. 
The optimal value of $\lambda$ naturally depends on the underlying distribution. Overall, selecting $\lambda$ in the neighborhood of 1 is generally acceptable when minimizing the negative log-likelihoods associated with the conditional radial model or the joint model, while a value of $\lambda=20$ is suitable when minimizing the negative log-likelihood associated with the angular model. These penalty strength values are used henceforth.

\subsection{Probability estimation}\label{sec:prob-est}
%\textcolor{gray}{Describe how probabilities are obtained. MCMC for angles.}
We perform extrapolation using a sampling-based approach in a similar manner to \cite{wadsworth2024statistical}. First, $n^\ast$ samples are drawn from $\bm{W}\mid\left\{R>\hat{r}_\tau(\bm{W})\right\}$, with each one used to draw a conditional sample from the truncated gamma distribution  $R\mid\left\{\bm{W}=\bm{w},R>\hat{r}_\tau(\bm{w})\right\}$.
Once sampled, the exceedance angles $\bm{w}^{\ast}_1,\dots,\bm{w}^{\ast}_{n^{\ast}}$ and radii $r^{\ast}_1,\dots,r^{\ast}_{n^{\ast}}$ are multiplied, resulting in samples $\bm{x}^{\ast}_1,\dots,\bm{x}^{\ast}_{n^{\ast}}$ from $\bm{X}\mid\left\{R>\hat{r}_\tau(\bm{W})\right\}$.
Given an extremal set $B\subset\left\{\bm{x}\in\mathbb{R}^d : \|\bm{x}\| >\hat{r}_{\tau}({\bm{x}}/{\|\bm{x}\|})\right\}$, probabilities can be estimated via
\begin{align*}
	\widehat{\Pr}(\bm{X}\in B) =& \,\,\widehat{\Pr}(\bm{X}\in B\mid R>r_\tau(\bm{W}))\widehat{\Pr}(R>r_\tau(\bm{W}))\\
	=& \left[\frac{1}{n^{\ast}}\sum\limits_{i=1}^{n^{\ast}}\mathbbm{1}_{ B}(\bm{x}^{\ast}_i)\right]\left[\frac{1}{n}\sum\limits_{i=1}^{n}\mathbbm{1}_{(\hat{r}_{\tau}(\bm{w}_i),\infty)}(r_i)\right]\, .
\end{align*}The second term should be near $1-\tau$, but we use the Monte-Carlo estimate to offset any potential bias the kernel density estimate of the threshold, i.e.,  in selecting the $n^\ast$ exceedance indices $i\in\{1,\dots,n\}$ such that $r_i>\hat{r}_\tau(\bm{w}_i)$.
In our approach, we have the option of a semiparametric model for $\bm{W}\mid\left\{R>\hat{r}_\tau(\bm{W})\right\}$ with joint density $f_{\tilde{\bm{W}}}$. 
To sample from $f_{\tilde{\bm{W}}}$, we use Metropolis-Hastings MCMC with a beta or Dirichlet proposal density. The performance of this MCMC is assessed in Supplement~\ref{supp:angle-MCMC-performance}.

\subsection{Model performance and diagnostics}\label{sec:diagnostics}

Several measures of goodness-of-fit for multivariate extremes align well with our piecewise-linear method. 
\cite{wadsworth2024statistical} assess %also assesses 
the performance of a fitted truncated gamma model for exceedance radii $R\mid\left\{\bm{W}=\bm{w},R>\hat{r}_\tau(\bm{w})\right\}$ through probability-probability (PP) and quantile-quantile (QQ) plots.
Plots of the fitted limit set boundary are also useful, since the shape should broadly correspond to that of the scaled sample clouds. 
\cite{simpson2020determining} present methodology for estimating the collection of sets $\mathcal{C}$ experiencing simultaneous extremes. The methodology depends on several tuning parameters, but provides helpful insight into possible structures, that is independent of gauge function estimation.
To that end, plots of the limit set boundary can be compared to findings based on the \cite{simpson2020determining} coefficients to determine whether or not we accurately capture the extremal dependence structure of a dataset.
For $d\leq 3$, plotting the limit set boundary is straightforward. In higher dimensions, one needs to project the gauge functions down to $d=3$ via minimization over the $d-3$ components. Let $J\subset\left\{1,\dots,d\right\}$ be an index set of size $\left|J\right|=3$, and $J^c$ its complement. One can plot the unit level set of the projection
\begin{equation}
	\label{eq:gauge-proj}
	g_J(\bm{x}_{J};\bm{\theta}) = \min_{\bm{x}_{J^c}\in\mathbb{R}_+^{d-3}}g(\bm{x};\bm{\theta})
\end{equation}
For example, when $d=4$, we plot four unit level sets minimizing over each of the four variables individually. In order to perform these minimizations, we opt to evaluate $g(\bm{x};\bm{\theta})$ on a mesh of $\bm{x}_{J^c}$ values in $[0,1]^{d-3}$, and take the minimum value, which we found to be less computationally expensive than other optimization methods. The size of the mesh is important: too small, and the resulting minima may be incorrect; too large, and the computation time will be high. In practice, a mesh of length $50^{d-3}$ was used in $d>3$. Furthermore, all fitted gauge functions are bounded using Algorithm~\ref{alg:bound-fit} in Supplement~\ref{supp:bounding-alg}, so the reduced domain $[0,1]^{d-3}$ is sufficient rather than minimizing over the entire $\mathbb{R}_+^{d-3}$ space.

Another potential goodness-of-fit measure is in the use of return-level curves \citep{papastathopoulos2025statistical}, sometimes also referred to as a version of ``environmental contours" \citep{simpson2024inference}. Given a return period $T$, the return-level curve defines a lower-bound of an open set such that we expect to see proportion $T^{-1}$ points lying beyond this curve. As we only model above the threshold $\hat{r}_{\tau}(\bm{w})$, we consider $T$ such that $1-T^{-1}\geq\tau$. The return curve in the truncated gamma setting is then defined as
\begin{align*}
	\mathcal{R}(T) =& \left\{\bm{x}\in\mathbb{R}^d_+\mid \bm{x}=F_{\small{\text{Ga}}}^{-1}\left[1-T^{-1};d,g(\bm{w},\bm{\theta})\right]\bm{w},\,\bm{w}\in\mathcal{S}_{d-1}\right\}.
\end{align*}
The full derivation of this expression is given in Supplement~\ref{supp:return-lvl}. Once such a curve is obtained, comparing the proportion of exceedances of $\mathcal{R}(T)$ in our data to the expected value of $T^{-1}$ is one way to assess the predictive performance of the piecewise-linear model. 

A check of how well our model captures the  upper joint tail dependence structure of the data can be assessed via estimates of the tail dependence coefficient
\begin{align*}
	\chi_C(u) = \left(\frac{1}{1-u}\right)\Pr\left(F_X(X_j)>u\:,\:j\in C\:,\:C\subseteq\left\{1,\dots,d\right\}\right),
\end{align*}
for sufficiently high values of $u<1$, where $F_X$ is the distribution function common to all margins. An empirical estimate of $\chi_C(u)$ can be compared to a model-based estimate, obtained via the methods in Section~\ref{sec:prob-est}, where the extremal set considered is $B_u^C=\left\{\bm{x}\in\mathbb{R}^d_+\mid x_{j}>F_X^{-1}(u)\:,\:j\in C\right\}$. 
The value of $u$ used needs to be high enough such that $B_u^C \subset\left\{\bm{x}\in\mathbb{R}^d_+\mid \left\|\bm{x}\right\|\geq \hat{r}_{\tau}({\bm{x}}/{\left\|\bm{x}\right\|})\right\}$. The boundary of this region in $\mathbb{R}^d_+$ is $\bm{x}=\hat{r}_{\tau}(\bm{w})\bm{w}$, $\bm{w}={\bm{x}}/{\left\|\bm{x}\right\|}\in\mathcal{S}_{d-1}$.
For any coordinate $x_j$, we want $u$ such that $\hat{r}_{\tau}(\bm{w})w_j>F_X^{-1}(u)$ for all $j\in C$, i.e., $\min_{j\in C}x_j = \min_{j\in C}\hat{r}_{\tau}(\bm{w})w_j>F_X^{-1}(u)$.
To find the maximum point on this boundary in each coordinate we consider $\max_{\bm{w}\in\mathcal{S}_{d-1}}\hat{r}_{\tau}(\bm{w})w_j$.
The minimum value of $u$ at which we can start estimating $\chi_C(u)$ is therefore given by
$
	u_0 = F_X\left(\max_{\bm{w}\in\mathcal{S}_{d-1}} \min_{j\in C} \: \hat{r}_{\tau}(\bm{w}) w_j\right).
$
In practice, $u_0$ can be obtained by taking the maximum over a mesh of values $\bm{w}\in\mathcal{S}_{d-1}$.
To assess the fit of the angular density $f_{\tilde{\bm{W}}}$, the empirical distribution of exceedance angles $\bm{W}\mid\left\{R>\hat{r}_{\tau}(\bm{W})\right\}$ can be compared to the fitted distribution through density plots ($d=2$), or comparing marginal histograms of the sample with those obtained by simulation from the fitted density ($d>2$).

A key preprocessing step in any geometric extremes modeling procedure is the estimation of the threshold $r_{\tau}(\bm{w})$ for level $\tau\in[0,1]$ close to 1. In dimensions $d=2,3$, one could simply plot the threshold curve $\hat{r}_{\tau}(\bm{w})\bm{w}$ for a fine mesh of angles $\bm{w}\in\mathcal{S}_{d-1}$. A good threshold will approximately follow the shape of the data. Visualization tools are not available in dimensions $d>3$, and projections to $d=3$ are therefore needed. To do this, we define a 1-homogeneous function using the threshold estimate $\hat{r}_{\tau}$: $\zeta(\bm{x})={\left\|\bm{x}\right\|}/{\hat{r}_{\tau}({\bm{x}}/{\left\|\bm{x}\right\|})}$. This can be viewed as the gauge function of the star-shaped quantile set at level $\tau$ \citep{papastathopoulos2025statistical}. Similarly to limit set projections, defined through \eqref{eq:gauge-proj}, we can also project the quantile set through a minimization operation. 
That is, given a subset $J\subset\{1,\dots,d\}$, $|J|=3$ and $\bm{x}_J\in\mathbb{R}^3$, we compute the projection
$$
    \zeta_J(\bm{x}_J) = \min_{\bm{x}_{J^c} \in\mathbb{R}_+^{d-3}}\zeta(\bm{x}).
$$
In practice, this minimization task can be computed on a grid of values for 
$\bm{x}_{J^c}$, or can be done using the \texttt{optim} function in \texttt{R}. Plotting the surface ${\bm{w}}/{\zeta_J(\bm{w})}$ for a mesh of angles $\bm{w}\in\mathcal{S}_2$ gives an interpretable projection of higher-dimensional thresholds to the space $\mathbb{R}^3$.

\section{Simulation studies}
\label{sec:sim}

% An overarching goal in multivariate extreme value inference is estimation of $\Pr\left(\bm{X}\in B\right)$, where $B\subset\mathbb{R}^d$ is an extremal set generally lying outside the range of the data. The parametric geometric approach of \cite{wadsworth2024statistical} showed greater accuracy and flexibility in estimating $\Pr\left(\bm{X}\in B\right)$ compared to competing methods. 
We consider the probability estimation problem outlined in Section~\ref{sec:prob-est} for different extremal sets $B$ in dimensions $d=2$ and $d=3$. We compare probability estimates obtained using the form of the true gauge as the rate of the truncated gamma, with parametric estimation of its parameters, and with those obtained using the piecewise-linear gauge. Because the parametric approach uses knowledge of the true gauge function and the piecewise-linear approach does not, the results here are intended to compete with those of \cite{wadsworth2024statistical}, not to outperform them.
In the simulation studies, we consider distributions \eqref{distn:log1}--\eqref{distn:mix}, exhibiting a variety of extremal dependence structures.

For each distribution, we generate $n=5000$ observations, and use the KDE approach with Gaussian kernel to estimate the quantiles $r_{\tau}(\bm{w})$ for $\tau=0.95$,
and $h_{\bm{W}}(\cdot)$ selected using the adaptive method discussed in Section~\ref{sec:performanceevaluation}.
% and bandwidths $h_{R}=0.05$, $h_{\bm{W}}=0.05$.
After obtaining  $\hat{r}_{\tau}(\bm{w})$, we first fit parametric models via maximization of $L_{R\mid \bm{W}}$ with knowledge of the true gauge function and use the empirical distribution of $\bm{W}\mid\left\{R>\hat{r}_{\tau}(\bm{W})\right\}$ for probability estimation, as is done in  \cite{wadsworth2024statistical}.
For the piecewise-linear model, we consider six options for model fitting:
\begin{enumerate}[SS1:]
	\setlength\itemsep{1.9pt}
	\item \label{enumerate:SS1} $R\mid\left\{\bm{W},R>\hat{r}_{\tau}(\bm{W})\right\}$ unbounded; empirical distribution for $\bm{W}\mid\left\{R>\hat{r}_{\tau}(\bm{W})\right\}$.
	\item \label{enumerate:SS2} $R\mid\left\{\bm{W},R>\hat{r}_{\tau}(\bm{W})\right\}$ bounded; empirical distribution for $\bm{W}\mid\left\{R>\hat{r}_{\tau}(\bm{W})\right\}$.
	\item \label{enumerate:SS3} $R\mid\left\{\bm{W},R>\hat{r}_{\tau}(\bm{W})\right\}$ unbounded; model for $\bm{W}\mid\left\{R>\hat{r}_{\tau}(\bm{W})\right\}$.
	\item \label{enumerate:SS4} $R\mid\left\{\bm{W},R>\hat{r}_{\tau}(\bm{W})\right\}$ bounded; model for $\bm{W}\mid\left\{R>\hat{r}_{\tau}(\bm{W})\right\}$.
	\item \label{enumerate:SS5} Unbounded joint model for $(R,\bm{W})\mid\left\{R>\hat{r}_{\tau}(\bm{W})\right\}$.
	\item \label{enumerate:SS6} Bounded joint model for $(R,\bm{W})\mid\left\{R>\hat{r}_{\tau}(\bm{W})\right\}$.
	\end{enumerate}
The term ``bounded'' refers to ensuring $\max(G_{\textsc{pwl}})=1$ via Algorithm~\ref{alg:bound-fit} in Supplement~\ref{supp:bounding-alg} during model fitting. Likelihoods for $R\mid\bm{W}$, $\bm{W}$, and $(R,\bm{W})$ are given in Section~\ref{sec:lik}.
All six settings are considered when performing model fitting.
In all cases we use penalized fitting on regular grid of reference angles.
For each fitted model, $n^{\ast}=50,000$ exceedance observations are generated and probabilities $\Pr(\bm{X}\in B_i)$, $i=1,2,3$, are estimated and compared with the true probabilities. This procedure is repeated 200 times.

%\subsection{Dimension $d=2$}\label{sec:2dsim}

\begin{figure}[htpb!]
    \centering
%    \begin{subfigure}{0.7\textwidth}
%    	\centering
     	\includegraphics[width=0.7\textwidth]{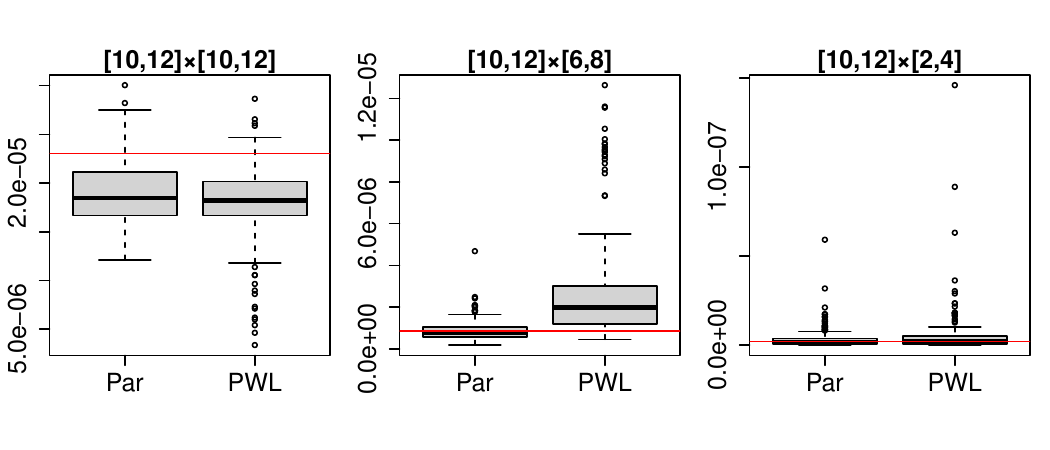}
%    	\caption{distribution \eqref{distn:log1}}
%    \end{subfigure}\hspace{1.5cm}
%    \begin{subfigure}{0.7\textwidth}
%    	\centering
     	\includegraphics[width=0.7\textwidth]{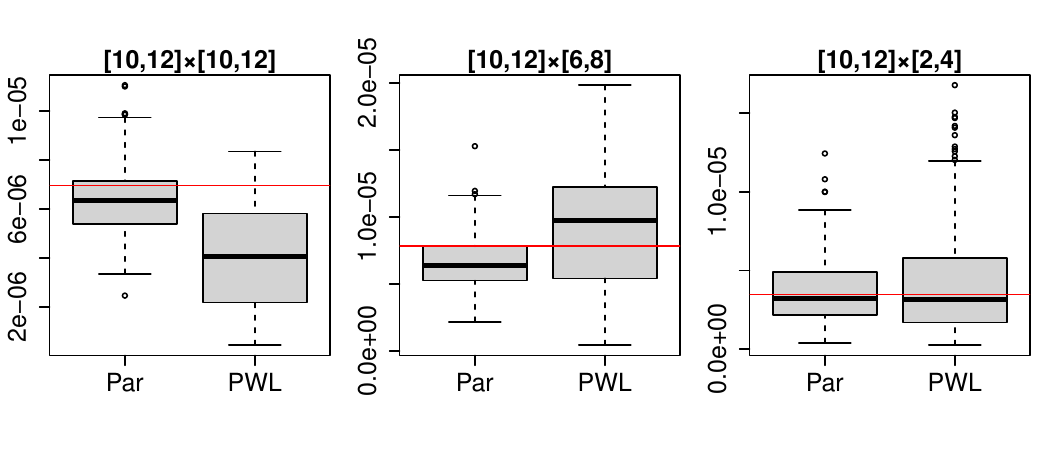}
%    	\caption{distribution \eqref{distn:log2}}
%    \end{subfigure}
%    \begin{subfigure}{0.7\textwidth}
%    	\centering
     	\includegraphics[width=0.7\textwidth]{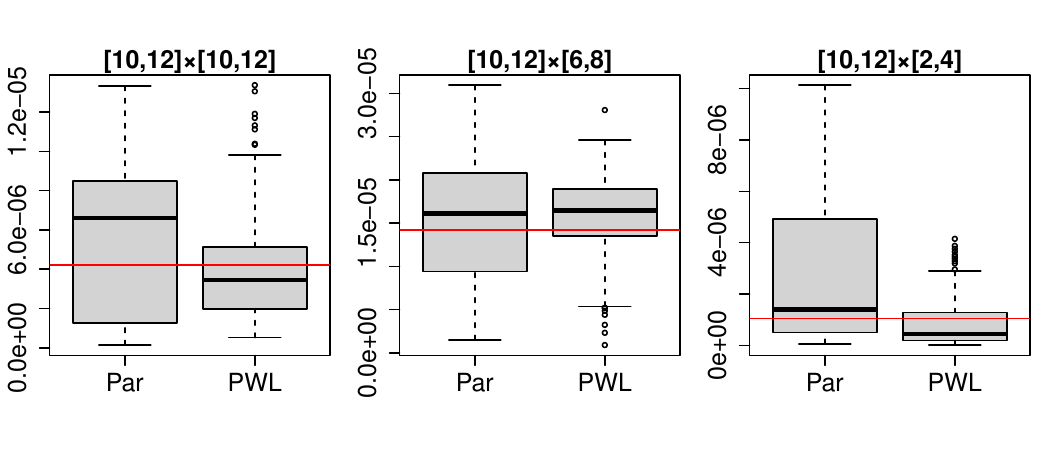}
%    	\caption{distribution \eqref{distn:gauss}}
%    \end{subfigure}\hspace{1.5cm}
%    \begin{subfigure}{0.7\textwidth}
%    	\centering
     	\includegraphics[width=0.7\textwidth]{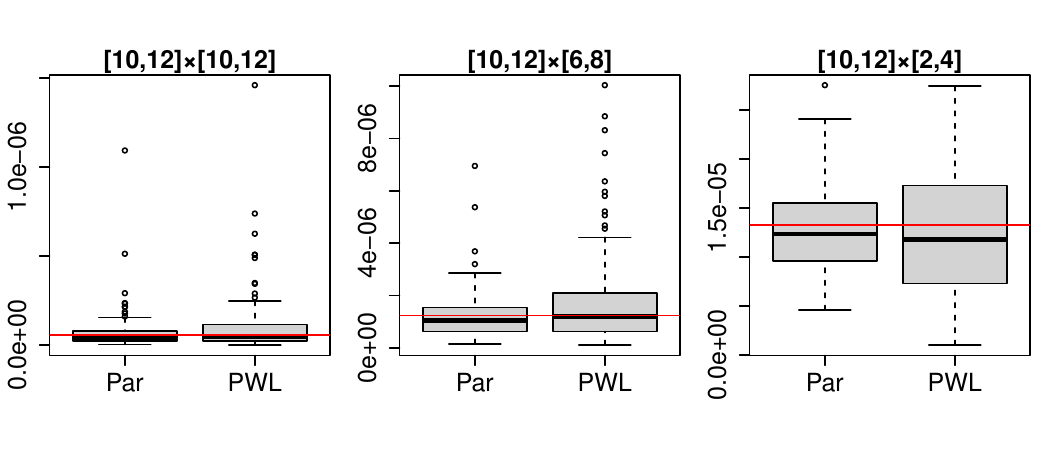}
%    	\caption{distribution \eqref{distn:invlog}}
%    \end{subfigure}
    \caption[\protect{$d=2$ simulation study probability estimates associated with distributions \protect\eqref{distn:log1}--\protect\eqref{distn:invlog} across replicated model fits.}]{$d=2$ simulation study probability estimates associated with distributions \protect\eqref{distn:log1}--\protect\eqref{distn:invlog} (top to bottom) across replicated model fits. ``Par" refers to modeling  with knowledge of the true parametric gauge function, ``PWL" is semiparametric modeling  using the piecewise-linear approach. Solid line is the true probability.}
    \label{fig:d2-simstudy-main-boxplots}
\end{figure}

We first consider the bivariate distributions \eqref{distn:log1}--\eqref{distn:invlog}, with $B_1=[10,12]\times[10,12]$, $B_2=[10,12]\times[6,8]$, and $B_3=[10,12]\times[2,4]$.
When comparing probability estimates to the true values, SS1--SS4 perform similarly, with a slight preference for SS2 and SS4.
The similarity of these shows that the angular fit from maximizing $L_W$ performs as well as its empirical counterpart. 
SS5 and SS6 tend to show more bias as the $\bm{W}$ distribution impacts the estimation of $g_{\small{\textsc{pwl}}}$.
The probability estimates from SS4 are displayed in the boxplots in Figure~\ref{fig:d2-simstudy-main-boxplots}, demonstrating that our semiparametric approach is comparable to the parametric method despite using no knowledge of the underlying distribution.
A full summary of all possible $d=2$ model fits is presented in Supplement~\ref{supp:sim}.

\begin{figure}[htpb!]
    \centering
%    \begin{subfigure}{0.7\textwidth}
%    	\centering
     	\includegraphics[width=0.7\textwidth]{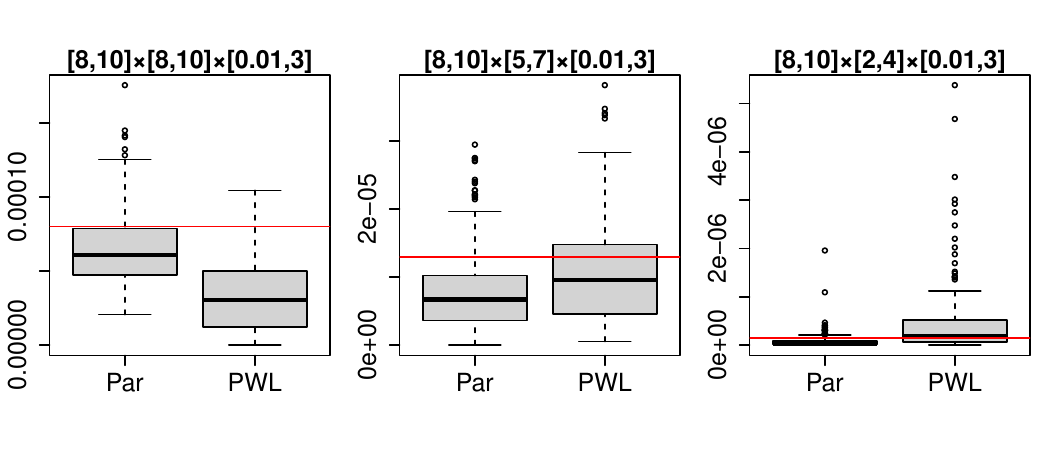}
%    	\caption{distribution \eqref{distn:alog1}}
%    \end{subfigure}\hspace{1.5cm}
%    \begin{subfigure}{0.7\textwidth}
%    	\centering
     	\includegraphics[width=0.7\textwidth]{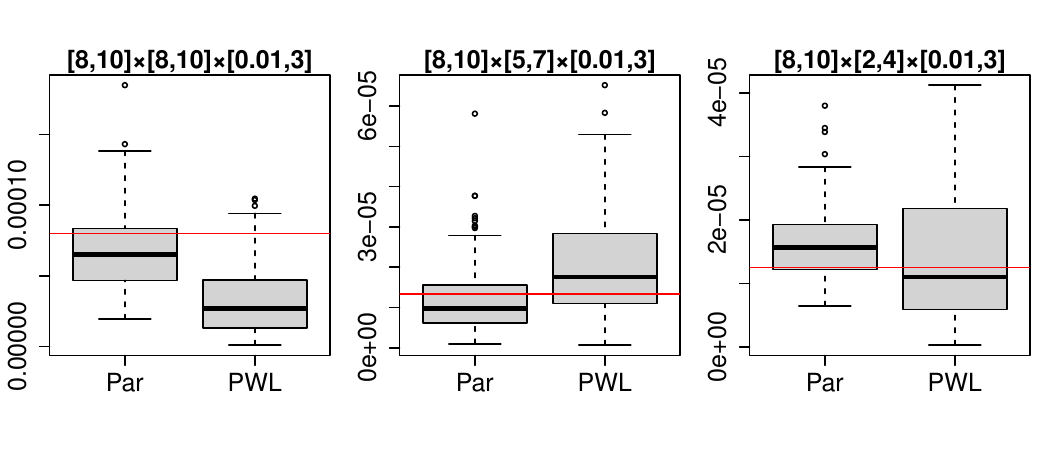}
%    	\caption{distribution \eqref{distn:alog2}}
%    \end{subfigure}
%    
%    \begin{subfigure}{0.7\textwidth}
%    	\centering
     	\includegraphics[width=0.7\textwidth]{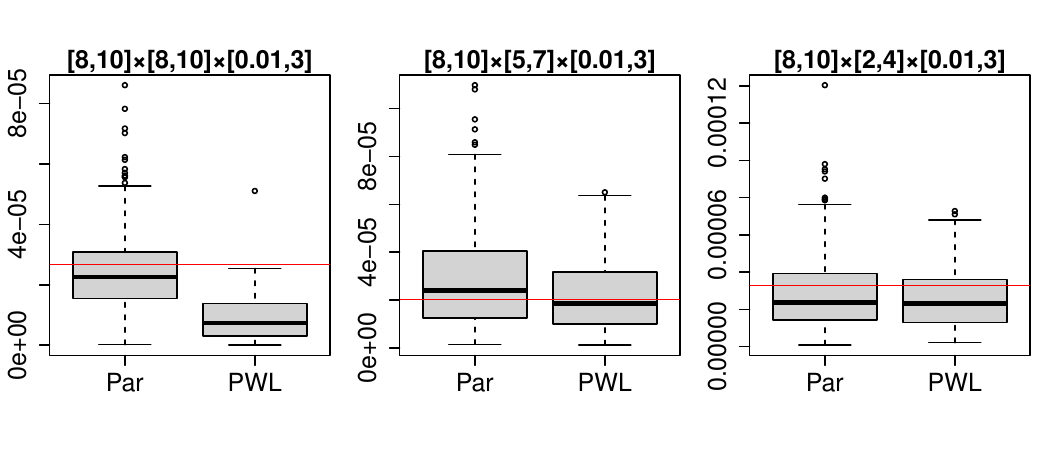}
%    	\caption{distribution \eqref{distn:mix}}
%    \end{subfigure}
    \caption[\protect{$d=3$ extremal probability estimates associated with distributions \protect\eqref{distn:alog1}--\protect\eqref{distn:mix} across replicated model fits.}]{$d=3$ extremal probability estimates associated with distributions \protect\eqref{distn:alog1}--\protect\eqref{distn:mix} (top to bottom) across replicated model fits. ``Par" refers to modeling  with knowledge of the true parametric gauge function, ``PWL" is semiparametric modeling  using the piecewise-linear approach. Solid line is the true probability.}
    \label{fig:d3-simstudy-main-boxplots}
\end{figure}

A similar conclusion can me made from the trivariate data generated from distributions \eqref{distn:alog1}--\eqref{distn:mix}. Here, the extremal regions are defined as $B_1=[8,10]\times[8,10]\times[0.01,3]$, $B_2=[8,10]\times[5,7]\times[0.01,3]$, and $B_3=[8,10]\times[2,4]\times[0.01,3]$.
In comparing probability estimates to the true values, it was also found that the model fits from the SS4 set-up were largely comparable to the parametric method, but without knowledge of the underlying distribution (see Figure~\ref{fig:d3-simstudy-main-boxplots}).
% found that  while SS4 performed slightly worse overall to the other simulation study setups, this approach is still largely comparable to the parametric method, but without knowledge of the underlying distribution (see Figure~\ref{fig:d3-simstudy-main-boxplots}).
A full summary of all possible $d=3$ model fits is again presented in Supplement~\ref{supp:sim}.

Figure~\ref{fig:ss-gauges-main} displays fitted unit level sets of the bounded piecewise-linear gauge function obtained using Algorithm~\ref{alg:bound-fit} when performing inference on the distribution of $R\mid\left\{\bm{W},R>\hat{r}_{\tau}(\bm{W})\right\}$ (setting SS2 and SS4). From this, we see that parameter values across 200 fits produce limit sets that are flexible enough to capture a wide range of tail behaviors. We remark the slight difficulty in capturing asymptotic dependence in distribution \eqref{distn:log2}, or in capturing one of the mixing components of distribution \eqref{distn:mix}, but overall the results indicate reasonable performance. The limit set estimates for the remaining simulation study setups are displayed in Supplement~\ref{supp:sim}.

\begin{figure}[htpb!]
    \centering
    \includegraphics[width=0.23\textwidth]{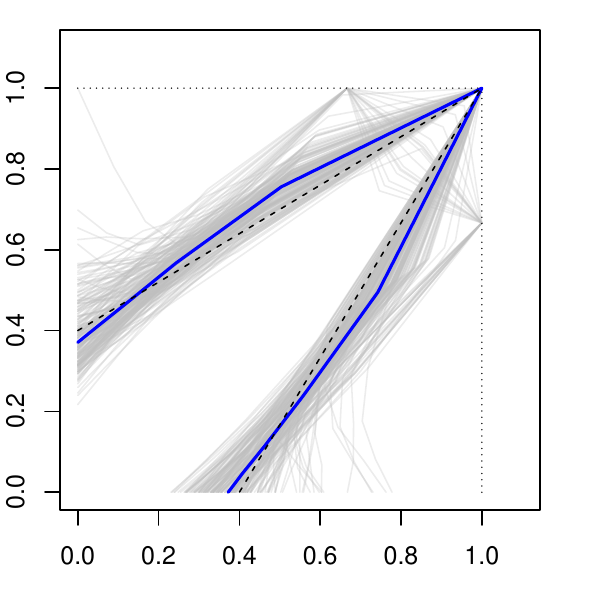}
    \includegraphics[width=0.23\textwidth]{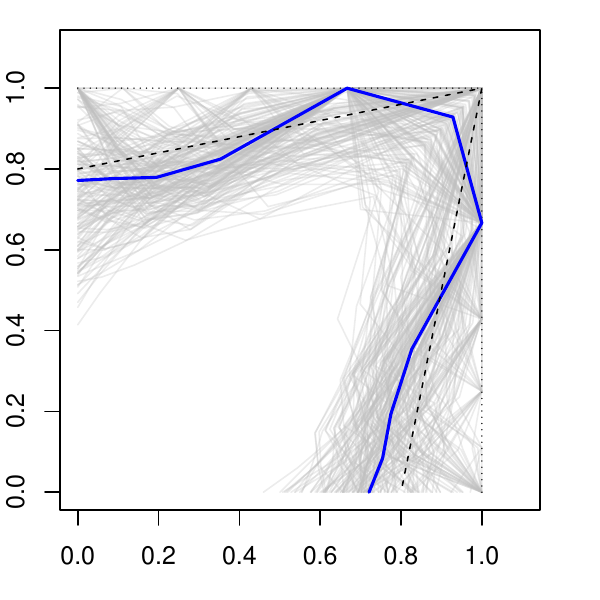}
    \includegraphics[width=0.23\textwidth]{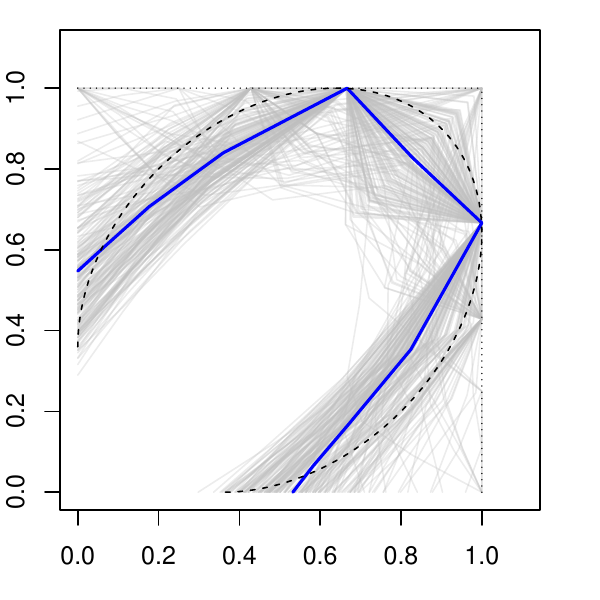}
    \includegraphics[width=0.23\textwidth]{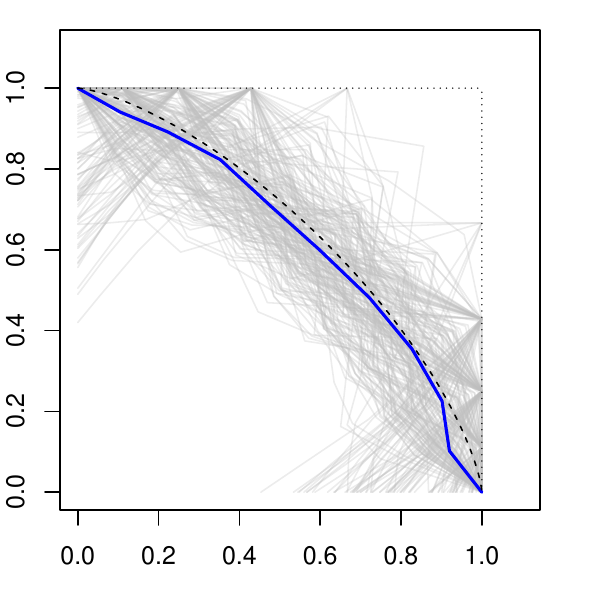}
    
    \includegraphics[width=0.3\textwidth]{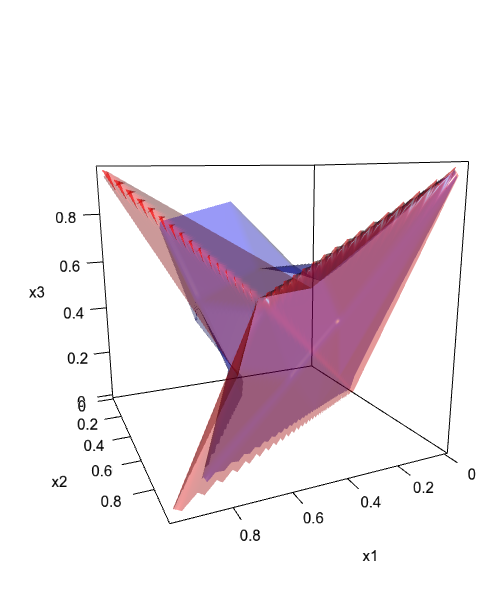}
    \includegraphics[width=0.3\textwidth]{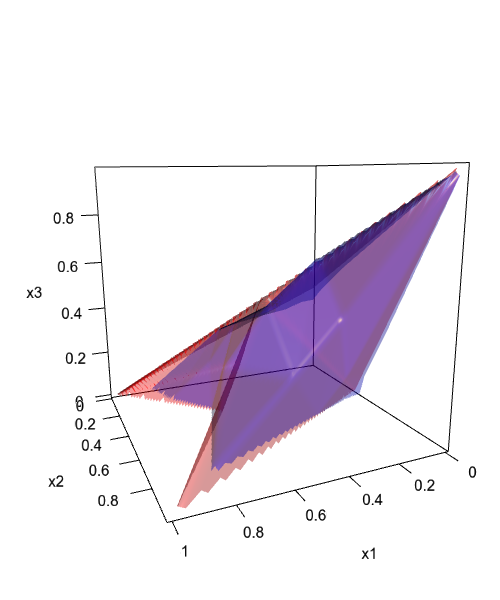}
    \includegraphics[width=0.3\textwidth]{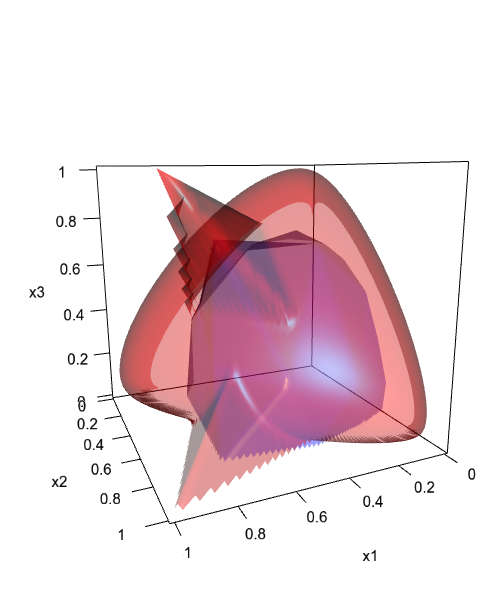}
    \caption[\protect{Estimates of the gauge function unit level sets on data generated from distributions \protect\eqref{distn:log1}--\protect\eqref{distn:mix}.}]{Top row: 200 estimates of the gauge function unit level sets (in grey) on data generated from distributions \protect\eqref{distn:log1}--\protect\eqref{distn:invlog} (left to right), with the gauge function at median parameter values in blue, and the true unit level set given by the dashed line. 
    Bottom row: Gauge function unit level sets evaluated at median parameter values (in blue) with the true unit level set (in red) on data generated from distributions \protect\eqref{distn:alog1}--\protect\eqref{distn:mix} (left to right).}
    \label{fig:ss-gauges-main}
\end{figure}

\newpage

\section{Application to air pollution measurements}
\label{sec:pollution}

We consider air pollution measurements from the Automatic Urban and Rural Network (AURN), a UK-based air quality monitoring network. 
For the North Kensington site in London, we gather hourly measurements from April 1996 to June 2024 for carbon monoxide (CO, ${\text{mg}}/{\text{m}^3}$), nitrogen dioxide ($\text{NO}_2$, $\mu{\text{g}}/{\text{m}^3}$), particulate matter with an aerodynamic diameter of $10\,\mu\text{m}$ or less (PM10, $\mu{\text{g}}/{\text{m}^3}$), and nitric oxide (NO, $\mu{\text{g}}/{\text{m}^3}$). These are labeled 1, 2, 3, and 4 for brevity.
We take the daily maxima over the 247,296 hourly measurements to avoid daily trends, and only consider measurements from October to April, inclusive, to reduce seasonal trends. Any measurements with missing data are excluded. The final dataset has $n=5584$ observations. There is no negative association between these measurements, so the margins are standardized to exponential using the empirical distribution function below the $0.95$ marginal threshold and a generalized Pareto distribution function fitted above this threshold, as outlined in \cite{coles1991modelling}.

\begin{figure}[htpb!]
    \centering
     \includegraphics[width=0.24\textwidth,trim={30 30 10 30},clip]{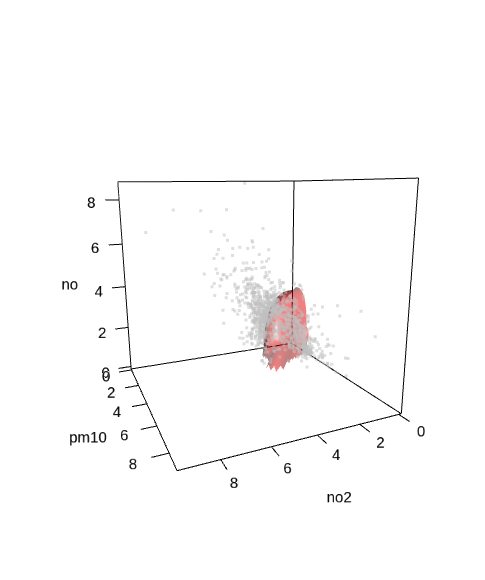}
    \includegraphics[width=0.24\textwidth,trim={30 30 10 30},clip]{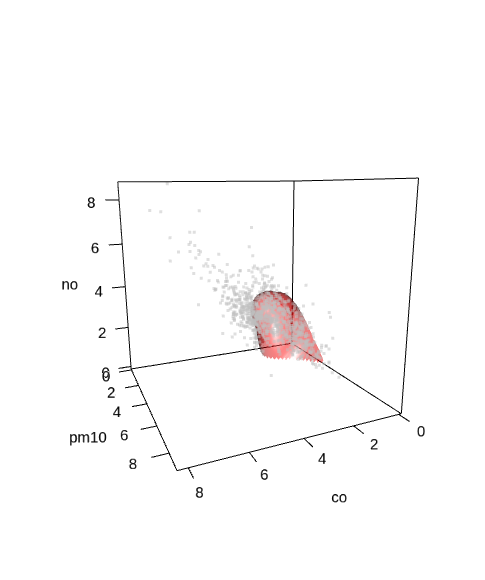}
    \includegraphics[width=0.24\textwidth,trim={30 30 10 30},clip]{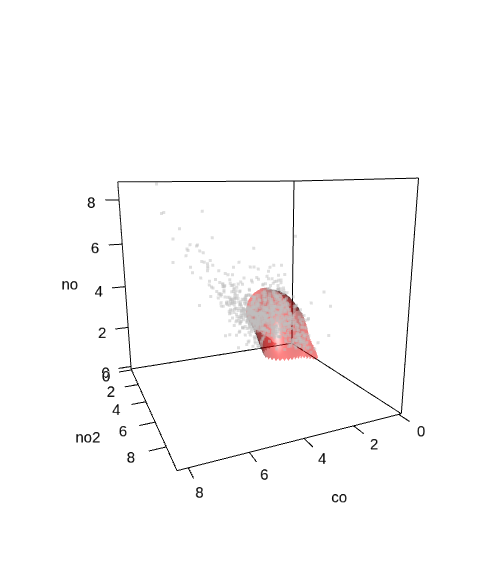}
    \includegraphics[width=0.24\textwidth,trim={30 30 10 30},clip]{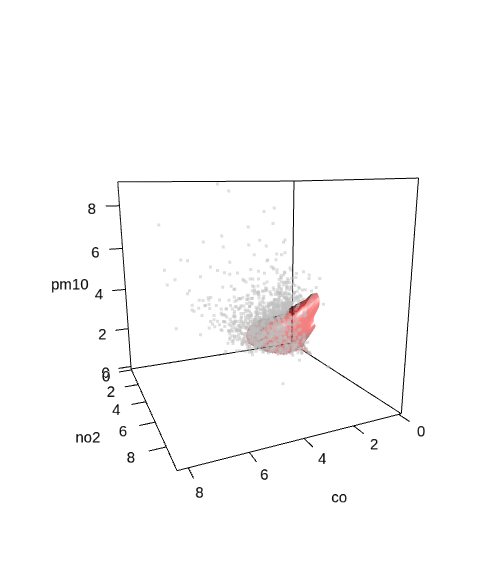}
    
    \includegraphics[width=0.24\textwidth]{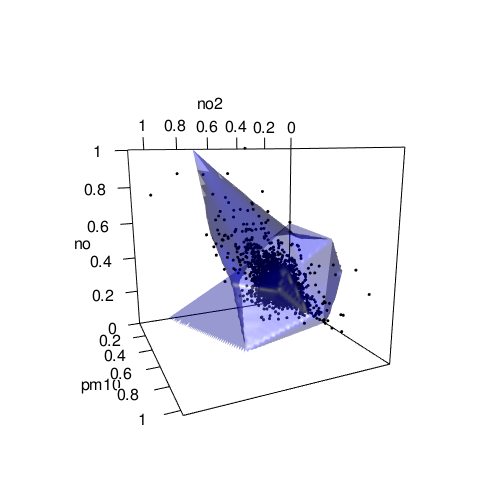}
    \includegraphics[width=0.24\textwidth]{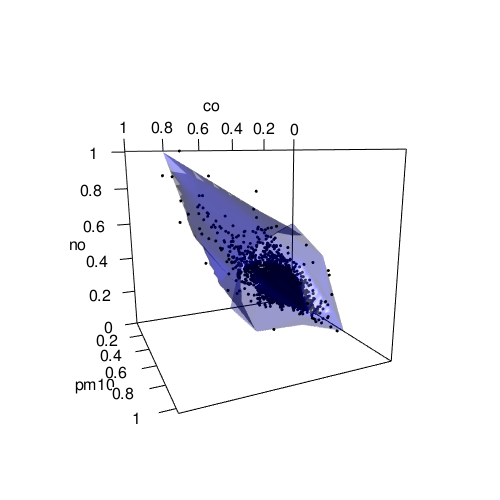}
    \includegraphics[width=0.24\textwidth]{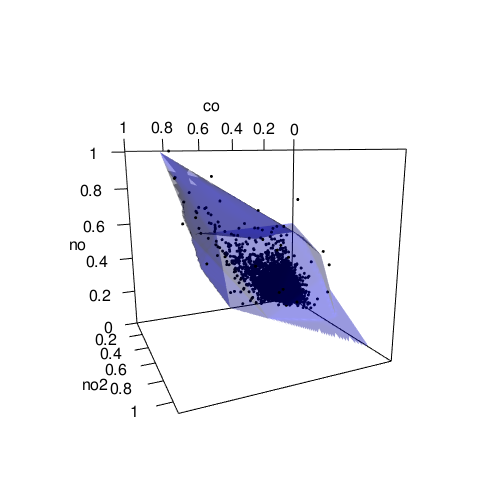}
    \includegraphics[width=0.24\textwidth]{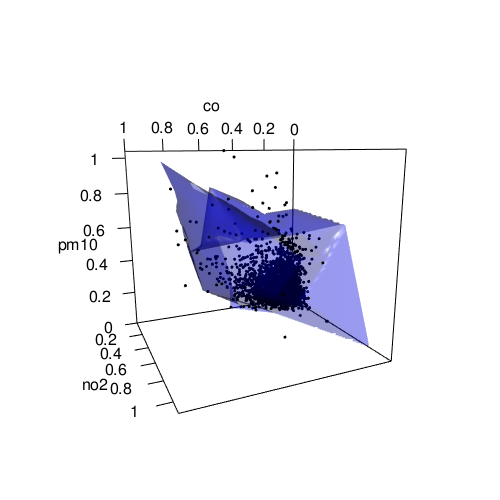}
    \caption{Top row: Projections of the threshold surface $\hat{r}_{\tau}(\bm{w})\bm{w}$ (in red) with data standardized to exponential margins (in gray). Bottom row: Projections of the estimated unit level set of $g_{\small{\textsc{pwl}}}$ (in blue) on 4-dimensional air pollution data with $\log(n)$-scaled data (in black). Gauge functions are projected to 3-dimensions using equation \protect\eqref{eq:gauge-proj} with $J=\left\{1\right\},\left\{2\right\},\left\{3\right\},\left\{4\right\}$ (left to right).}
    \label{fig:d4-pollution-thresh-gauge-proj}
\end{figure}

The radial threshold $r_{\tau}(\bm{w})$, $\bm{w}\in\mathcal{S}_3$, is estimated at $\tau=0.70$, to increase the amount of exceedance data, while still focusing primarily on high values. We note that lower thresholds have been used in \cite{simpson2024inference} and \cite{murphybarltrop2024MLgauges} without inducing large biases.
The angular bandwidth was fixed to $h_{\bm{W}}=0.075$. 
This is slightly different to the adaptive method used in simulation studies, but we felt that this helped eliminate excessive noise in the threshold estimate.
A visualization of the four-dimensional threshold projected to three dimensions is displayed in Figure~\ref{fig:d4-pollution-thresh-gauge-proj}.
A triangulation was obtained by first considering the vertices $\bm{e}_1,\bm{e}_2,\bm{e}_3,\bm{e}_4\in\mathbb{R}^4$, the center of the $\mathcal{S}_3$ simplex, $({1}/{4},{1}/{4},{1}/{4},{1}/{4})^{\top}$, and the center of the subfaces $\mathcal{S}_2,\mathcal{S}_1$. Once obtained, additional reference angles were placed in the center of
each triangle from the resulting initial Delaunay triangulation. The final result is $N = 39$ reference angles.
We fit the conditional radial component $R\mid\left\{\bm{W}=\bm{w},R>\hat{r}_{\tau}(\bm{w})\right\}$ by minimizing the associated penalized negative log-likelihood with penalty strength $\lambda=1$ and bounding using Algorithm~\ref{alg:bound-fit} in Supplement~\ref{supp:bounding-alg}. The corresponding angular model $\bm{W}\mid\left\{R>\hat{r}_{\tau}(\bm{W})\right\}$ was fit separately, with $\lambda=20$.
% fitted limit set boundaries, $\chi_C(u)$ plots, return level set boundaries, PP, and QQ plots

\cite{simpson2020determining} introduce coefficients to suggest which combinations of the components of a random vector can obtain extreme values simultaneously.
Using these coefficients, we observe that all four variables are likely to grow large simultaneously, while PM10 can likely achieve large values when the remaining pollutants are jointly small. 
The projected three-dimensional unit level sets displayed in Figure~\ref{fig:d4-pollution-thresh-gauge-proj} appear able to capture the joint tail dependence, with perhaps slight difficulty capturing the behavior of PM10 when the remaining variables are jointly small (i.e., we do not have exactly $g_{\textsc{pwl}}(\gamma_1,\gamma_2,1,\gamma_3)=1$ for $\gamma_1,\gamma_2,\gamma_3<1$). However, the estimate is not far off, and we are able to capture this behavior in a $d=3$ fit, as is shown in Supplement~\ref{supp:pollution}.
%During experimentation, we found that limit set estimates associated with bounded models were able to capture this behavior; however, their performance in other diagnostics, such as $\chi_C(u)$ plots, were worse than the unbounded fit.
Also by our findings with the \cite{simpson2020determining} coefficients, values of $\chi_C(u)$ are expected to be positive for all values of $u\in[0,1]$ for any collection of variables $C\subseteq\left\{1,2,3,4\right\}$. This is demonstrated by our estimated values, displayed in Figure~\ref{fig:d4-pollution-chi} and in Figure~\ref{fig:supp-d4-pollution-chi} in Supplement~\ref{supp:pollution}, showing a general agreement with the corresponding empirical values.
Note that probability estimates were obtained from extremal samples using our fitted angular model; therefore, good probability estimates indicate a well-fitted angular model.
Figure~\ref{fig:d4-pollution-return} in Supplement~\ref{supp:pollution} shows good agreement between estimated and true return levels $T\in[10,1000]$, while the PP and QQ plots in Figure~\ref{fig:d4-pollution-PPQQ} generally show reasonable agreement between the fitted model and the truncated gamma distribution, with some small deviation.
In Supplement~\ref{supp:pollution} we also present a three-dimensional fit to a subset of the pollutants for comparison, showing agreement with the four-dimensional findings in detecting extremal tail behavior.

\begin{figure}[htpb!]
    \centering
    \includegraphics[width=0.28\textwidth]{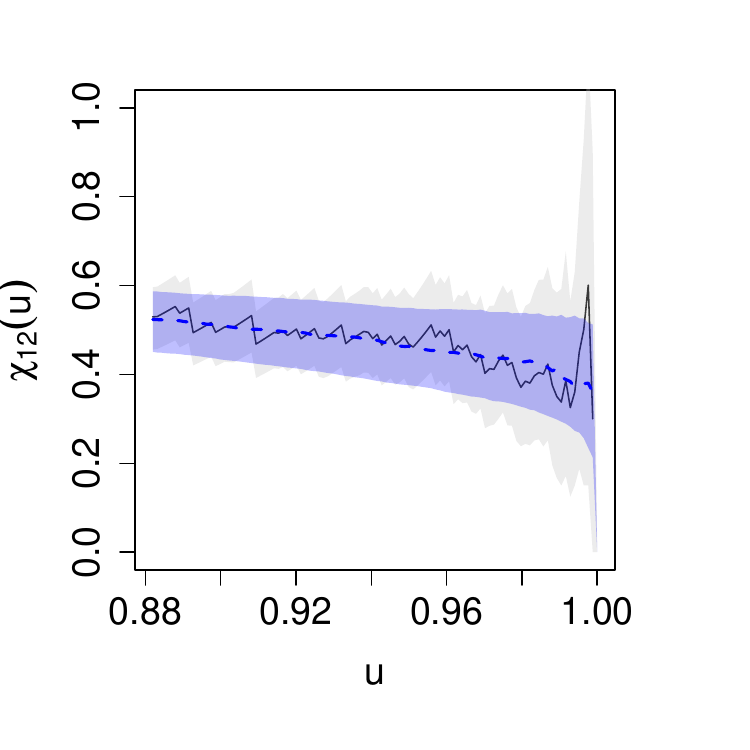}
	\includegraphics[width=0.28\textwidth]{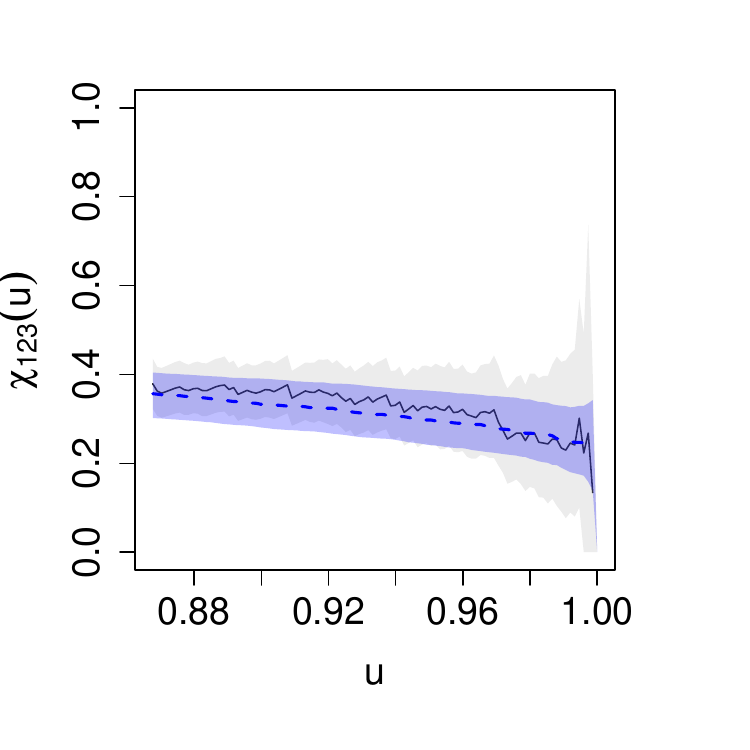}
	\includegraphics[width=0.28\textwidth]{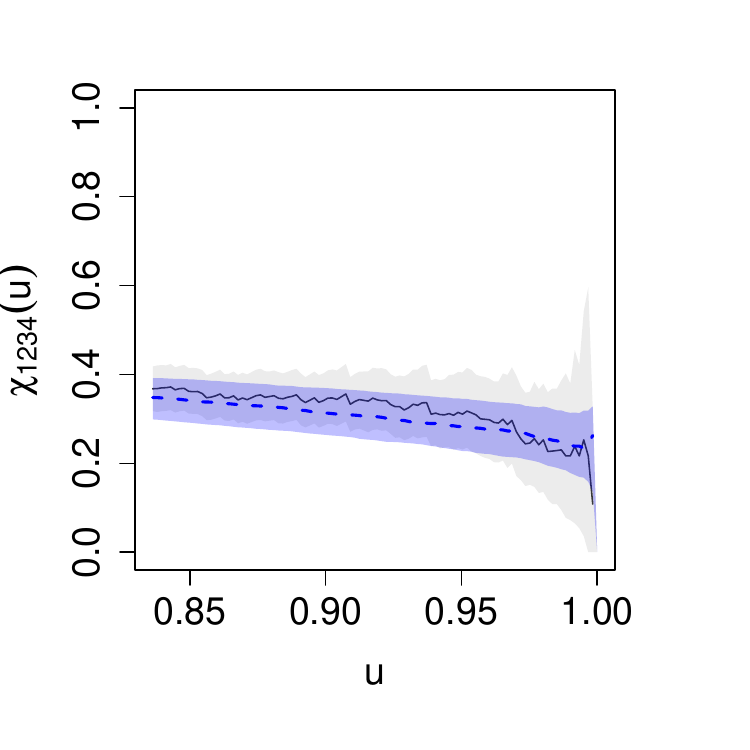}
    \caption[Model-based and empirical $\chi_C(u)$ plots with $C=\left\{1,2\right\}$, $\left\{1,2,3\right\}$, and $\left\{1,2,3,4\right\}$ for the pollution dataset.]{Model-based and empirical $\chi_C(u)$ plots with $C=\left\{1,2\right\}$, $\left\{1,2,3\right\}$, and $\left\{1,2,3,4\right\}$ for the pollution dataset. Solid lines are empirical values, and dashed lines are estimated using the piecewise-linear model. Shaded regions represent 7-day 95\% block bootstrap confidence intervals.}
    \label{fig:d4-pollution-chi}
\end{figure}

\section{Concluding remarks}
\label{sec:conc}
In this work, we aimed to bypass the current difficulties in semiparametric modeling  of multivariate extreme values through the geometric approach by proposing a simple piecewise-linear construction. 
Furthermore, this piecewise-linear construction allows for easy computation of the volume of its limit set, which in turn allows for efficient fitting of an angular model. Its calculation relies on standard operations of linear algebra, rendering it quick to evaluate and perform estimation on, including the use of gradient-based penalties.
Simulation studies show that our semiparametric method is comparable to parametric methods using the true model forms, but without knowledge of the underlying distribution of the dataset.

A difficulty in our proposed method is selecting the reference angles at which parameters are to be estimated. To avoid choosing these angles, we selected a regular grid on the simplex $\mathcal{S}_{d-1}$ for $d=2,3$, while opting for a sparser approach with $d\geq 4$.
In unreported results, we were able to fit our model in dimension $d=5$, but faced difficulties with choice of reference angles. 
Future avenues of work include the development of a method for selecting only important reference angles with a view to performing modeling in higher dimensions, or a way of eliminating unimportant parameters through a different type of penalization, leading to models with fewer parameters but potentially higher predictive performance.

We close with some further comments on possible future work.
Piecewise linear gauge functions may be considered crude since they cannot provide smooth estimates of $g$, if this feature was desired. One may consider ways to smooth the estimates; in a different context \cite{muhlenstadt2011kernel} developed a kernel-based approach to a function estimation problem for which they previously used a piecewise linear solution. In Section~\ref{sec:quantiles}, we used kernel-based methods for threshold estimation, and a version of this could also be considered for gauge function estimation.

\cite{murphybarltrop2025nonstationary} recently extended statistical methods for limit set estimation to a non-stationary setting. Our approach could be adapted to such cases through covariate-dependent parameterization of the $\bm{\theta}$ values, while time-varying kernel methods such as that of \cite{harvey2012kernel} may be used for non-stationary threshold estimation.

\subsection*{Acknowledgments, data availability, and code.}
AURN data was imported using the \texttt{importAURN} function within the \texttt{openair} package in \texttt{R}, but is publicly available at \url{https://uk-air.defra.gov.uk/networks/network-info?view=aurn}. 
Code to reproduce the figures and data analyses in this article can be found at
\url{https://github.com/ryancampbell514/PWLExtremes}, while functionality for fitting piecewise-linear models is embedded in the R package \texttt{geometricMVE}, available at
\url{https://github.com/jennywadsworth/geometricMVE}. 
The authors are grateful to Sylwia Bathe-Taylor for identifying an error in an earlier version of the manuscript, and to referees for their constructive comments.

\section*{Funding}
We are grateful for EPSRC (UK) DTP funding EP/W523811/1, NSERC postgraduate scholarships, doctoral program (Canada), FRQNT doctoral training scholarships (Québec), and the PEPR Risques COCHAIR grant ANR-22-EXIR-0002 from the French National Research Agency supporting Ryan Campbell, and EPSRC grant EP/X010449/1 supporting Jennifer Wadsworth.

\renewcommand{\thesection}{\Alph{section}}
\setcounter{section}{0}

\renewcommand{\thesubsection}{A.\arabic{subsection}}
\setcounter{subsection}{0}

\section*{Appendix}

\renewcommand{\thefigure}{A.\arabic{figure}}
\setcounter{figure}{0}

\renewcommand{\theequation}{A.\arabic{equation}}
\setcounter{equation}{0}

\renewcommand{\thetable}{A.\arabic{table}}
\setcounter{table}{0}

\subsection{On the error bounds of multivariate piecewise-linear interpolation}\label{app:pwl-error-bounds}

We address the approximation error of a gauge function $g$ by the piecewise linear $g_{\small{\textsc{pwl}}}$. 
Under mild conditions, Theorem 4.4.4 of \cite{brenner2008mathematical} ensures that
\begin{equation}\label{eq:pwl-error-bound}
    |g_{\small{\textsc{pwl}}}(\bm{w})-g(\bm{w})| = O(h),
\end{equation}
where $h>0$ is the diameter of the segment of the triangulation that contains $\bm{w}$, defined by
$$
    h=\sup_{\bm{w}_i,\bm{w}_j\in\triangle^{(k)}}\left\|\bm{w}_i-\bm{w}_j\right\|
$$
for some choice of norm.
As $N$ gets arbitrarily large, $h$ gets arbitrarily small.

A requirement for Theorem 4.4.4 of \cite{brenner2008mathematical} is that $g$ and its partial derivatives up to order $\geq1$ have finite norms (i.e., is in a Sobolev space). A particular case of this is Lipschitz continuity. In this case we can derive a bound in terms of the Lipschitz constants for $g$, denoted $K_g$, and $g_{\small{\textsc{pwl}}}$, denoted $K_{\small{\textsc{pwl}}}$.

Consider $g$ on $\mathcal{S}_{d-1}$. Lipschitz continuity gives
\[
\left|g(\bm{w}_1) - g(\bm{w}_2)\right| \leq K_g\|\bm{w}_1-\bm{w}_2\|.
\]
A continuous piecewise-linear function on a compact domain is also Lipschitz, that is:
\[
\left|g_{\textsc{pwl}}(\bm{w}_1) - g_{\textsc{pwl}}(\bm{w}_2)\right| \leq K_{\textsc{pwl}}\|\bm{w}_1-\bm{w}_2\|.
\]
 The Lipschitz constant $K_{\small{\textsc{pwl}}}$ will generally change with $N$.
Let $\bm{w}_k$ be a node of the Delaunay triangulation, at which $\theta_k = 1/g(\bm{w}_k)$ and hence $g_{\textsc{pwl}} (\bm{w}_k) = g(\bm{w}_k)$. Now consider the approximation error at arbitrary $\bm{w}$, where $\bm{w}_k$ is a vertex of the triangle on which $\bm{w}$ lies:
\begin{align}
    \left|g(\bm{w}) - g_{\textsc{pwl}}(\bm{w})\right| &= \|g(\bm{w}) - g(\bm{w}_k) + g(\bm{w}_k)- g_{\textsc{pwl}}(\bm{w})\|\notag\\
    &\leq \|g(\bm{w}) - g(\bm{w}_k)\| + \|g(\bm{w}_k)- g_{\textsc{pwl}}(\bm{w})\|\notag\\
       &= \|g(\bm{w}) - g(\bm{w}_k)\| + \|g_{\textsc{pwl}}(\bm{w}_k)- g_{\textsc{pwl}}(\bm{w})\|\notag\\
    &\leq K_g \|\bm{w}-\bm{w}_k\| + K_{\textsc{pwl}} \|\bm{w}-\bm{w}_k\|\notag\\
    &= (K_g + K_{\textsc{pwl}})\|\bm{w}-\bm{w}_k\|.\label{eq:g-pwl-lipscitz-bound}
\end{align}
Since $\bm{w}$ belongs to the triangle with vertex $\bm{w}_k$, we have that $\|\bm{w}-\bm{w}_k\| \leq \sup\|\bm{w}_i-\bm{w}_j\|$, where the supremum is over all vertices belonging to the same triangle.

We note that in practice, an estimate of the piecewise-linear gauge function will be inferior to the approximation of $g$ with $\theta_k = 1/g(\bm{w}_k)$. Consistent estimation of the parameters $\theta_k$ will, as a minimum, require the sample size of the threshold exceedances to grow at a faster rate than the number of nodes $N$ in the Delaunay triangulation.

We further note that there exist tighter bounds for the interpolation error than the one presented in \eqref{eq:pwl-error-bound} and \eqref{eq:g-pwl-lipscitz-bound}; however, they rely on $g$ having continuous higher-order derivatives (see Theorem 4.4.4 of \citeay{brenner2008mathematical}). This is not guaranteed to be the case with gauge functions, with an example being  $g(\bm{x})=\max(\bm{x})$.

\subsection{Volume of $G_{\small{\textsc{PWL}}}$}
\label{app:vol-proof}

\begin{proof}[Proof of Proposition \ref{prop:vol}]

The limit set $G_{\small{\textsc{pwl}}}$ with a piecewise-linear boundary is a union of subregions defined by parameters and reference angles. Therefore, its volume can be decomposed in the following manner:
\begin{align*}
	\text{vol}(G_{\small{\textsc{pwl}}}) =& \sum\limits_{k=1}^{M} \text{vol}\left(\bm{\theta}^{(k)}\triangle^{(k)}\right),
\end{align*}
where $\bm{\theta}^{(k)}\triangle^{(k)}\subset\mathbb{R}^d$ is the region with vertices at $\theta^{(k)}_1 \bm{w}^{\star(k),1},\dots,\theta^{(k)}_d \bm{w}^{\star(k),d}$, and the origin if working in the positive orthant.
Any point $\bm{x}\in\bm{\theta}^{(k)}\triangle^{(k)}$ can be written as 
$$
	\bm{x} = \sum\limits_{j=1}^d \theta^{(k)}_j a_j \bm{w}^{\star(k),j}\,\,;\,a_j\geq0\,\forall\, j,\,\sum\limits_{j=1}^d a_j=1
$$
where 
$$
	\theta^{(k)}_j\bm{w}^{\star(k),j} = \begin{pmatrix}
		\theta^{(k)}_1\bm{w}^{\star(k),1} & \theta^{(k)}_2\bm{w}^{\star(k),2} & \hdots & \theta^{(k)}_d\bm{w}^{\star(k),d}
	\end{pmatrix}\bm{e}_j.
$$
Let $\textbf{M}^{(k)}=\begin{pmatrix}
		\theta^{(k)}_1\bm{w}^{\star(k),1} & \theta^{(k)}_2\bm{w}^{\star(k),2} & \hdots & \theta^{(k)}_d\bm{w}^{\star(k),d}
\end{pmatrix}$ be the change of basis matrix, then we can write 
$$
	\bm{x} = \textbf{M}^{(k)}\sum\limits_{j=1}^d a_j \bm{e}_j\,\,;\,a_j\geq0\,\forall\, j,\,\sum\limits_{j=1}^d a_j=1.
$$
Note that $\sum_{j=1}^d a_j \bm{e}_j$ defines any point in the simplex $\mathcal{S}_{d-1}$; therefore, it follows that
\begin{align*}
	\text{vol}\left(\bm{\theta}^{(k)}\triangle^{(k)}\right) 
	= \int\limits_{\bm{\theta}^{(k)}\triangle^{(k)}}d\bm{x} =& \left|\det\left(\textbf{M}^{(k)}\right)\right|\int\limits_{\mathcal{S}_{d-1}}d\bm{u}\\
	=& \left|\det\begin{pmatrix}
		\theta^{(k)}_1\bm{w}^{\star(k),1} & \theta^{(k)}_2\bm{w}^{\star(k),2} & \hdots & \theta^{(k)}_d\bm{w}^{\star(k),d}
	\end{pmatrix}\right|{1}/{d!},
\end{align*}
where ${1}/{d!}$ is the volume of the $d-1$-dimensional simplex \citep{simplex-vol}.

\end{proof}

\bibliography{bibliography.bib}

\newpage

\begin{center}
    {\LARGE\bf Supplementary Material}
\end{center}

% \renewcommand{\thesection}{\Alph{section}}
% \setcounter{section}{0}
% \renewcommand{\thefigure}{S.\arabic{figure}}
% \setcounter{figure}{0}
% \renewcommand{\theequation}{S.\arabic{equation}}
% \setcounter{equation}{0}
% \renewcommand{\thetable}{S.\arabic{table}}
% \setcounter{table}{0}
% \renewcommand{\thesection}{\Alph{section}}
% \setcounter{section}{0}
% \renewcommand{\thefigure}{S.\arabic{figure}}
% \setcounter{figure}{0}
% \renewcommand{\theequation}{S.\arabic{equation}}
% \setcounter{equation}{0}
% \renewcommand{\thetable}{S.\arabic{table}}
% \setcounter{table}{0}
% 1. Use the built-in appendix command to safely switch to A, B, C...
\appendix
\renewcommand{\thesection}{\Alph{section}}
\setcounter{section}{0}
\renewcommand{\thefigure}{S.\arabic{figure}}
\setcounter{figure}{0}
\renewcommand{\theequation}{S.\arabic{equation}}
\setcounter{equation}{0}
\renewcommand{\thetable}{S.\arabic{table}}
\setcounter{table}{0}

\section{Catalog of considered multivariate distributions}\label{supp:copulas}

\subsection{Logistic}\label{supp:copulas-log}
The joint distribution function in standard Fr\'{e}chet margins is given by
\begin{equation*}
	F_F(\bm{x};\alpha) = \exp\left\{-V(\bm{x};\alpha)\right\}\:,\:\:\:\: V(\bm{x};\alpha) = \left(\sum\limits_{j=1}^d x_j^{-{{1}/{\alpha}}}\right)^\alpha,
\end{equation*}
where $V$ is the $-1$-homogeneous exponent function, and $\alpha\in(0,1]$ controls the strength of dependence. The corresponding parametric gauge function is obtained by differentiating $F_F$ to obtain the joint density $f_F$, performing a change of variables to standard exponential margins, and taking the limit \eqref{eq:gauge-limit} to obtain
\begin{equation*}
	g(\bm{x};\alpha) = \frac{1}{\alpha}\left(\sum\limits_{j=1}^d x_j\right) +\left(1-\frac{d}{\alpha}\right)\min\left\{x_1,\dots,x_d\right\}.
\end{equation*}

\subsection{Gaussian}\label{supp:copulas-gauss}

Consider the matrix $\Sigma_{ij}=\rho_{ij}$, $i,j\in\left\{1,\dots,d\right\}$ where $\rho_{ij}=\text{Corr}(X_i,X_j)>0$. The gauge function is obtained by performing a change of variables of a joint multivariate normal density with covariance matrix $\Sigma$ in standard normal margins to standard exponential margins, then taking the limit \eqref{eq:gauge-limit}. This results in the following
\begin{equation}\label{eq:gauss-gauge}
	g(\bm{x};\Sigma) = \sqrt{\bm{x}}^\top\Sigma^{-1}\sqrt{\bm{x}}.
\end{equation}
In equation \eqref{eq:gauss-gauge}, all operations performed on vectors are done componentwise. The case with some $\rho_{ij}<0$ is given for Laplace margins in equation \eqref{eq:gauss-gauge-Lap}.

\subsection{Inverted logistic}\label{supp:copulas-invlog}
The inverted logistic distribution is most simply presented by its joint survival function in standard exponential margins,
\begin{equation*}
	\bar{F}_E(x,y;\alpha) = \exp\left\{-V({1}/{\bm{x}};\alpha)\right\}\:,\:\:\:\: V({1}/{\bm{x}};\alpha) = \left(\sum\limits_{j=1}^d x_j^{{{1}/{\alpha}}}\right)^\alpha,
\end{equation*}
where $\alpha\in(0,1]$ controls the rate at which the $d$ marginal variables grow large together.
The corresponding parametric gauge function is obtained by differentiating $\bar{F}_E$ to obtain the joint density $f_E$ and taking the limit \eqref{eq:gauge-limit} to obtain
\begin{equation*}
	g(\bm{x};\alpha) = V({1}/{\bm{x}};\alpha).
\end{equation*}

\subsection{Asymmetric logistic}\label{supp:copulas-alog}

The joint distribution function in Fr\'{e}chet margins is given by
\begin{equation*}
	F_F(\bm{x};\bm{\alpha}) = \exp\left\{-V(\bm{x};\bm{\alpha})\right\},
\end{equation*}
where $V$ is a prespecified $-1$-homogeneous exponent function, depending on the desired dependence structure. In general, $V$ is given by
\begin{equation}\label{eq:alog-V-fn}
	V(\bm{x};\bm{\phi},\bm{\alpha}) = \sum_{C \in P_D}\phi_C \left(\sum\limits_{j\in C} x_j^{-{{1}/{\alpha_C}}}\right)^{\alpha_C},
\end{equation}
where $P_D$ is the power set of indices $D=\left\{1,\dots,d\right\}$, 
$$
	\phi_C = \begin{cases}
		1 &;\:\text{variables}\:C\:\text{can grow large simultaneously.}\\
		0 &;\:\text{otherwise}
	\end{cases},
$$
and $\alpha_C\in(0,1]$ controls the dependence in group $C$. 
Additional parameters are required in \eqref{eq:alog-V-fn} to make the margins standard Fr\'{e}chet, though these do not affect the limiting gauge function and so are omitted.
Here, the gauge function is obtained by differentiating $F_F$ to obtain the joint density $f_F$, performing a change of variables to standard Gumbel margins. 
%by considering limiting behavior of $r\bm{w}$ as $r\rightarrow\infty$ for simplicity, and taking the limit \eqref{eq:gauge-limit}. 
The standard Gumbel and standard exponential distribution are asymptotically equivalent in $\mathbb{R}_+^d$, and therefore we can obtain the gauge function in the usual way (see \cite{wadsworth2024statistical}). 
The true gauge is given by
\begin{equation*}
	g(\bm{x};\bm{\alpha}) = \min_{\pi\in\Pi}\min_{C\in\mathcal{C}_s^+\,:\,s\in\pi}\left[\sum\limits_{s\in\pi}\left(\sum\limits_{j\in s}\frac{x_j}{\alpha_{C}} + \left(1-\frac{|s|}{\alpha_{C}}\right)\min_{j\in C}x_j\right)\right]
\end{equation*}
where $\Pi$ is the set of all partitions of $D$, where $\mathcal{C}_s^+$ denotes a collection of indices corresponding to a group {obtaining simultaneous extremes}, i.e., when $\phi_{s}=1$.

\subsection{Mixture model}\label{supp:copulas-mix}

Consider a mixture model with exponential margins whose joint density is given by
\begin{equation*}
	f(\bm{x};\bm{\theta}_1,\bm{\theta}_2) = p f_1(\bm{x};\bm{\theta}_1) + (1-p) f_2(\bm{x};\bm{\theta}_2),
\end{equation*}
where $f_1$ and $f_2$ are joint densities with respective parameters $\bm{\theta}_1$ and $\bm{\theta}_2$ in exponential margins, and $p\in(0,1)$, with $p=0.5$ throughout this work. The corresponding gauge function is given by
$$
	g(\bm{x};\bm{\theta}_1,\bm{\theta}_2) = \min\left\{g_1(\bm{x};\bm{\theta}_1),g_2(\bm{x};\bm{\theta}_2)\right\},
$$
where $g_1$ and $g_2$ are the gauge functions corresponding to the joint densities of $f_1$ and $f_2$, respectively.

\section{Truncated gamma return level sets}\label{supp:return-lvl}

For a given return period $T$, set $\mathcal{R}(T)\subset\mathbb{R}^d$ to be the corresponding return level set. For $T={1}/{(1-\tau)}$, where $\tau$ is the level at which we \rc{estimate the threshold}, we have
$$
	\mathcal{R}({1}/{(1-\tau)}) = \left\{\bm{x}\in\mathbb{R}^d_+\middle| \bm{x}=r_\tau(\bm{w})\bm{w},\,\bm{w}\in\mathcal{S}_{d-1}\right\}.
$$
For general $T$, we have
\begin{align*}
	\Pr\left(R\geq r_{1-T^{-1}}(\bm{w})\right) =& T^{-1}.
\end{align*}
\noindent It follows that, for $T\geq (1-\tau)^{-1}$,
\begin{align*}
	&\Pr\left(R\geq r_{1-T^{-1}}(\bm{w})\middle| \bm{W}=\bm{w}\right) \\
	&=
	\Pr\left(R\geq r_{1-T^{-1}}(\bm{w})\middle| R\geq r_{\tau}(\bm{w}) , \bm{W}=\bm{w}\right)\Pr\left(R\geq r_{\tau}(\bm{w}) \middle| \bm{W}=\bm{w}\right)\\
	&= \frac{\bar{F}_{\small{\text{Ga}}}\left[r_{1-T^{-1}}(\bm{w});d,g(\bm{w})\right]}{\bar{F}_{\small{\text{Ga}}}\left[r_{\tau}(\bm{w});d,g(\bm{w})\right]} \times \bar{F}_{\small{\text{Ga}}}\left[r_{\tau}(\bm{w});d,g(\bm{w})\right]\\
	&= T^{-1},
\end{align*}
%Deconditioning,
%\begin{align*}
%	\Pr\left(R\geq r_{1-T^{-1}}(\bm{w})\middle| R\geq r_{\tau}(\bm{w})\right) =& (1-\tau)^{-1}T^{-1}
%\end{align*}
%Using the assumption of truncated Gamma radial values over the threshold $r_\tau$ and solving for $r_{1-T^{-1}}(\bm{w})$,
%$$
%	r_{1-T^{-1}}(\bm{w}) = F_{\small{\text{Ga}}}^{-1}\left[1-(1-\tau)^{-1}T^{-1};d,g(\bm{w},\bm{\theta})\right]
%$$
where $F_{\small{\text{Ga}}}^{-1}\left[\cdot;d,g(\bm{w},\bm{\theta})\right]$ is the quantile function corresponding to the gamma distribution with shape parameter $d$ and rate parameter $g(\bm{w},\bm{\theta})$. 
Therefore, the $1-T^{-1}$ radial quantile is given by $$r_{1-T^{-1}}(\bm{w})=F_{\small{\text{Ga}}}^{-1}\left[1-T^{-1};d,g(\bm{w},\bm{\theta})\right],$$ and the return level set is therefore given by the curve
\begin{align*}
	\mathcal{R}(T) =& \left\{\bm{x}\in\mathbb{R}^d_+\middle| \bm{x}=F_{\small{\text{Ga}}}^{-1}\left[1-T^{-1};d,g(\bm{w},\bm{\theta})\right]\bm{w},\,\bm{w}\in\mathcal{S}_{d-1}\right\}.
\end{align*}

\section{\rc{Threshold} estimation}\label{supp:quants}

\begin{figure*}[h!]
    \centering
%    \begin{subfigure}{0.22\textwidth}
%        \centering
        \includegraphics[width=0.22\textwidth]{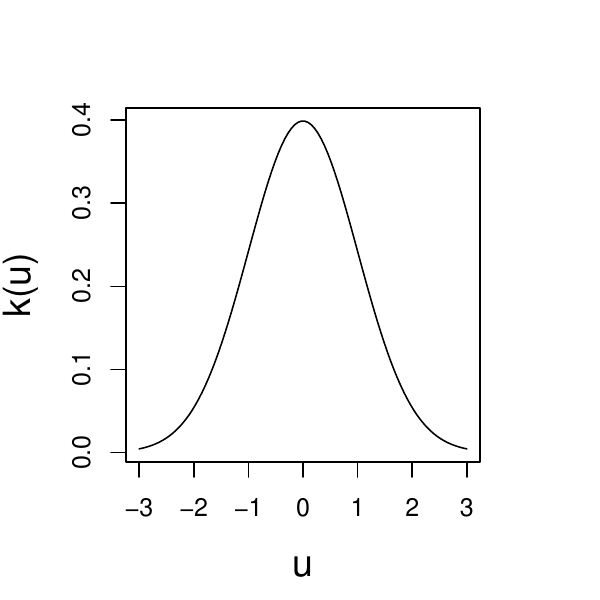}
%        \caption{}
%    \end{subfigure}%
%    ~ 
%    \begin{subfigure}{0.22\textwidth}
%        \centering
        \includegraphics[width=0.22\textwidth]{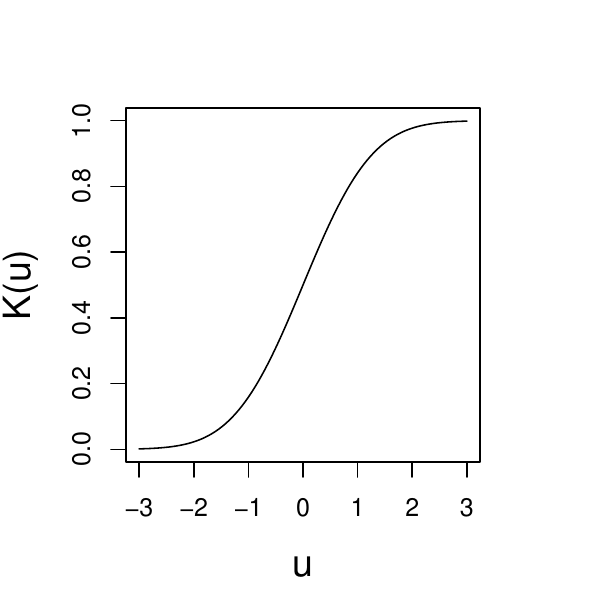}
%        \caption{}
%    \end{subfigure}%
%    ~ 
%    \begin{subfigure}{0.22\textwidth}
%        \centering
        \includegraphics[width=0.22\textwidth]{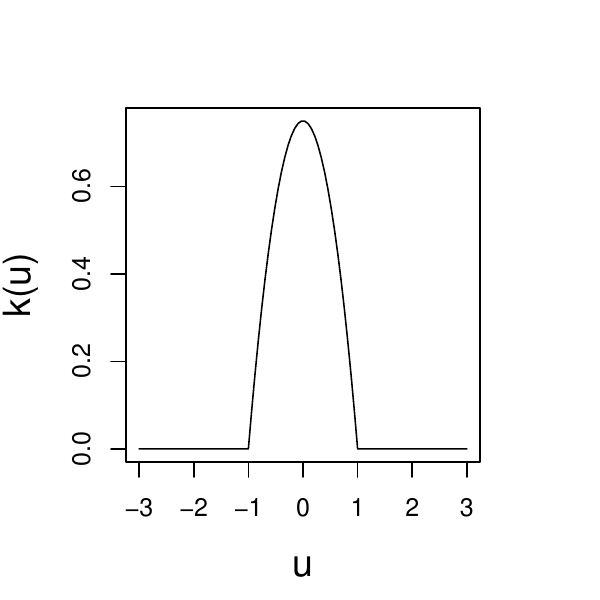}
%        \caption{}
%    \end{subfigure}
%    ~ 
%    \begin{subfigure}{0.22\textwidth}
%        \centering
        \includegraphics[width=0.22\textwidth]{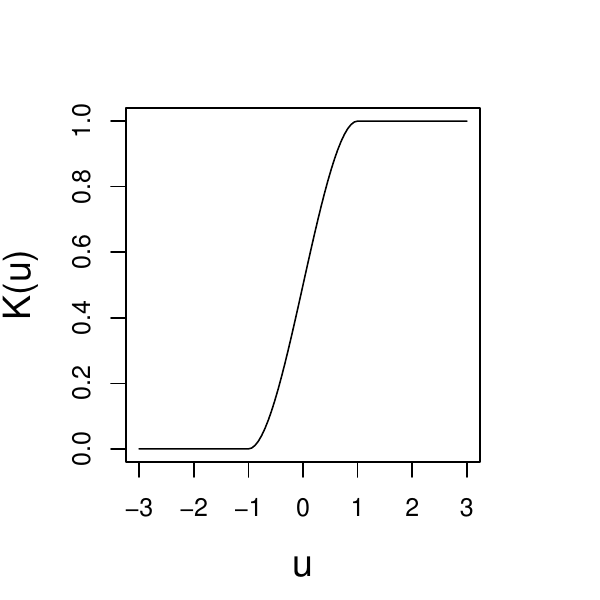}
%        \caption{}
%    \end{subfigure}
	\caption{Left to right: Gaussian PDF, Gaussian CDF, Epanechnikov PDF, and Epanechnikov CDF}
    \label{fig:kernels}
\end{figure*}

The kernel density estimation (KDE) approach of obtaining \rc{thresholds} introduced in Section~\ref{sec:quantiles} is an important step before obtaining maximum likelihood estimates of parameters of our piecewise-linear model.
Given the compact domain $\mathcal{S}_{d-1}$, the use of compactly-supported kernels may be of interest.
Here, we compare the Gaussian kernel to the Epanechnikov kernel \citep{epanechnikov1969non}, which has a bounded support and is defined by $$k(u)=\mathbbm{1}_{(-1,1)}(u)\frac{3}{4}(1-u^2)$$ and $$K(u)=\begin{cases}0 &;u\leq -1\\ \frac{3}{4}\left[u(1-\frac{1}{3}u^2) + \frac{2}{3}\right] &; -1<u<1 \\ 1&;u\geq1\end{cases}.$$
Both kernels' associated densities (PDFs) and distribution functions (CDFs) are shown in \autoref{fig:kernels}.

%\subsection{Choosing the angular bandwidth}\label{supp:quants-hW}

\begin{figure}[h!]
    \centering
%    \begin{subfigure}{0.3\textwidth}
%        \centering
        \includegraphics[width=0.3\textwidth]{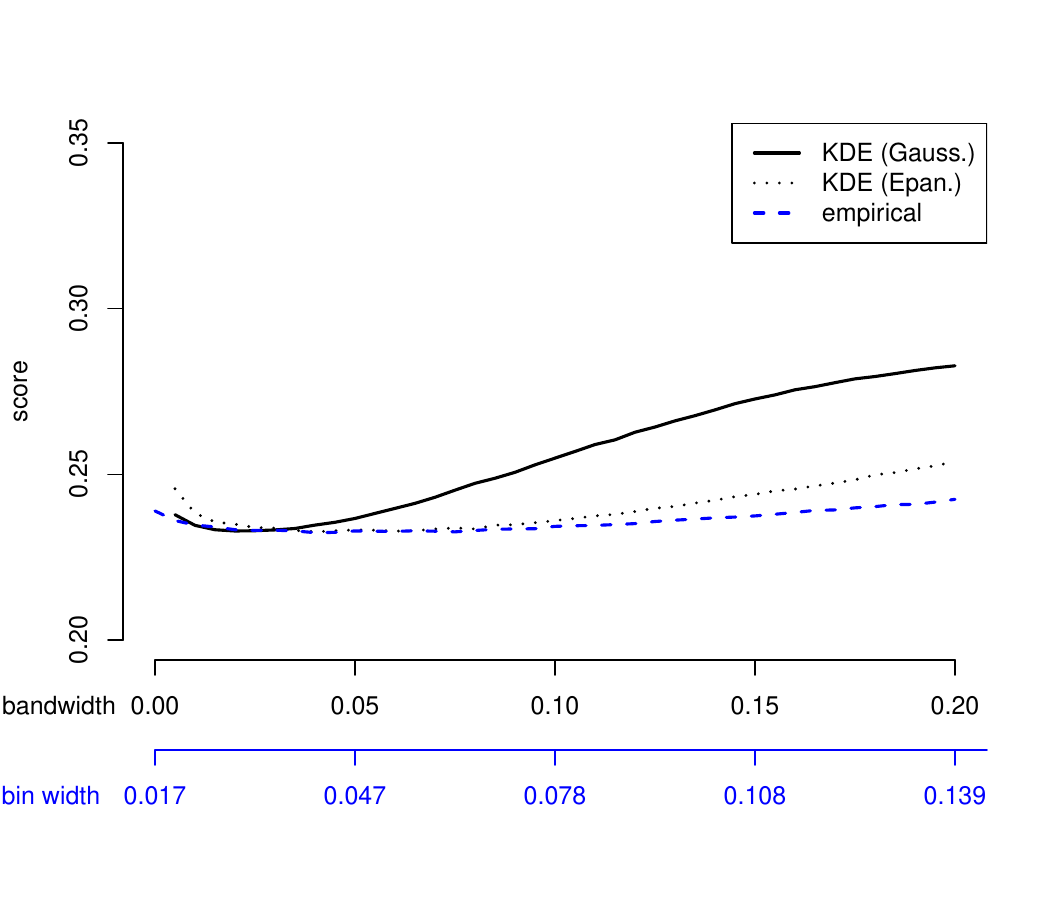}
%        \caption{}
%        \label{fig:d2-quant-scores-log1}
%    \end{subfigure}
%%    \hspace{1cm}
%    \begin{subfigure}{0.3\textwidth}
%        \centering
        \includegraphics[width=0.3\textwidth]{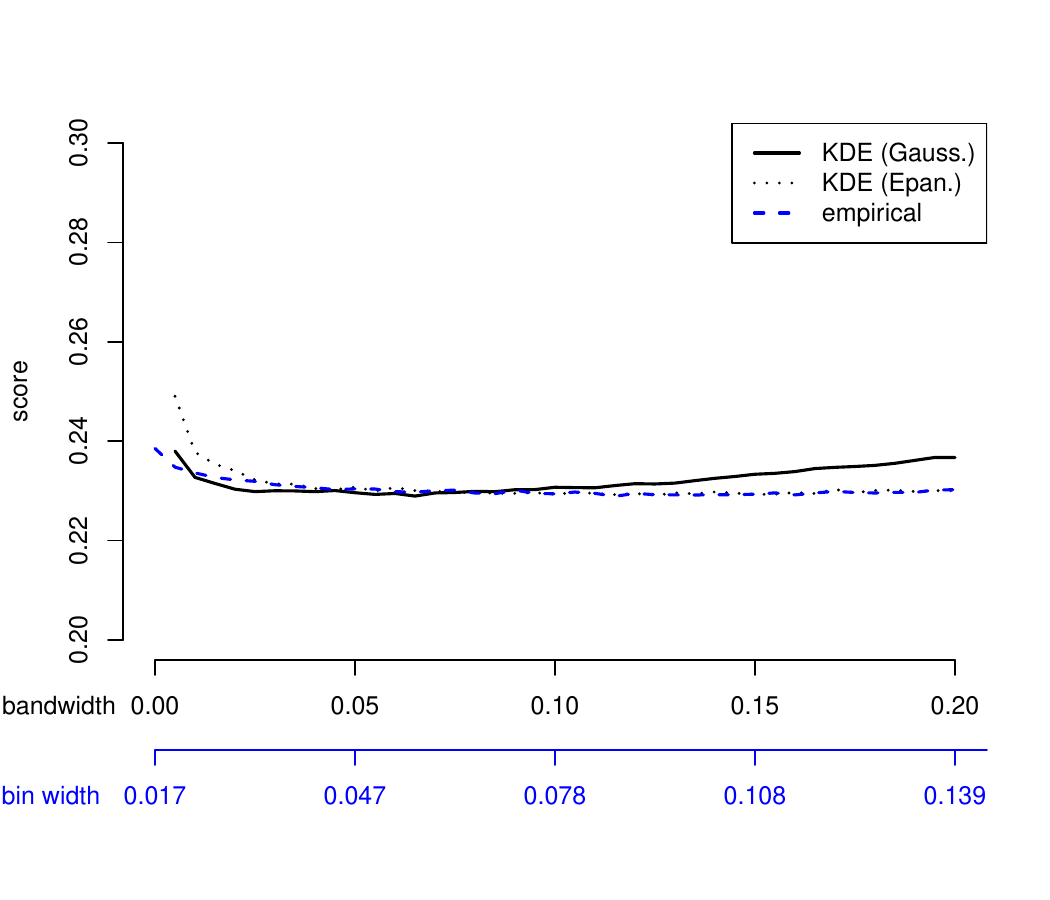}
%        \caption{}
%        \label{fig:d2-quant-scores-log2}
%    \end{subfigure}
%    \begin{subfigure}{0.3\textwidth}
%        \centering
        \includegraphics[width=0.3\textwidth]{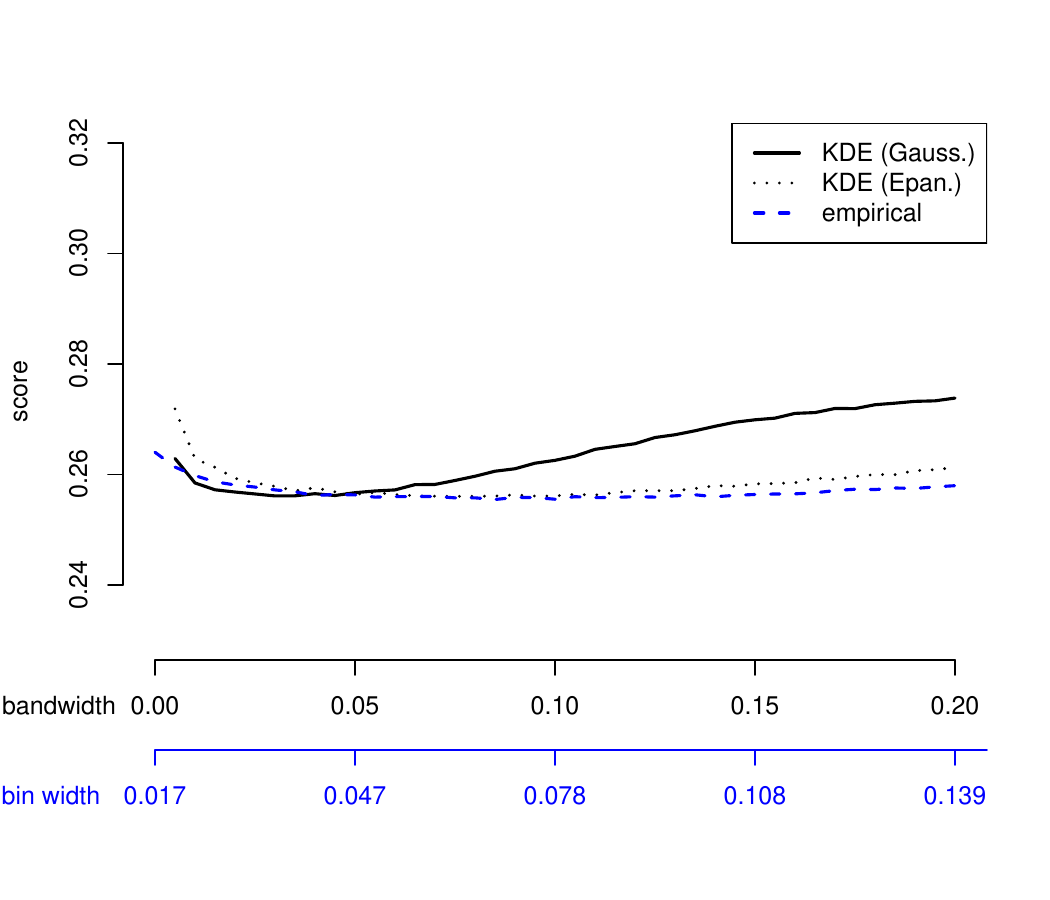}
%        \caption{}
%        \label{fig:d2-quant-scores-gauss}
%    \end{subfigure}
%%    \hspace{1cm}
%    \begin{subfigure}{0.3\textwidth}
%        \centering
        \includegraphics[width=0.3\textwidth]{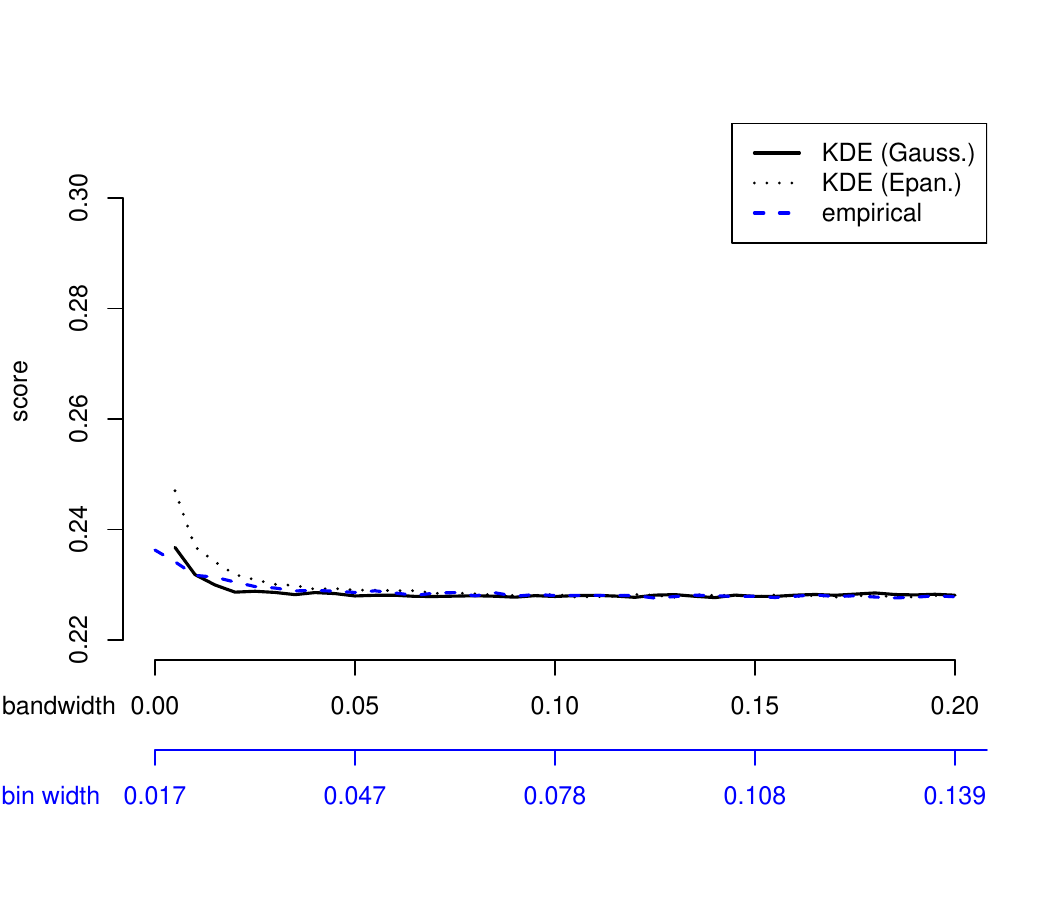}
%        \caption{}
%        \label{fig:d2-quant-scores-invlog}
%    \end{subfigure}
%    \begin{subfigure}{0.3\textwidth}
%        \centering
        \includegraphics[width=0.3\textwidth]{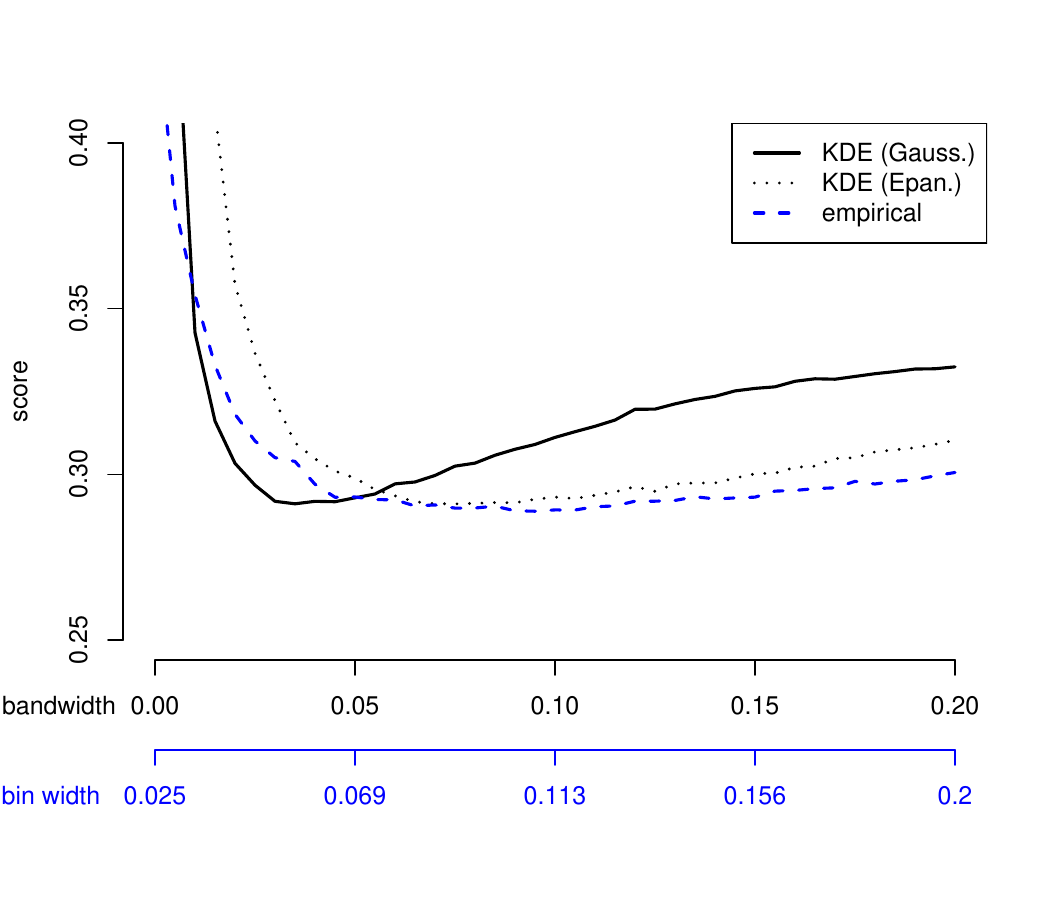}
%        \caption{}
%        \label{fig:d3-quant-scores-alog1}
%    \end{subfigure}
%%    \hspace{1cm}
%    \begin{subfigure}{0.3\textwidth}
%        \centering
        \includegraphics[width=0.3\textwidth]{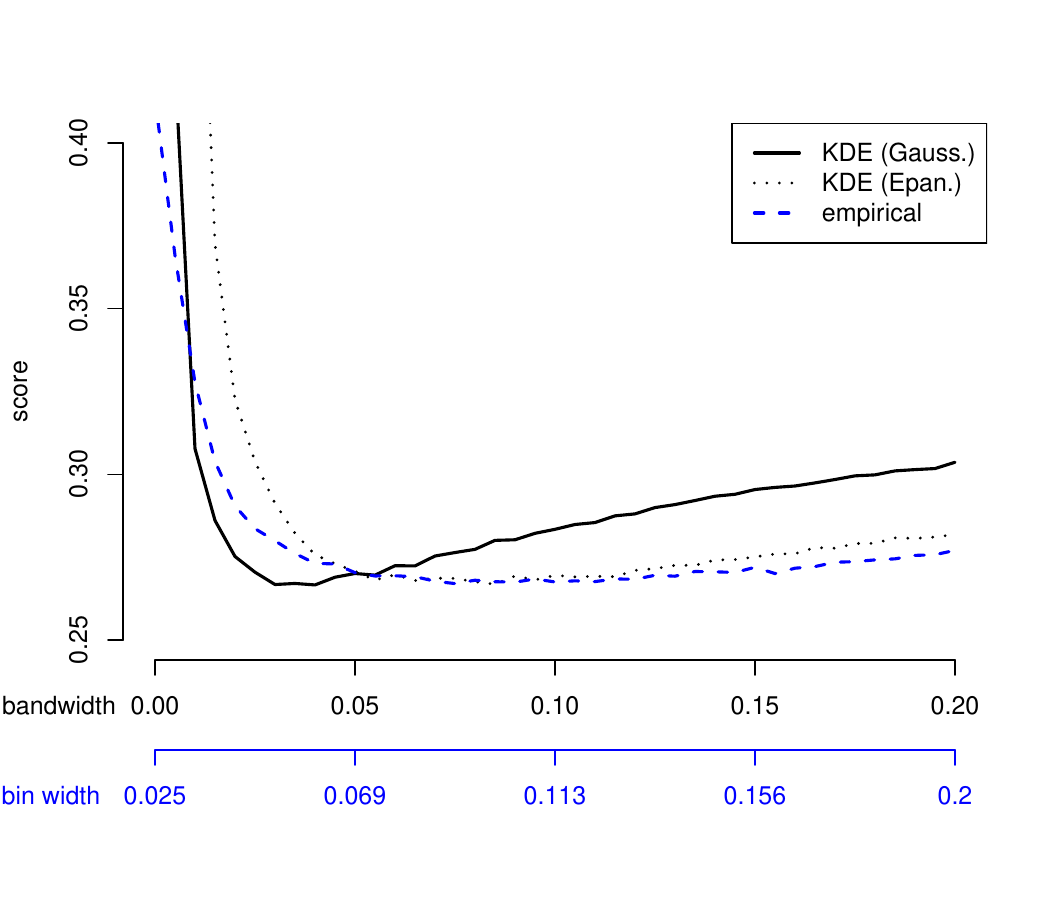}
%        \caption{}
%        \label{fig:d3-quant-scores-alog2}
%    \end{subfigure}
    
%    \begin{subfigure}{0.3\textwidth}
%        \centering
        \includegraphics[width=0.3\textwidth]{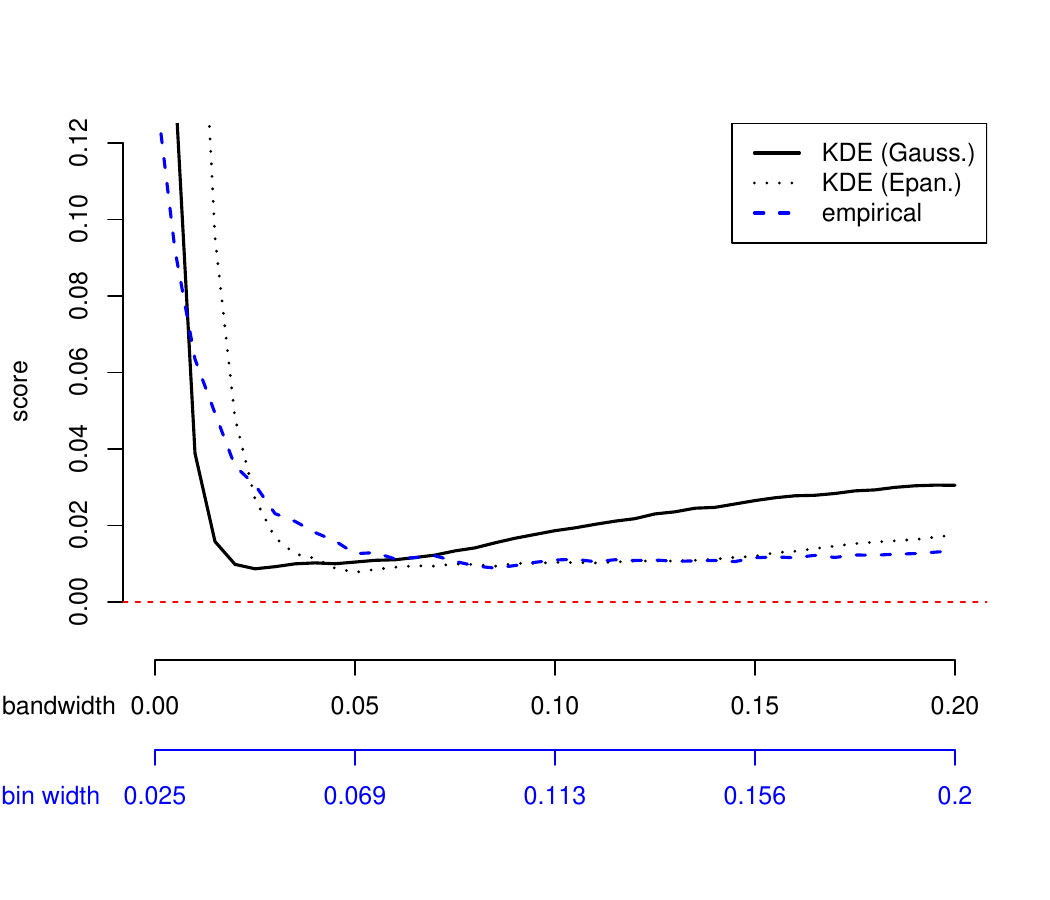}
%        \caption{}
%        \label{fig:d3-quant-scores-mix}
%    \end{subfigure}
	\caption[\protect{Median \rc{threshold} estimation scores at $\tau=0.95$ for data generated from distributions \protect\eqref{distn:log1}--\protect\eqref{distn:gauss}, \protect\eqref{distn:mix}-- \protect\eqref{distn:alog2}, and  \protect\eqref{distn:mix}.}]{Median \rc{threshold} estimation scores at $\tau=0.95$ for data generated from distributions \protect\eqref{distn:log1}--\protect\eqref{distn:gauss} (top row), \protect\eqref{distn:mix}-- \protect\eqref{distn:alog2} (middle row), and  \protect\eqref{distn:mix} (bottom row). \rc{Thresholds} are estimated using KDE with the Gaussian and the Epanechnikov kernels \rc{for both radii and angles with fixed $h_{R}$ and increasing $h_{\bm{W}}$}, and empirically with increasing bin width.}
    \label{fig:quant-scores}
\end{figure}

\begin{figure}[h]
    \centering
%    \begin{subfigure}{0.234\textwidth}
%        \centering
        \includegraphics[width=0.234\textwidth]{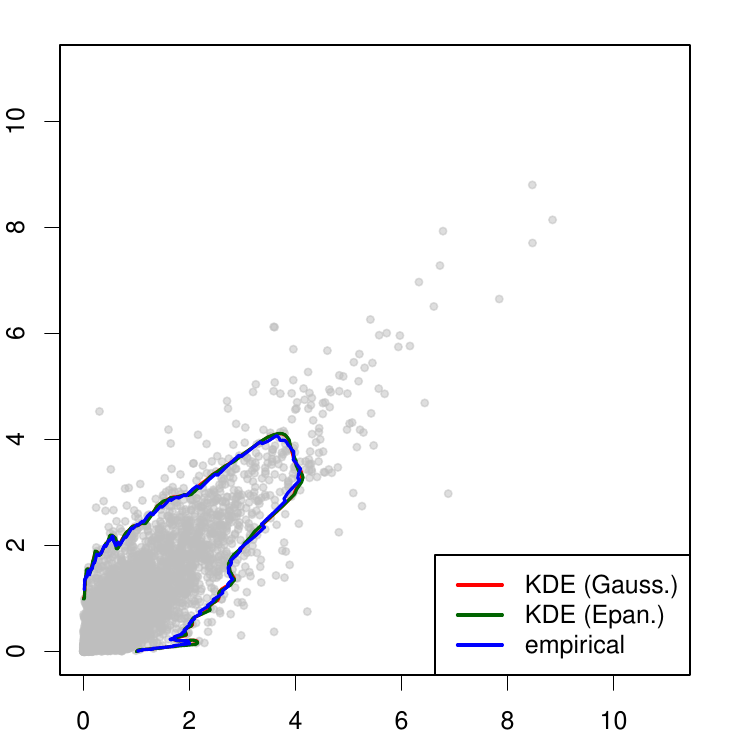}
%        \caption{}
%    \end{subfigure}%
%    ~ 
%    \begin{subfigure}{0.234\textwidth}
%        \centering
        \includegraphics[width=0.234\textwidth]{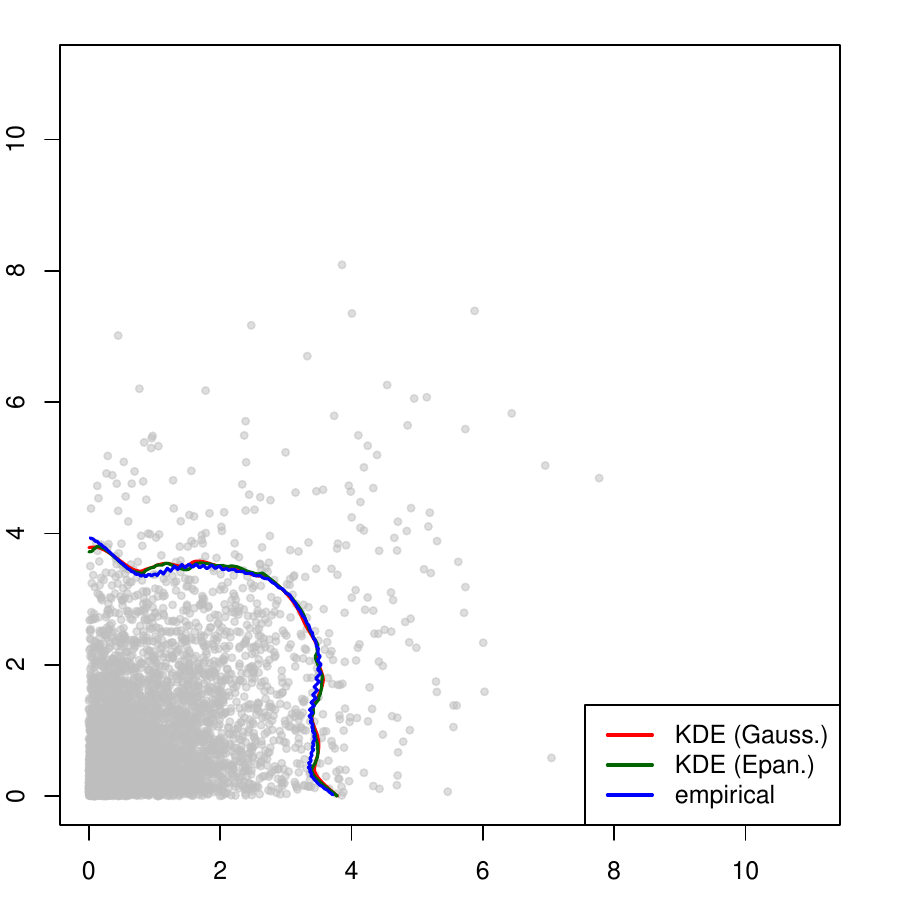}
%        \caption{}
%    \end{subfigure}%
%    ~
%    \begin{subfigure}{0.234\textwidth}
%        \centering
        \includegraphics[width=0.234\textwidth]{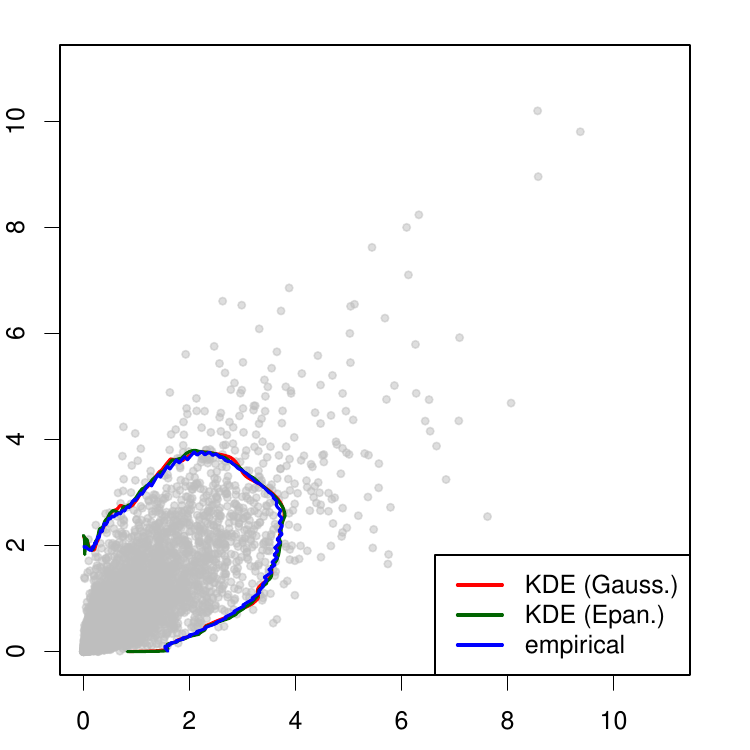}
%        \caption{}
%    \end{subfigure}
%    ~ 
%    \begin{subfigure}{0.234\textwidth}
%        \centering
        \includegraphics[width=0.234\textwidth]{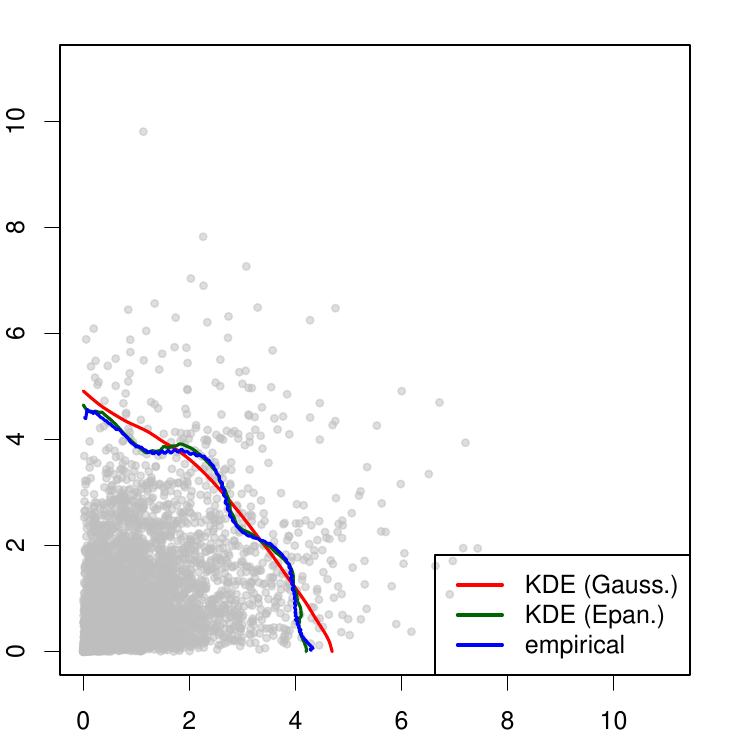}
%        \caption{}
%    \end{subfigure}
	\caption[\protect{Left to right: empirical and KDE-based threshold estimates for $d=2$ on datasets \protect\eqref{distn:log1}--\protect\eqref{distn:invlog} with $\tau=0.95$.}]{Left to right: empirical and KDE-based threshold estimates for $d=2$ on datasets \protect\eqref{distn:log1}--\protect\eqref{distn:invlog} with $\tau=0.95$. The Gaussian and Epanechnikov kernels use bandwidth values $h_{R}=0.05$ and optimal values for $h_W$ and amount of bin overlap governed by the scores in Figure \protect\ref{fig:quant-scores}.}
    \label{fig:QRex-d2}
\end{figure}

\begin{figure}[h!]
    \centering
%    \begin{subfigure}{0.25\textwidth}
%        \centering
        \includegraphics[width=0.25\textwidth]{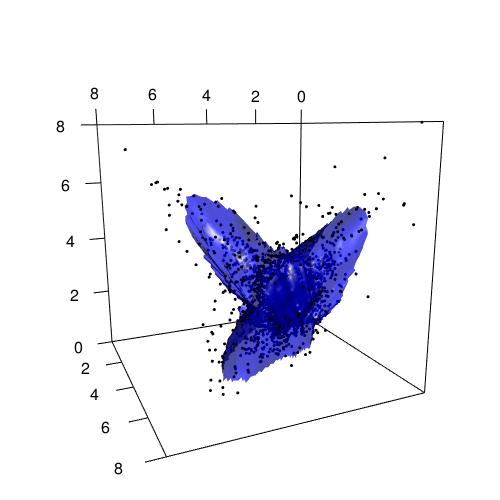}
%        \caption{}
%        \label{fig:d3-best-quant-alog1-emp}
%    \end{subfigure}
%%    \hspace{1cm}
%    \begin{subfigure}{0.25\textwidth}
%        \centering
        \includegraphics[width=0.25\textwidth]{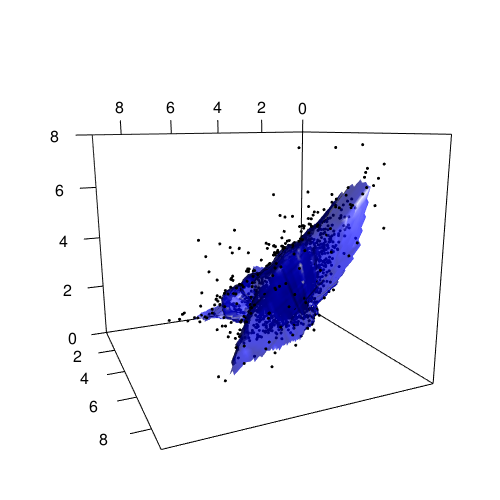}
%        \caption{}
%        \label{fig:d3-best-quant-alog2-emp}
%    \end{subfigure}
%    \begin{subfigure}{0.25\textwidth}
%        \centering
        \includegraphics[width=0.25\textwidth]{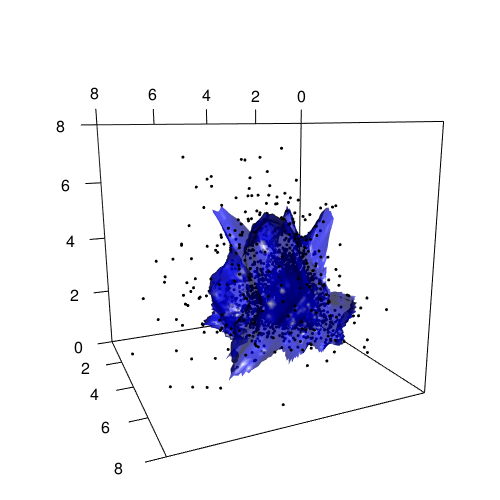}
%        \caption{}
%        \label{fig:d3-best-quant-mix-emp}
%    \end{subfigure}    
    
%    \begin{subfigure}{0.25\textwidth}
%        \centering
        \includegraphics[width=0.25\textwidth]{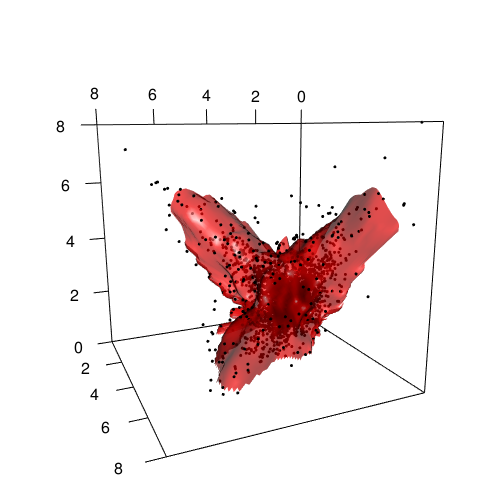}
%        \caption{}
%        \label{fig:d3-best-quant-alog1-KDE}
%    \end{subfigure}
%%    \hspace{1cm}
%    \begin{subfigure}{0.25\textwidth}
%        \centering
        \includegraphics[width=0.25\textwidth]{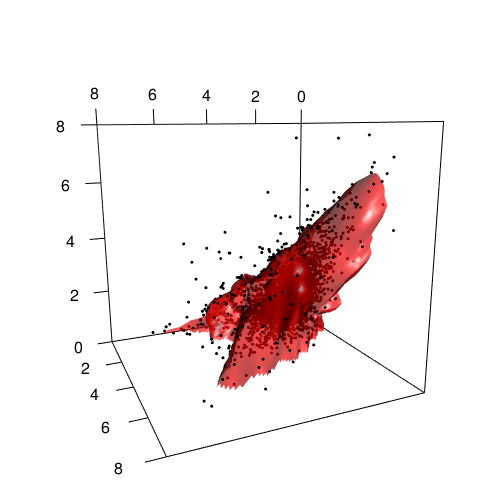}
%        \caption{}
%        \label{fig:d3-best-quant-alog2-KDE}
%    \end{subfigure}
%    \begin{subfigure}{0.25\textwidth}
%        \centering
        \includegraphics[width=0.25\textwidth]{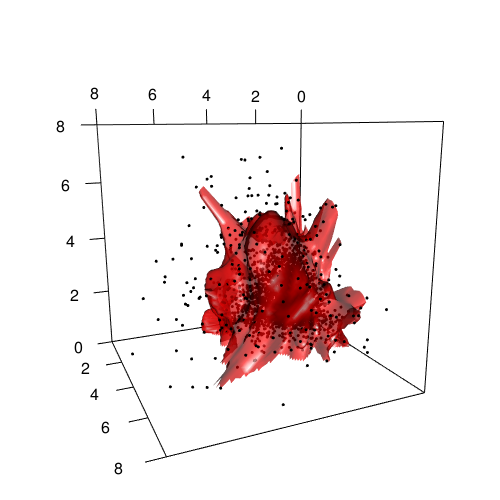}
%        \caption{}
%        \label{fig:d3-best-quant-mix-KDE}
%    \end{subfigure}
    
%    \begin{subfigure}{0.25\textwidth}
%        \centering
        \includegraphics[width=0.25\textwidth]{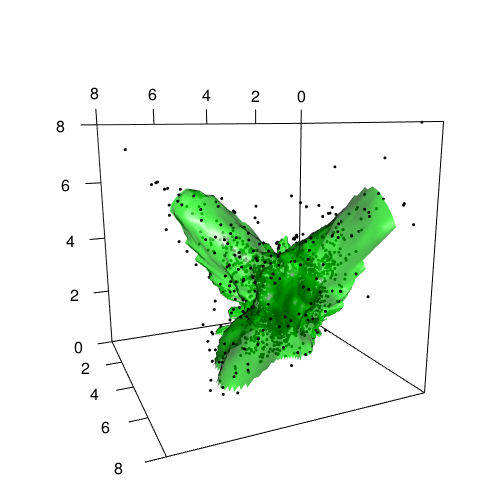}
%        \caption{}
%        \label{fig:d3-best-quant-alog1-KDE}
%    \end{subfigure}
%%    \hspace{1cm}
%    \begin{subfigure}{0.25\textwidth}
%        \centering
        \includegraphics[width=0.25\textwidth]{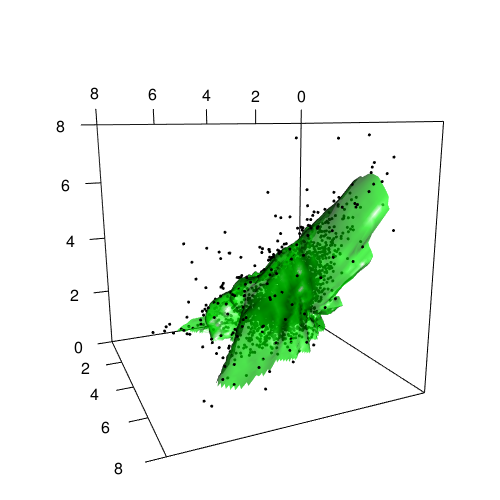}
%        \caption{}
%        \label{fig:d3-best-quant-alog2-KDE}
%    \end{subfigure}
%    \begin{subfigure}{0.25\textwidth}
%        \centering
        \includegraphics[width=0.25\textwidth]{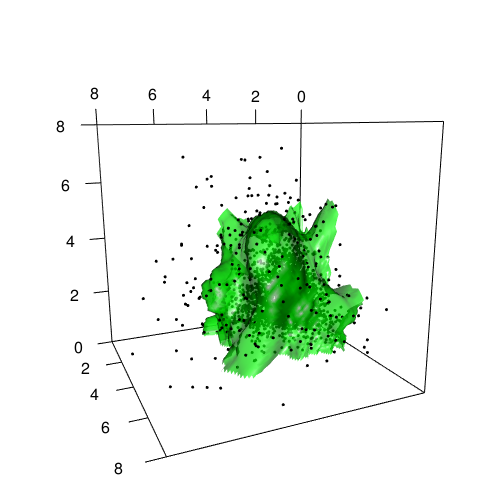}
%        \caption{}
%        \label{fig:d3-best-quant-mix-KDE}
%    \end{subfigure}
	\caption[\protect{Empirical, Gaussian KDE, and Epanechnikov KDE threshold estimates for $d=3$ on datasets \protect\eqref{distn:alog1}--\protect\eqref{distn:mix} (left to right) with $\tau=0.95$.}]{Left to right: empirical (top row), Gaussian KDE (middle row), and Epanechnikov KDE (bottom row) threshold estimates for $d=3$ on datasets \protect\eqref{distn:alog1}--\protect\eqref{distn:mix} (left to right) with $\tau=0.95$. The Gaussian and Epanechnikov kernels use bandwidth values $h_{R}=0.05$ and optimal values for $h_W$ and bin overlap governed by the scores in Figure~\ref{fig:quant-scores}.}
    \label{fig:QRex-d3}
\end{figure}

Using the \rc{threshold} score proposed in Section~\ref{sec:quantiles}  with $K=5$, we assess how well our KDE-based \rc{threshold} estimation performs as the smoothing hyperparameter $h_{\bm{W}}$ increases when using the Gaussian and Epanechnikov kernels.
Figure~\ref{fig:quant-scores} shows these scores for $d=2$ and $d=3$ datasets \eqref{distn:log1}--\eqref{distn:mix} at the $\tau=0.95$, comparing to the empirical binning method of \rc{threshold} estimation as the amount of overlapping increases. 
{
Note that for simplicity there is no boundary correction for either kernel in our KDE methodology. This would be important for actual density estimation, but we are simply searching for approximate high \rc{thresholds} of $R\mid\bm{W}$.
}
%For Gaussian KDE, the optimal values of $h_{\bm{W}}$ based on median scoring for distributions \eqref{distn:log1}--\eqref{distn:mix} were found to be 
%0.02, 0.055, 0.035, 0.14, 0.06, 0.05, 0.055, respectively.
%For Epanechnikov KDE, the optimal values of $h_{\bm{W}}$ for distributions \eqref{distn:log1}--\eqref{distn:mix} were found to be 
%0.06, 0.065, 0.045, 0.175, 0.105, 0.095, 0.135, respectively.
%For the empirical method, the optimal bin overlap amount was for distributions \eqref{distn:log1}--\eqref{distn:mix} were found to be 
%0.095, 0.122, 0.108, 0.20, 0.143, 0.134, 0.161, respectively.
Because of its bounded support, the Epanechnikov kernel was found to be computationally cheaper to evaluate compared to the Gaussian kernel in the $d=2$ case. 
However, \rc{threshold} estimates are not visually better than using the Gaussian kernel, as is shown in Figure~\ref{fig:QRex-d2}, and the Epanechnikov kernel is more computationally expensive to evaluate in $\texttt{R}$ than the Gaussian kernel in the $d\geq3$ setting, as the Gaussian kernel has an efficient multivariate evaluation using functions in the \texttt{mvtnorm} package in \texttt{R}, while the Epanechnikov setting requires taking products of the univariate kernels. For this reason, we continue using the Gaussian kernel exclusively in KDE \rc{threshold} estimation for dimensions $d\geq3$.
Figures~\ref{fig:QRex-d2} and~\ref{fig:QRex-d3} display the \rc{threshold} boundaries $\hat{r}_{\tau}(\bm{w})\bm{w}$ for $\bm{w}\in\mathcal{S}_{d-1}$ and $\tau=0.95$ at the specified adopted smoothing parameters for datasets \eqref{distn:log1}--\eqref{distn:mix}. 
%In the KDE approach, these are $h_{\bm{W}}=0.05$ for $d=2,3$.
\rc{Threshold} score values at optimal levels of $h_{\bm{W}}$ are shown in Table~\ref{tab:quant-scores}.
Results show that the KDE approach results in estimation quality similar to the empirical method in dimensions $d=2,3$. 
%Given the results in Figures \ref{fig:QRex-d2} and \ref{fig:QRex-d3}, we proceed with $h_{\bm{W}}=0.05$ for all new datasets, as this seems to be close to optimal across a range of distributions. 

\begin{table}[t!]
\begin{center}
\begin{tabular}{c|ccccccc}
Dataset & \eqref{distn:log1} & \eqref{distn:log2} & \eqref{distn:gauss} & \eqref{distn:invlog} & \eqref{distn:alog1} & \eqref{distn:alog2} & \eqref{distn:mix} \\ \hline
KDE score       & 0.2329 & 0.2296 & 0.2562 & 0.2282 & 0.2919 & 0.2656 & 0.3119 \\ \hline
empirical score & 0.2324 & 0.2291 & 0.2557 & 0.2280 & 0.2890 & 0.2643 & 0.3125
\end{tabular}
\end{center}
\caption{Median scores of repeated \rc{threshold} estimates for the KDE approach with Gaussian kernel and the empirical binning approach of \cite{wadsworth2024statistical} at their optimal smoothing hyperparameters. \label{tab:quant-scores}
}
\end{table}
%\vspace{-0.7cm}

\begin{figure}[h!]
    \centering
%    \begin{subfigure}{0.3\textwidth}
%        \centering
        \includegraphics[width=0.3\textwidth]{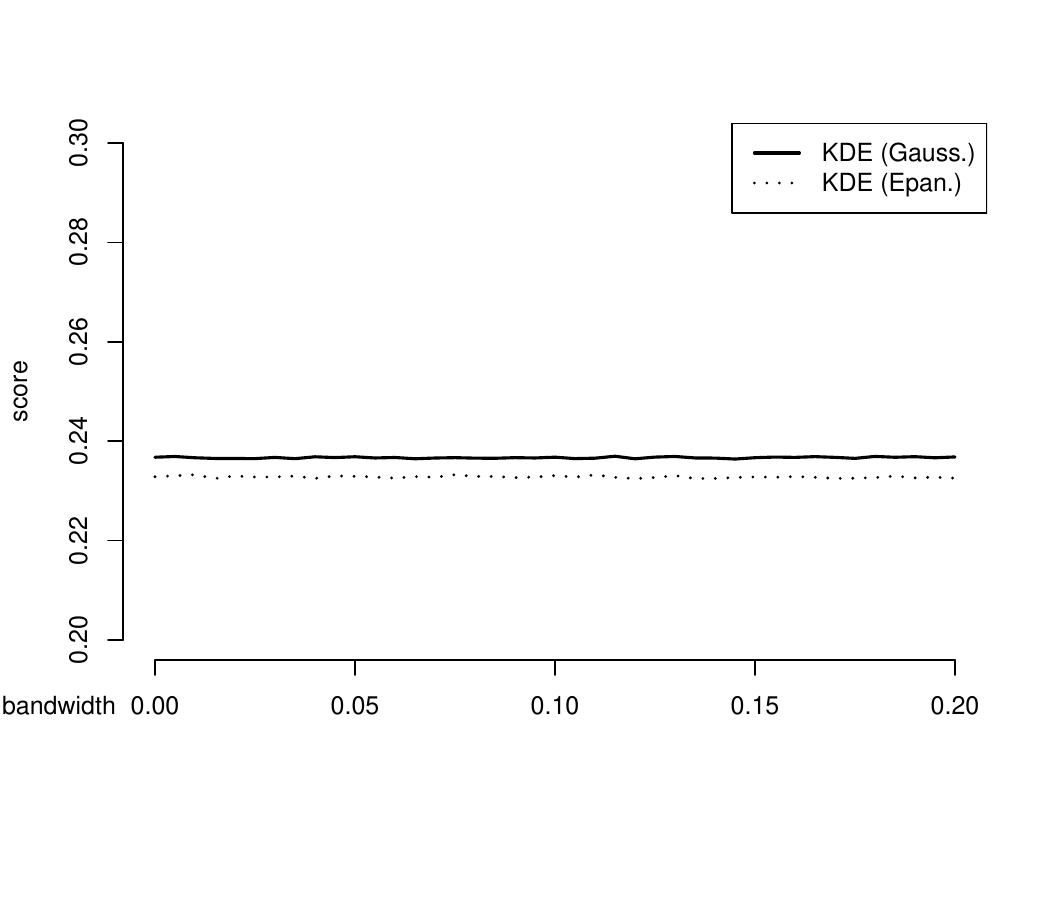}
%        \caption{}
%        \label{fig:d2-quant-scores-log1-hR}
%    \end{subfigure}
%%    \hspace{1cm}
%    \begin{subfigure}{0.3\textwidth}
%        \centering
        \includegraphics[width=0.3\textwidth]{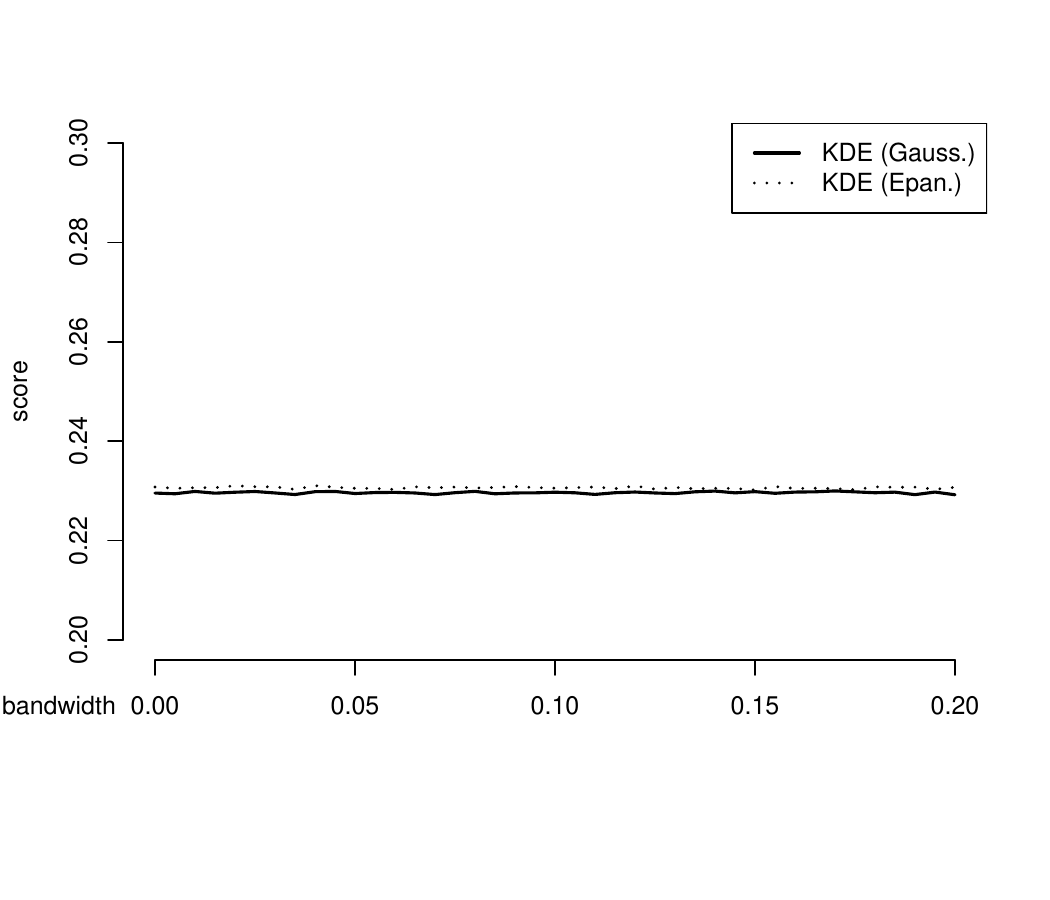}
%        \caption{}
%        \label{fig:d2-quant-scores-log2-hR}
%    \end{subfigure}
%    \begin{subfigure}{0.3\textwidth}
%        \centering
        \includegraphics[width=0.3\textwidth]{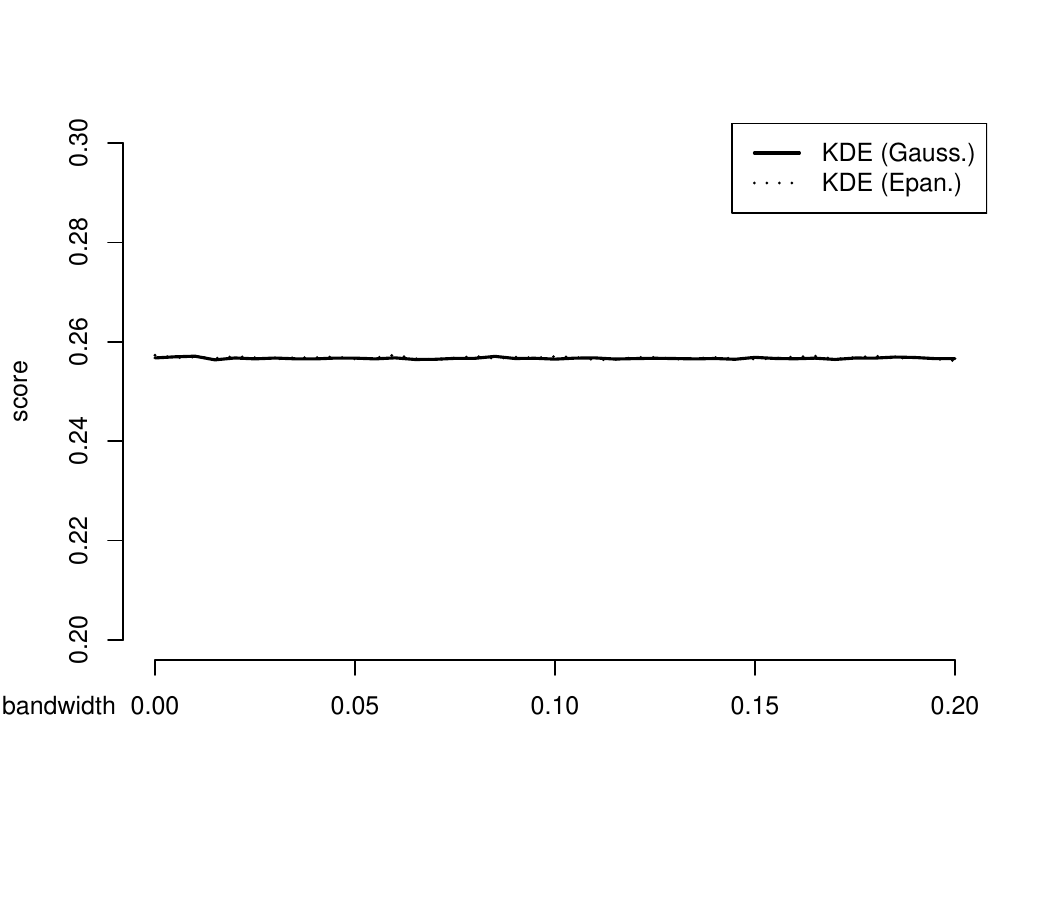}
%        \caption{}
%        \label{fig:d2-quant-scores-gauss-hR}
%    \end{subfigure}
%%    \hspace{1cm}
%    \begin{subfigure}{0.3\textwidth}
%        \centering
        \includegraphics[width=0.3\textwidth]{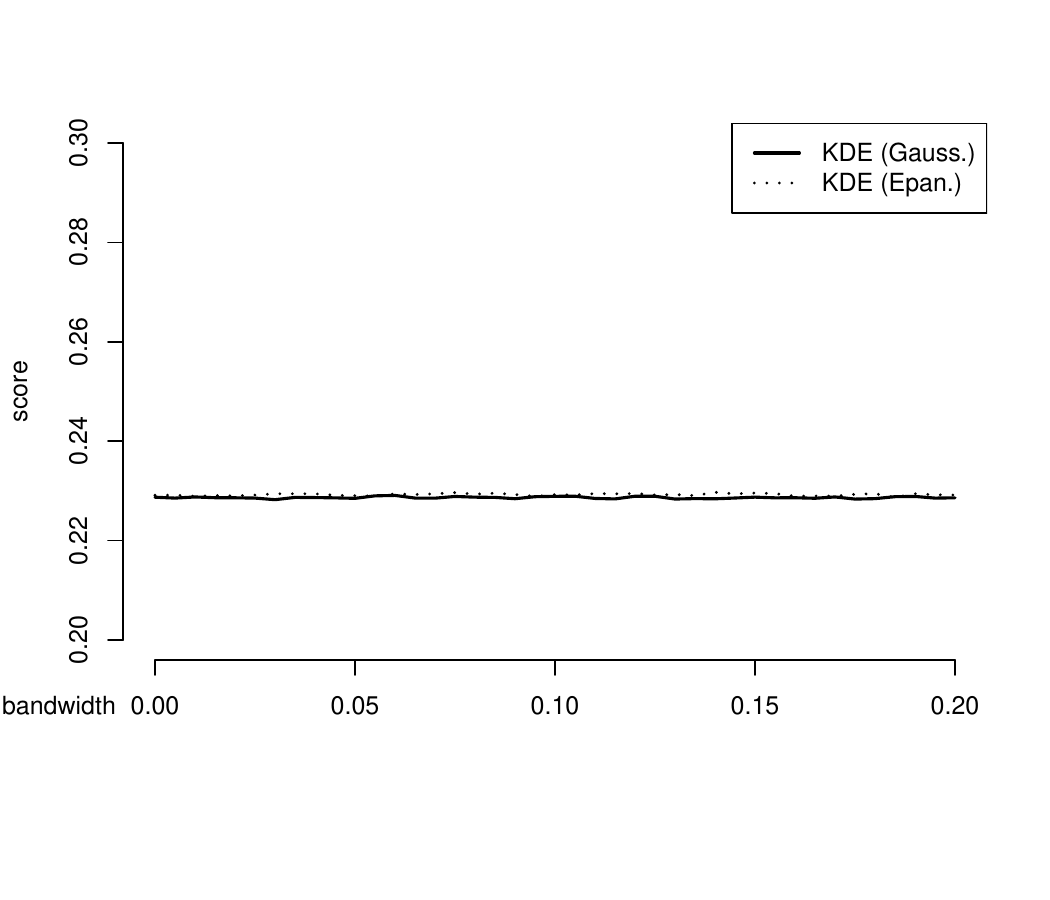}
%        \caption{}
%        \label{fig:d2-quant-scores-invlog-hR}
%    \end{subfigure}
%    \begin{subfigure}{0.3\textwidth}
%        \centering
        \includegraphics[width=0.3\textwidth]{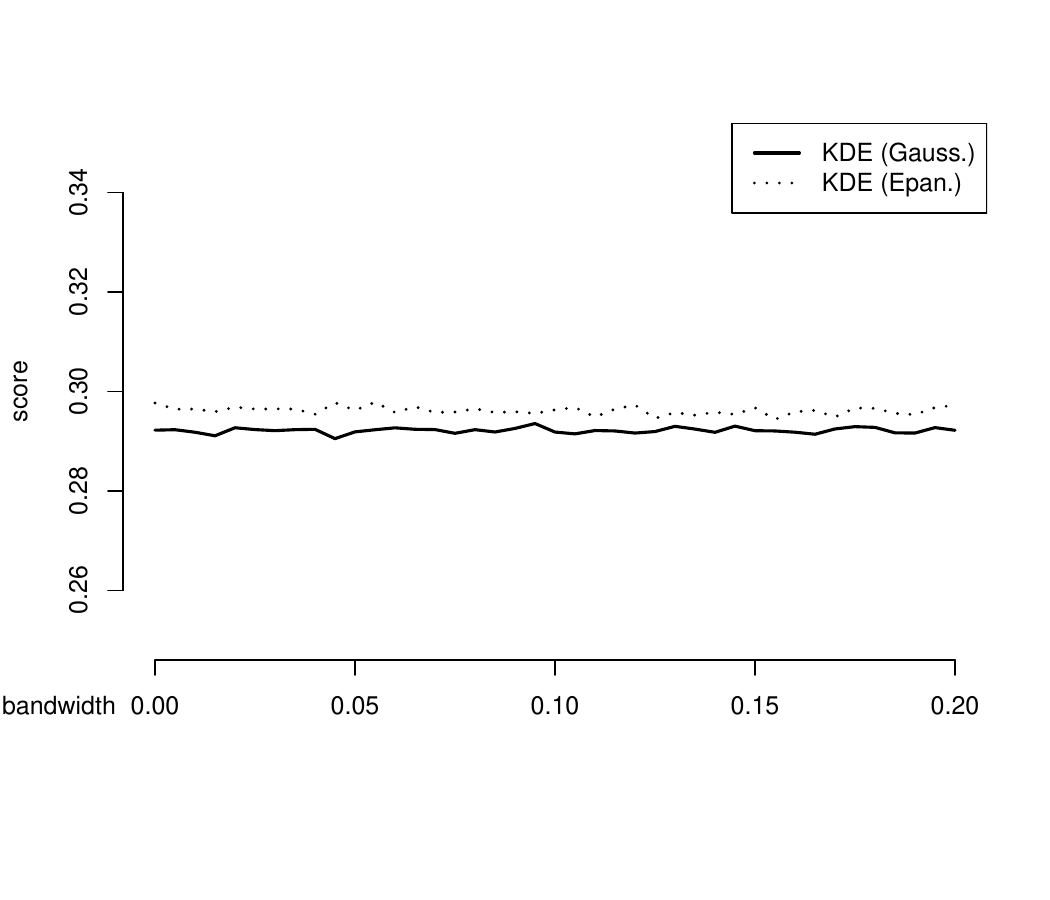}
%        \caption{}
%        \label{fig:d3-quant-scores-alog1-hR}
%    \end{subfigure}
%%    \hspace{1cm}
%    \begin{subfigure}{0.3\textwidth}
%        \centering
        \includegraphics[width=0.3\textwidth]{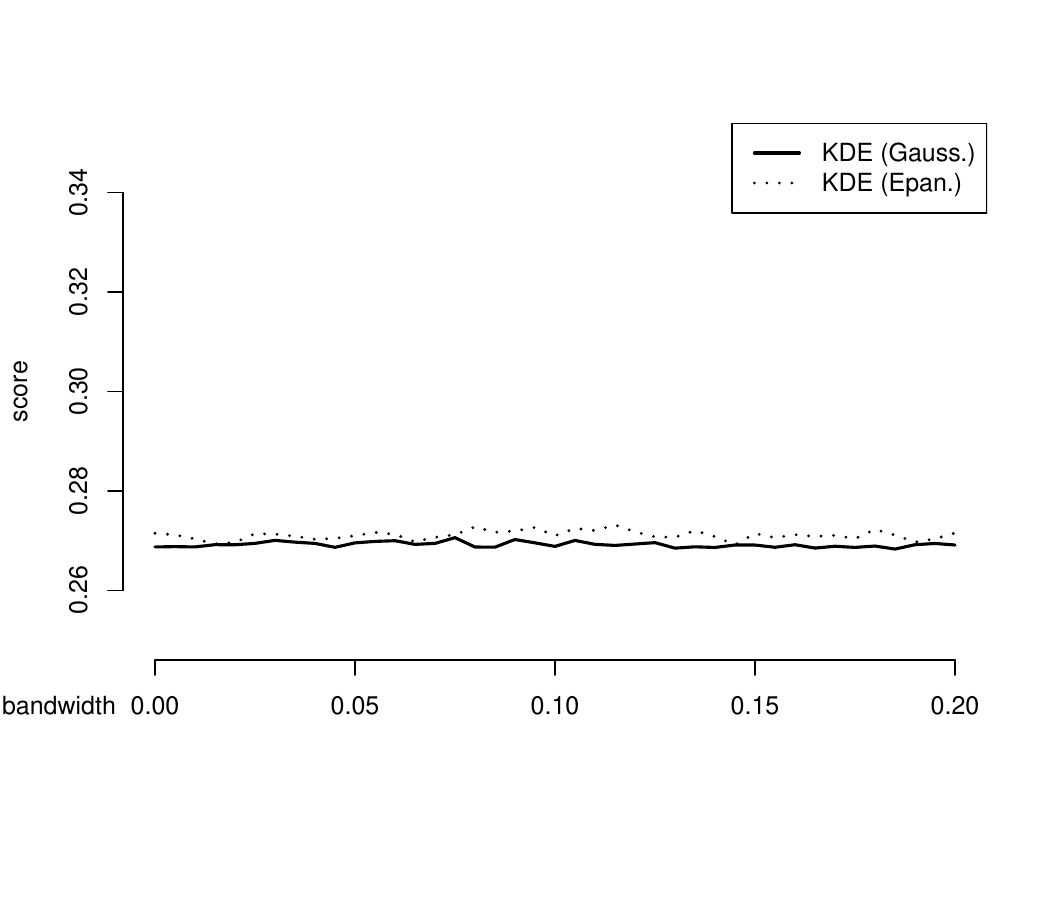}
%        \caption{}
%        \label{fig:d3-quant-scores-alog2-hR}
%    \end{subfigure}
    
%    \begin{subfigure}{0.3\textwidth}
%        \centering
        \includegraphics[width=0.3\textwidth]{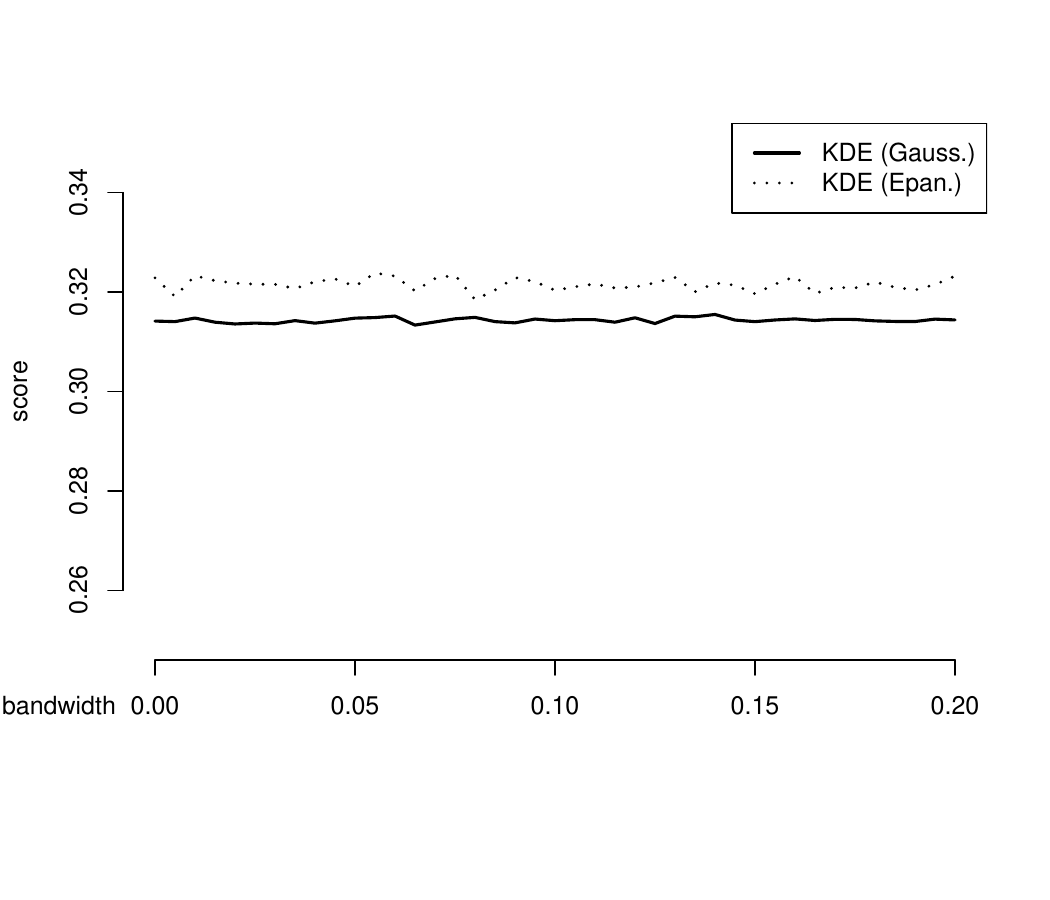}
%        \caption{}
%        \label{fig:d3-quant-scores-mix-hR}
%    \end{subfigure}
	\caption[\protect{Median KDE \rc{threshold} estimation scores at $\tau=0.95$ as $h_R$ varies with fixed $h_{\bm{W}}=0.05$ for data generated from distributions \protect\eqref{distn:log1}--\protect\eqref{distn:gauss}, \protect\eqref{distn:mix}-- \protect\eqref{distn:alog2}, and  \protect\eqref{distn:mix}.}]{Median KDE \rc{threshold} estimation scores at $\tau=0.95$ as $h_R$ varies with fixed $h_{\bm{W}}=0.05$ for data generated from distributions \protect\eqref{distn:log1}--\protect\eqref{distn:gauss} (top row), \protect\eqref{distn:mix}-- \protect\eqref{distn:alog2} (middle row), and  \protect\eqref{distn:mix} (bottom row).
    \rc{Thresholds are estimated using KDE with the Gaussian and the Epanechnikov kernels for both radii and angles with increasing $h_{R}$ and fixed $h_{\bm{W}}$.}
    }
    \label{fig:quant-scores-hR}
\end{figure}

The empirical binning method can lead to empty regions in higher dimensions, leading to the inability to estimate \rc{thresholds} in the entire $\mathcal{S}_{d-1}$ simplex. In the empirical method, given a new angle $\bm{w}\in\mathcal{S}_{d-1}$, a local average of \rc{threshold} values $\hat{r}_{\tau}(\bm{w}_i)$ already estimated are taken for angles $\bm{w}_i$ neighboring $\bm{w}$. Therefore, we are entirely dependent on the radial \rc{threshold} values of the dataset. If there is insufficient data to estimate the radial \rc{threshold} on a given dataset, then one may not be able to estimate radial \rc{thresholds} at new angles. This problem does not arise in the KDE approach, i.e., we can evaluate $\hat{r}_{\tau}(\bm{w})$ for all $\bm{w}\in\mathcal{S}_{d-1}$.
We note that the \rc{thresholds} performance score of distributions \eqref{distn:log2} and \eqref{distn:invlog} are near-independent of the amount of smoothing applied to the \rc{threshold} estimation procedure. 
For these two distributions, $r_{\tau}(\bm{w})$ does not depend strongly on $\bm{w}$ for this value of $\tau$.
This threshold is therefore easy to estimate regardless of the amount of smoothing applied. 
%The findings for these distributions are therefore not indicative of the threshold behavior of general bivariate data.
Different dependence structures have different optimal bandwidths, but $h_{\bm{W}}=0.05$ is close to optimal in all cases. In accompanying code, users can instead allow for automatic selection of $h_{\bm{W}}$ using $K$-fold cross-validation scoring on the check function $S(h_{\bm{W}})$ defined in Section~\ref{sec:quantiles}.
Figure~\ref{fig:quant-scores-hR} demonstrates that, for a fixed value of $h_{\bm{W}}$, varying the radial bandwidth $h_R$ has no effect on the quality of radial \rc{threshold} estimates.

\clearpage

%\subsection{Choosing the radial bandwidth}\label{supp:quants-hR}

\clearpage
\section{Algorithm: gauge function unit level set bounding}\label{supp:bounding-alg}

The following describes the bounded model fitting algorithm, introduced in Section~\ref{sec:lik} and implemented throughout the simulation studies and real-data applications in Sections~\ref{sec:sim} and~\ref{sec:pollution}, respectively.

\begin{algorithm}[h!]
\doublespacing
\caption{Bounding the piecewise-linear limit set during model fitting}\label{alg:bound-fit}
\KwInput{$r_i,\bm{w}_i$, threshold estimates $\hat{r}_{\tau}(\bm{w}_i)$, likelihood $L_{R,\bm{W}}$ or $L_{R\mid\bm{W}}$, reference angle set $\left\{\bm{w}^{\star 1},\dots,\bm{w}^{\star N}\right\}$}
$\widehat{\bm{\theta}} \gets \mathrm{argmax}_{\bm{\theta}}L_{\bullet}(\bm{\theta};r_{1:n},\bm{w}_{1:n})$\;
$\mathcal{F} \gets \emptyset$\;
\While{$\max \left({w^{\star i}_j}/{g(\bm{w}^{\star i};\widehat{\bm{\theta}})}\right)\neq 1 $ $\forall$ $i=1,\dots,N$ and $j=1,\dots,d$}{
  % $\mathcal{F} \gets \mathcal{F}\bigcup\left\{i\in\left\{1,\dots,N\right\} : {w^{\star i}_j}/{g(\bm{w}^{\star i};\widehat{\bm{\theta}})} > {w^{\star \left\{1,\dots,N\right\}\setminus i}_j}/{g(\bm{w}^{\star \left\{1,\dots,N\right\}\setminus i};\widehat{\bm{\theta}})} ,j\in\left\{1,\dots,d\right\}\right\} $\;
  \begin{align*}
      \mathcal{F} \gets &\mathcal{F}\bigcup\left\{i\in\left\{1,\dots,N\right\} : {w^{\star i}_j}/{g(\bm{w}^{\star i};\widehat{\bm{\theta}})} > {w^{\star \left\{1,\dots,N\right\}\setminus i}_j}/{g(\bm{w}^{\star \left\{1,\dots,N\right\}\setminus i};\widehat{\bm{\theta}})} ,\right.\\
      &\qquad\quad\left.j\in\left\{1,\dots,d\right\}\right\} \text{;}
  \end{align*}
  % \vspace{-0.8cm}
  $\widehat{\bm{\theta}}_{\mathcal{F}}\gets{\widehat{\bm{\theta}}_{\mathcal{F}}}/{g(\bm{w}^{\star \mathcal{F}};\widehat{\bm{\theta}})}$\;
  $\widehat{\bm{\theta}}_{-\mathcal{F}} \gets \mathrm{argmax}_{\bm{\theta}_{-\mathcal{F}}}L_{\bullet}(\bm{\theta}_{-\mathcal{F}};r_{1:n},\bm{w}_{1:n},\bm{\theta}_{\mathcal{F}})$
}
\KwReturn{Scaled parameter estimates $\widehat{\bm{\theta}}\in\mathbb{R}^N_+$}
\end{algorithm}
%\spacingset{1}

\newpage

\section{Angular fit examples}\label{supp:angle-MCMC-performance}
\subsection{Gaussian distribution \eqref{distn:gauss}, $d=2$}
\begin{figure*}[h]
    \centering
    \includegraphics[width=0.8\textwidth]{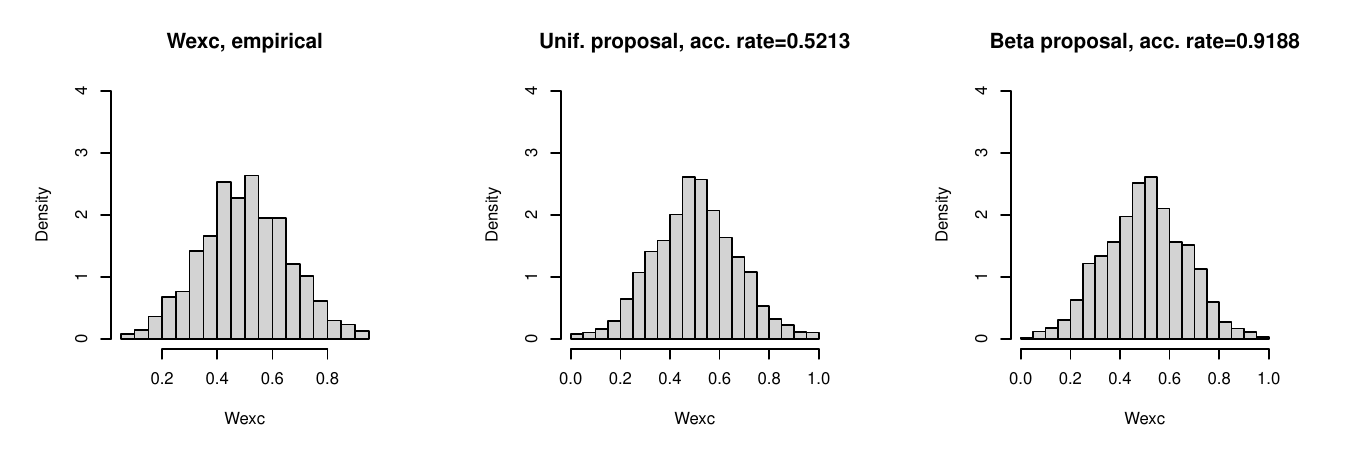}
	\caption{Left: empirical distribution of $W\mid\left\{R>\hat{r}_{\tau}(W)\right\}$; center/right: samples drawn via MCMC on fitted model, using uniform (center) and beta (right) proposals.}
    \label{fig:W-bimodal-d2-MCMC}
\end{figure*}

We model the angles of data generated from a bivariate Gaussian distribution with correlation $\rho=0.8$, using the density $f_{\tilde{\bm{W}}}(\bm{w})={g_{\small{\textsc{pwl}}}(\bm{w})^{-d}}/{\{d\text{vol}(G_{\small{\textsc{pwl}}})\}}$ and $g_{\small{\textsc{pwl}}}$ specified piecewise-linearly using equation \eqref{eq:gauge-full-w}.
Included in Figure~\ref{fig:d2-d3-fW-ex-main} are estimates of the density $f_{\tilde{\bm{W}}}$ for increasing number of parameters $N$. Figure~\ref{fig:W-bimodal-d2-MCMC} shows samples from this fitted density using MCMC with a uniform proposal density and a beta density whose parameters were fitted using the exceedance angles of the dataset. The beta proposal is preferred, as the MCMC acceptance rate is much higher, leading to a more efficient sampling algorithm.

% should separate these...

\clearpage
\subsection{Mixture model \eqref{distn:mix}, $d=3$}

\begin{figure*}[h]
    \centering
%    \begin{subfigure}{0.22\textwidth}
%        \centering
        \includegraphics[width=0.22\textwidth]{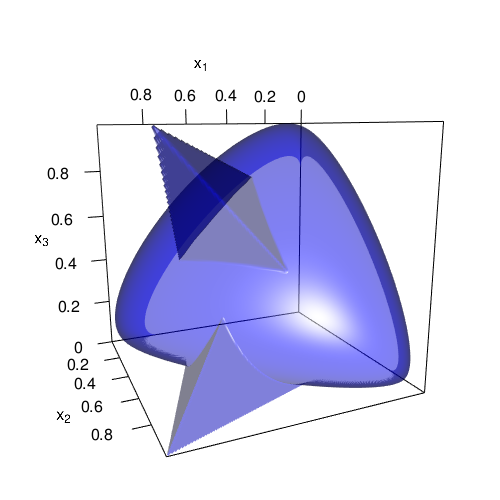}
%        \caption{}
%    \end{subfigure}%
%    ~ 
%    \begin{subfigure}{0.22\textwidth}
%        \centering
        \includegraphics[width=0.22\textwidth]{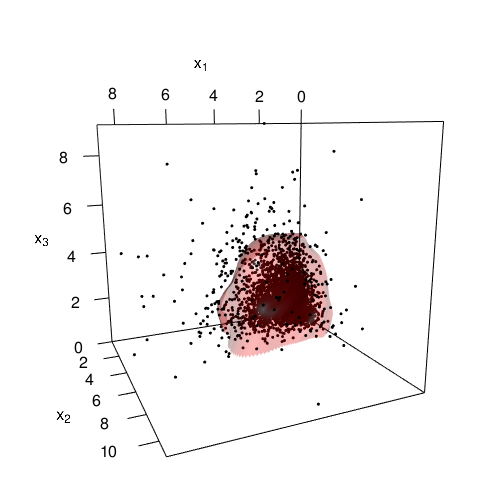}
%        \caption{}
%    \end{subfigure}%
%    ~ 
%    \begin{subfigure}{0.22\textwidth}
%        \centering
        \includegraphics[width=0.22\textwidth]{d3_mix_fWfit.pdf}
%        \caption{}
%    \end{subfigure}
    
%    \begin{subfigure}{0.22\textwidth}
%        \centering
        \includegraphics[width=0.22\textwidth]{d3_mix_Wexc_samp.pdf}
%        \caption{}
%    \end{subfigure}%
%    ~ 
%    \begin{subfigure}{0.22\textwidth}
%        \centering
        \includegraphics[width=0.22\textwidth]{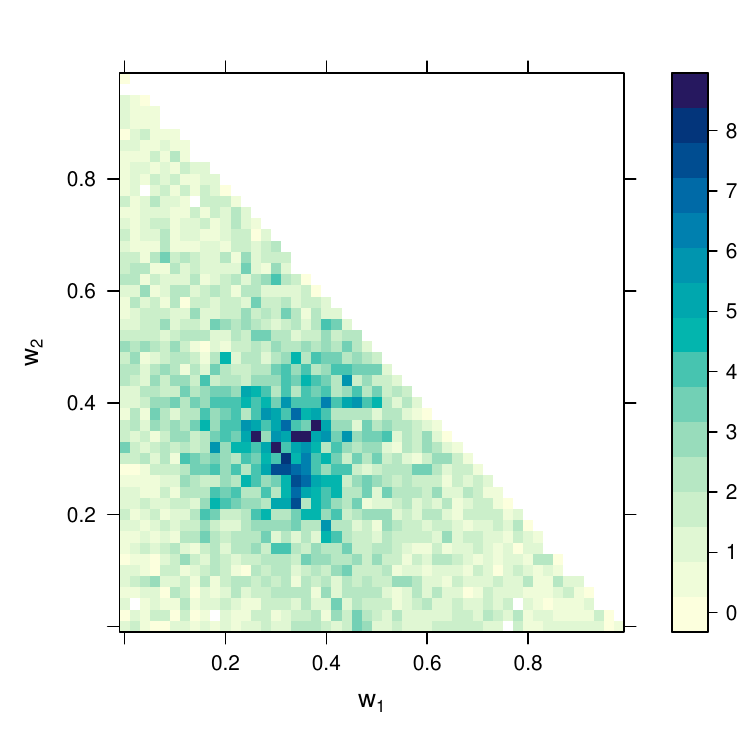}
%        \caption{}
%    \end{subfigure}%
%    ~ 
%    \begin{subfigure}{0.22\textwidth}
%        \centering
        \includegraphics[width=0.22\textwidth]{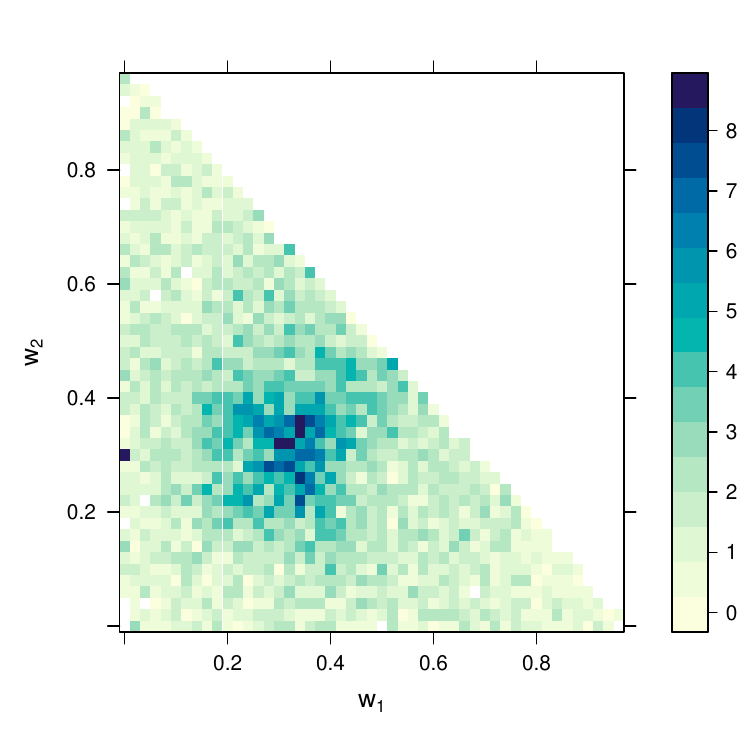}
%        \caption{}
%    \end{subfigure}
	\caption[$d=3$ mixture model example]{$d=3$ mixture model example: Top-left: Unit level set of the gauge function, $g$. Top-middle: Dataset and estimated high \rc{thresholds}. Top-right: Fit of $f_{\tilde{\bm{W}}}(\bm{w})$ with reference angles overlaid. Bottom-left: histogram of $\bm{W}\mid\left\{R>\hat{r}_{\tau}(\bm{W})\right\}$. Bottom-middle and bottom-right: Histogram densities of an MCMC sample of exceedance angles using the fitted $f_{\tilde{\bm{W}}}$ using a uniform and a Dirichlet proposal, respectively.}
    \label{fig:W-mix-d3-ex}
\end{figure*}

\begin{figure*}[h]
    \centering
    \includegraphics[width=0.8\textwidth]{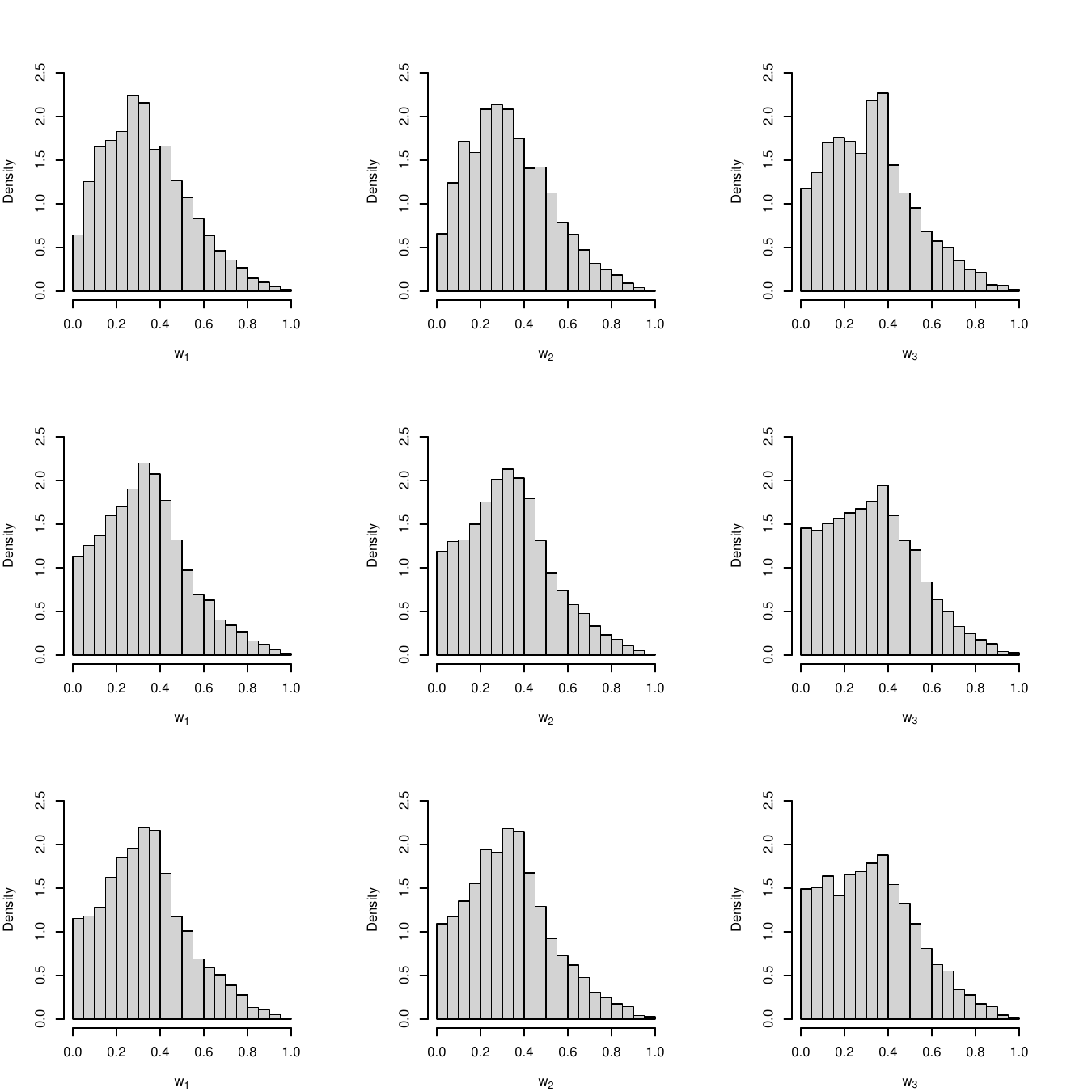}
	\caption[$d=3$ mixture model example.]{$d=3$ mixture model example: Marginal sample of exceedance angles using the empirical distribution of exceedance angles (top row), MCMC on the fitted density $f_{\tilde{\bm{W}}}$ with uniform proposal (middle row) and Dirichlet proposal (bottom row).}
    \label{fig:W-mix-d3-MCMC-marginal}
\end{figure*}

Using data sampled from the mixture distribution \eqref{distn:mix}, we fit the exceedance angle model $f_{\tilde{\bm{W}}}$. Figure~\ref{fig:W-mix-d3-ex} includes the estimated density plotted on the $\mathcal{S}_2$ simplex. Samples obtained using the uniform and a fitted Dirichlet proposal show reasonable agreement with the underlying true angular distribution, as suggested by agreement with the empirical density. Marginal samples as seen in Figure~\ref{fig:W-mix-d3-MCMC-marginal} show that both the uniform and Dirichlet proposals reasonably capture the behavior of the underlying angular distribution.

\clearpage
\newpage
\section{Choosing the gradient penalty strength, $\lambda$}\label{supp:pen-str}
\begin{figure}[h!]
    \centering
%    \begin{subfigure}{0.15\textwidth}
%        \centering
        \includegraphics[width=0.15\textwidth]{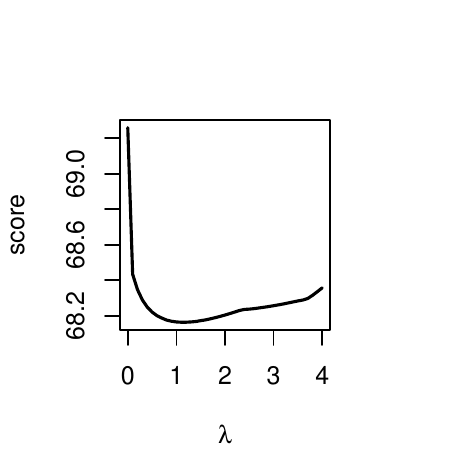}
%    \end{subfigure}
%    \begin{subfigure}{0.15\textwidth}
%        \centering
        \includegraphics[width=0.15\textwidth]{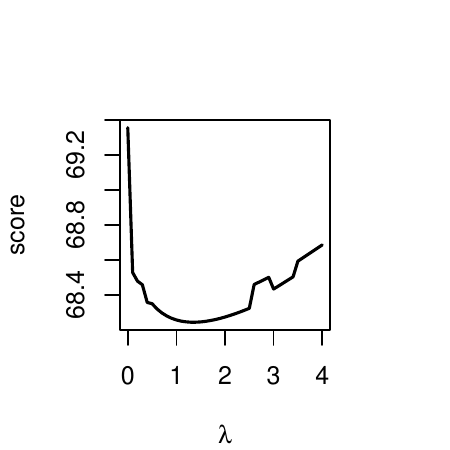}
%    \end{subfigure}
%    \begin{subfigure}{0.15\textwidth}
%        \centering
        \includegraphics[width=0.15\textwidth]{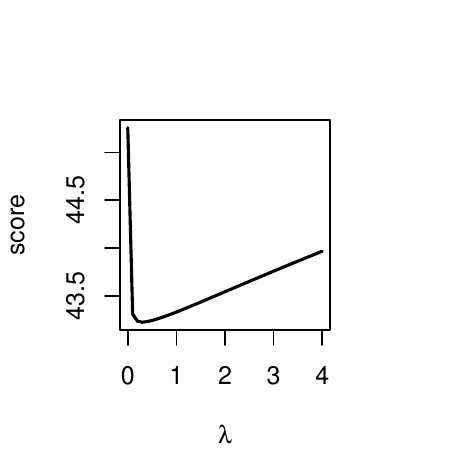}
%    \end{subfigure}
%    \begin{subfigure}{0.15\textwidth}
%        \centering
        \includegraphics[width=0.15\textwidth]{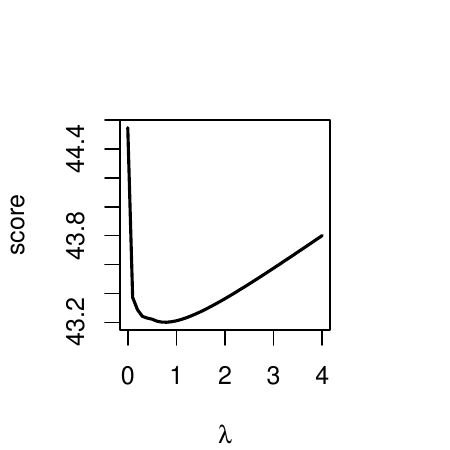}
%    \end{subfigure}
%	\begin{subfigure}{0.15\textwidth}
%        \centering
        \includegraphics[width=0.15\textwidth]{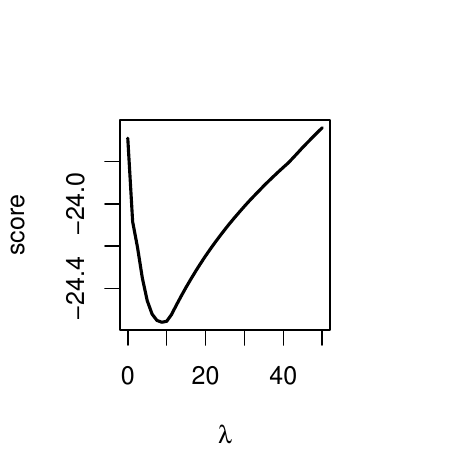}
%    \end{subfigure}
    
%    \begin{subfigure}{0.15\textwidth}
%        \centering
        \includegraphics[width=0.15\textwidth]{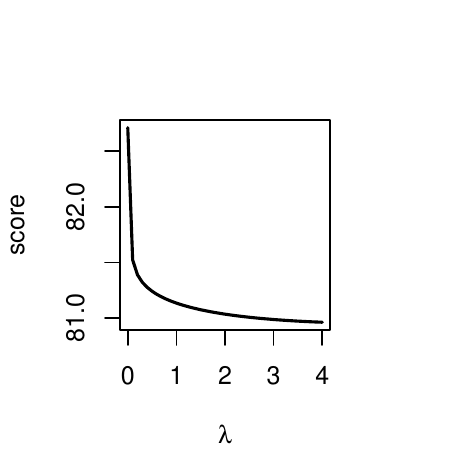}
%    \end{subfigure}
%    \begin{subfigure}{0.15\textwidth}
%        \centering
        \includegraphics[width=0.15\textwidth]{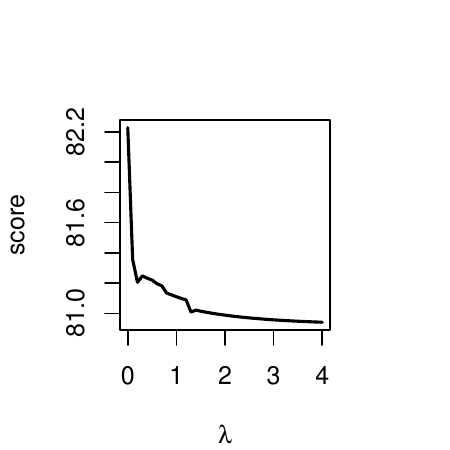}
%    \end{subfigure}
%    \begin{subfigure}{0.15\textwidth}
%        \centering
        \includegraphics[width=0.15\textwidth]{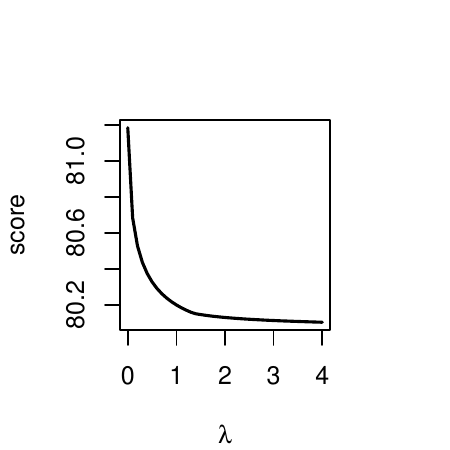}
%    \end{subfigure}
%    \begin{subfigure}{0.15\textwidth}
%        \centering
        \includegraphics[width=0.15\textwidth]{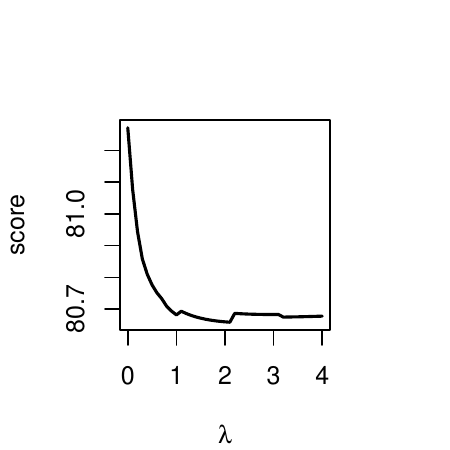}
%    \end{subfigure}
%	\begin{subfigure}{0.15\textwidth}
%        \centering
        \includegraphics[width=0.15\textwidth]{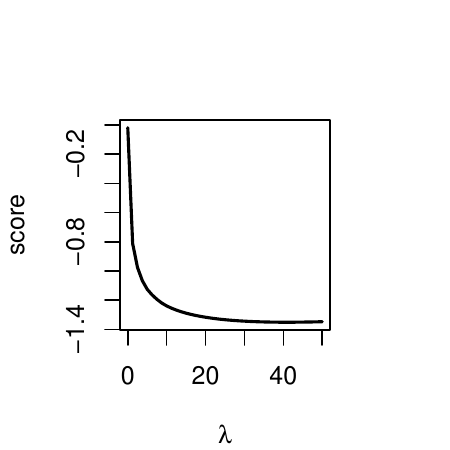}
%    \end{subfigure}

%    \begin{subfigure}{0.15\textwidth}
%        \centering
        \includegraphics[width=0.15\textwidth]{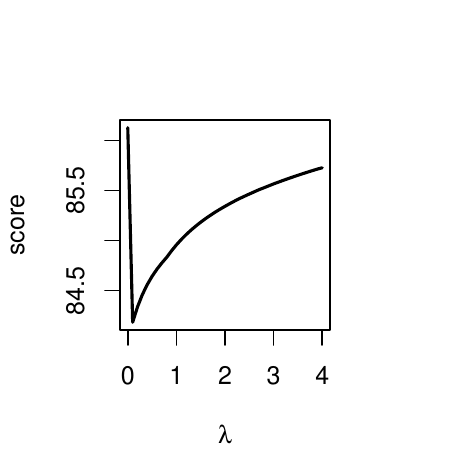}
%    \end{subfigure}
%    \begin{subfigure}{0.15\textwidth}
%        \centering
        \includegraphics[width=0.15\textwidth]{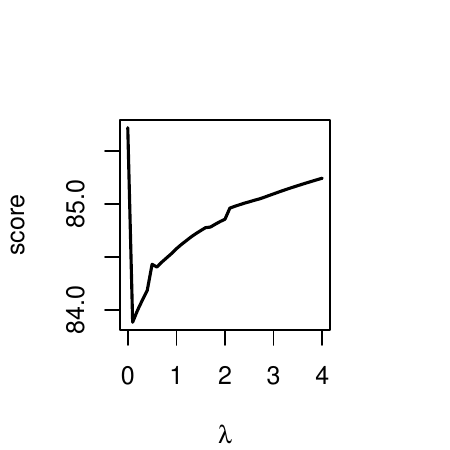}
%    \end{subfigure}
%    \begin{subfigure}{0.15\textwidth}
%        \centering
        \includegraphics[width=0.15\textwidth]{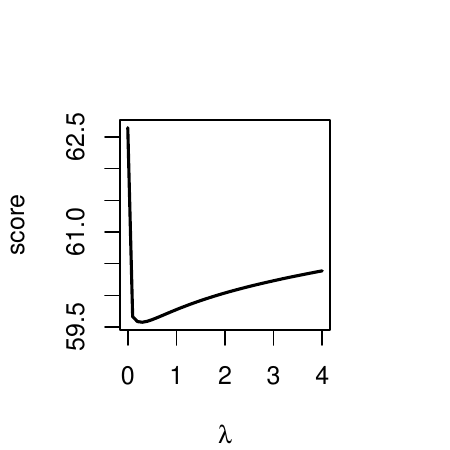}
%    \end{subfigure}
%    \begin{subfigure}{0.15\textwidth}
%        \centering
        \includegraphics[width=0.15\textwidth]{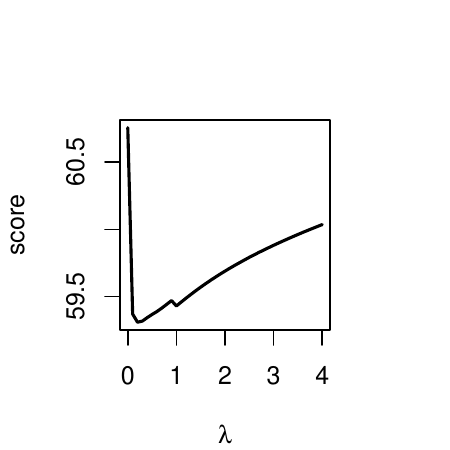}
%    \end{subfigure}
%	\begin{subfigure}{0.15\textwidth}
%        \centering
        \includegraphics[width=0.15\textwidth]{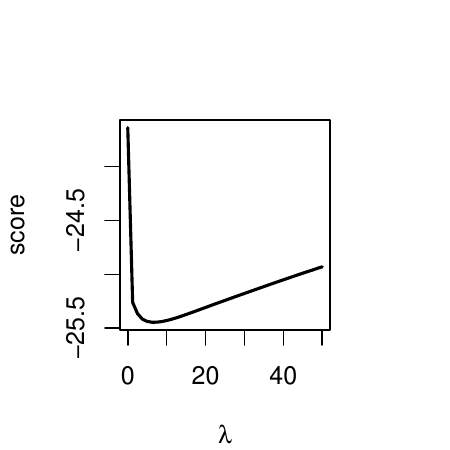}
%    \end{subfigure}
    
%    \begin{subfigure}{0.15\textwidth}
%        \centering
        \includegraphics[width=0.15\textwidth]{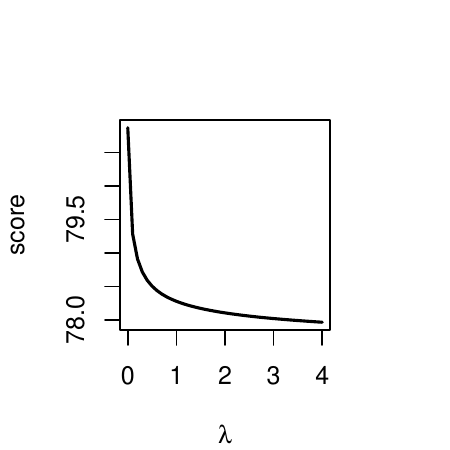}
%    \end{subfigure}
%    \begin{subfigure}{0.15\textwidth}
%        \centering
        \includegraphics[width=0.15\textwidth]{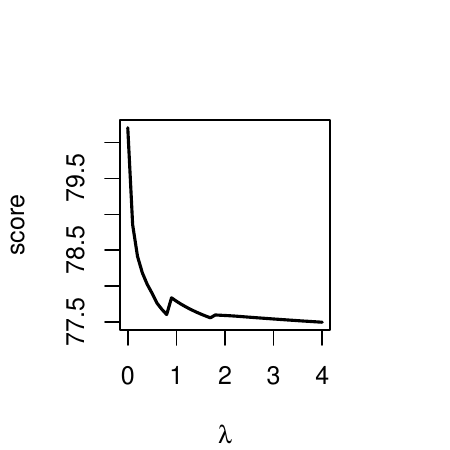}
%    \end{subfigure}
%    \begin{subfigure}{0.15\textwidth}
%        \centering
        \includegraphics[width=0.15\textwidth]{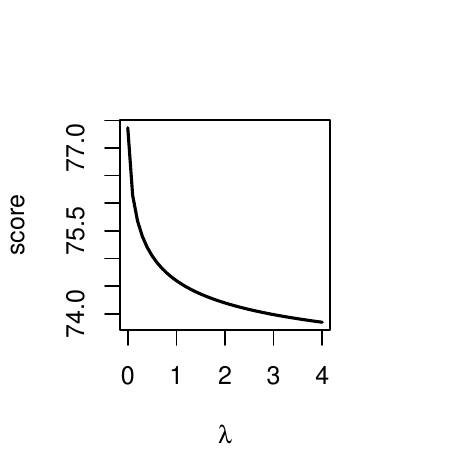}
%    \end{subfigure}
%    \begin{subfigure}{0.15\textwidth}
%        \centering
        \includegraphics[width=0.15\textwidth]{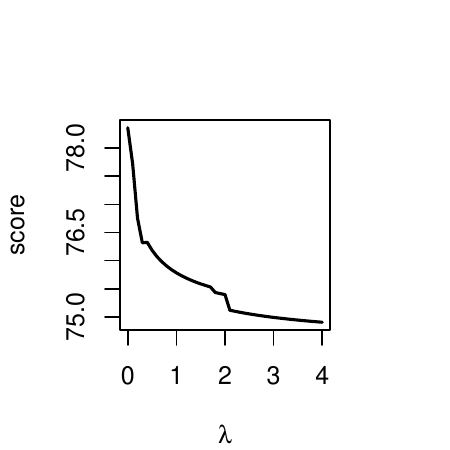}
%    \end{subfigure}
%	\begin{subfigure}{0.15\textwidth}
%        \centering
        \includegraphics[width=0.15\textwidth]{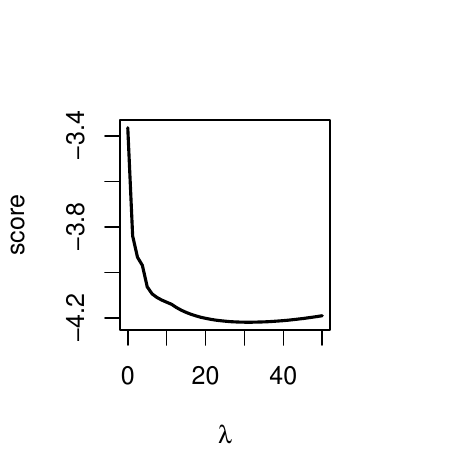}
%    \end{subfigure}
    
%    \begin{subfigure}{0.15\textwidth}
%        \centering
        \includegraphics[width=0.15\textwidth]{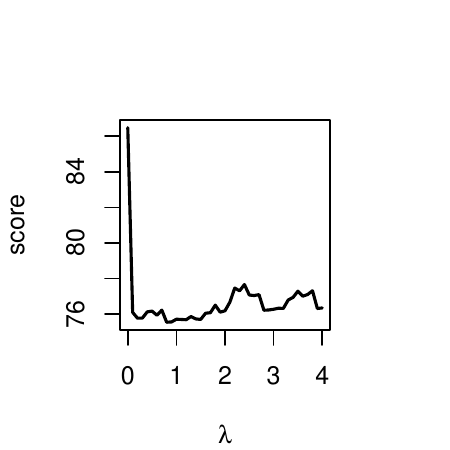}
%    \end{subfigure}
%    \begin{subfigure}{0.15\textwidth}
%        \centering
        \includegraphics[width=0.15\textwidth]{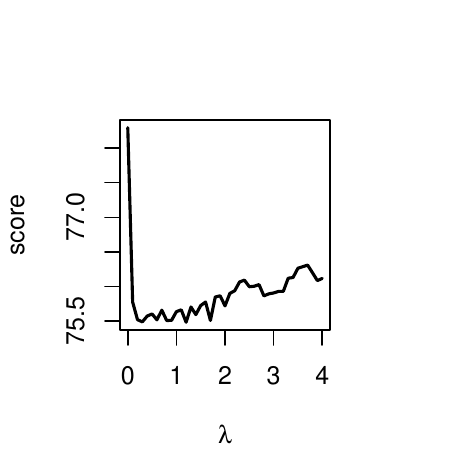}
%    \end{subfigure}
%    \begin{subfigure}{0.15\textwidth}
%        \centering
        \includegraphics[width=0.15\textwidth]{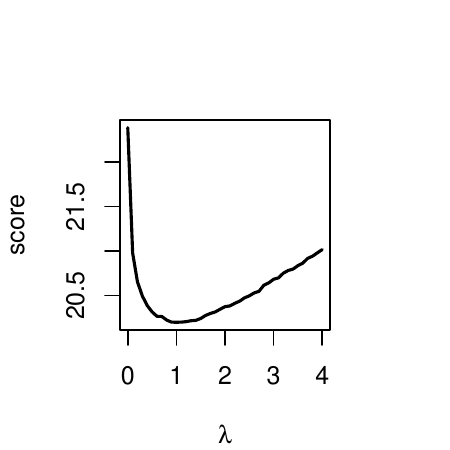}
%    \end{subfigure}
%    \begin{subfigure}{0.15\textwidth}
%        \centering
        \includegraphics[width=0.15\textwidth]{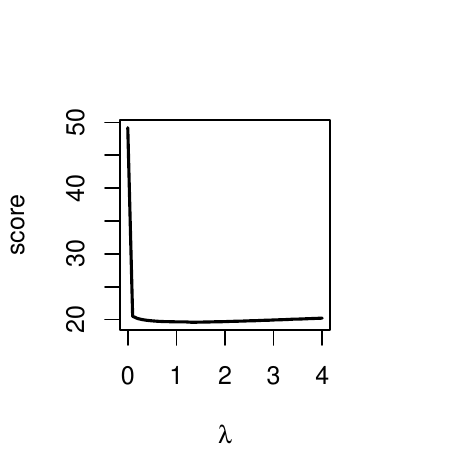}
%    \end{subfigure}
%	\begin{subfigure}{0.15\textwidth}
%        \centering
        \includegraphics[width=0.15\textwidth]{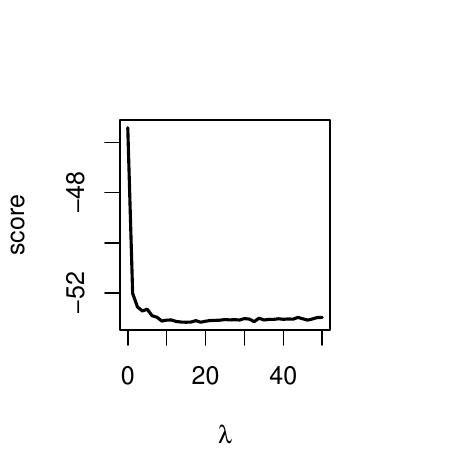}
%    \end{subfigure}
    
%    \begin{subfigure}{0.15\textwidth}
%        \centering
        \includegraphics[width=0.15\textwidth]{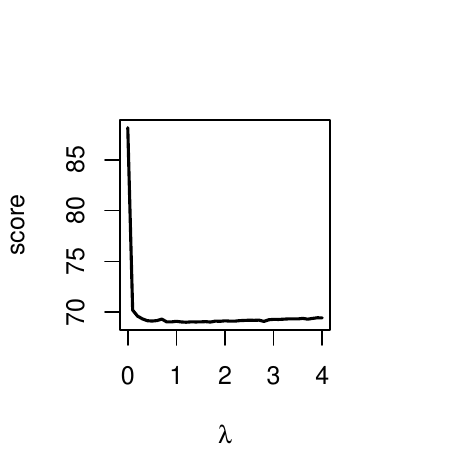}
%    \end{subfigure}
%    \begin{subfigure}{0.15\textwidth}
%        \centering
        \includegraphics[width=0.15\textwidth]{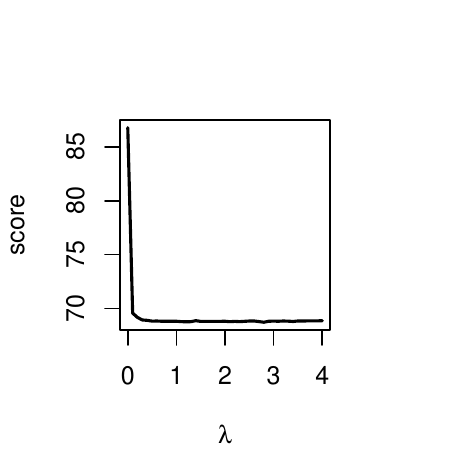}
%    \end{subfigure}
%    \begin{subfigure}{0.15\textwidth}
%        \centering
        \includegraphics[width=0.15\textwidth]{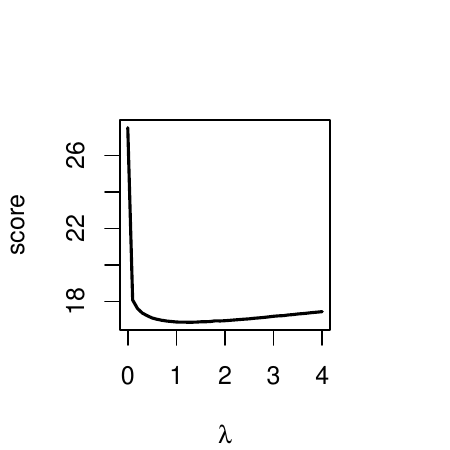}
%    \end{subfigure}
%    \begin{subfigure}{0.15\textwidth}
%        \centering
        \includegraphics[width=0.15\textwidth]{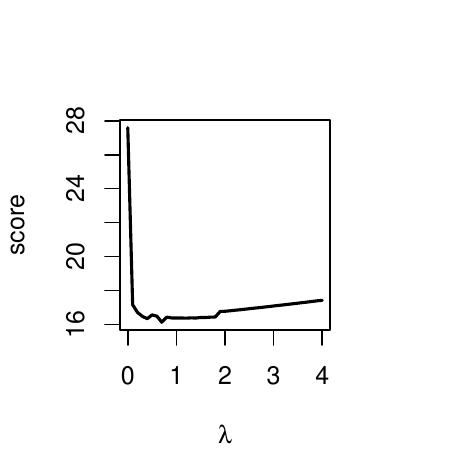}
%    \end{subfigure}
%	\begin{subfigure}{0.15\textwidth}
%        \centering
        \includegraphics[width=0.15\textwidth]{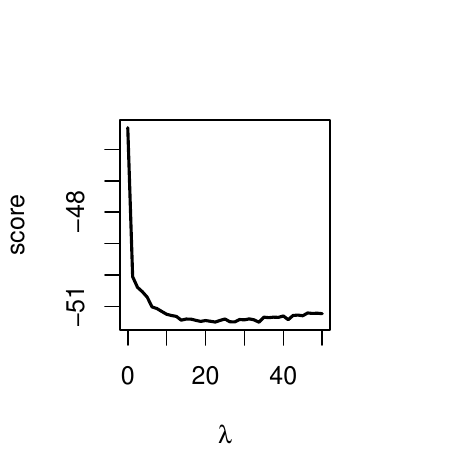}
%    \end{subfigure}
    
%    \begin{subfigure}{0.15\textwidth}
%        \centering
        \includegraphics[width=0.15\textwidth]{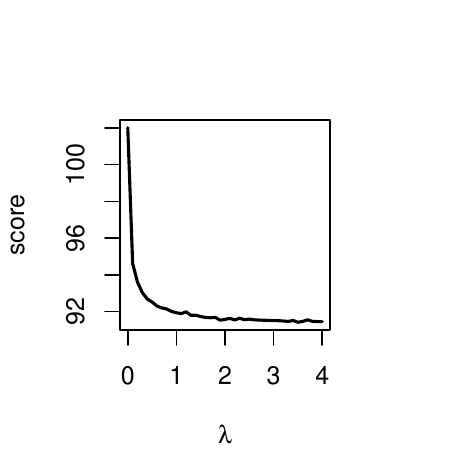}
%    \end{subfigure}
%    \begin{subfigure}{0.15\textwidth}
%        \centering
        \includegraphics[width=0.15\textwidth]{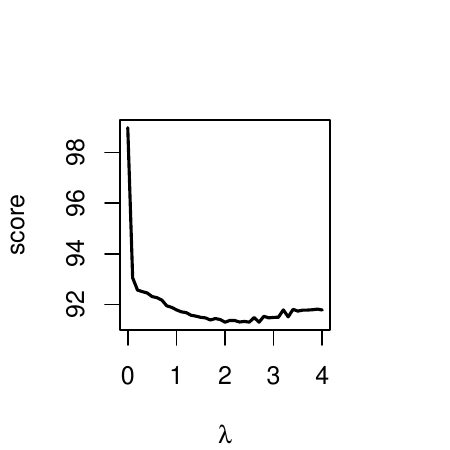}
%    \end{subfigure}
%    \begin{subfigure}{0.15\textwidth}
%        \centering
        \includegraphics[width=0.15\textwidth]{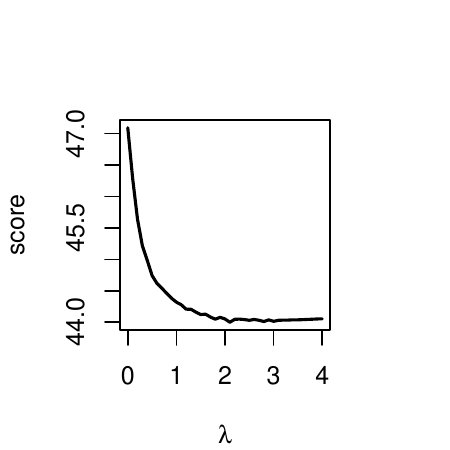}
%    \end{subfigure}
%    \begin{subfigure}{0.15\textwidth}
%        \centering
        \includegraphics[width=0.15\textwidth]{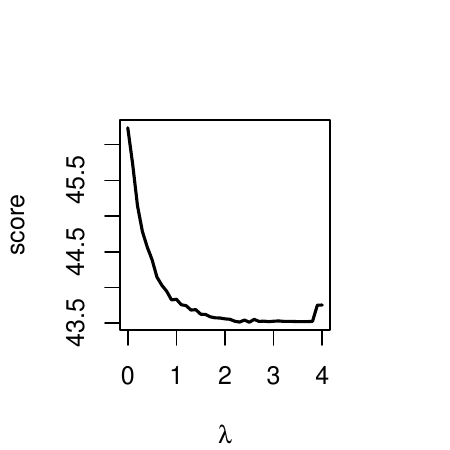}
%    \end{subfigure}
%	\begin{subfigure}{0.15\textwidth}
%        \centering
        \includegraphics[width=0.15\textwidth]{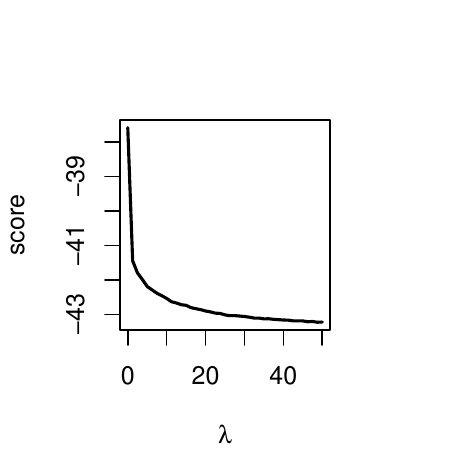}
%    \end{subfigure}
    
    \caption{Cross-validation scores for models (left to right, see text) fitted using data generated from distributions \protect\eqref{distn:log1}--\protect\eqref{distn:mix} (top to bottom) for varying $\lambda$.}\label{fig:pen-score-plots}.
\end{figure}

In Section~\ref{sec:ref-angle-choice}, a gradient penalty is introduced. To select the best penalty strength $\lambda\geq0$, a $K$-fold cross-validation score on negative log-likelihood (NLL) is implemented. This score is based on negative log-likelihood values at parameters obtained from a fitting set and evaluated on a hold-out set, and is explicitly given by
\begin{equation*}
	\text{CV}(\lambda) = \frac{1}{K}\sum\limits_{k=1}^K -\log L_{\bullet}\left(\widehat{\bm{\theta}}_{\lambda,k};r_{1:n_k},\bm{w}_{1:n_k}\right)\:,
\end{equation*}
where $-\log L_{\bullet}$ is the negative log-likelihood associated with one of the three likelihood functions introduced in Section~\ref{sec:lik}, $\widehat{\bm{\theta}}_{\lambda,k}$ is the vector of parameter values obtained from the $k^{\text{th}}$ fitting set of datapoints with gradient penalty with strength $\lambda\geq0$, and $(r_{1:n_k},\bm{w}_{1:n_k})$ are held-out evaluation radii and angle data.
Figure~\ref{fig:pen-score-plots} shows these scores with $K=4$ and $n=5000$ on a mesh of $\lambda$ values. From top to bottom, the score is a median value obtained across 20 datasets from distributions \eqref{distn:log1}--\eqref{distn:mix}. Columns 1--2 correspond to fitting the radial model conditioned on angles by maximizing $L_{R\mid\bm{W}}$ unbounded and bounded using Algorithm~\ref{alg:bound-fit} in Supplement~\ref{supp:bounding-alg}, respectively. Columns 3--4 correspond to jointly fitting the radial-angular model by maximizing $L_{R,\bm{W}}$ unbounded and bounded using Algorithm~\ref{alg:bound-fit} in Supplement~\ref{supp:bounding-alg}, respectively. Column 5 corresponds to fitting the angular model by maximizing $L_{\bm{W}}$. 

	As expected, the optimal degree of smoothing $\lambda$ depends on the underlying dependence structure of the data. In bivariate data (rows 1--4 in Figure~\ref{fig:pen-score-plots}), Gaussian data from distribution \eqref{distn:gauss} requires the least smoothing as the true limit set is curved and therefore changes in the gradient of each segment are desirable. For distributions such as \eqref{distn:log2} and \eqref{distn:invlog} (rows 2 and 4 in Figure~\ref{fig:pen-score-plots}), a flatter limit set is desired, meaning a higher $\lambda$ is preferred. To account for this change in optimal smoothing hyperparameter, the code in our GitHub repository allows the user to not specify $\lambda$, and a $K$-fold cross-validation scoring procedure on $-\log L_{\bullet}$ is performed instead. In three-dimensions (rows 5--7 in Figure~\ref{fig:pen-score-plots}), a good middle-ground would be $\lambda=1$ for the radial and joint models, while a $\lambda$ value of around 20 seems appropriate when fitting the angular model on its own.

\clearpage
\newpage
\section{Simulation studies}\label{supp:sim}

\begin{figure}[h!]
    \centering
     \includegraphics[width=0.35\textwidth]{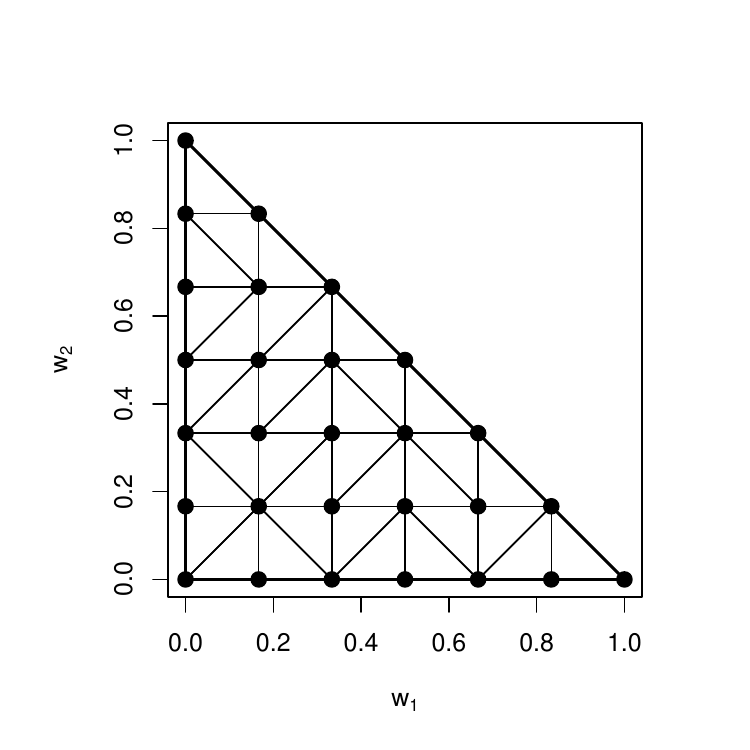}
    \caption[A mesh of $N=28$ reference angles (points) and the resulting Delaunay triangulation (solid lines) on the $\mathcal{S}_2$ simplex.]{A mesh of $N=28$ reference angles (points) and the resulting Delaunay triangulation (solid lines) on the $\mathcal{S}_2$ simplex. This setting is used throughout all $d=3$ model fits in simulation studies and in real-data examples.}
    \label{fig:d3-DT}
\end{figure}

As described in Section~\ref{sec:sim}, there are six settings for fitting the piecewise-linear model for inference on multivariate extremes, labelled SS1--SS6. Here, we outline results for all possible settings for distributions \eqref{distn:log1}--\eqref{distn:mix} in dimensions $d=2,3$, and compare probability estimates and estimates of the unit level set boundary when using $N=11$ equally-spaced scalar-valued reference angles from 0 to 1, inclusive. In each case, 200 datasets of length $n=5000$ are generated. In each replication, $r_{\tau}(\bm{w})$ is first estimated at the $\tau=0.95$ threshold, using the KDE approach with Gaussian kernel, and a probability estimate is obtained for each extremal region. 
Furthermore, in the $d=2$ setting we plot all 200 limit set boundaries in gray, the limit set boundary obtained at the median parameter values in blue, and the true limit set boundary using a black-dashed line. For $d=3$, the piecewise-linear limit set boundary is plotted in blue at median parameter values across the 200 replications, and the true limit set boundary is plotted in red. Here, we use $N=28$ reference angles, and the resulting triangulation of $\mathcal{S}_2$ is presented in Figure~\ref{fig:d3-DT}.
Based on the penalty strength findings in Supplement~\ref{supp:pen-str}, we fix the penalty strength $\lambda=1$ for the radial and joint fits, while setting $\lambda=20$ for the angular model fit.

In summary, for $d=2$, settings SS1,3 and SS2,4 produce similar results, showing no particular advantage or disadvantage of modeling  the angles separately compared to the corresponding empirical distribution. The joint fitting procedures of SS5,6 lead to good estimates in the case of distribution \eqref{distn:log1}, but an increased bias in the probability estimates when $f_{\tilde{\bm{W}}}$ is not well-approximated by the same $g$ in used in the radial model (distributions \eqref{distn:log2}--\eqref{distn:invlog}).
Similarly, SS5 and SS6 induce more bias in the $d=3$ distributions, while SS1,3 and SS2,4 perform similarly well overall.
These findings can be seen in Tables~\ref{tab:2d-SS} and \ref{tab:3d-SS}, which present root mean squared error (RMSE) of the log-probability estimates. In it, we see that the piecewise-linear model fitting method competes well across dimensions two and three and across different distributions exhibiting a variety of extremal dependence properties. 
We note that SS5 and SS6 often have the lowest RMSEs due to reduced variance, but we place high value on unbiasedness. Therefore, for as an overall well-performing model, we report the bounded model fit presented in the setup of SS4 in the main body of this paper.

\subsection{Simulation studies, $d=2$}\label{supp:d2-sim}
\begin{figure}[h]
    \centering
    \includegraphics[width=0.7\textwidth]{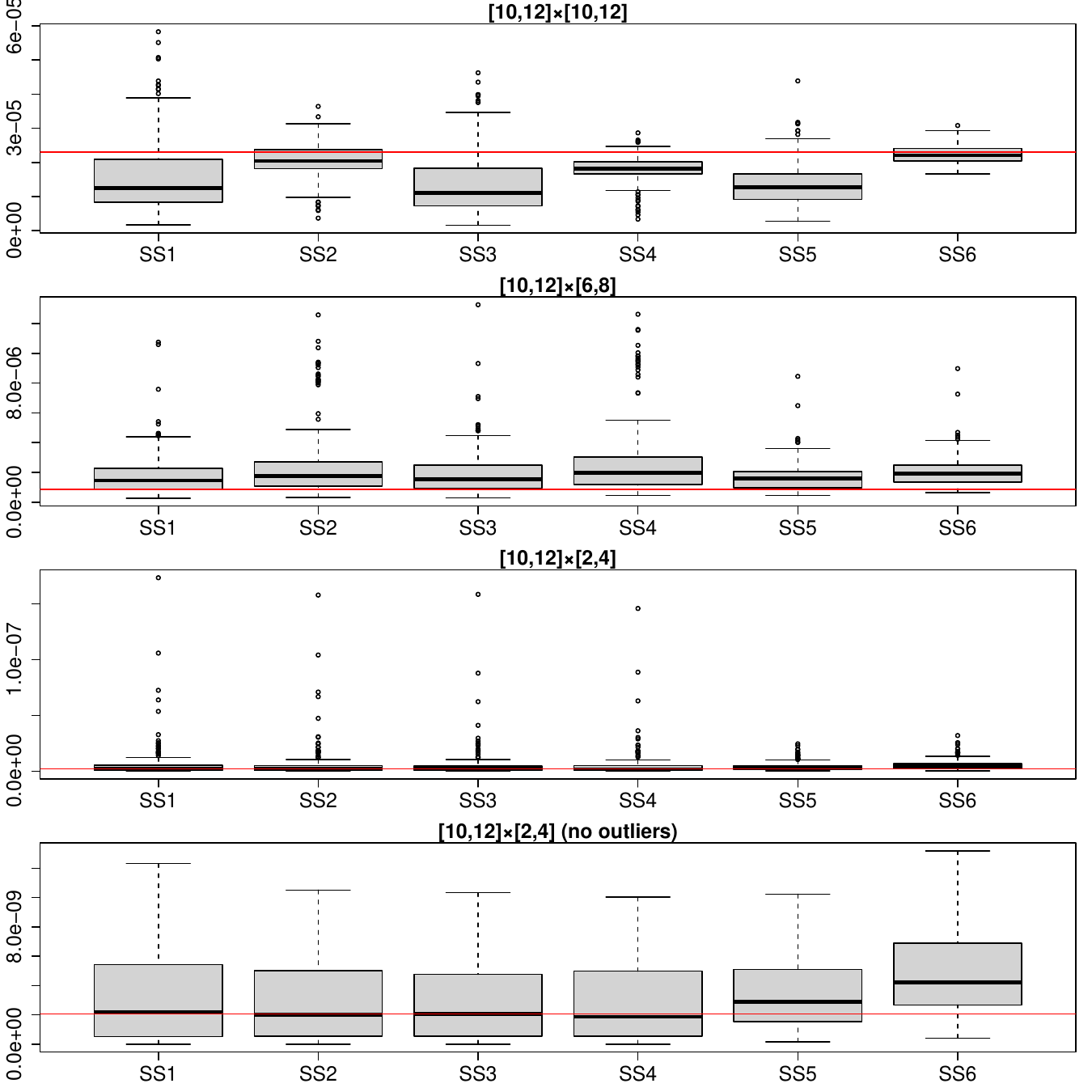}
    \caption[\protect{Probability estimates for distribution \protect\eqref{distn:log1}.}]{Probability estimates for distribution \protect\eqref{distn:log1}. True values shown by the solid line. {Regions of interest $B_i$, $i=1,2,3$, is given in the title, with results for $B_3$ given with and without outliers for clarity.}}
\end{figure}
\begin{figure}[h]
    \centering
%    \begin{subfigure}{0.17\textwidth}
%        \centering
        \includegraphics[width=0.17\textwidth]{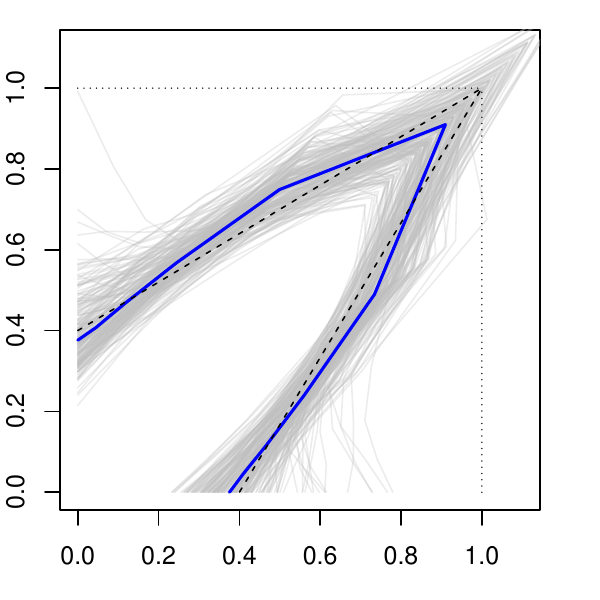}
%        \caption{SS1}
%    \end{subfigure}%
%    ~ 
%    \begin{subfigure}{0.17\textwidth}
%        \centering
        \includegraphics[width=0.17\textwidth]{SimStudy2d_log_strongdep_SS2_unitg.pdf}
%        \caption{SS2}
%    \end{subfigure}%
%    ~ 
%    \begin{subfigure}{0.17\textwidth}
%        \centering
        \includegraphics[width=0.17\textwidth]{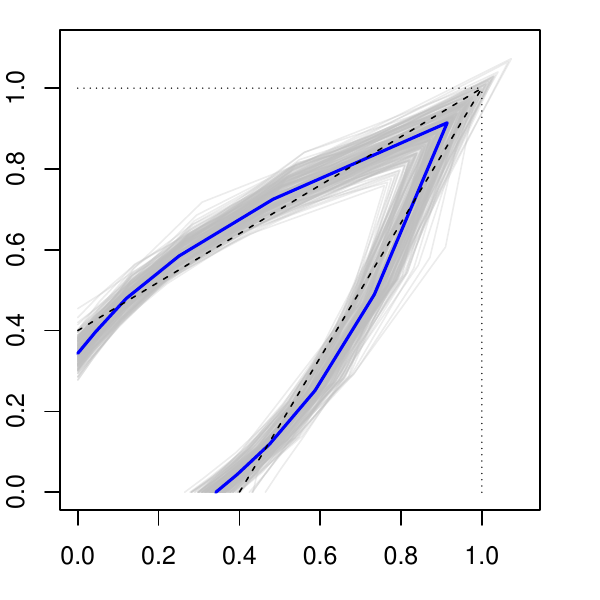}
%        \caption{SS5}
%    \end{subfigure}%
%    ~ 
%    \begin{subfigure}{0.17\textwidth}
%        \centering
        \includegraphics[width=0.17\textwidth]{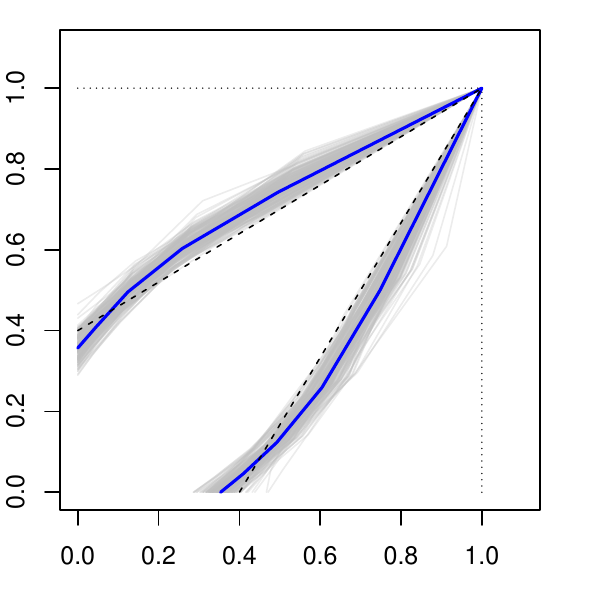}
%        \caption{SS6}
%    \end{subfigure}
%    ~ 
%    \begin{subfigure}{0.19\textwidth}
%        \centering
        \includegraphics[width=0.19\textwidth]{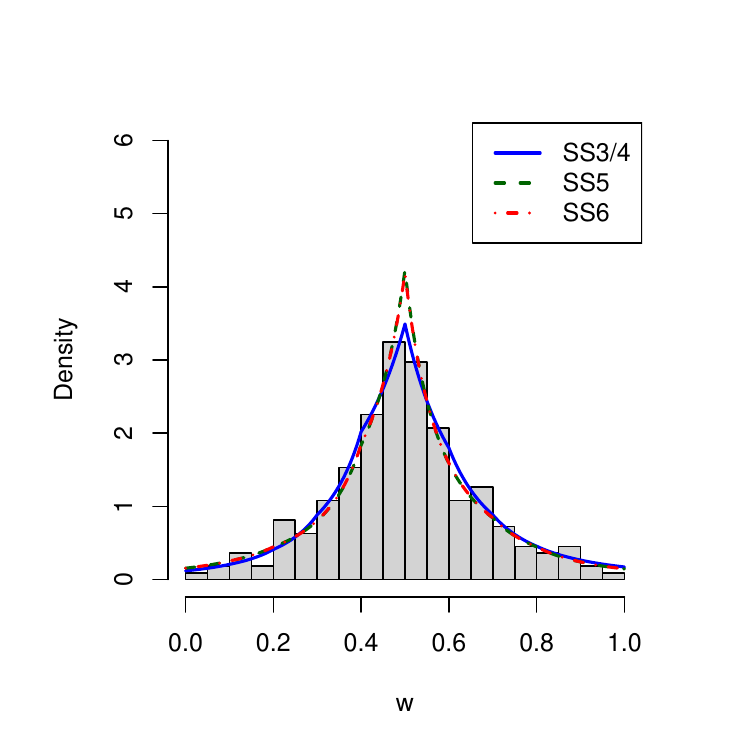}
%        \caption{}
%    \end{subfigure}
	\caption{(a)--(d): Estimates of the unit level set of $g$ for distribution \protect\eqref{distn:log1}. (e) Estimated angular density $f_{W}$, with a sample histogram from one sample.}
\end{figure}

\clearpage
\begin{figure}[h]
    \centering
    \includegraphics[width=0.7\textwidth]{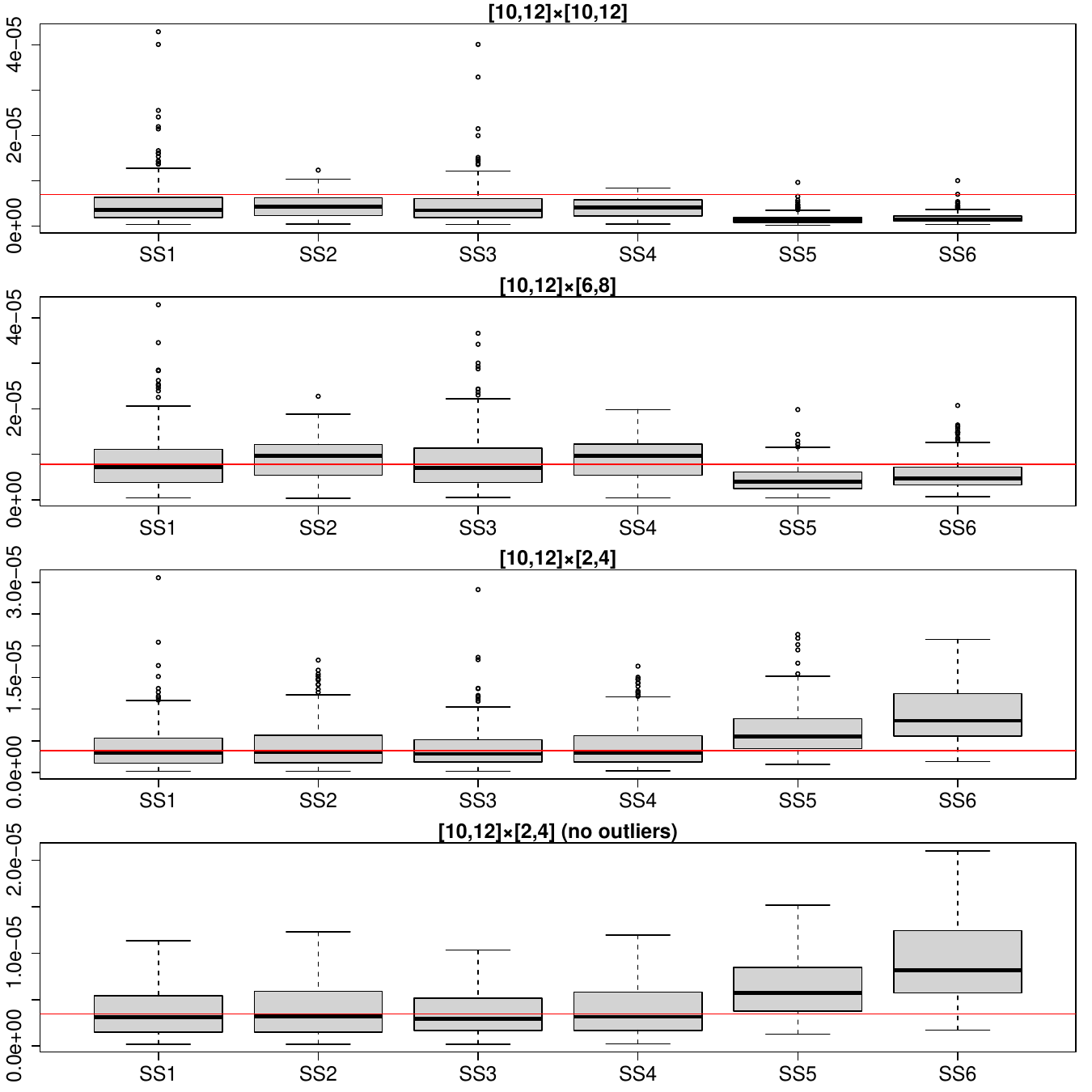}
    \caption[\protect{Probability estimates for distribution \protect\eqref{distn:log2}.}]{Probability estimates for distribution \protect\eqref{distn:log2}. True values shown by the solid line. Regions of interest $B_i$, $i=1,2,3$, is given in the title, with results for $B_3$ given with and without outliers for clarity.}
\end{figure}
\begin{figure}[h]
    \centering
%    \begin{subfigure}{0.17\textwidth}
%        \centering
        \includegraphics[width=0.17\textwidth]{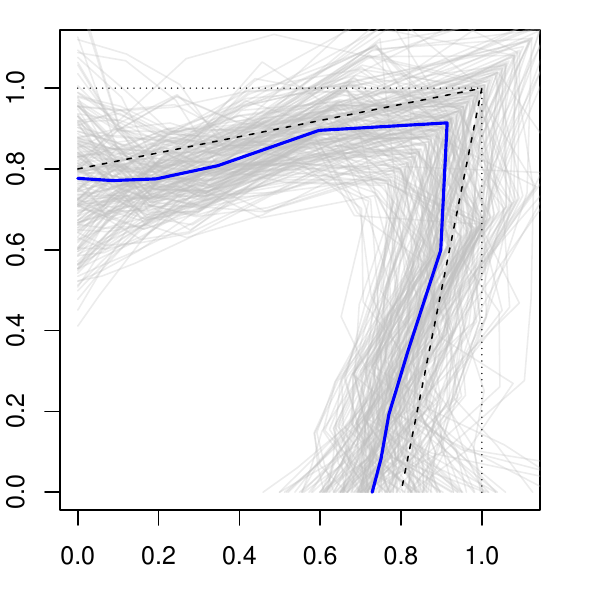}
%        \caption{SS1}
%    \end{subfigure}%
%    ~ 
%    \begin{subfigure}{0.17\textwidth}
%        \centering
        \includegraphics[width=0.17\textwidth]{SimStudy2d_log_weakdep_SS2_unitg.pdf}
%        \caption{SS2}
%    \end{subfigure}%
%    ~ 
%    \begin{subfigure}{0.17\textwidth}
%        \centering
        \includegraphics[width=0.17\textwidth]{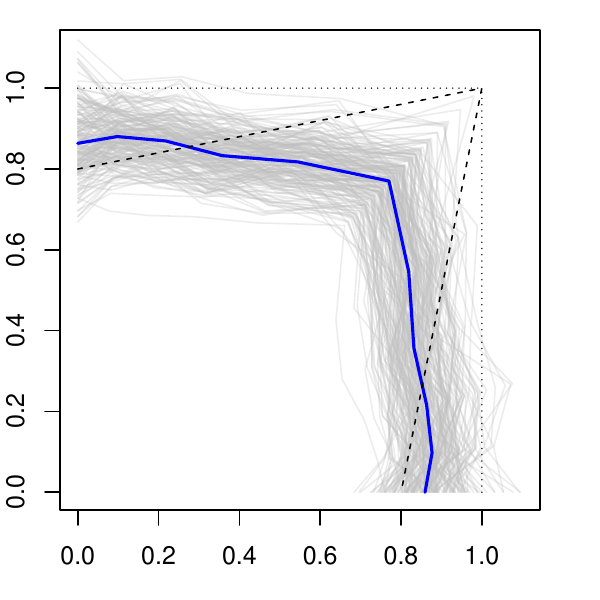}
%        \caption{SS5}
%    \end{subfigure}%
%    ~ 
%    \begin{subfigure}{0.17\textwidth}
%        \centering
        \includegraphics[width=0.17\textwidth]{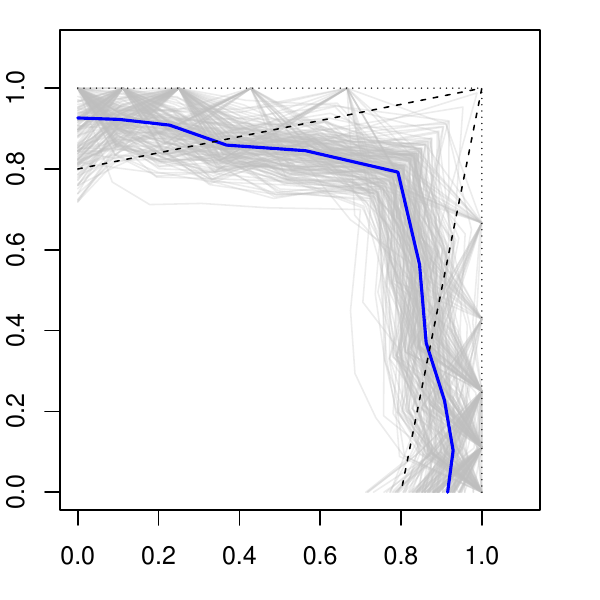}
%        \caption{SS6}
%    \end{subfigure}
%     ~ 
%    \begin{subfigure}{0.19\textwidth}
%        \centering
        \includegraphics[width=0.19\textwidth]{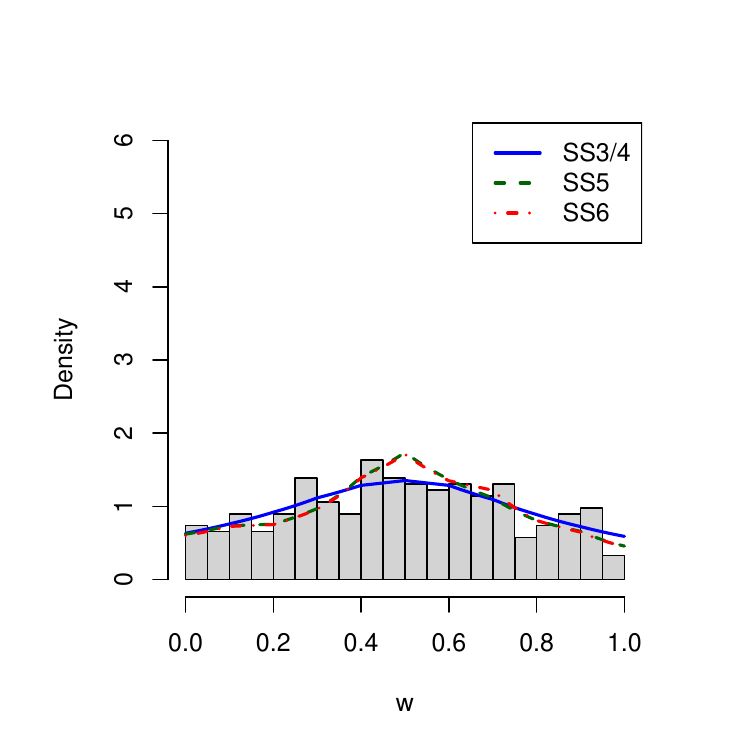}
%        \caption{}
%    \end{subfigure}
	\caption{Estimates of the unit level set of $g$ for distribution \protect\eqref{distn:log2}. Right: Estimated angular density $f_{W}$, with a sample histogram from one sample.}
\end{figure}
\clearpage
\begin{figure}[h]
    \centering
    \includegraphics[width=0.7\textwidth]{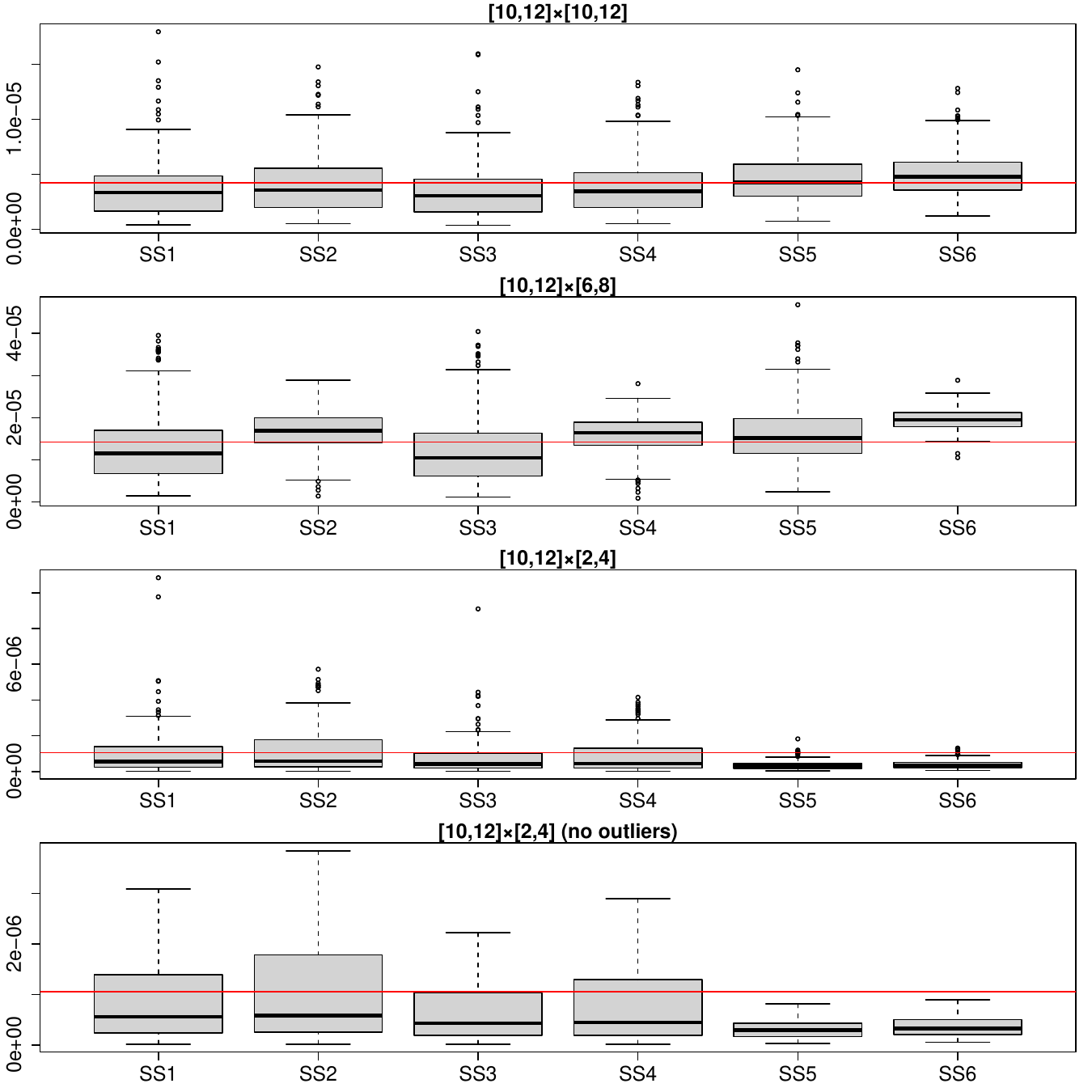}
    \caption[\protect{Probability estimates for distribution \protect\eqref{distn:gauss}.}]{Probability estimates for distribution \protect\eqref{distn:gauss}. True values shown by the solid line. Regions of interest $B_i$, $i=1,2,3$, is given in the title, with results for $B_3$ given with and without outliers for clarity.}
\end{figure}
\begin{figure}[h]
    \centering
%    \begin{subfigure}{0.17\textwidth}
%        \centering
        \includegraphics[width=0.17\textwidth]{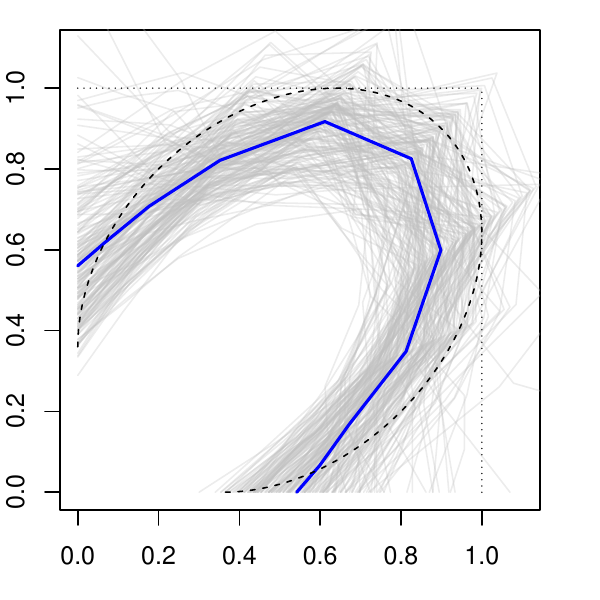}
%        \caption{SS1}
%    \end{subfigure}%
%    ~ 
%    \begin{subfigure}{0.17\textwidth}
%        \centering
        \includegraphics[width=0.17\textwidth]{SimStudy2d_gauss_SS2_unitg.pdf}
%        \caption{SS2}
%    \end{subfigure}%
%    ~ 
%    \begin{subfigure}{0.17\textwidth}
%        \centering
        \includegraphics[width=0.17\textwidth]{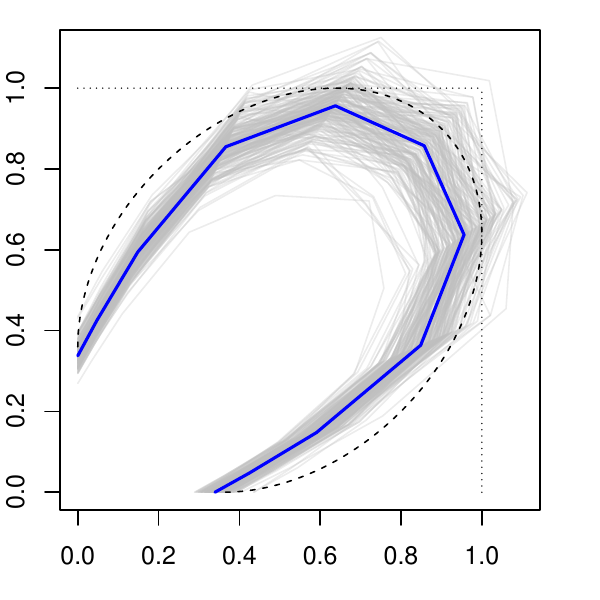}
%        \caption{SS5}
%    \end{subfigure}%
%    ~ 
%    \begin{subfigure}{0.17\textwidth}
%        \centering
        \includegraphics[width=0.17\textwidth]{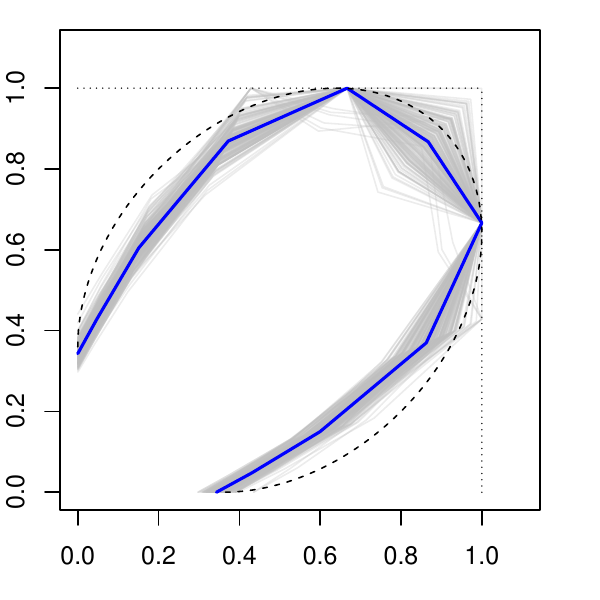}
%        \caption{SS6}
%    \end{subfigure}
%    ~ 
%    \begin{subfigure}{0.19\textwidth}
%        \centering
        \includegraphics[width=0.19\textwidth]{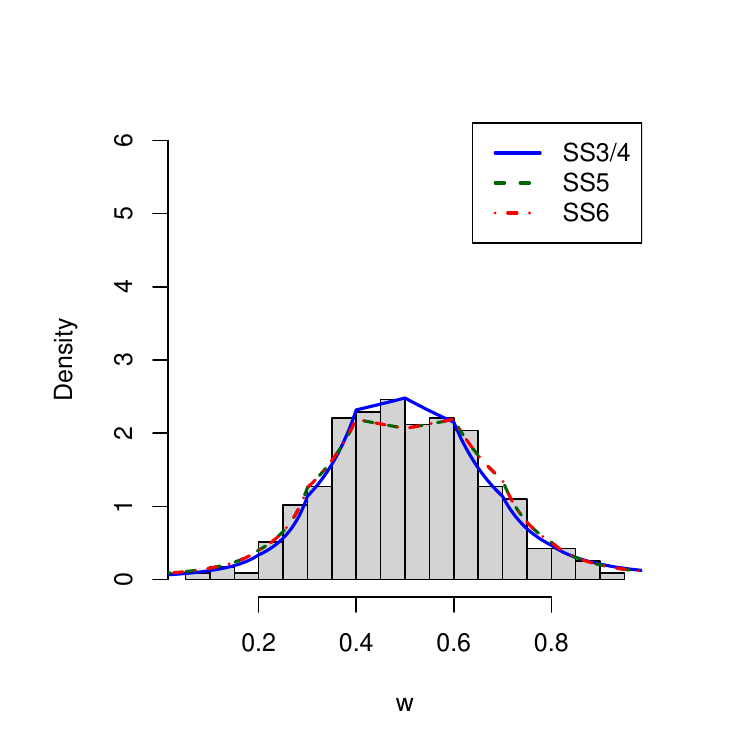}
%        \caption{}
%    \end{subfigure}
	\caption{Estimates of the unit level set of $g$ for distribution \protect\eqref{distn:gauss}. Right: Estimated angular density $f_{W}$, with a sample histogram from one sample.}
\end{figure}

\clearpage
\begin{figure}[h]
    \centering
    \includegraphics[width=0.7\textwidth]{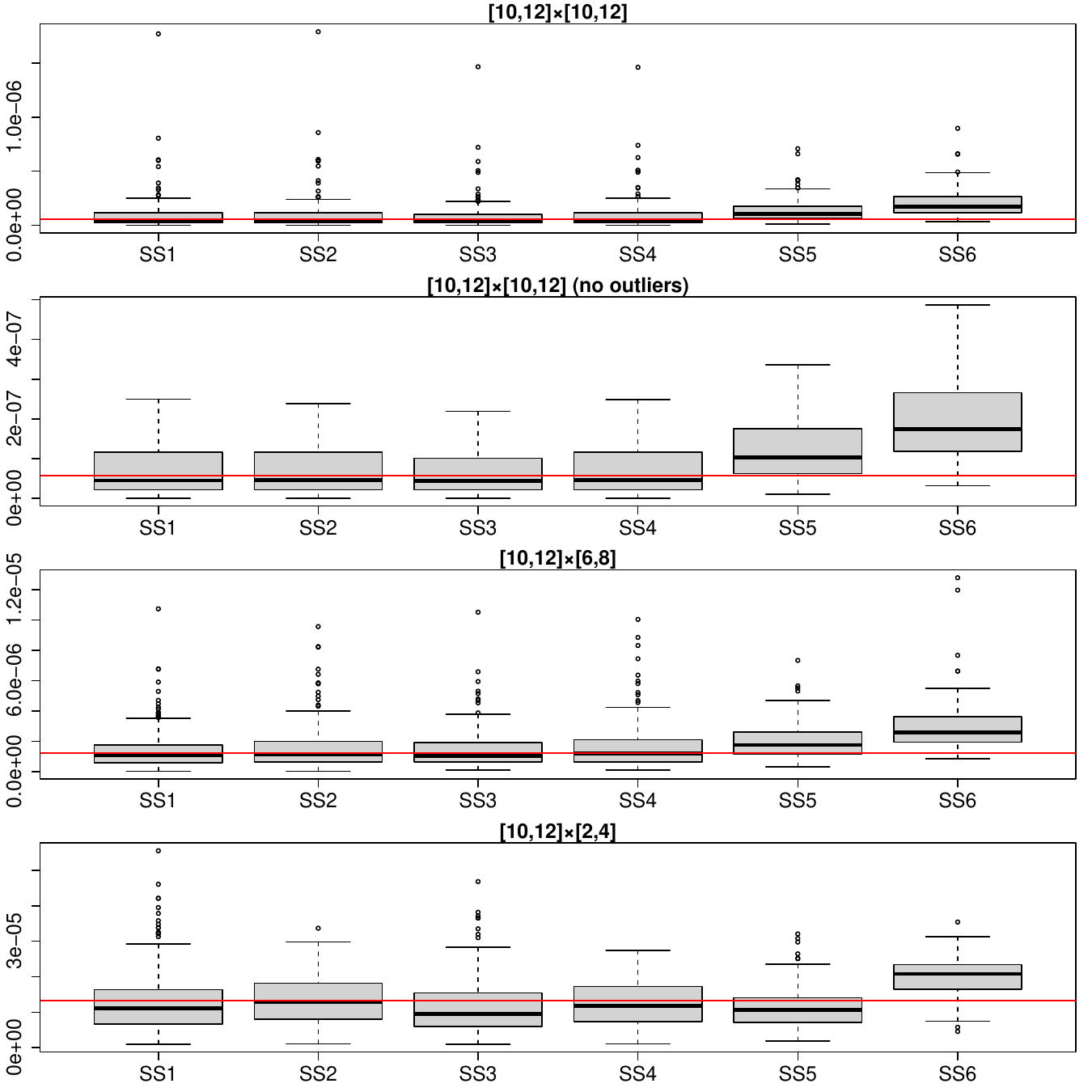}
    \caption[\protect{Probability estimates for distribution \protect\eqref{distn:invlog}.}]{Probability estimates for distribution \protect\eqref{distn:invlog}. True values shown by the solid line. Regions of interest $B_i$, $i=1,2,3$, is given in the title, with results for $B_1$ given with and without outliers for clarity.}
\end{figure}
\begin{figure}[h]
    \centering
%    \begin{subfigure}{0.17\textwidth}
%        \centering
        \includegraphics[width=0.17\textwidth]{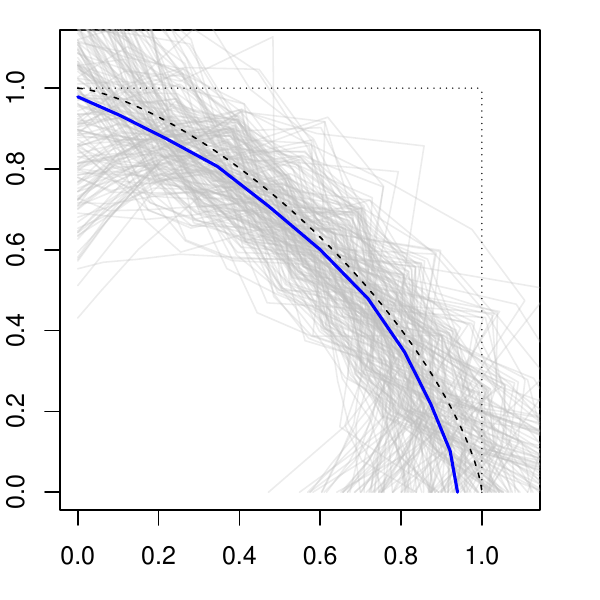}
%        \caption{SS1}
%    \end{subfigure}%
%    ~ 
%    \begin{subfigure}{0.17\textwidth}
%        \centering
        \includegraphics[width=0.17\textwidth]{SimStudy2d_invlog_SS2_unitg.pdf}
%        \caption{SS2}
%    \end{subfigure}%
%    ~ 
%    \begin{subfigure}{0.17\textwidth}
%        \centering
        \includegraphics[width=0.17\textwidth]{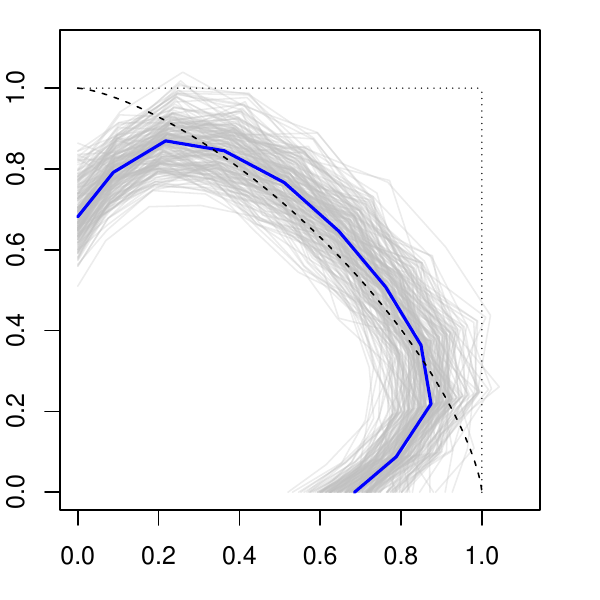}
%        \caption{SS5}
%    \end{subfigure}%
%    ~ 
%    \begin{subfigure}{0.17\textwidth}
%        \centering
        \includegraphics[width=0.17\textwidth]{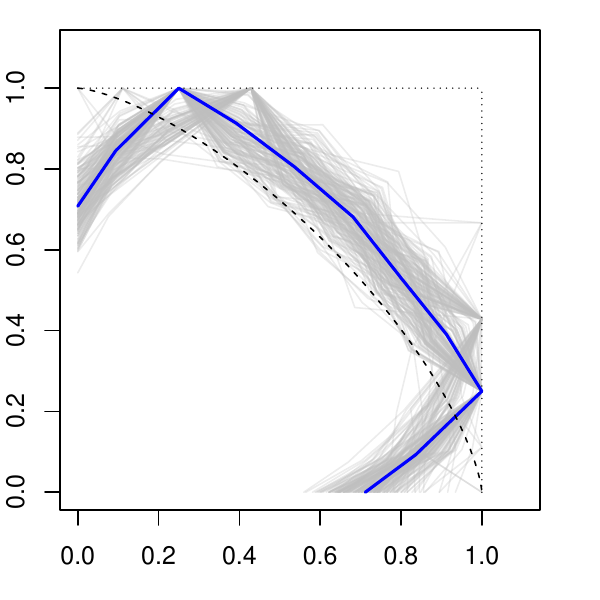}
%        \caption{SS6}
%    \end{subfigure} 
%    ~ 
%    \begin{subfigure}{0.19\textwidth}
%        \centering
        \includegraphics[width=0.19\textwidth]{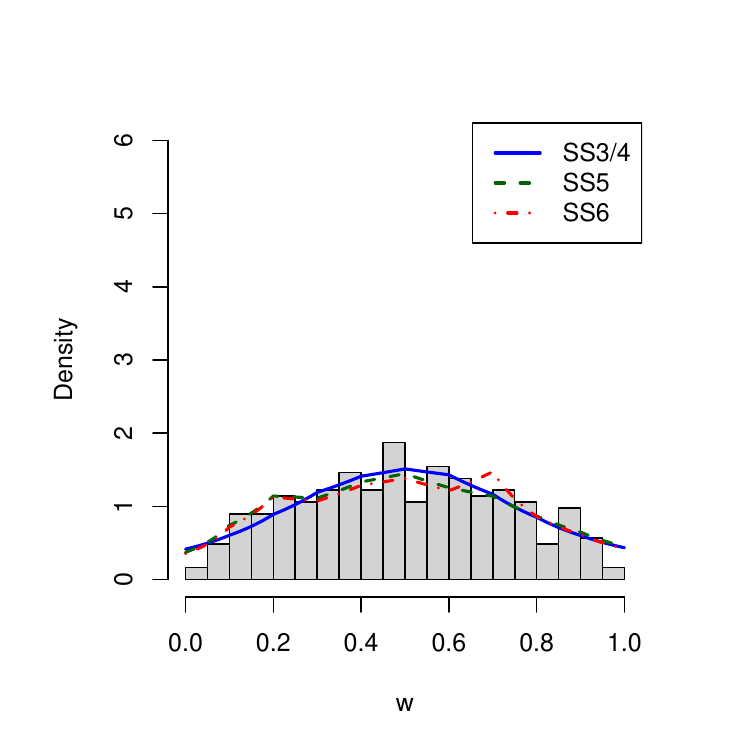}
%        \caption{}
%    \end{subfigure}
	\caption{Estimates of the unit level set of $g$ for distribution \protect\eqref{distn:invlog}. Right: Estimated angular density $f_{W}$, with a sample histogram from one sample.}
\end{figure}

\clearpage

\input{RMSEtable_SS2d}

\clearpage
\subsection{Simulation studies, $d=3$}\label{supp:d3-sim}
\begin{figure}[h]
    \centering
    \includegraphics[width=0.7\textwidth]{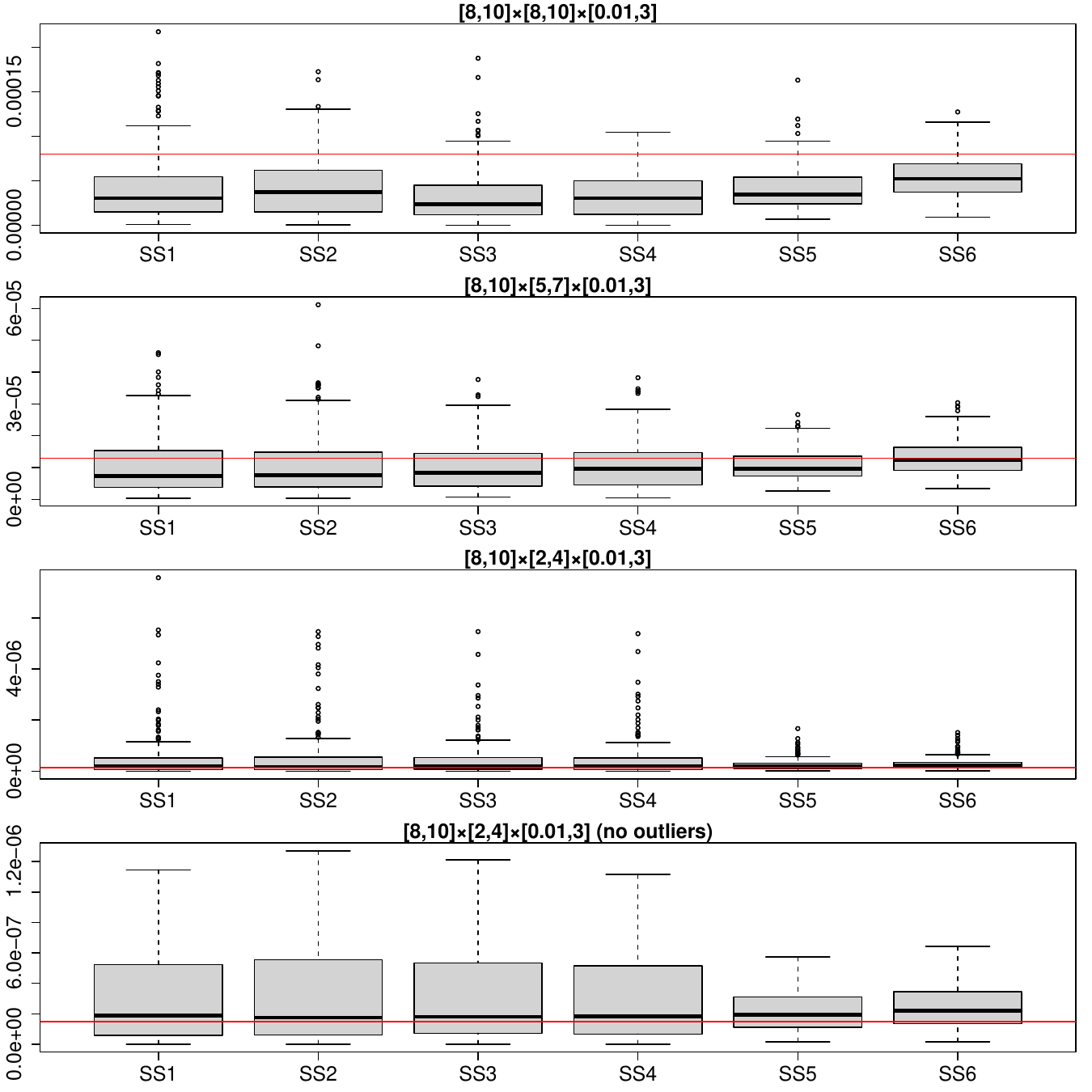}
    \caption[\protect{Probability estimates for distribution \protect\eqref{distn:alog1}.}]{Probability estimates for distribution \protect\eqref{distn:alog1}. True values shown by the solid line. Regions of interest $B_i$, $i=1,2,3$, is given in the title, with results for $B_3$ given with and without outliers for clarity.}
\end{figure}
\begin{figure}[h]
    \centering
%    \begin{subfigure}{0.2\textwidth}
%        \centering
        \includegraphics[width=0.2\textwidth]{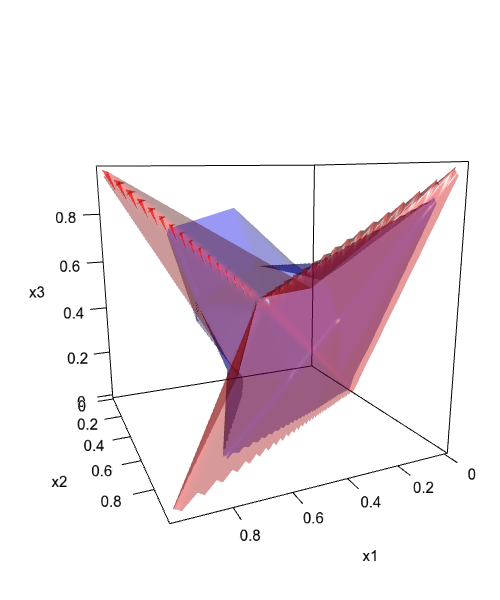}
%        \caption{SS1}
%    \end{subfigure}%
%    ~ 
%    \begin{subfigure}{0.2\textwidth}
%        \centering
        \includegraphics[width=0.2\textwidth]{alog1_SS2.png}
%        \caption{SS2}
%    \end{subfigure}%
%    ~ 
%    \begin{subfigure}{0.2\textwidth}
%        \centering
        \includegraphics[width=0.2\textwidth]{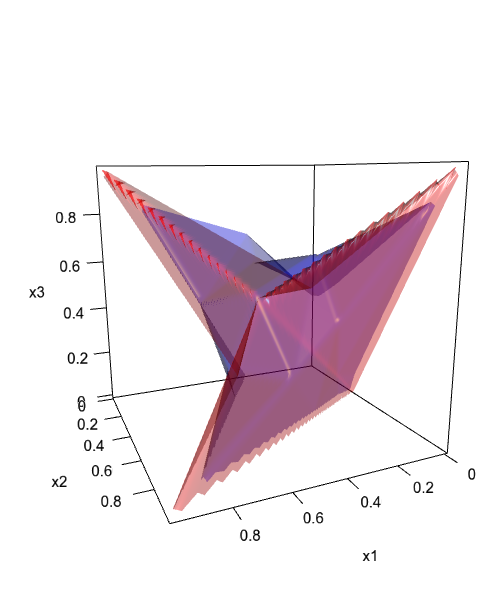}
%        \caption{SS5}
%    \end{subfigure}%
%    ~ 
%    \begin{subfigure}{0.2\textwidth}
%        \centering
        \includegraphics[width=0.2\textwidth]{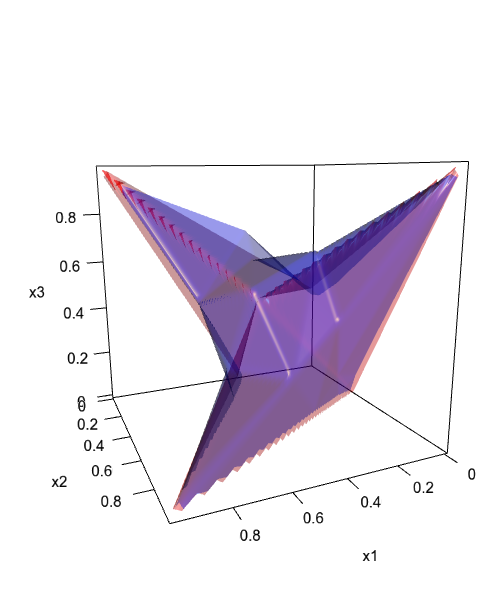}
%        \caption{SS6}
%    \end{subfigure}
	\caption{Blue: median estimates of the unit level set of $g$ for distribution \protect\eqref{distn:alog1}. Red: true unit level set.}
\end{figure}

\begin{figure}[h]
    \centering
    \includegraphics[width=0.7\textwidth]{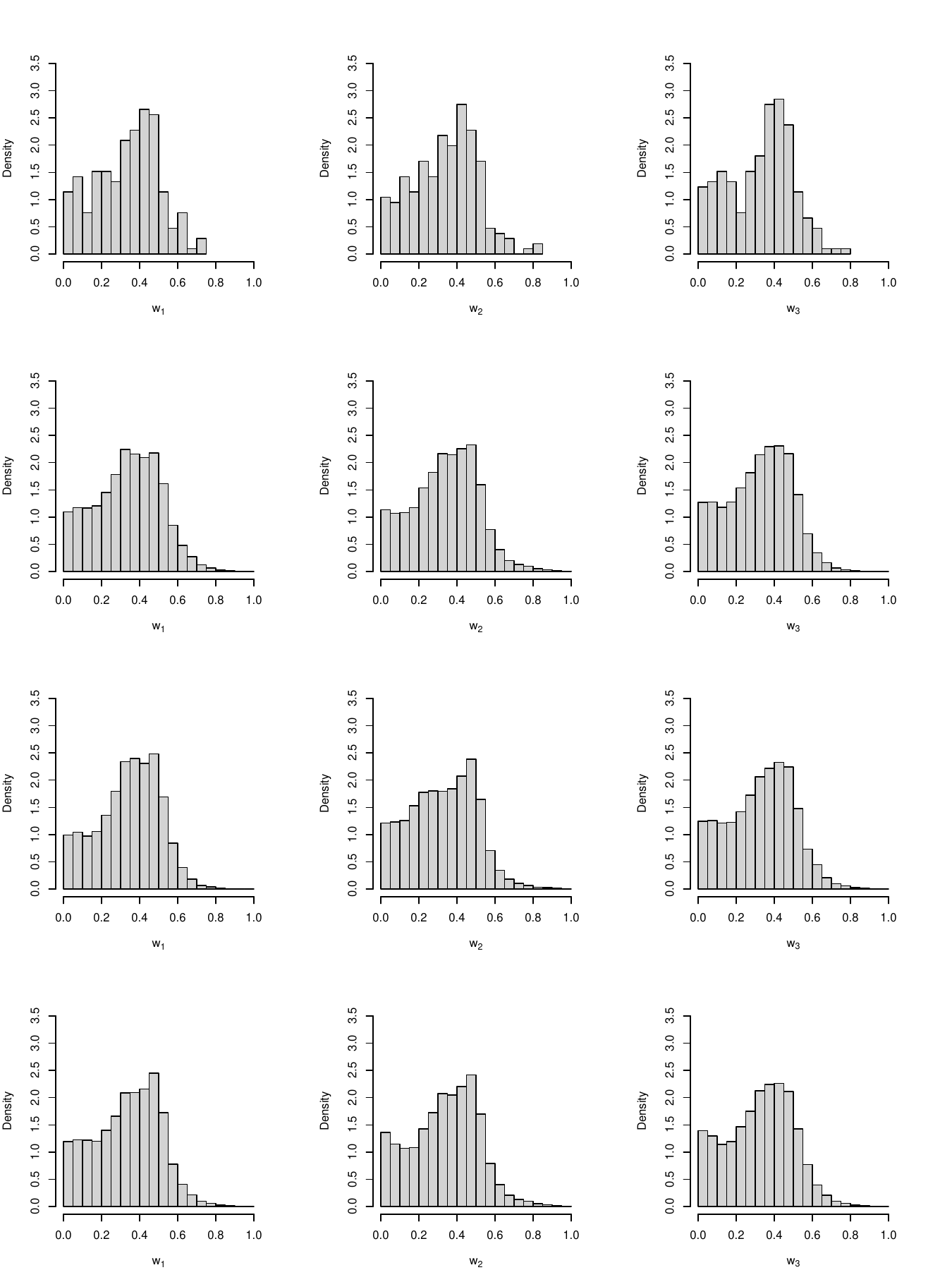}
    \caption{Top row: Marginal sample of exceedance angles generated of a sample from distribution \protect\eqref{distn:alog1}. Rows 2--4: Marginal MCMC samples of exceedance angles from SS3/4, SS5, and SS6, respectively.}
\end{figure}

\begin{figure}[h]
    \centering
    \includegraphics[width=0.7\textwidth]{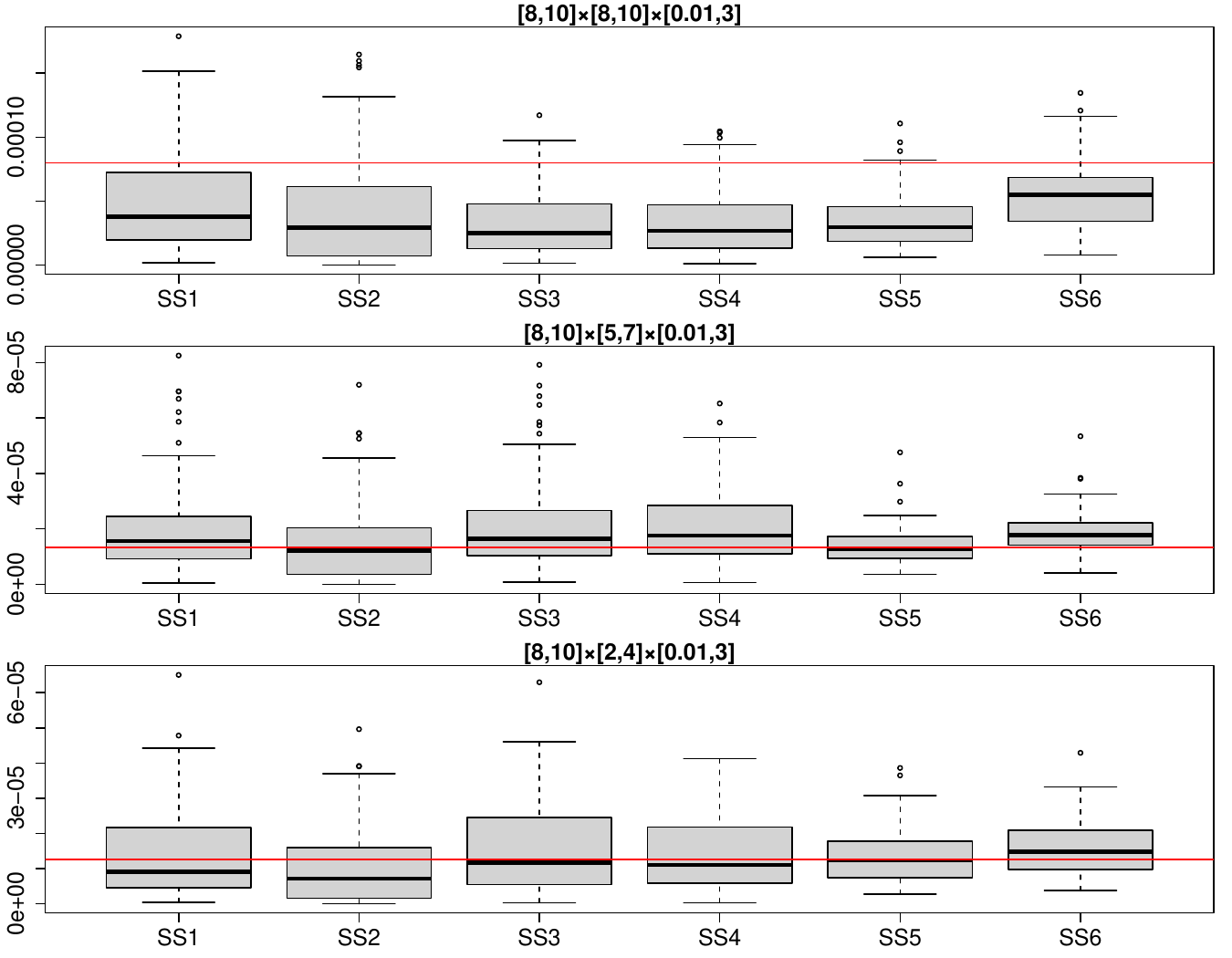}
    \caption[\protect{Probability estimates for distribution \protect\eqref{distn:alog2}.}]{Probability estimates for distribution \protect\eqref{distn:alog2}. True values shown by the solid line. Regions of interest $B_i$, $i=1,2,3$, is given in the title.}
\end{figure}
\begin{figure}[h!]
    \centering
%    \begin{subfigure}{0.2\textwidth}
%        \centering
        \includegraphics[width=0.2\textwidth]{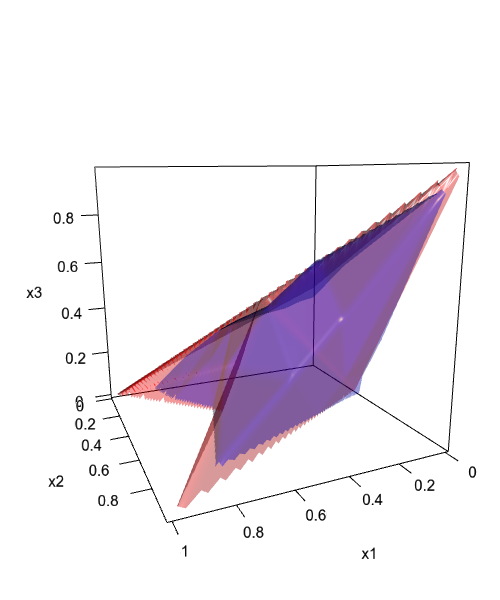}
%        \caption{SS1}
%    \end{subfigure}%
%    ~ 
%    \begin{subfigure}{0.2\textwidth}
%        \centering
        \includegraphics[width=0.2\textwidth]{alog2_SS2.png}
%        \caption{SS2}
%    \end{subfigure}%
%    ~ 
%    \begin{subfigure}{0.2\textwidth}
%        \centering
        \includegraphics[width=0.2\textwidth]{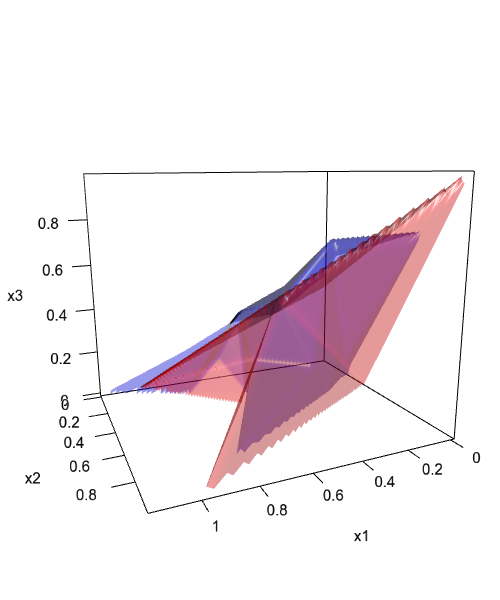}
%        \caption{SS5}
%    \end{subfigure}%
%    ~ 
%    \begin{subfigure}{0.2\textwidth}
%        \centering
        \includegraphics[width=0.2\textwidth]{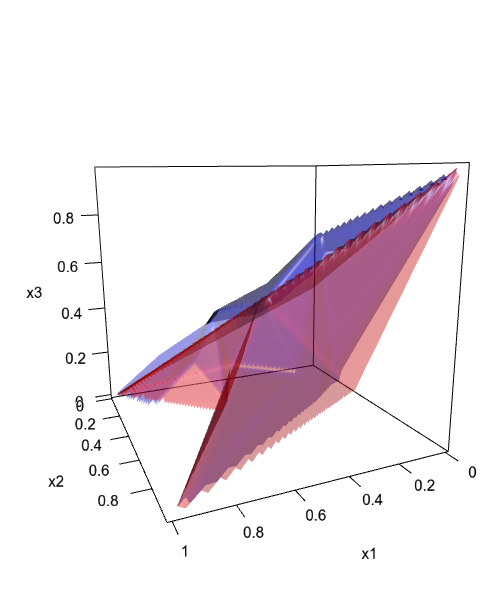}
%        \caption{SS6}
%    \end{subfigure}
	\caption{Blue: median estimates of the unit level set of $g$ for distribution \protect\eqref{distn:alog2}. Red: true unit level set.}
\end{figure}

\begin{figure}[h]
    \centering
    \includegraphics[width=0.7\textwidth]{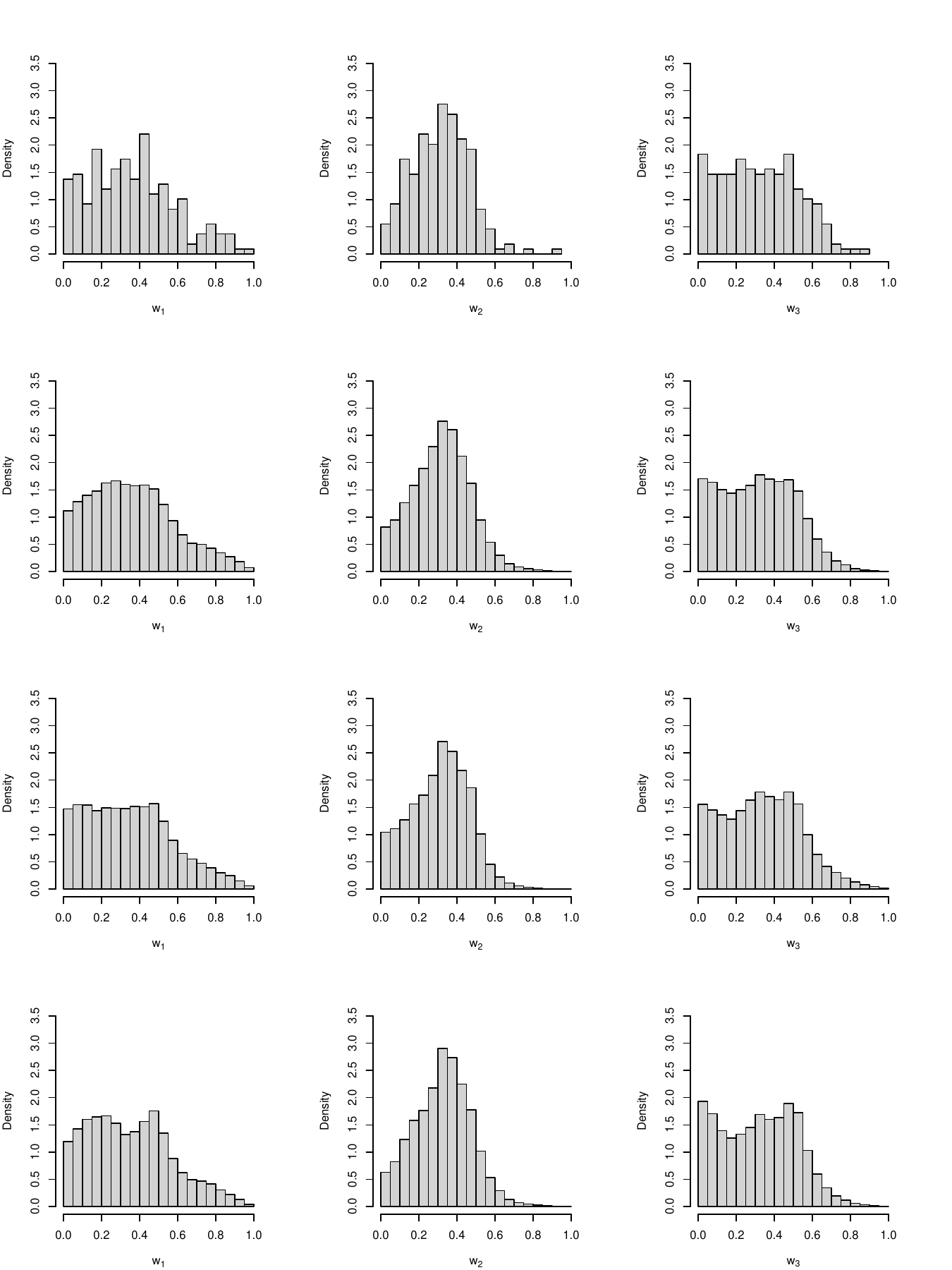}
    \caption{Top row: Marginal sample of exceedance angles generated of a sample from distribution \protect\eqref{distn:alog2}. Rows 2--4: Marginal MCMC samples of exceedance angles from SS3/4, SS5, and SS6, respectively.}
\end{figure}

\begin{figure}[h]
    \centering
    \includegraphics[width=0.7\textwidth]{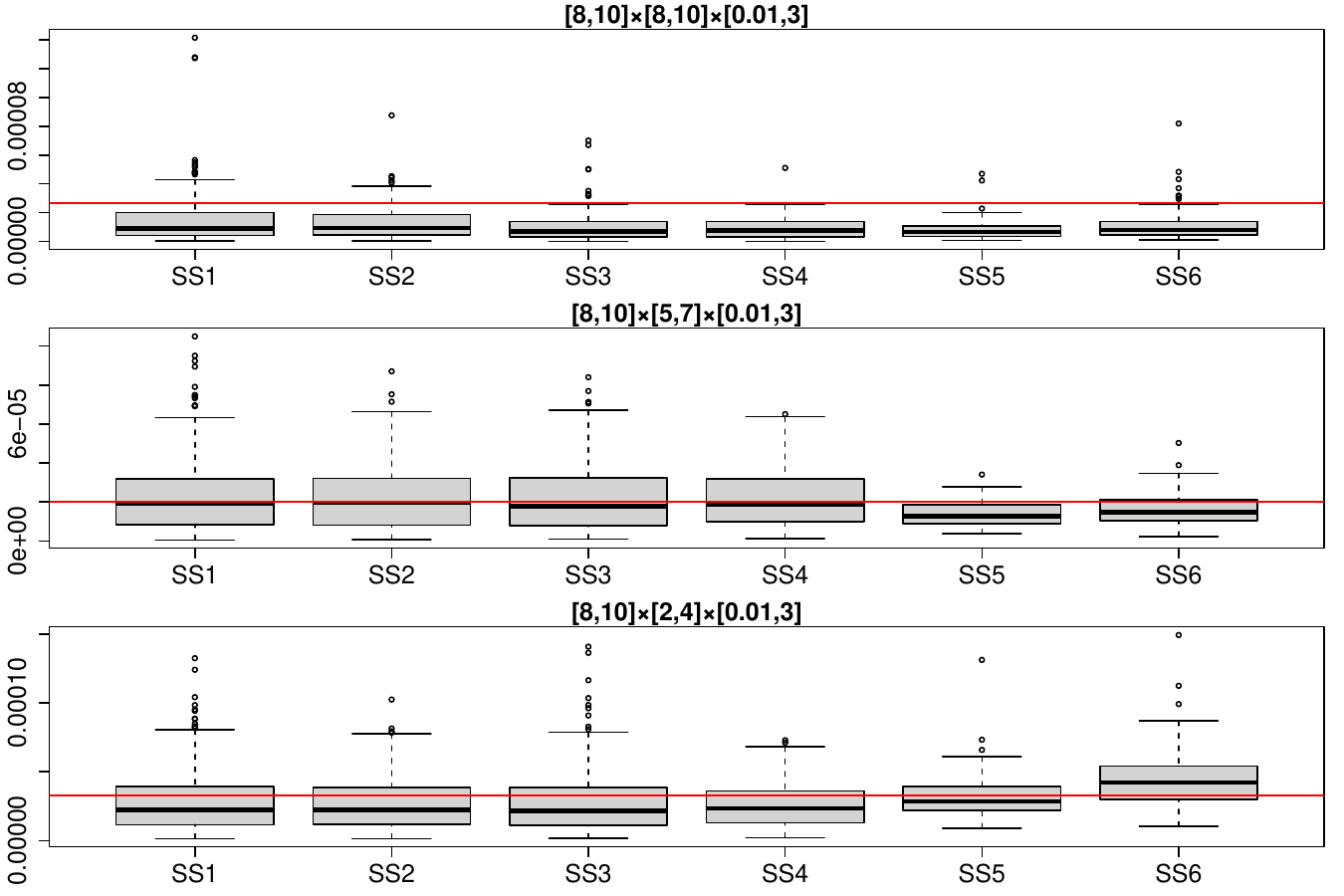}
    \caption[\protect{Probability estimates for distribution \protect\eqref{distn:mix}.}]{Probability estimates for distribution \protect\eqref{distn:mix}. True values shown by the solid line.Regions of interest $B_i$, $i=1,2,3$, is given in the title.}
\end{figure}
\begin{figure}[h!]
    \centering
%    \begin{subfigure}{0.2\textwidth}
%        \centering
        \includegraphics[width=0.2\textwidth]{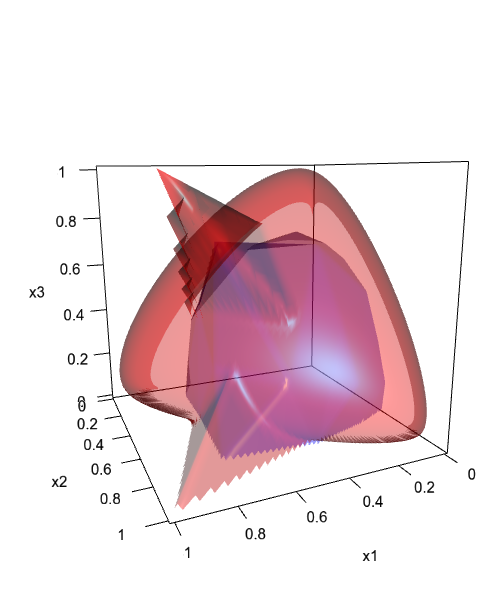}
%        \caption{SS1}
%    \end{subfigure}%
%    ~ 
%    \begin{subfigure}{0.2\textwidth}
%        \centering
        \includegraphics[width=0.2\textwidth]{mix_SS2.png}
%        \caption{SS2}
%    \end{subfigure}%
%    ~ 
%    \begin{subfigure}{0.2\textwidth}
%        \centering
        \includegraphics[width=0.2\textwidth]{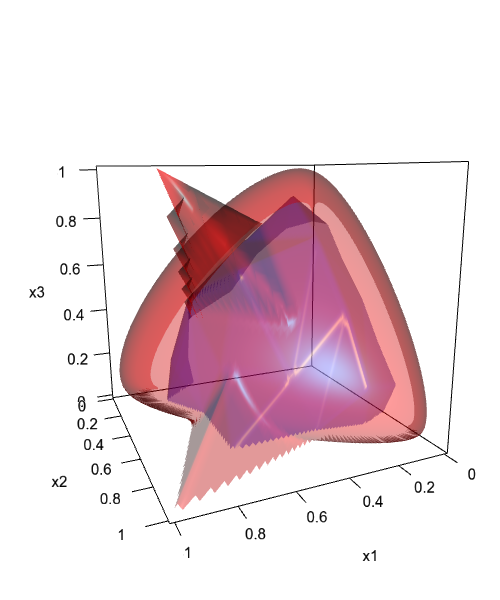}
%        \caption{SS5}
%    \end{subfigure}%
%    ~ 
%    \begin{subfigure}{0.2\textwidth}
%        \centering
        \includegraphics[width=0.2\textwidth]{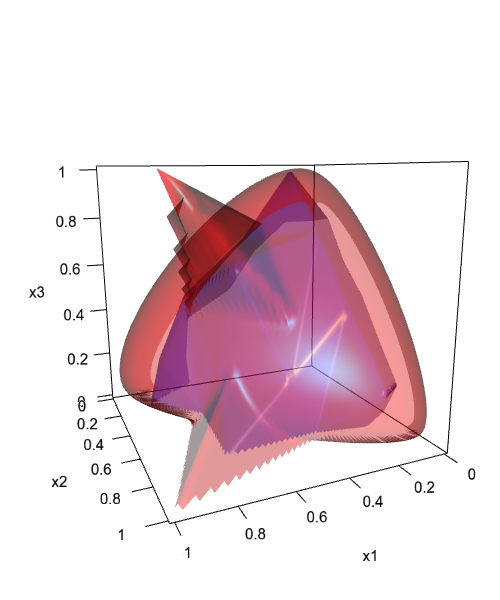}
%        \caption{SS6}
%    \end{subfigure}
	\caption{Blue: median estimates of the unit level set of $g$ for distribution \protect\eqref{distn:mix}. Red: true unit level set.}
\end{figure}

\begin{figure}[h]
    \centering
    \includegraphics[width=0.7\textwidth]{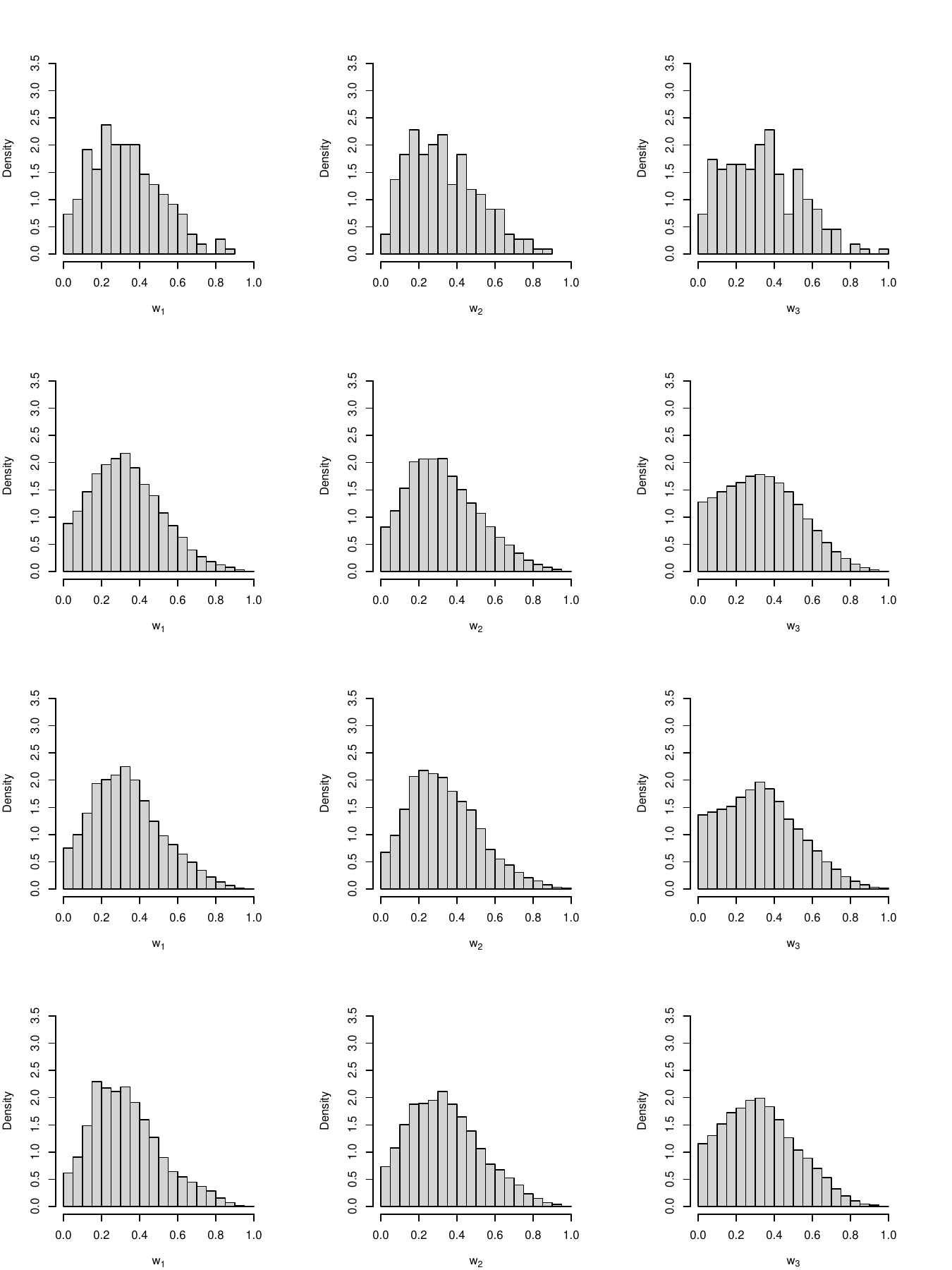}
    \caption{Top row: Marginal sample of exceedance angles generated of a sample from distribution \protect\eqref{distn:mix}. Rows 2--4: Marginal MCMC samples of exceedance angles from SS3/4, SS5, and SS6, respectively.}
\end{figure}

\clearpage

\input{RMSEtable_SS3d}

\clearpage
\newpage
\section{Additional pollution data results}\label{supp:pollution}

\subsection{Setting 1: $d=4$}\label{supp:pollution-d4}

\begin{figure}[h!]
    \centering
    \includegraphics[width=0.24\textwidth]{d4pollution_chi_plot_12.pdf}
    \includegraphics[width=0.24\textwidth]{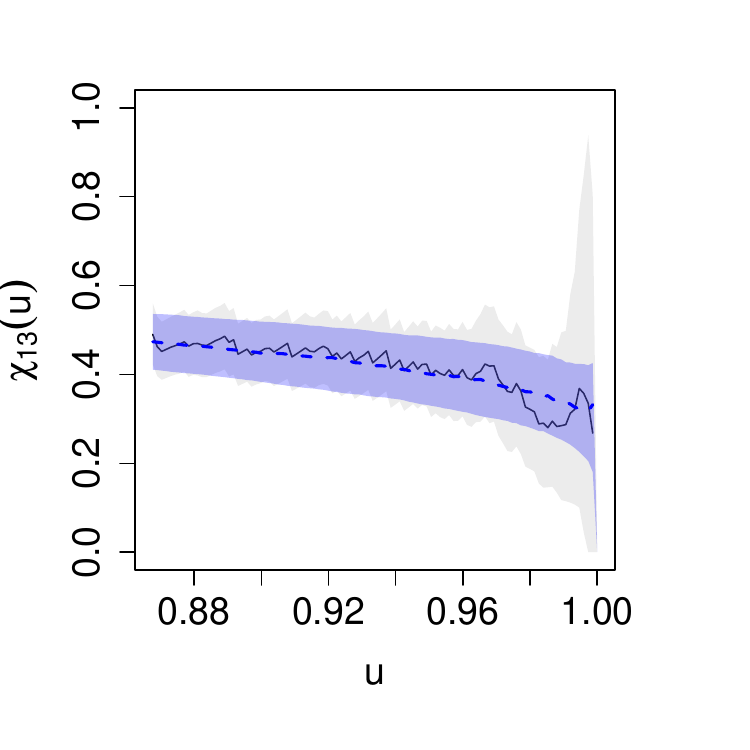}
    \includegraphics[width=0.24\textwidth]{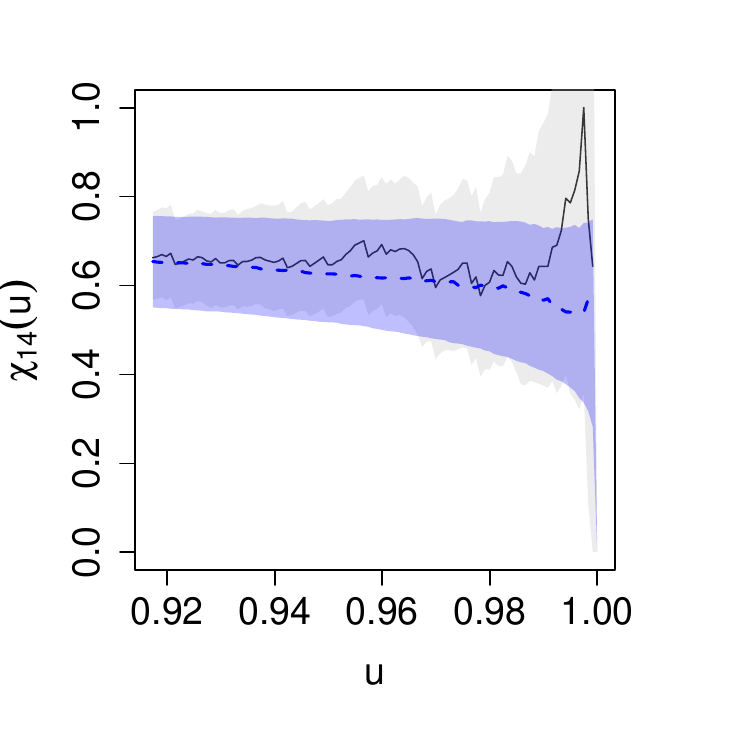}
    \includegraphics[width=0.24\textwidth]{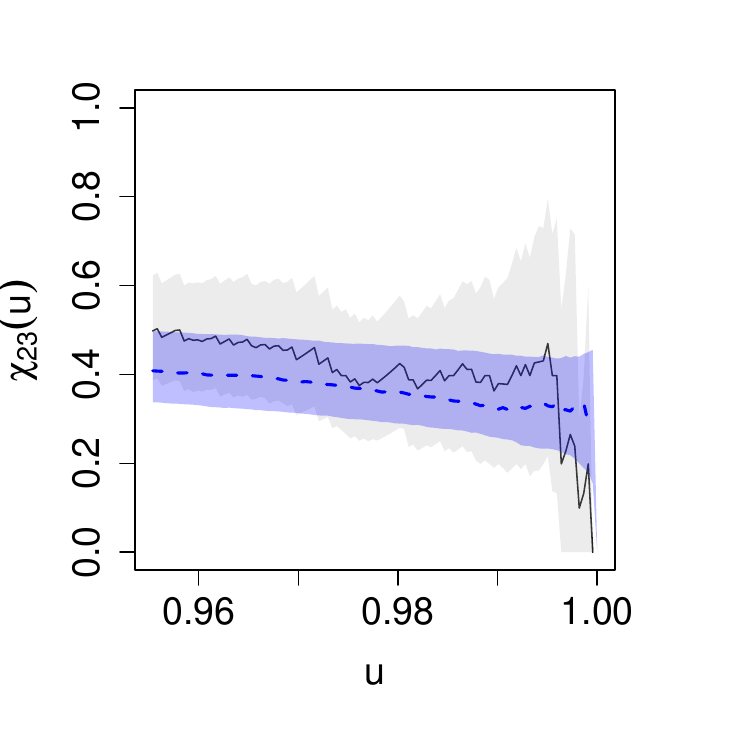}
    \includegraphics[width=0.24\textwidth]{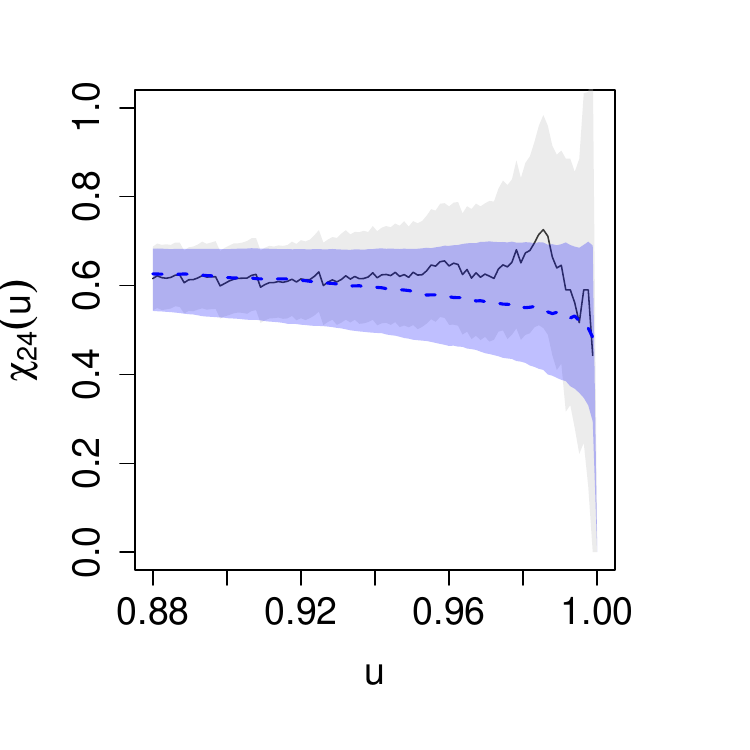}
    \includegraphics[width=0.24\textwidth]{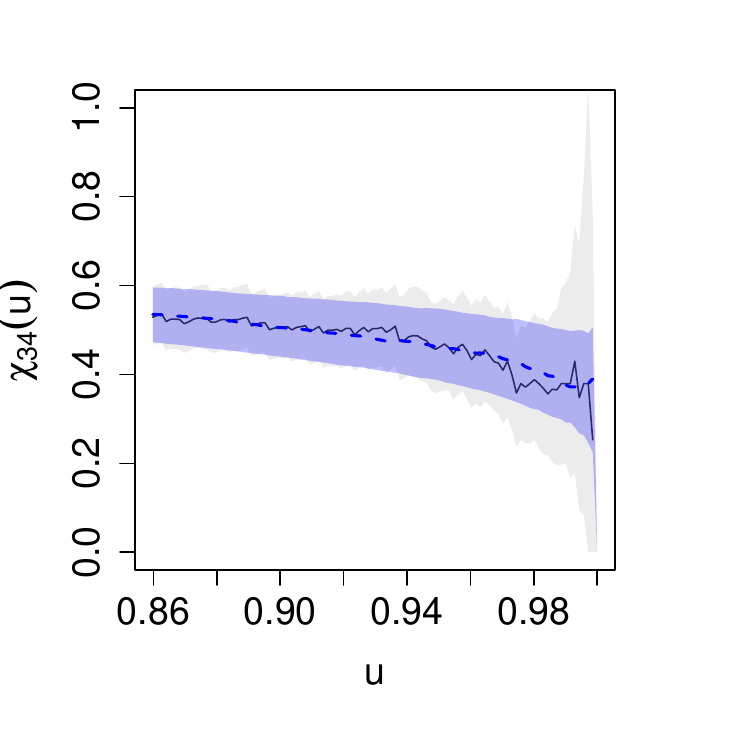}
    \includegraphics[width=0.24\textwidth]{d4pollution_chi_plot_123.pdf}
    \includegraphics[width=0.24\textwidth]{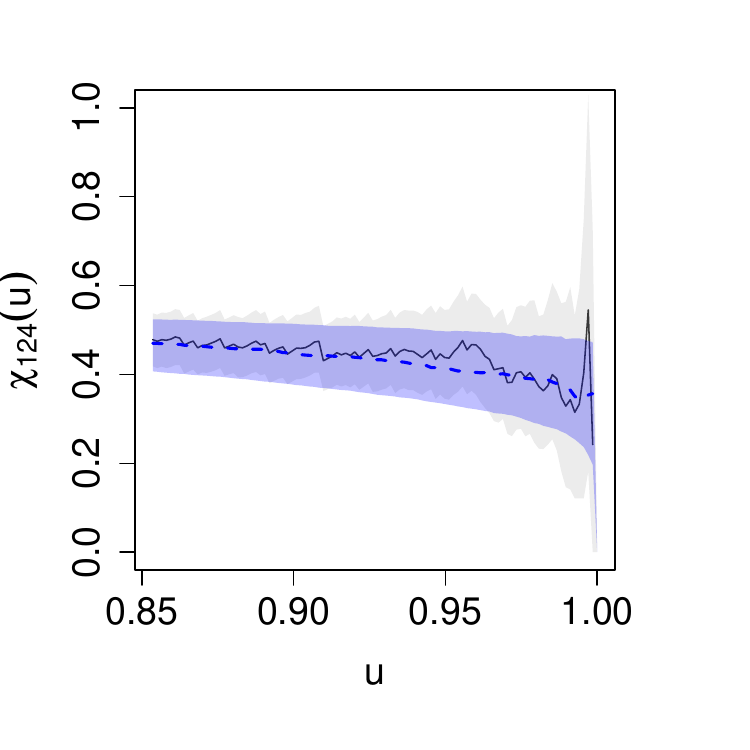}
    \includegraphics[width=0.24\textwidth]{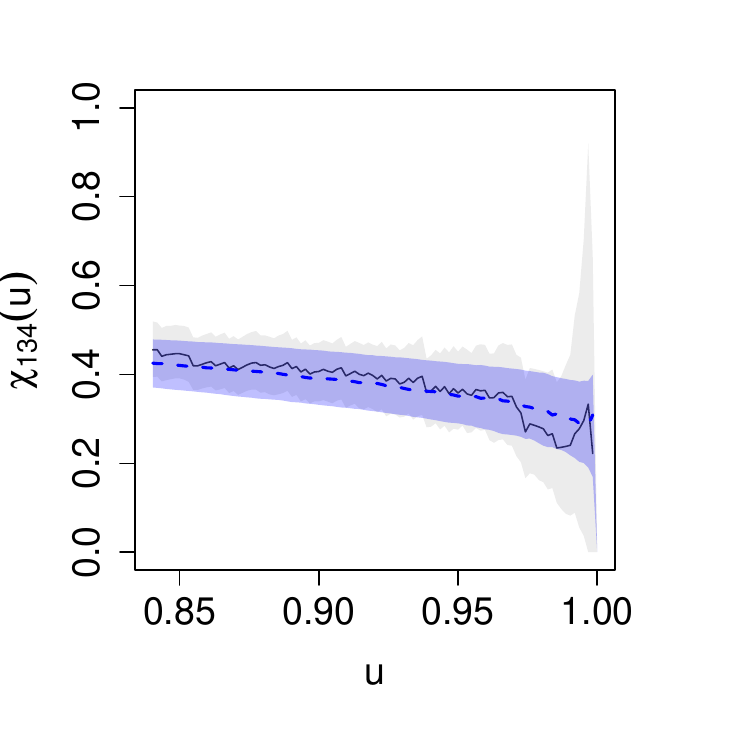}
    \includegraphics[width=0.24\textwidth]{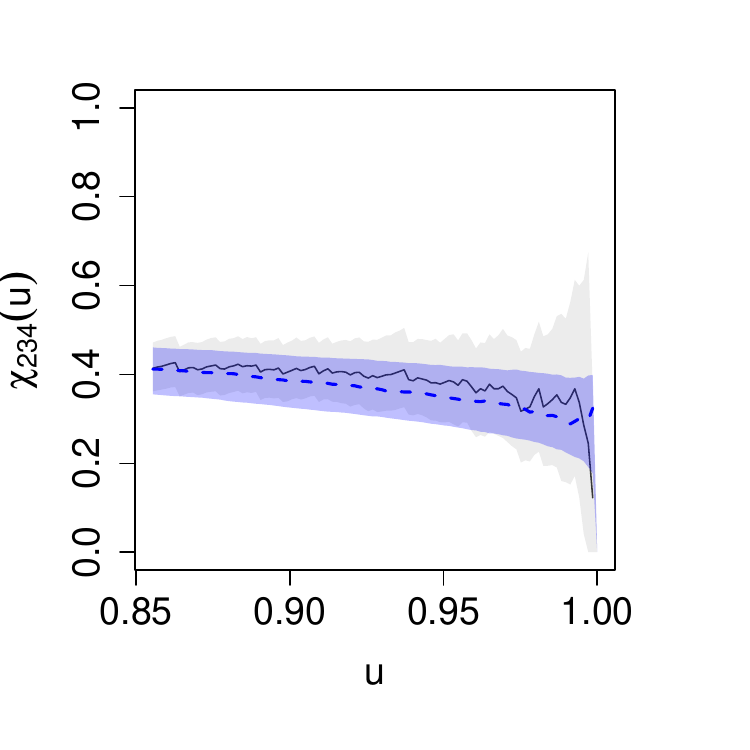}
    \includegraphics[width=0.24\textwidth]{d4pollution_chi_plot_1234.pdf}
    \caption[$\chi_C(u)$ plots estimated empirically (solid line) and piecewise-linearly (dashed line) on the $d=4$ pollution dataset.]{$\chi_C(u)$ plots estimated empirically (solid line) and piecewise-linearly (dashed line) on the $d=4$ pollution dataset. Black solid lines are empirical values, and blue dashed lines are estimated using the piecewise-linear model. Shaded regions represent 7-day 95\% block bootstrap confidence intervals.}
    \label{fig:supp-d4-pollution-chi}
\end{figure}

Here, we present additional diagnostic plots for the $d=4$ pollution data example from Section~\ref{sec:pollution}, in which we aim to estimate the tail behavior of the pollutants CO, NO, PM10, and $\text{NO}_2$. In Figure~\ref{fig:supp-d4-pollution-chi}, we show plots of $\chi_C(u)$ for high values of $u$ for all combinations of indices $C\subseteq\left\{1,2,3,4\right\}$. Values of $\chi_C(u)$ are estimated empirically and with a truncated gamma model \eqref{eq:trunc-gam-cond} with gauge function estimated piecewise-linearly. %, with 95\% 7-day block bootstrap confidence intervals. 
We see strong agreement between the empirical values and those from the fitted model. Furthermore, all model estimates capture the asymptotic positive association between the variables in $\mathcal{C}$, as was suggested by tools introduced in \cite{simpson2020determining}. This demonstrates the ability of both the fitted radial and angular models to capture the extremal dependence of the four pollutants.

\begin{figure}[h!]
    \centering
    \includegraphics[width=0.4\textwidth]{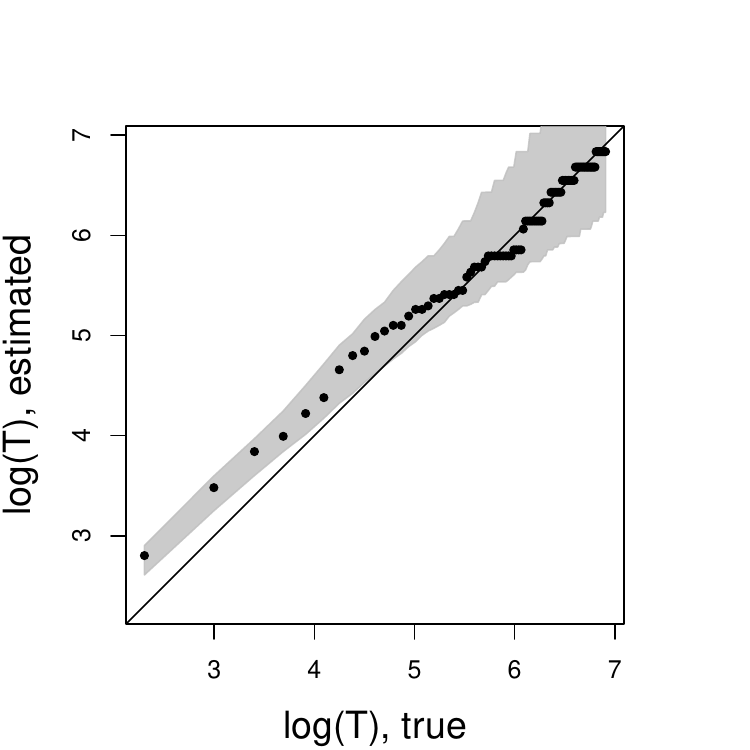}
    \caption{$d=4$ pollution fitted piecewise-linear gauge estimated return periods (log-scale) compared to true values $T\in\left\{10,20,\dots,1000\right\}$.}
    \label{fig:d4-pollution-return}
\end{figure}

\begin{figure}[h!]
    \centering
    \includegraphics[width=0.4\textwidth]{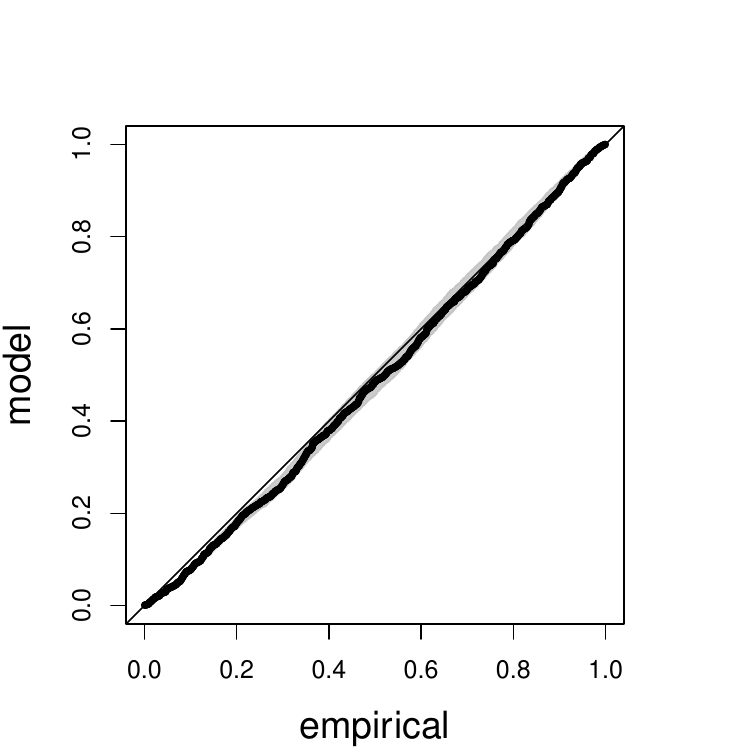}
    \includegraphics[width=0.4\textwidth]{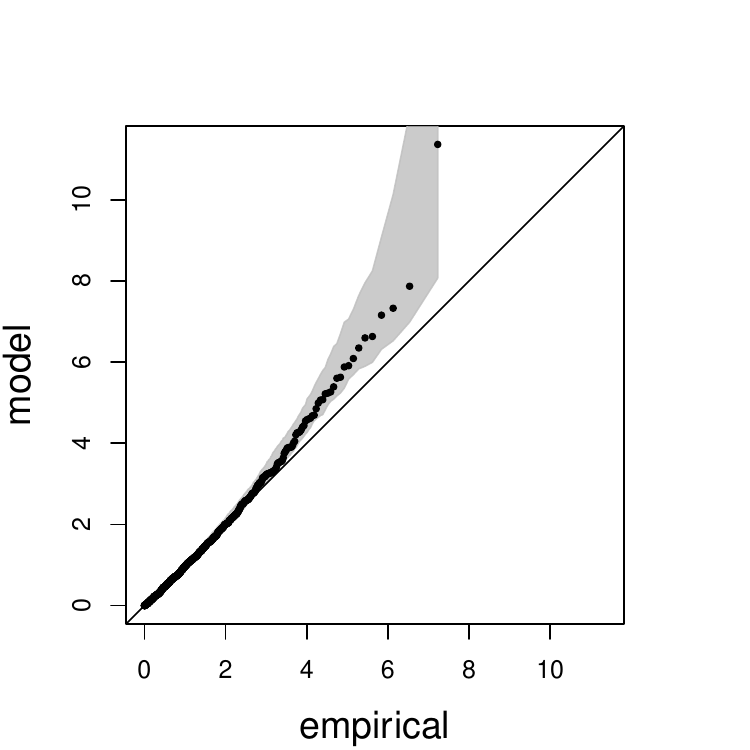}
    \caption[$d=4$ pollution PP and QQ plots.]{$d=4$ pollution PP and QQ plots, with 95\% 7-day block bootstrap confidence intervals in grey.}
    \label{fig:d4-pollution-PPQQ}
\end{figure}

Figure~\ref{fig:d4-pollution-return} shows accurate estimated return level periods corresponding to curves $\mathcal{R}(T)$ against $T$ on the $\log$-scale. Estimates are formed by counting the proportion of points exceeding the computed boundary and taking the reciprocal as an estimate of the return period $T$, with good matching to the true values. 
The PP and QQ plots in Figure~\ref{fig:d4-pollution-PPQQ} show the fitted model for exceedance radii $R\mid\left\{\bm{W}=\bm{w},R>\hat{r}_{\tau}(\bm{W})\right\}$ is in general agreement with the theoretical truncated gamma model \eqref{eq:trunc-gam-cond}, further validating our proposed modeling approach.
%The tolerance intervals on these plots are those obtained through the beta distribution of uniform order statistics of an independent sample; therefore, they are likely to be overly narrow for temporally dependent data.

\clearpage
\subsection{Setting 2: $d=3$}\label{supp:pollution-d3}

\begin{figure}[h!]
    \centering
    \includegraphics[width=0.4\textwidth]{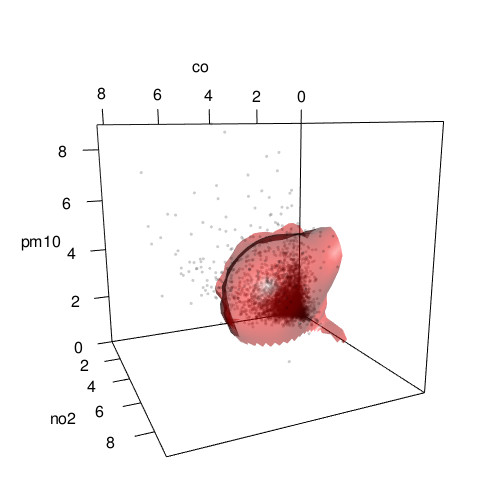}
    \includegraphics[width=0.4\textwidth]{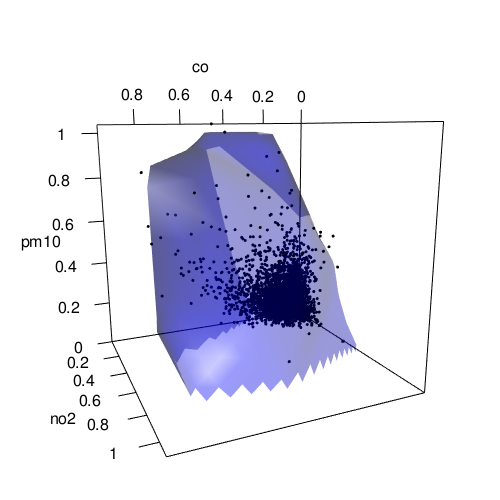}
    \caption{Left: $d=3$ pollution dataset, with the radial threshold $r_{0.95}(\bm{w})$. Right: estimated gauge function unit level set with $\log(n)$-scaled data.}
    \label{fig:d3-pollution-quants-gauge}
\end{figure}

We now consider the three-dimensional setting of modeling  measurements of CO, $\text{NO}_2$, and PM10, encoded as in the four-dimensional setting as variables 1, 2, and 3.
%\textcolor{red}{TO DO: is SS2 really the best fit? Needs refining.}
The first step in modeling  is to obtain $\hat{r}_{\tau}(\bm{w})$, $\bm{w}\in\mathcal{S}_2$, using the KDE approach of Section~\ref{sec:quantiles} with a Gaussian kernel with angular bandwidth $h_{\bm{W}}=0.075$. 
Here, we take $\tau=0.95$ as the level associated with our threshold.
% Compared to the four-dimensional fits, we obtained good model fits using this higher threshold; however, we found that values of $\tau$ larger than 0.90 resulted in a piecewise-linear model whose parameters varied too much even with a large amount of penalization.
  Figure~\ref{fig:d3-pollution-quants-gauge} shows the resulting threshold curve $\hat{r}_{\tau}(\bm{w})\bm{w}$ for values $\bm{w}\in\mathcal{S}_2$.
Next, we obtain a triangulation using a grid of $N=28$ reference angles at $\left\{0,{1}/{6},\dots,1\right\}^3$ within the simplex $\mathcal{S}_2$, displayed in Figure~\ref{fig:d3-DT} in Supplement~\ref{supp:sim}.
%After fitting all possible piecewise-linear models (i.e., the configurations outlined in SS1 to SS6 outlined in Section \ref{sec:sim}), the available diagnostics suggest that fitting the bounded conditional radial model for $R\mid\left\{\bm{W}=\bm{w},R>r_{\tau}(\bm{w})\right\}$ and using the empirical distribution of exceedance angles $\bm{W}\mid\left\{R>r_{\tau}(\bm{W})\right\}$ was the best performing model. A penalty of $\lambda=0.005$ for the radial model was selected based on NLL cross-validation scores.
As in the $d=4$ setting, we fit the conditional radial model for $R\mid\left\{\bm{W}=\bm{w},R>\hat{r}_{\tau}(\bm{w})\right\}$ with penalty strength $\lambda=1$ with bounding using Algorithm~\ref{alg:bound-fit} in Supplement~\ref{supp:bounding-alg}, along with the angular model for $\bm{W}\mid\left\{R>\hat{r}_{\tau}(\bm{W})\right\}$ with penalty strength $\lambda=20$.

\begin{figure}[h!]
    \centering
    \includegraphics[width=0.24\textwidth]{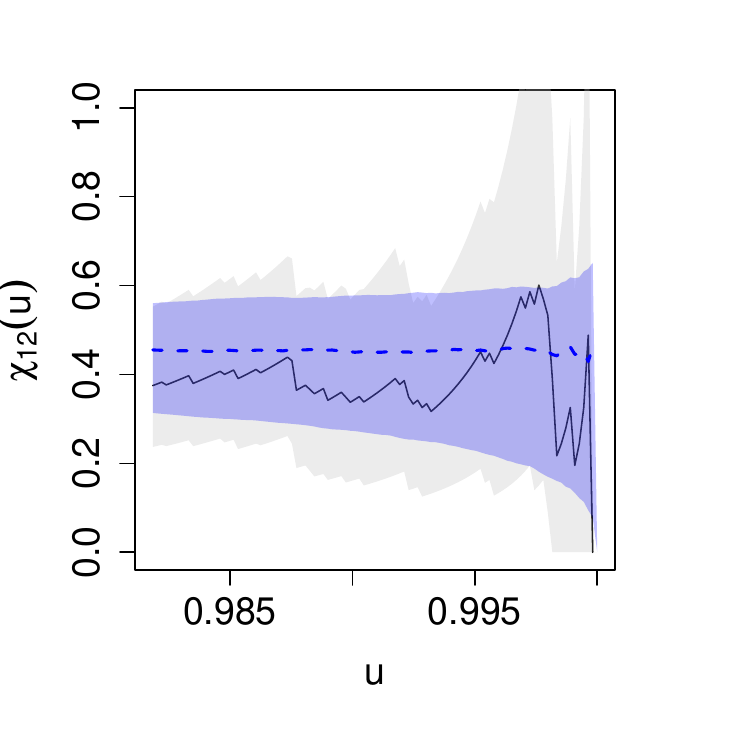}
    \includegraphics[width=0.24\textwidth]{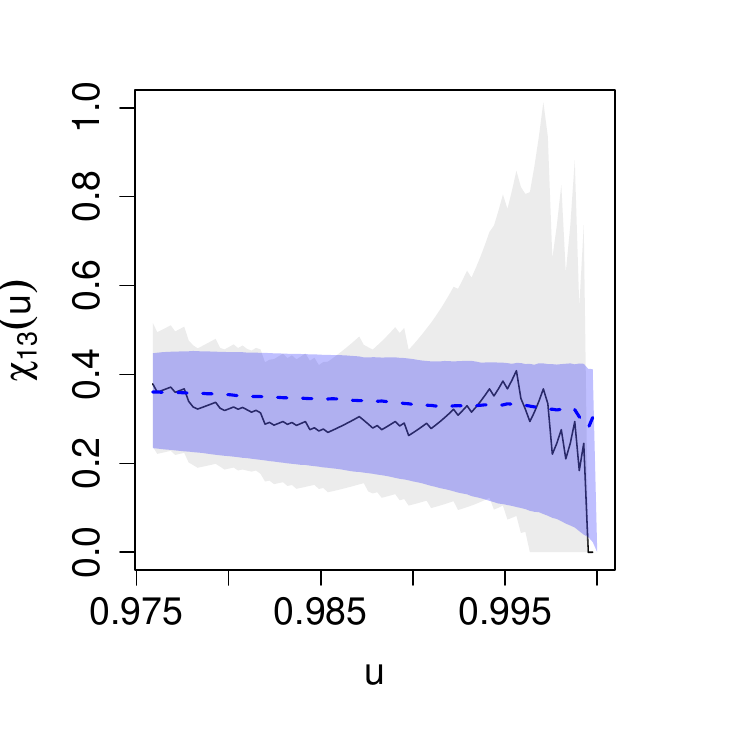}
    \includegraphics[width=0.24\textwidth]{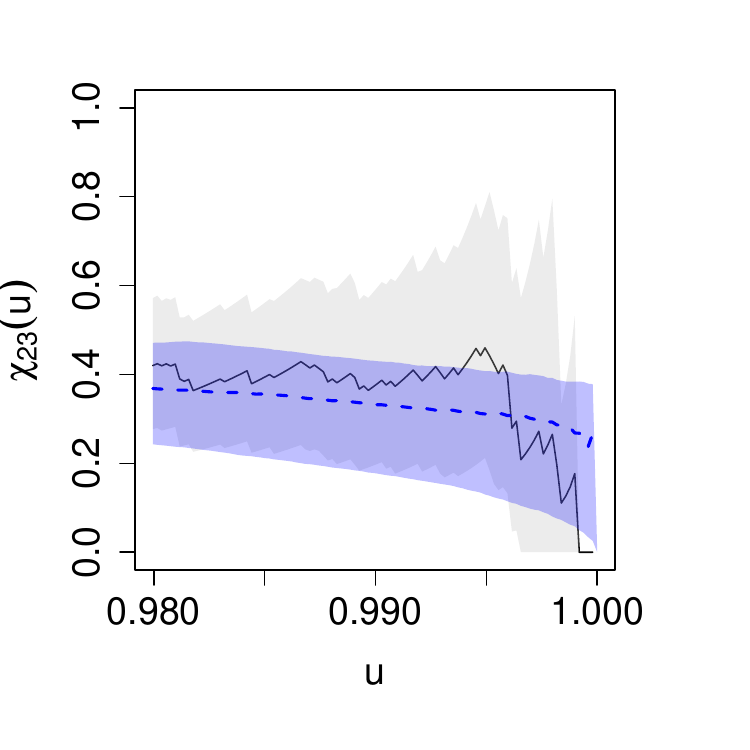}
    \includegraphics[width=0.24\textwidth]{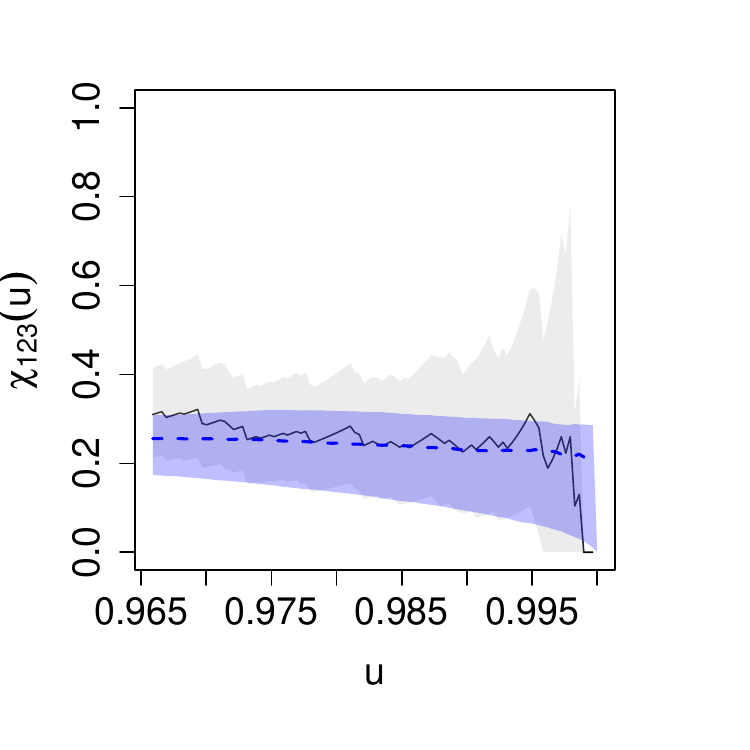}
    \caption[$\chi_C(u)$ plots estimated empirically (solid line) and piecewise-linearly (dashed line) on the $d=3$ pollution dataset.]{$\chi_C(u)$ plots estimated empirically (solid line) and piecewise-linearly (dashed line) on the $d=3$ pollution dataset. % with 7-day block bootstrap 95\% confidence intervals. 
    From left to right: $C=\left\{1,2\right\}, \left\{1,3\right\}, \left\{2,3\right\}, \left\{1,2,3\right\}$. Variables 1, 2, and 3 correspond to pollutants CO, $\text{NO}_2$, and PM10, respectively. Shaded regions represent 7-day 95\% block bootstrap confidence intervals.}
    \label{fig:d3-pollution-chi}
\end{figure}

\begin{figure}[h!]
    \centering
%    \begin{subfigure}{0.24\textwidth}
%    	\centering
     	\includegraphics[width=0.24\textwidth]{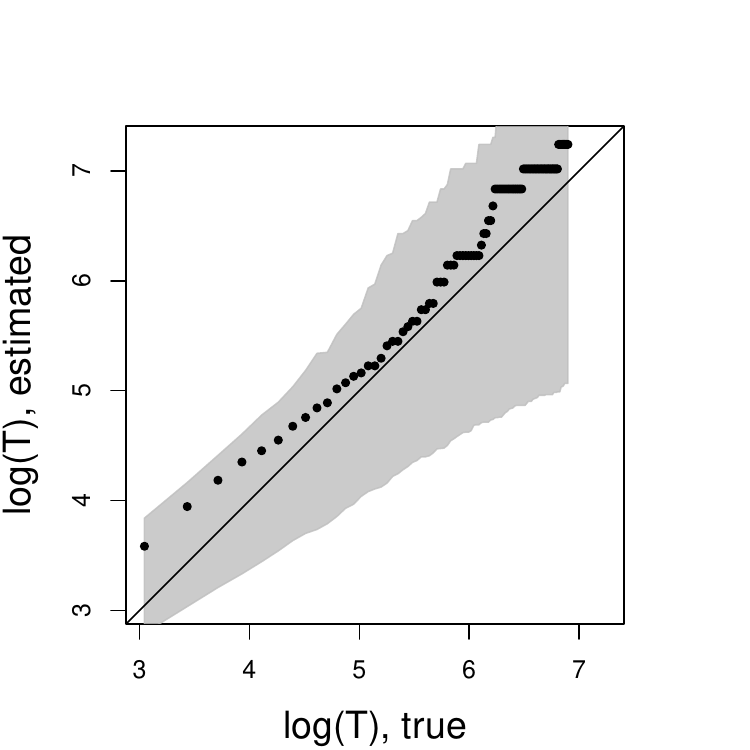}
%    	\caption{}
%    	\label{fig:d3-pollution-return-Ts}
%    \end{subfigure}
%    \begin{subfigure}{0.24\textwidth}
%    	\centering
     	\includegraphics[width=0.24\textwidth]{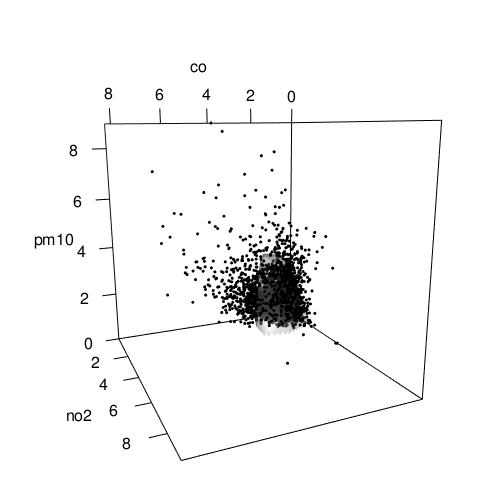}
%    	\caption{}
%    	\label{fig:d3-pollution-return1}
%    \end{subfigure}
%    \begin{subfigure}{0.24\textwidth}
%    	\centering
     	\includegraphics[width=0.24\textwidth]{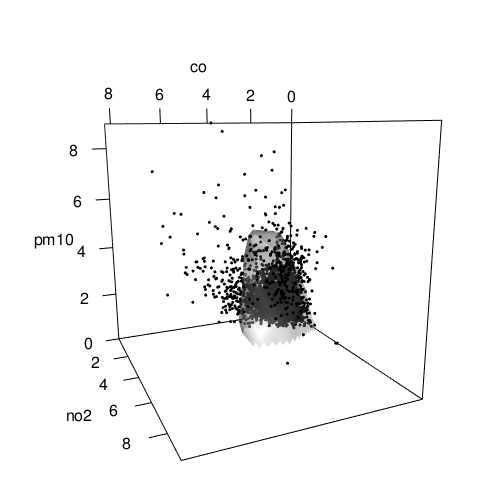}
%    	\caption{}
%    	\label{fig:d3-pollution-return2}
%    \end{subfigure}
%    \begin{subfigure}{0.24\textwidth}
%    	\centering
     	\includegraphics[width=0.24\textwidth]{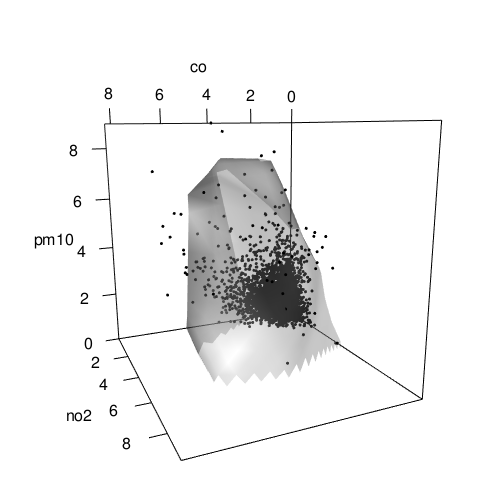}
%    	\caption{}
%    	\label{fig:d3-pollution-return3}
%    \end{subfigure}
    \caption{Left: $d=3$ pollution fitted piecewise-linear gauge estimated return periods (log-scale), compared to true values $T\in\left\{10,20,\dots,1000\right\}$. 95\% 7-day block bootstrap confidence intervals are shown in grey. Center-left--right: $T=50, 100, 1000$ day return level set boundaries.}
    \label{fig:d3-pollution-return}
\end{figure}

\begin{figure}[h!]
    \centering
     \includegraphics[width=0.4\textwidth]{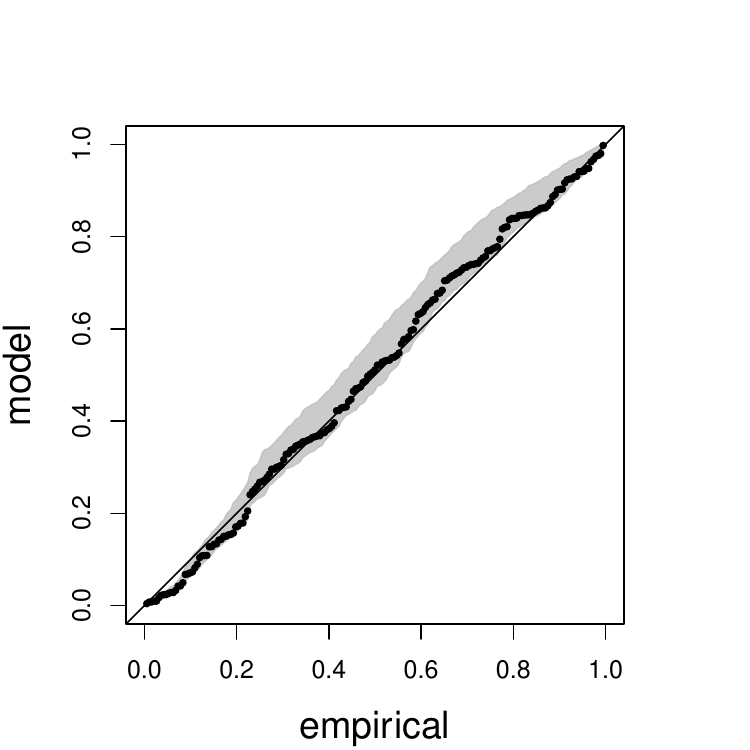}
     \includegraphics[width=0.4\textwidth]{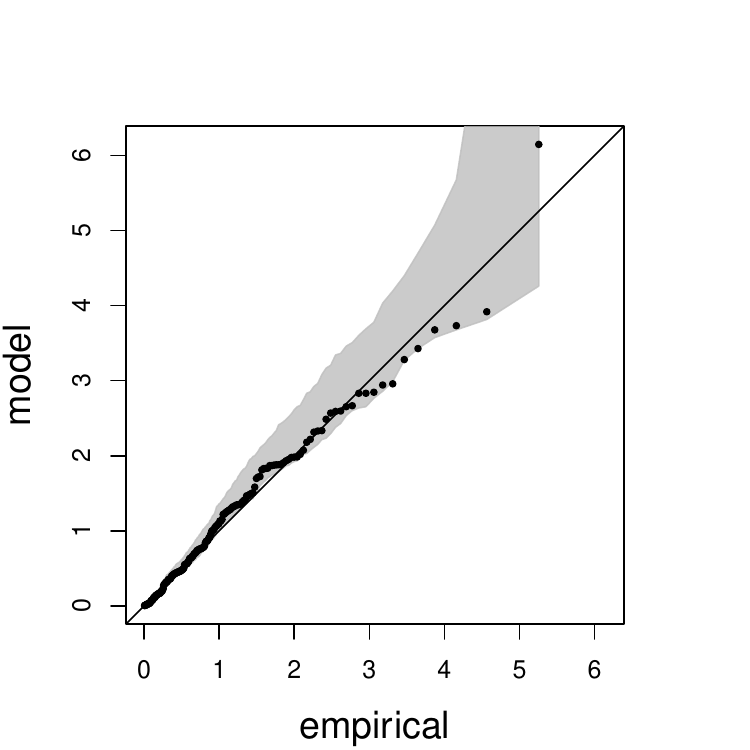}
    \caption[$d=3$ pollution PP and QQ plots.]{$d=3$ pollution PP and QQ plots, with 95\% 7-day block bootstrap confidence intervals in grey.}
    \label{fig:d3-pollution-PP-QQ}
\end{figure}

In employing methods from \cite{simpson2020determining}, it was estimated that all three variables can obtain large values simultaneously, while PM10 can grow large when CO and $\text{NO}_2$ are both small, i.e., $\mathcal{C}=\left\{\left\{3\right\},\left\{1,2,3\right\}\right\}$. The resulting limit set boundary in~\ref{fig:d3-pollution-quants-gauge} is in agreement with this, since $g(1,1,1)=1$ and $g(\gamma_1,\gamma_2,1)=1$ for $\gamma_1=0.673$ and $\gamma_2=0.3471$.
%This aligns with the four-dimensional findings, where we found that all variables grow large together. 
The results from the methods in \cite{simpson2020determining} also imply that values of $\chi_C(u)$ are expected to be positive for all values of $u\in[0,1]$ for any collection of variables $C\subseteq\left\{1,2,3\right\}$.
The $\chi_C(u)$ plots of Figure~\ref{fig:d3-pollution-chi} also indicate this possibility, and show that our piecewise-linear model is in close agreement with the empirical estimates, demonstrating good capability of the angular and radial models in capturing the extremal behavior of the data.
Figure~\ref{fig:d3-pollution-return} shows that our model accurately estimates return periods, with three different return-level sets also displayed.
%Figure \ref{fig:d3-pollution-return} shows some slight overestimation in return periods, but the values are generally close overall.
Furthermore, the PP and QQ plot in Figure~\ref{fig:d3-pollution-PP-QQ} show that the fitted truncated gamma model for $R\mid\left\{\bm{W}=\bm{w},R>\hat{r}_{\tau}(\bm{w})\right\}$ agrees with the theoretical model.
%The fitted angular density in Figure \ref{fig:d3-pollution-angles} shows agreement with the empirical sample, though there isn't much data in the empirical sample to make a fair assessment.

\clearpage
\newpage
\section{Piecewise-linear models for data in standard Laplace margins}\label{supp:laplace}

\begin{figure}[h!]
    \centering
    \includegraphics[width=0.4\textwidth]{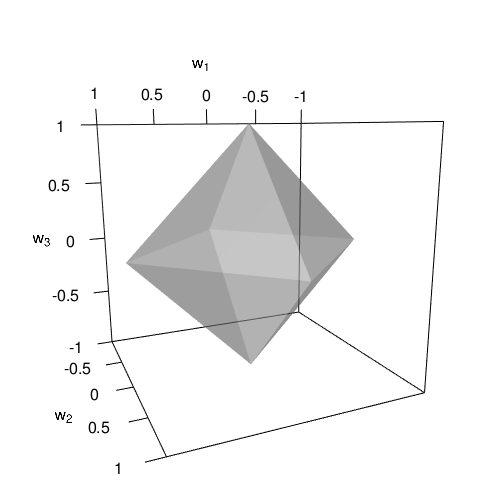}
    \caption{The simplex $\mathcal{S}^{(+,-)}_2$ is given by the boundary of the above surface plot.}
    \label{fig:SL2}
\end{figure}

In this section, we outline adaptations to the methodology for working in standard Laplace, rather than standard exponential, margins. For $d=2$, this is straightforward, and we demonstrate the methodology.
For $d>2$, it is more complex to represent the $(d-1)$-dimensional $L_1$ manifold $\mathcal{S}^{(+,-)}_{d-1}=\left\{\bm{x}\in\mathbb{R}^d \middle| \left\|\bm{x}\right\|_1=1\right\}$ in $\mathbb{R}^{d-1}$.
Figure~\ref{fig:SL2} gives a demonstration of $\mathcal{S}_2^{(+,-)}$ in three dimensions. From this figure alone, the task of projecting to $(d-1)$-dimensions,  implementing the Delaunay triangulation on this lower-dimensional space, and recovering $d$-dimensional vectors from their lower-dimensional $L_1$ representation can be seen as difficult when not restricting to the positive orthant, and is left to future work. 

\cite{mackay2024modeling} provide a useful $L_1$-based decomposition of bivariate copies of $\bm{X}=(X_1,X_2)^\top$ when the margins $X_1,X_2$ follow the standard Laplace distribution,
$$
	(R,W) = \left(\left|X_1\right|+\left|X_2\right|,\varepsilon\left(\frac{X_2}{\left|X_1\right|+\left|X_2\right|}\right)\left(1-\frac{X_1}{\left|X_1\right|+\left|X_2\right|}\right)\right)\in\mathbb{R}_+\times [-2,2),
$$
where $\varepsilon(u)=1$ when $u\geq0$ and $\varepsilon(u)=-1$ otherwise. In this setting, we can recover the corresponding Cartesian vectors using
\begin{equation}\label{eq:cos1sin1trnasf}
(X_1,X_2)^\top = R\left(\frac{1-|W|}{\left|1-|W|\right| + \left|1-|W-1|\right|},\frac{1-|W-1|}{\left|1-|W|\right| + \left|1-|W-1|\right|}\right)^\top\in\mathbb{R}^2.
\end{equation}
With this representation, it is possible to use the piecewise-linear framework outlined in this paper. 
Take, for example, the bivariate Gaussian distribution with standard Gaussian margins and correlation $\rho<0$. When transforming to standard exponential margins, the limit set cannot be defined on the axes through a continuous gauge. However, in standard Laplace margins, the gauge function is well-defined in its $\mathbb{R}^2$ domain.
In standard Laplace margins, this gauge function is given in general $d$-dimensions by
\begin{equation}\label{eq:gauss-gauge-Lap}
	g(\bm{x};\Sigma) = \left(\text{sign}(\bm{x})\sqrt{\left|\bm{x}\right|}\right)^\top\Sigma^{-1}\left(\text{sign}(\bm{x})\sqrt{\left|\bm{x}\right|}\right).
\end{equation}
In equation \eqref{eq:gauss-gauge-Lap}, all operations performed on vectors are done componentwise.

\begin{figure}[h!]
    \centering
%    \begin{subfigure}{0.22\textwidth}
%        \centering
        \includegraphics[width=0.22\textwidth]{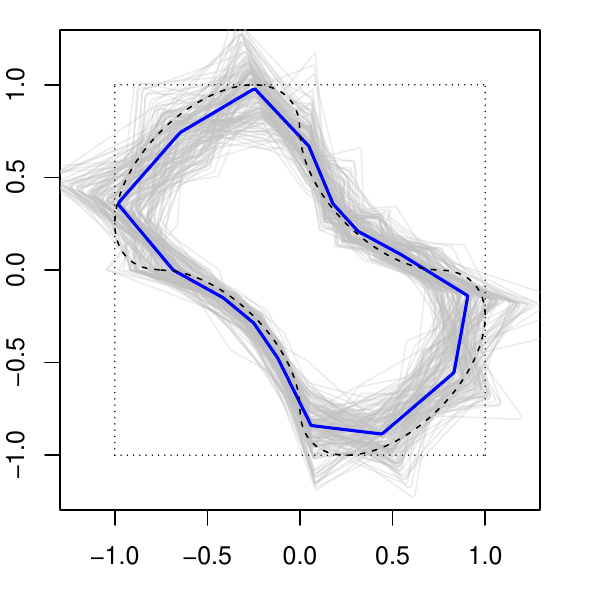}
%        \caption{SS1/3}
%    \end{subfigure}%
%    ~ 
%    \begin{subfigure}{0.22\textwidth}
%        \centering
        \includegraphics[width=0.22\textwidth]{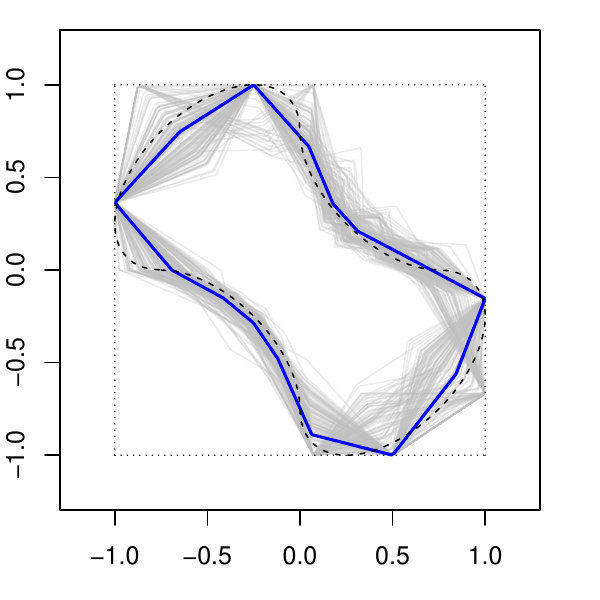}
%        \caption{SS2/4}
%    \end{subfigure}%
%    ~ 
%    \begin{subfigure}{0.22\textwidth}
%        \centering
        \includegraphics[width=0.22\textwidth]{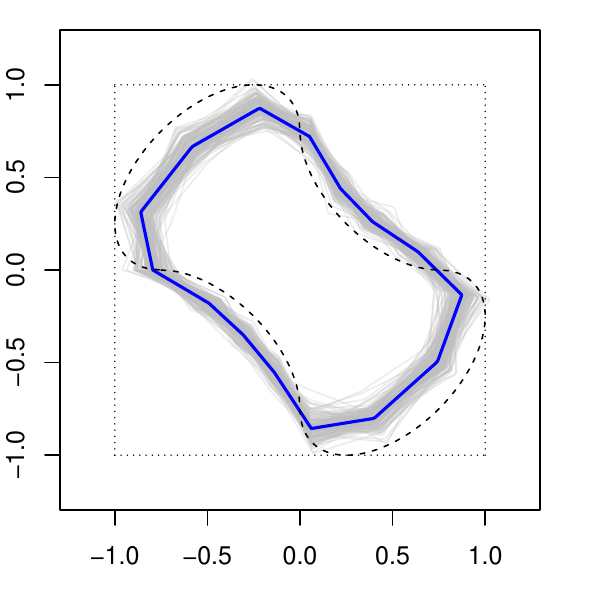}
%        \caption{SS5}
%    \end{subfigure}%
%    ~ 
%    \begin{subfigure}{0.22\textwidth}
%        \centering
        \includegraphics[width=0.22\textwidth]{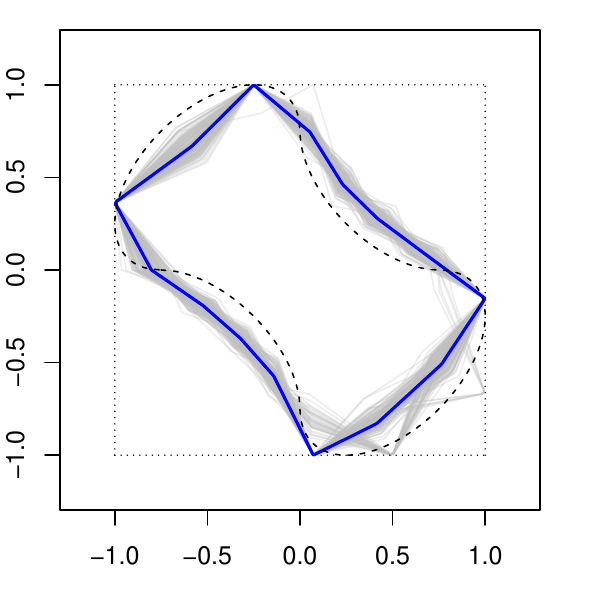}
%        \caption{SS6}
%    \end{subfigure}
	\caption[\protect{200 estimates of the unit level set of $g_{\small{\textsc{pwl}}}$ on model fits using the setup SS1, SS2, SS5, and SS6 (left to right) for the Gaussian distribution in standard Laplace margins.}]{200 estimates of the unit level set of $g_{\small{\textsc{pwl}}}$ on model fits using the setup SS1, SS2, SS5, and SS6 (left to right) for the Gaussian distribution in standard Laplace margins, with median value given by the solid line. The true unit level set is given by the  dashed line.}
    \label{fig:lapgauss-gaugefits}
\end{figure}

\begin{figure}[h!]
    \centering
    \includegraphics[width=0.4\textwidth]{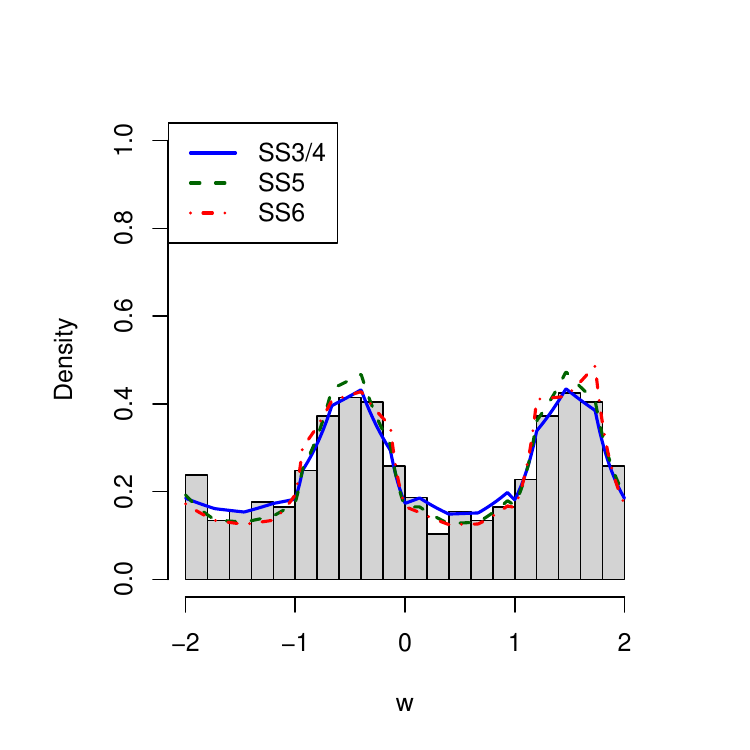}
    \caption{Estimate of the angular density for various simulation study setups, with an empirical angular density given by the histogram.}
    \label{fig:lapgauss-fWfits}
\end{figure}

\begin{figure}[h!]
    \centering
    \includegraphics[width=0.58\textwidth]{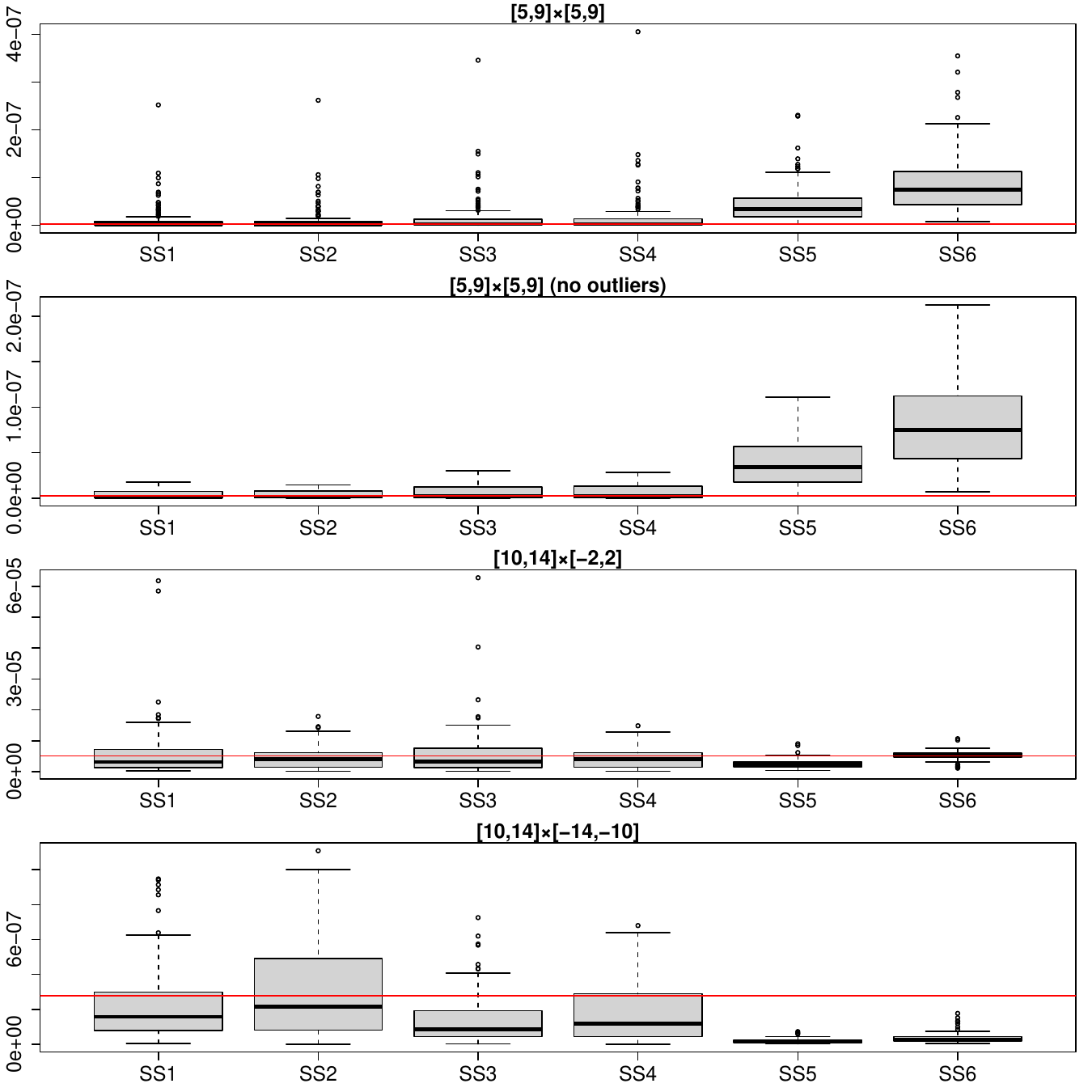}
    \caption[Probability estimates for bivariate Gaussian distribution in standard Laplace margins.]{Probability estimates for bivariate Gaussian distribution in standard Laplace margins. True values given by the solid line. The region of interest is given in the figure title, and results for region $B_1$ are shown with and without outliers for clarity.}
    \label{fig:lapgauss-2d-SS}
\end{figure}

We illustrate the estimation of this gauge function piecewise-linearly in a simulation study with data generated from the bivariate Gaussian distribution with standard Laplace margins and correlation $\rho=-0.5$. We generate $n=5000$ datapoints, and perform KDE-based \rc{threshold} estimation at $\tau=0.90$. After defining a regular grid of $N=15$ reference angles from $[-2,2)$, including $-2$, we fit the models SS1--SS6 outlined in Section~\ref{sec:sim}, and draw samples from these models to estimate the probability of lying in the regions $B_1=[5,9]\times[5,9]$, $B_2=[10,14]\times[-2,2]$, and $B_3=[10,14]\times[-14,-10]$. This is repeated 200 times. The estimated unbounded and bounded limit set boundaries from modeling  $R|\left\{W=w,R>\hat{r}_{\tau}(w)\right\}$ and $(R,W)|\left\{R>\hat{r}_{\tau}(W)\right\}$ are displayed in Figure~\ref{fig:lapgauss-gaugefits}, where good agreement with the true gauge functions is shown. Angular models in the setting of SS3/4, SS5, and SS6 are shown in Figure~\ref{fig:lapgauss-fWfits}. All models show good agreement with a histogram of exceedance angles. Probability estimates are displayed in Figure~\ref{fig:lapgauss-2d-SS}, where models in the settings SS2 and SS4 perform best overall.
A further adjustment needs to be taken when sampling the angles $W$ from the density $f_{\tilde{\bm{W}}}$ introduced in Section~\ref{sec:angular-model}. In our MCMC algorithm, a beta proposal distribution is used. Sampled angles need to be shifted to the $[-2,2)$ domain using the transformation $W^\prime = 4W-2$ before proceeding to sampling radii and obtaining extremal points using  \eqref{eq:cos1sin1trnasf}.

\end{document}

%% file: RMSEtable_SS2d.tex
\begin{table}[h!]
\begin{center}
\begin{tabular}{c|c|c|c|c|c}
\multirow{2}{*}{Region} & \multirow{2}{*}{Setup} & \multicolumn{4}{|c}{Dataset} \\ 
 & & \protect\eqref{distn:log1} & \protect\eqref{distn:log2} & \protect\eqref{distn:gauss} & \protect\eqref{distn:invlog} \\ \hline
& par       & 0.2670 & 0.2524 & 1.1132 & 0.9905 \\ 
& SS1       & 0.9001 & 1.1045 & 0.8375 & 1.2622 \\ 
B1: & SS2   & 0.3364 & 0.9425 & 0.7518 & 1.2646 \\
$[10,12]\times[10,12]$ & SS3       
            & 0.9809 & 1.1401 & 0.8603 & 1.2347 \\ 
& SS4       & 0.3946 & 0.9869 & 0.7593 & 1.2262 \\ 
& SS5       & 0.7972 & 1.8985 & 0.5214 & 0.9567 \\ 
& SS6       & 0.1294 & 1.6248 & 0.4319 & 1.2256 \\ \hline
& par       & 0.4907 & 0.3959 & 0.5417 & 0.7207 \\ 
& SS1       & 0.8687 & 0.8010 & 0.7437 & 0.9127 \\ 
B2: & SS2   & 1.0807 & 0.6947 & 0.4268 & 0.9170 \\ 
$[10,12]\times[6,8]$ & SS3       
            & 0.9165 & 0.7845 & 0.7764 & 0.8888 \\ 
& SS4       & 1.1378 & 0.6716 & 0.4440 & 0.8957 \\ 
& SS5       & 0.7644 & 0.9377 & 0.4445 & 0.6534 \\ 
& SS6       & 0.8928 & 0.7413 & 0.3405 & 0.8807 \\ \hline
& par       & 1.4067 & 0.6098 & 1.3380 & 0.3382 \\ 
& SS1       & 1.8054 & 0.9160 & 1.3650 & 0.7787 \\ 
B3:& SS2    & 1.7825 & 0.9142 & 1.3567 & 0.6357 \\ 
$[10,12]\times[2,4]$ & SS3
            & 1.7323 & 0.8924 & 1.5058 & 0.8082 \\ 
& SS4       & 1.7132 & 0.8898 & 1.4609 & 0.6460 \\ 
& SS5       & 1.0190 & 0.7359 & 1.5618 & 0.6046 \\ 
& SS6       & 1.0860 & 0.9765 & 1.3691 & 0.4877 \\ 
\end{tabular}
\end{center}
\caption{RMSE across the 200 log-probability estimates for $d=2$ simulation studies. \label{tab:2d-SS}
}
\end{table}

%% file: RMSEtable_SS3d.tex
\begin{table}[h!]
\begin{center}
\begin{tabular}{c|c|c|c|c}
\multirow{2}{*}{Region} & \multirow{2}{*}{Setup} & \multicolumn{3}{|c}{Dataset} \\ 
 & & \protect\eqref{distn:alog1} & \protect\eqref{distn:alog2} & \protect\eqref{distn:mix} \\ \hline
& par       & 0.4718 & 0.4689 & 0.7422  \\ 
& SS1       & 1.5314 & 1.1990 & 1.6711  \\ 
B1: & SS2   & 1.4664 & 1.2166 & 1.6724  \\ 
$[8,10]\times[8,10]\times[0.01,3]$ & SS3      
            & 1.7604 & 1.5459 & 2.0096  \\ 
& SS4       & 1.6946 & 1.5332 & 2.0664  \\ 
& SS5       & 1.0075 & 1.2129 & 1.7306  \\ 
& SS6       & 0.7353 & 0.8186 & 1.4942  \\ \hline
& par       & 1.2837 & 0.7504 & 0.9414  \\ 
& SS1       & 1.1307 & 0.8785 & 1.0231  \\ 
B2: & SS2   & 1.1208 & 0.8625 & 0.9689 \\ 
$[8,10]\times[5,7]\times[0.01,3]$ & SS3      
            & 1.0140 & 0.8304 & 0.9906  \\ 
& SS4       & 0.9875 & 0.8094 & 0.9107  \\ 
& SS5       & 0.5664 & 0.4643 & 0.7013  \\ 
& SS6       & 0.4669 & 0.4675 & 0.6321  \\ \hline
& par       & 2.0299 & 0.3831 & 0.8872   \\ 
& SS1       & 1.5820 & 1.0670 & 1.0320  \\ 
B3:& SS2    & 1.5801 & 1.0461 & 0.9804 \\ 
$[8,10]\times[2,4]\times[0.01,3]$ & SS3  
            & 1.4419 & 1.0380 & 1.0352   \\ 
& SS4       & 1.4746 & 0.9895 & 0.9212  \\ 
& SS5       & 0.8932 & 0.5867 & 0.4768   \\ 
& SS6       & 0.9401 & 0.5358 & 0.5046  \\ 
\end{tabular}
\end{center}
\caption{RMSE across the 200 log-probability estimates for $d=3$ simulation studies. \label{tab:3d-SS}}
\end{table}